\documentclass[iop]{emulateapj}
\usepackage{graphicx}
\usepackage{amsmath}

\slugcomment{Submitted to The Astrophysical Journal}

\gdef\deg{$^{\circ}$}
\gdef\kms{km s$^{-1}$}

\gdef\imacs{{\it IMACS}}

\def\etal{\hbox{et al.}}

\gdef\ltsima{$\scriptscriptstyle \; \buildrel < \over \sim \;$}
\gdef\simlt{\lower.3ex\hbox{\ltsima}}

\gdef\gtsima{$\scriptscriptstyle \; \buildrel > \over \sim \;$}
\gdef\simgt{\lower.3ex\hbox{\gtsima}}

\gdef\about{\raise.3ex\hbox{$\scriptscriptstyle \sim $}}

\def\gs{\mathrel{\raise0.35ex\hbox{$\scriptstyle >$}\kern-0.6em 
\lower0.40ex\hbox{{$\scriptstyle \sim$}}}}
\def\ls{\mathrel{\raise0.35ex\hbox{$\scriptstyle <$}\kern-0.6em 
\lower0.40ex\hbox{{$\scriptstyle \sim$}}}}

\def\etal{\hbox{et al.}}
\def\OII{\hbox{[O II]}$\,\,$}

\def\Ha{\hbox{H$\alpha$}$\,$}
\def\Hb{\hbox{H$\beta$}$\,\,$}
\def\Hd{\hbox{H$\delta$}$\,\,$}

\def\Msun{\rm{\hbox{M$_{\odot}$}}}           % Solar masses
\def\ang{\hbox{$\,$\AA}}
\def\kms{\rm{\hbox{\,km s$^{-1}$}}}
\def\24m{\hbox{24\,$\micron$}$\,$}
\def\10-18{\hbox{$\times~10^{-18}$}}
\def\deg{\hbox{$^{\circ}$}}
\def\Lgal{\hbox{L$_{gal}$}}
\def\Ntot{\hbox{N$_{tot}$}}
\def\Mgal{\hbox{M$_{gal}$}}
\def\tauenc{\hbox{$\tau_{enc}$}}

\def\R200{\hbox{$R_{200}$}}
\def\Rnorm{\hbox{$Rcl/R_{200}$}}

\shortauthors{Dressler \etal\ }
\shorttitle{Structure and Spectral Types in the ICBS}

\begin{document}

\title{The IMACS Cluster Building Survey: \break II. Spectral Evolution of Galaxies in the Epoch of  Cluster Assembly
\footnote{T\lowercase{his paper includes data gathered with the 6.5 meter \uppercase{M}agellan \uppercase{T}elescopes 
located at \uppercase{L}as \uppercase{C}ampanas \uppercase{O}bservatory, \uppercase{C}hile.}}}

\author{
Alan Dressler\altaffilmark{1},
Augustus Oemler, Jr.\altaffilmark{1},
Bianca M.~Poggianti\altaffilmark{2},
Michael D.~Gladders\altaffilmark{3,4},
Louis Abramson\altaffilmark{3,4},
Benedetta Vulcani\altaffilmark{2,5}
}
\altaffiltext{1}{The Observatories of the Carnegie Institution for Science, 813 Santa Barbara St., Pasadena, CA 91101, USA}
\altaffiltext{2}{INAF-Astronomical Observatory of Padova, Italy}
\altaffiltext{3}{Department of Astronomy and Astrophysics, University of Chicago, 5640 S. Ellis Ave., Chicago, IL 60637, USA}
\altaffiltext{4}{Kavli Institute for Cosmological Physics, University of Chicago, 5640 S. Ellis Ave., Chicago, IL 60637, USA}
\altaffiltext{5}{Kavli Institute for the Physics and Mathematics of the Universe, University of Tokyo, Kashiwa, 277-8582, Japan}

\begin{abstract}
The \imacs\ Cluster Building Survey (ICBS) provides spectra of $\sim$2200 galaxies $0.31<z<0.54$ in 
5 rich clusters ($R\ls5$ Mpc) and the field.  Infalling, dynamically cold groups with tens of members 
account for approximately half of the supercluster population, contributing to a growth in cluster mass of 
$\sim$100\% by today. The ICBS spectra distinguish non-starforming (PAS) and poststarburst (PSB) from 
starforming galaxies --- continuously  starforming (CSF) or starbursts, (SBH or SBO), identified by anomalously 
strong \Hd absorption or \OII emission.  For the infalling cluster groups and similar field groups, we 
find a correlation between PAS+PSB fraction and group mass, indicating substantial ``preprocessing" through 
\emph{quenching} mechanisms that can turn starforming galaxies into passive galaxies without the unique 
environment of rich clusters.  SBH + SBO starburst galaxies are common, and they maintain an approximately 
constant ratio (SBH+SBO)/CSF $\approx$ 25\% in all environments --- from field, to groups, to rich clusters.  
Similarly, while PSB galaxies strongly favor denser environments, PSB/PAS $\approx$ 10-20\% for all environments. 
This result, and their timescale $\tau\sim$ 500 Myr, indicates that starbursts are not signatures of a quenching 
mechanism that produces the majority of passive galaxies.  We suggest instead that starbursts and poststarbursts 
signal minor mergers and accretions, in starforming and passive galaxies, respectively, and that the principal 
mechanisms for producing passive systems are (1) early major mergers, for elliptical galaxies, and (2) later, less 
violent processes --- such as starvation and tidal stripping, for S0 galaxies.  
\end{abstract}


\section{Introduction}

The plummeting rate of cosmic star formation --- a factor of 10 in the last 7 Gyr --- would have been wholly unexpected 
by astronomers in the 1960s and 1970s as they took the first steps in the young field of galaxy evolution.  Baade's (1944) 
resolution of the stellar disk in M31 pointed to two distinct populations --- one ancient, one ongoing.  The only star 
formation history available at the time, the stellar fossil record in the Milky Way, consisted of an exclusively old 
population of globular clusters and stellar halo, and a contemporary disk supporting a low, steady level of star formation 
and heavy element production (Twarog 1980).  When considered together, the picture emerged of a remote epoch of 
vigorous star formation and rapid chemical evolution that soon decayed into a long, uneventful puttering of these 
processes, for at least the last 5 Gyr.  By inference, other large galaxies like our own also formed early, and aged slowly.

Optical astronomers were not the first to discover how wrong this picture was --- until the 1970s, their instruments 
were incapable of pushing photometric or spectroscopic measurements even a billion years back in cosmic time. Instead, 
radio astronomers first glimpsed the fireworks of a younger universe.  Radio galaxies, powered by supermassive 
black-hole accretion, were detectable 10 Gyr in the past: their strong evolution, and that of the quasars they 
pointed to, signaled a period of great activity 8-11 Gyr ago.  

Our first view of the sinking cosmic star formation rate (SFR) through optical observations of normal galaxies came from 
Butcher \& Oemler (1976), who carried out the first photometric study of galaxies in rich clusters at $z\sim0.5$.  Butcher 
\& Oemler found a substantial population of blue, starforming galaxies in an environment essentially purged of star formation 
by the present epoch.  Photometric studies in the 1980s that found rapid evolution of faint field galaxies out to $z\sim1$ 
confirmed that galaxies were anything but `puttering along' over this epoch.  

When spectroscopy --- the most powerful tool of optical astronomy --- could finally be applied to typical galaxies at 'cosmological 
distances,' an opportunity to understand this dynamic star formation history (SFH) had arrived.  The spectral features in galaxy 
integrated spectra are powerful diagnostics of the star formation history of galaxies: the presence and strength of specific spectral 
lines yield a wealth of information, especially regarding the galaxy current and recent star formation activity.  Galaxy spectra are, 
among other things, a valuable way to investigate how and why many galaxies that were actively forming stars at intermediate 
redshifts ($0.3<z<1$) have become passive by the present epoch. Butcher \& Oemler (1984) were the first to recognize that 
this phenomenon was not limited to clusters, but widespread; today it is known to be largely independent of environment --- seen 
in clusters, groups, and the general field.

The work we describe in this paper can trace its beginnings to these first early discoveries of the 1980s, in particular to 
Dressler \& Gunn's (1983) spectroscopy of cluster galaxies from the Butcher--Oemler study and spectra of 
intermediate-redshift clusters by Couch \& Sharples (1987).  With these first spectra came another surprise, the 
presence in large numbers of galaxies whose star formation had ended abruptly, a behavior that is seen only rarely 
today, in a few galaxies per thousand.   At first, with only the high-density \emph{cores} of rich clusters well studied,
these ``poststarburst" galaxies were known only as a cluster phenomenon, at a level of $\sim$10\% of the population ---
remarkable, for a class of objects with a time scale of $\tau<$1\,Gyr.  

But, as spectroscopic samples reached into the hundreds (Morphs: Dressler \etal\ 1997, Poggianti \etal\ 1999 -- P99; 
CNOC: Balogh \etal\ 1997, 1999; Couch \etal\ 1994; EDisCS: White et al. 2005, Poggianti \etal\ 2006, 2009a), a sufficient 
number of line-of-sight field galaxies at similar redshifts began to reveal that the poststarburst population dropped 
significantly in the field environment, but that a population of ``active" starburst galaxies exhibiting the spectral signatures
of poststarburst galaxies and ongoing star formation became common in its place (Dressler \etal\ 1999).  These observations 
strongly implicated some cluster-specific process as responsible for rapidly and efficiently quenching star formation in galaxies 
infalling into clusters, but the instrumentation needed to follow this conjecture has only recently become available on large 
telescopes.  For example, a recent study of cluster, group and field galaxies using the FORS2 spectrograph on VLT has 
revealed that some of the richest, distant groups share the high post-starburst fraction of clusters and, likely, the cluster 
quenching efficiency (Poggianti \etal\ 2009a).

Both Couch \& Sharples (1987) and Dressler \etal\ (1999) posited a simple sequential relation between starforming starbursts 
and poststarbursts, but the quantification of the former through low-resolution spectroscopy has remained challenging.  
Starforming galaxies are of course readily identifiable by the presence of emission lines in their spectra, but quantifying the 
level of star formation is non-trivial, both for optical studies, due to dust obscuration and the difficulty of separating a 
poststarburst component from strong (but non-bursting) star formation, and for IR studies, due to the limited sensitivity 
that only allows detection of galaxies with vigorous ongoing star formation.  Despite these difficulties, several works have 
attempted to identify starforming galaxies in the starburst phase in distant clusters (Dressler \etal\ 1999, 2004, 2009a,b; 
Poggianti \etal\ 2009a, Geach \etal\ 2009, 2011; Finn \etal\ 2010 --- to name a few), with most studies finding little or no 
enhancement of starburst activity in clusters compared to the general field at similar redshifts, but a clear enhancement in
moderate starburst activity in galaxies compared to the local universe. 

Even ignoring active starbursts, simply quantifying the evolution of the starforming fraction of galaxies with redshift as a function 
of environment (for example, group or cluster mass) has turned out to be challenging.  The first attempts have found that 
the evolution from z\,$\sim$\,0.8 to today is strongest in low-mass ($\sigma\sim$ 500\,\kms) clusters (Poggianti \etal\ 2006), 
paralleling the evolution of galaxy morphologies (Poggianti \etal\ 2009b). These results are consistent with a scenario in which 
the population of passive galaxies in clusters and groups today consists of galaxies that stopped forming stars at $z\gs2$ 
(the most massive --- primarily elliptical galaxies) and galaxies that have been quenched as a consequence of environmental 
effects (lower mass --- mostly S0 galaxies) following the hierarchical growth of structure (Poggianti \etal\ 2006).

The last ten years has seen increasing effort to obtain a better picture of the role of environment in quenching star 
formation. Detailed photometric and spectroscopic studies, extended to galaxy groups at intermediate redshifts, have 
underlined the importance of the group environment and the pre-processing occurring in group galaxies before they enter 
more massive clusters (Wilman \etal\ 2005, 2008). In groups, the fraction of starforming galaxies of a given galaxy mass 
is already lower than in similar mass galaxies in the field, with a population that is moving from the red sequence to the 
blue cloud (Balogh \etal\ 2011),  while the star formation rate in starforming galaxies appears to be unaffected by groups, 
suggesting a short quenching timescale (Vulcani \etal\ 2010; McGee \etal\ 2011).  In addition, larger area surveys have 
begun to investigate the outskirts and surrounding filaments of distant clusters (Ma \etal\ 2008; Fadda \etal\ 2008) and the 
effects of galaxy mergers and substructure on the star formation properties of galaxies (Ma \etal\ 2010; Hou \etal\ 2012), 
supporting the notion that the large population of poststarburst galaxies in clusters have been quenched as a consequence 
of some cluster-related mechanism, while starbursts are triggered in lower density environments such as cluster outskirts
(Geach \etal\ 2009, 2011).  However, wide-area spectroscopic surveys of distant clusters remain few, and further studies are
needed to link the evolution observed in the cluster cores with the processes active in the groups and filaments infalling to 
clusters, as well as in groups and lower density regions unrelated to clusters that comprise the general field.

\subsection{The \imacs\ Cluster Building Survey}

The direction of research since 2000 showed the need for a survey of wide-enough field to study the environs 
immediately outside the clusters --- the so-called supercluster population, to look for an evolutionary sequence 
associated with the infall of galaxies onto accreting clusters, with the goal of better elucidating the process(es) 
responsible for the starburst/poststarburst phenomenon.  An opportunity to make such a study came with the 
building of the Magellan-Baade telescope and installation of its Inamori-Magellan Areal Camera and Spectrograph 
--- \imacs (Dressler \etal\ 2011).  The combination of a large 6.5-m aperture and a field nearly 0.5\deg\ in diameter 
for multislit spectroscopy made \imacs\ an ideal instrument to obtain $\sim$1000 spectra of galaxies infalling 
to intermediate-redshift clusters.  

The \imacs\ Cluster Building Survey, ICBS, has in fact used the wide-field multislit spectroscopic capability of \imacs\ 
to map and study the galaxy populations of growing clusters at $z\sim0.4$.  One aim of the ICBS has been to learn 
whether the poststarburst population so prominent in the cores of intermediate-redshift rich clusters is solely or at least 
primarily associated with the dense environment of a cluster core, or whether instead this activity is more widespread and 
associated in some way with the infalling population that is building the cluster.  A second important goal was to 
clarify the relationship of the starburst-poststarburst phenomenon to the growth of passive galaxies in intermediate-redshift 
groups and clusters.  Because the idea was to follow the star formation histories of galaxies transitioning from the 
field to rich clusters, and to compare the field and group environment in the vicinity of a rich cluster to that of the general field, 
it was essential to target all galaxies in the \imacs\ field, without preselection, so that the field environment could also be 
studied and compared fairly to the supercluster environment involved in building the rich clusters. 

Selection of the rich clusters for the ICBS also called for an unconventional strategy.  Our goal was to study ``average" 
rich clusters, not the most massive clusters at the $z\sim0.5$ epoch, and to catch them at a typical time of their growth 
between $0<z<1$.  Both these considerations meant that X-ray luminosity should not be used as a primary selection
criteria, since this technique preferentially selects massive clusters that have thoroughly virialized, that is,
those that have completed their major growth phase.  Instead, we used the Red-Sequence Cluster Survey technique
developed by Gladders and Yee (2000), which identifies rich clusters in a search through color-magnitude space
for strong red sequences of passive (non-starforming) galaxies.  By restricting the search volume  search to only 
$3\times10^7$\,Mpc$^3$, we aimed at choosing a typical rich cluster, one that might grow into (at most) a Coma 
cluster by today.

In this paper we present classification and analysis of a large subset of the ICBS survey that includes spectrometric 
data on 1073 galaxies in five rich clusters covering the redshift range $0.31< z< 0.54$, and 1091 galaxies covering 
the same redshift range that are members of the ``field.''  As described in Oemler \etal\ (2013a, Paper 1), the spectra 
from the approximately 50 multislit-mask exposures with \imacs\ and \emph{LDSS} were measured for spectral features and 
indices that quantitatively discriminate the degree and character of star formation in these galaxies.  In addition to these 
data, separate measurements of \Ha fluxes and, for two of the four fields, Spitzer-MIPS 24$\mu$m fluxes were 
available to provide measures of star formation rates that are less sensitive to dust extinction.

Our methodology here has been two-fold: first, to assign spectral types to these galaxies based on the spectrophotometric 
measures, and second, to associate these with the larger-scale structures of groups and clusters in these fields.  If possible, 
we hoped to identify aspects of spectral evolution that can be connected to the environments in which they are found, whether
or not that link is causative.  Such information will help decide to what extent the star formation histories of these galaxies 
are influenced by their present environment. 

The division of the spectra in discrete classes is advantageous for investigating the questions we address here, as will be
discussed below.  However, the ICBS has produced a highly uniform set of spectra and photometric data set over a wide
interval of cosmic time $0.2<z<0.8$ that can well address the more general question of the star formation histories (SFHs) of
common field galaxies.  In Oemler \etal\ (2013b, Paper 3) we discuss the distributions of SFHs for galaxies over this period
of steep decline in the cosmic average SFR.  A principal conclusion of Paper 3 is that, as observed from the perspective
of galaxies at intermediate redshift, the SFR was in general rising up to that epoch for a small but significant fraction of the
population.  This result inspired us to reconsider traditional $\tau$-models of the star formation rate and to consider alternatives.
In Gladders \etal\ (Paper 4) we identify the \emph{lognormal} parameterization as the best simple parameterization of 
individual SFHs and, in fact, a remarkably good fit to the volume-averaged star formation rate density evolution over cosmic
time.  Paper 4 begins with the distribution of SFR and specific SFR (sSFR) for present-epoch galaxies, but shows that, using
the lognormal paradigm for the SFHs of individual galaxies, the prediction of the distribution SFH and sSFH for 
intermediate-redshift field galaxies is a very good match to what is found in Paper 3 for the ICBS data.  The implication
is that describing the SFHs of galaxies in lognormal form shows promise for understanding the history of star formation
back to very early epochs.

This paper is organized as follows.  In \S2 we describe the separation of the spectra into 5 spectral types and (a) the distributions 
of other fundamental galaxy properties for these types, and (b) the spatial distribution as described by the correlation of spectral 
type with local galaxy density and radial distance from a cluster center, and with the angular correlation function.  In \S3 we
use the redshift data in these fields to identify moderate-sized, cold groups infalling into clusters, and to identify comparable 
groups and filaments in the field, and we describe the basic properties of these groups.  In \S4 we use spectral types and 
spatial/structural information together to discuss the evidence for quenching processes that turn starforming into passive galaxies 
across the full range of environments from isolated galaxies to rich cluster cores, and explore the connection of starburst 
galaxies to the quenching process, developing a model that could account for the wide range of data that describe star 
formation histories across all galaxy environments.

\section{Star Formation Histories from Spectrophotometric Data}

\subsection{Division into five spectral types}

As described in Paper 1, the fields chosen for the ICBS were centered on putative rich clusters selected by the red-sequence 
method Gladders \& Yee (2000); catalogs from these photometric observations were used to select objects for spectroscopic 
observations.  A non-trivial combination of prioritizing objects by brightness and populating multislit masks available 
galaxies resulted in a sample that is approximately magnitude-limited at Sloan r = 22.5 with a tail to galaxies as faint as r = 23.5.  
An general description of the magnitude and position incompleteness of the spectroscopic sample compared to the photometric 
source catalog can be found in Paper 1.  Determinations of the incompleteness for specific sub-samples used in this paper
are described below in \S2.3.

Our redshift survey of more than 1000 galaxies per field revealed 6 rich clusters, with the RCS1102 and SDSS1500 fields each 
containing two.  In RCS1102, the serendipitous cluster at $z=0.2550$ is rich (157  members in our sample) and has a high velocity 
dispersion of $\sigma_0 \approx 930$ \kms, but the cluster is centered at the edge of our field or perhaps beyond, and the lower 
redshift means that key spectral features fell below the spectral window we used for most of the spectroscopy, so this cluster has 
been excluded from our sample.  Basic parameters for remaining 5 clusters, including sample properties that apply specifically to 
this paper, are given in Table \ref{Clusters}.  `Cluster' members were chosen in an interval of rest frame velocity $\pm3000$ \kms, 
which reasonably if not perfectly sequesters the virialized cluster and its supercluster from the field (see Figure 16 of Paper 1).  
A field sample was selected for each sky field that covers the same redshift range as the 5 clusters, $0.31<z<0.54$ --- excluding, 
for each field, the z-range of the cluster(s), referred to as the \emph{cl\_field}.

The spectra classes we use in this study are closely related to the \emph{k,k+a,a+k,e(a),e(b),e(n)} types we developed in
Dressler \etal\ (1997), but we have now replaced criteria based the detection of emission-lines with well-measured star formation
rates.  SFRs were calculated from measurements of \24m, \Ha, \Hb, and \OII luminosities, using new calibrations which we 
develop in Paper 1.  Not all measurements are available for each galaxy: \24m photometry is available for galaxies in 2 of our 
4 fields; \Hb and \OII fluxes are available for most galaxies, while \Ha fluxes were obtained for about one-half of the cluster 
members, but less than one-third of the field galaxies in the surrounding supercluster.  In order of preference we use (1) \24m 
plus \Ha, (2) \24m plus \OII, (3) \Ha, (4) \Hb, and (5) \OII in order to calculate star formation rates.  All of these calculations 
implicitly include the effects of extinction, either object-by-object in the case of \24m + optical measurements, or in the mean 
for calibrations relying on a single line.  More detail on the calculation of SFRs can be found in Paper 1.

The detection of Balmer absorption lines remains an key component in the classifications.  As first noted by Dressler \& Gunn 
(1983) and quantified by Couch \& Sharples (1987), and P99, strong Balmer absorption is usually indicative of a recent 
starburst in the star formation history of a galaxy, due to the prevalence of A stars with a $\sim$1 Gyr lifetime after O and 
B stars have evolved off the main sequence.  \Hd is particularly well suited for this measurement, and a strong detection of 
\Hd in what is otherwise an old K-giant type spectrum is a reliable sign of a poststarburst galaxy.\footnote{As explained in 
Paper 1, we use a modified measure of the \Hd index in the ICBS which improves the measurement of Balmer line strength 
by including measurement of the (H+H$\epsilon$)/K ratio.} However, for a galaxy in which star formation is \emph{ongoing}, 
strong Balmer absorption lines are also the result of vigorous star formation.  The ICBS has taken a first step in accounting 
for this effect by using well-studied local  samples to define a relation between [O II] equivalent width and \Hd absorption in 
the absence of starburst, as explained in Paper 1.  Using this, we define a $\Delta$H$\delta$ index which measures the 
Balmer-line strength in excess of that expected in a continuously star forming system to identify  with a starburst.

Finally, to better identify starbursts, we have revisited the issue of whether strong \OII emission alone can signal a starburst, as 
introduced in P99.  Again, using local samples, we have determined that the fixed limit in equivalent width $W_{eqw} > 40$\,\AA\ 
used previously must be refined by using a limit that is a function of a galaxy's specific star formation rate,
sSFR.

Extra attention was given to the of-order hundred objects with a marginal detection of star formation using one or more indicators, 
to determine which of these were in fact likely PAS galaxies. The spectra of these objects were examined closely to estimate the 
S/N of emission-lines [O II], H$\beta$, [O III], and H$\alpha$, and to determine whether the the line-strength of any of these 
automatically-detected features were at the sensitivity limit of the ICBS data. In general, if the starforming classification was 
based on a single determination of the SFR at the detection limit (a function of redshift), the classification was changed to PAS, 
while multiple detections of a non-zero SFR, even if at their detection limits, kept the galaxy in the  starforming category.  
Although the boundary between PAS and starforming cannot be sharp at the detection limit for these optical features and 
\24m\ flux, the application of this criteria essentially divided ambiguous cases between PAS and starforming (CSF, SBH, or SBO) 
at a SFR of 1 \Msun\ yr$^{-1}$, equivalent, for a galaxy in our sample, to a specific star formation rate, sSFR of  
$10^{-11}$ yr$^{-1}$.  The distribution of these quantities is shown in the following section.

Galaxies in the 5 cluster and 4 field samples were divided into five distinct spectral classes based on the SFR and the strength 
of Balmer absorption lines.  We define five spectral types as follows: (1) PAS (passive) --- SFR $=0.0 ~and~ \Delta$H$\delta \le 
0.0$; (2) CSF (continuously starforming) --- SFR $> 0.0 ~and~ \Delta$H$\delta \le 0.0 ~and~not$  \OII starburst;  (3) SBH 
(starburst from H$\delta$) --- SFR $> 0.0 ~and~ \Delta$H$\delta > 0.0 ~and~not$ \OII starburst; (4) PSB (poststarburst) --- SFR 
$=0.0 ~and~ \Delta$H$\delta > 0.0$; and (5) SBO --- starburst based on equivalent width of [O II], as described in Paper 1.
Plots of composite spectra representing the 5 spectral classes can be found in Abramson \etal\ (2013).

\subsection{Relating the new spectral classes to star formation histories} 

The new spectral classes are an attempt to better constrain SFHs for the galaxies of the class, however, all are not equally 
successful in providing an unambiguous SFH.  PAS galaxies, formerly `k type,' are the most clear cut --- they are galaxies 
where star formation has been below $\sim$1\,\Msun\,yr$^{-1}$ for at least a Gyr.  Likewise, PSB, formerly `k+a' or the 
stronger `a+k,' are galaxies in which a starburst of at least moderate size has occurred within 1 Gyr, but where the SFR
has fallen to $\ls$1\,\Msun\,yr$^{-1}$ at the epoch of observation. SBO, formerly e(b), is a reliable classification for an
in-progress starburst, modulo an AGN contamination ($\sim$10\%).  However, the [ OII]-equivalent-width criterion that
we use for identifying SBOs is affected by differential extinction between HII regions and the stellar continuum.  We have
sufficient data to produce a ``dust-free" \OII equivalent width for only a minority of the spectroscopic sample, so we have
chosen not to apply this correction, but a test sample indicates that the SBO class would approximately double (with
additions formerly classified as CSF) if this affect were fully taken into account.  In other words, an SBO is reliably a 
starburst, but there are a substantial number of other objects that should be included in this class.   CSF, formerly e(c), is 
similarly a $\sim$90\% reliable attribution for a continuously star forming (non-bursting) galaxy, where the $\sim$10\% 
contamination is by systems that are in fact starbursts.  Our inability to properly assign these to the SBH class is because 
the ICBS spectral resolution is insufficient to isolate Balmer emission in the cores of Balmer absorption lines.  This means that
genuine starbursts with SFRs $\gs$10 \Msun\,yr$^{-1}$ can have their H$\delta$ and H$\epsilon$ absorption lines ``filled
in'' by Balmer emission (to a degree dependent on the dust extinction for the HII regions), such that they are misclassified
as CSF.

Finally, the SBH class, formerly e(a) has a range of SFHs that attest to the difficulty of identifying a starburst in a galaxy 
with \emph{ongoing} star formation, as we mentioned in the Introduction.  P99 identified these as dusty starbursts, but
the ICBS data show that this description only applies to a minority of the class.  The strong Balmer absorption lines in these 
systems guarantee that they have undergone a sizable starburst within the last Gyr, but what is ambiguous with the 
ICBS data is whether the burst is of fairly long duration and still in progress, or the galaxy is observed at some later time, 
up to and including systems where the ``ongoing star formation" (that makes this galaxy an SBH rather than PSB) is 
essentially what it was before the burst.  Using a starburst criterion of sSFR $>1.4 \times 10^{-10}$ \Msun\,$yr^{-1}$, 
appropriate for a subsample of SBH galaxies identified as ``old" and with constant or declining rates of star 
formation over their history (those in groups, with stellar masses M $> 1.4 \times 10^{11}$ \Msun --- see Paper 3), we 
estimate that at least 30\% of the SBH systems are still in the starburst state, with the others in some later phase of 
starburst decline.

Refining the SFHs of these systems will require  additional data for galaxies in the CSF and SBH classes, something
we plan for a future study.  We are satisfied, however, that these 5 spectral classes we use here are sufficiently well defined
for the purposes at hand, and certain that the conclusions of this work are unaffected by the ambiguous SFHs for a
minority of galaxies in the CSF and SBH classes.

\subsection{Galaxy properties of 5 spectral types}

Before we consider the properties of this sample of ``cluster" galaxies, we need to remind the reader that the samples 
discussed here are from fields of diameter $\sim$0.5\deg--- about 5 Mpc in radius, approximately 5 times larger 
than are usually studied for either local or distant clusters of galaxies, where the typical volume of investigation is a sphere 
of radius $R\sim1$ Mpc (approximately the virial radius, and a few core radii).   Due to the higher density of the cluster
cores and the associated difficulty of covering all objects with multislits, our spectroscopic sample within 500 kpc of the 
cluster centers is undersampled compared to the full catalog by $\sim$20\%, with a range of 0\% to 40\% for 
the individual clusters.  (This incompleteness is unbiased with spectral type.)  What we will refer to as the ICBS 
``metacluster" sample is, then, a mixture of supercluster and cluster galaxies.  In \S4.2 we will define a division
between these two samples based on measurements of R$_{200}$, which we identify as the cluster virial radius. 
We will show in \S3 that for 4 of the 5 metaclusters we have analyzed, the supercluster is a region of infall that will 
substantially build the cluster during this epoch, adding hundreds of galaxies that will become a major part of the 
traditional cluster sample of a present-epoch cluster.

\begin{figure*}[h]

\centerline{
\hspace{-0.0in}
\includegraphics[width=2.8in, angle=90]{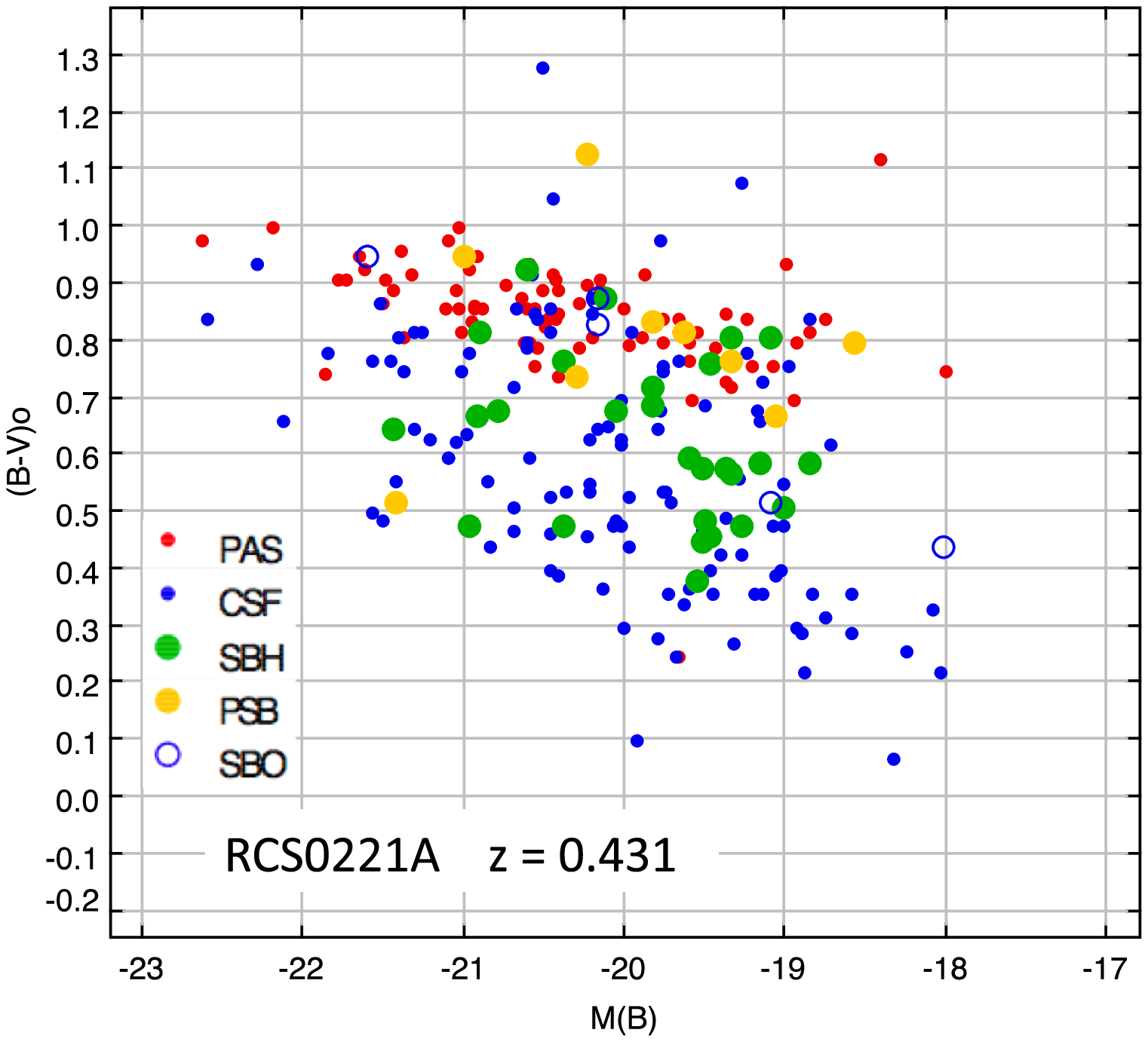}
%\qquad
%\hspace{0.65in}
\includegraphics[width=2.8in, angle=90]{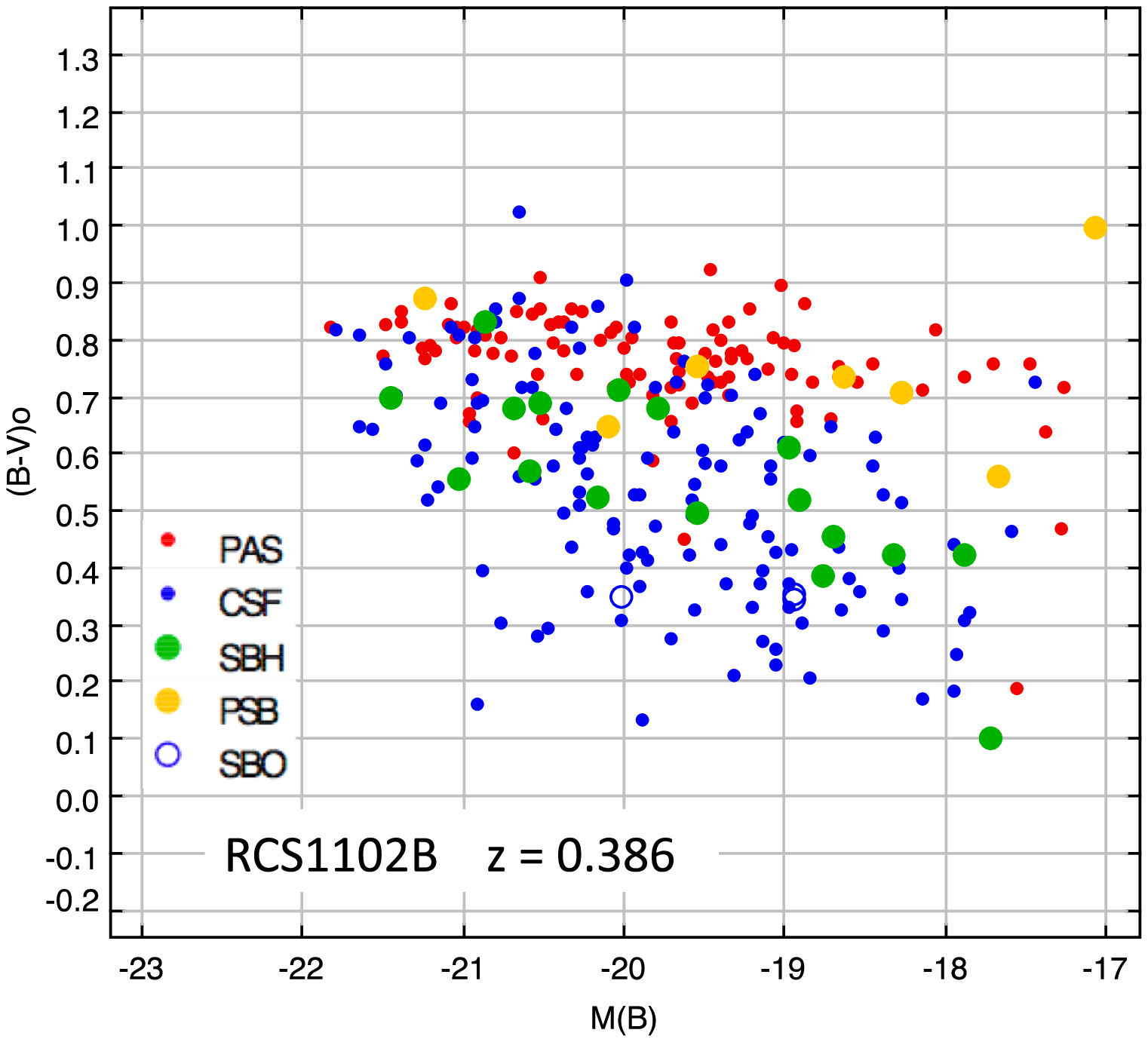}
}
\centerline{
\includegraphics[width=2.8in, angle=90]{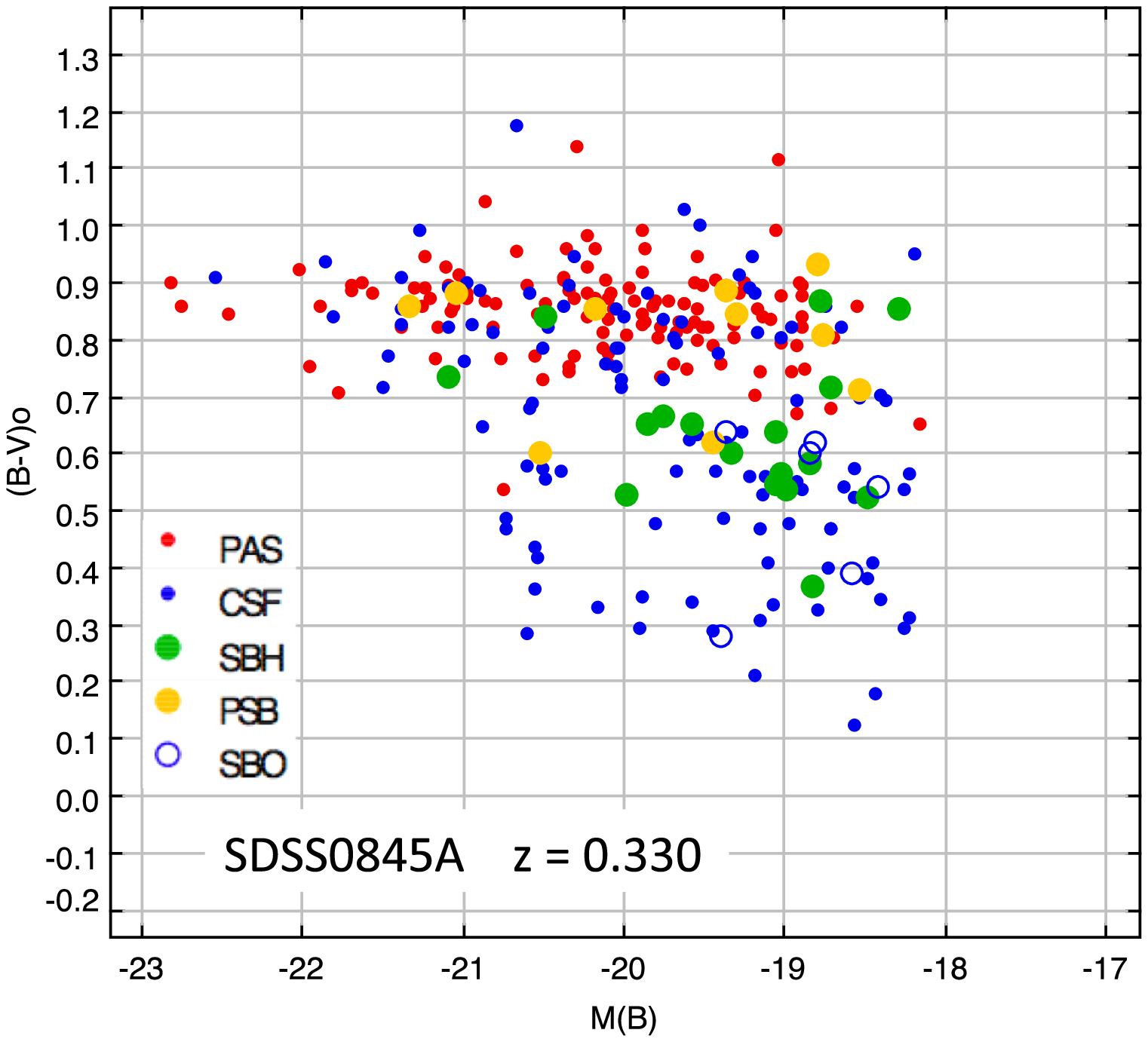}
\includegraphics[width=2.8in, angle=90]{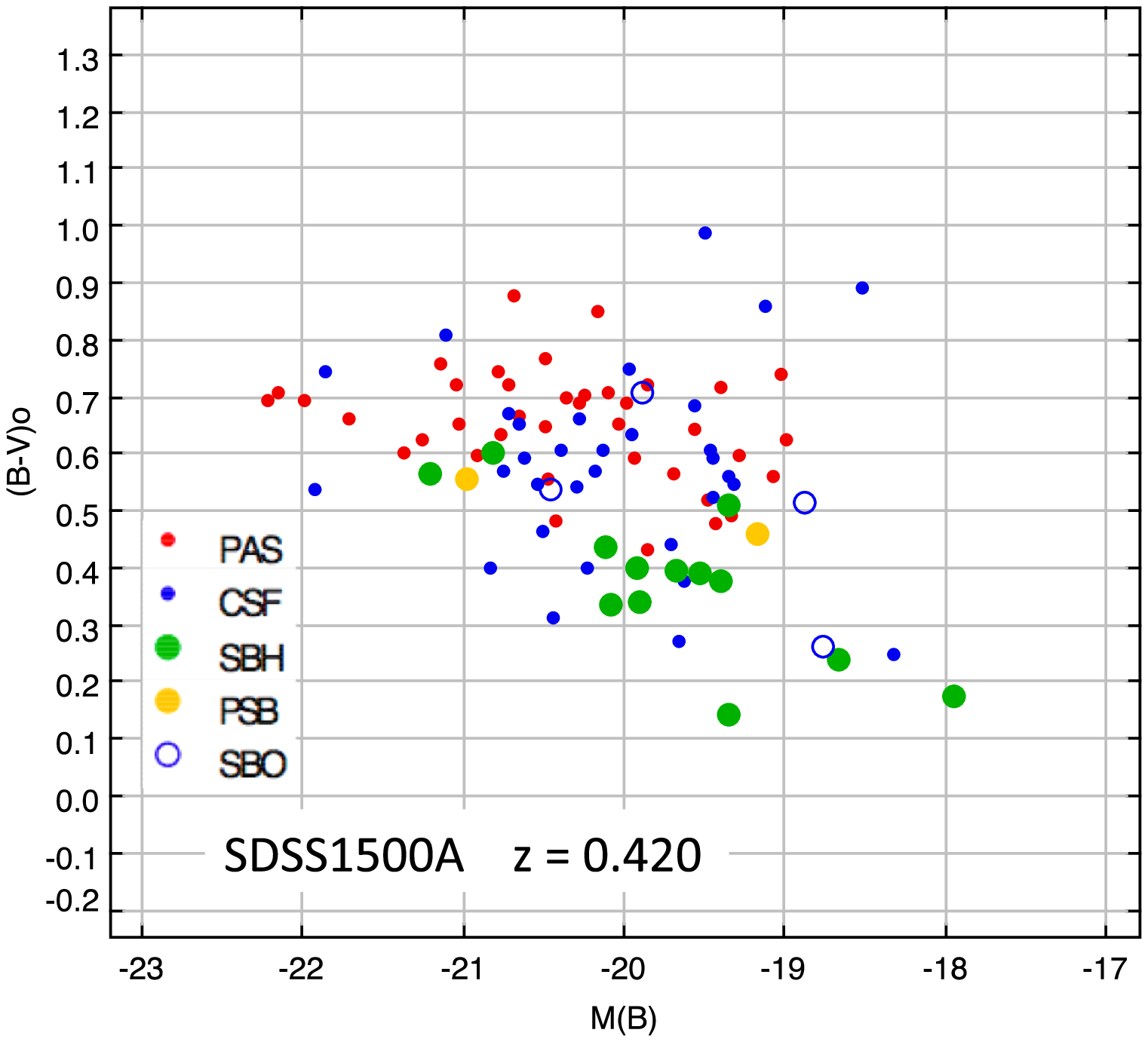}
}
\centerline{
\includegraphics[width=2.8in, angle=90]{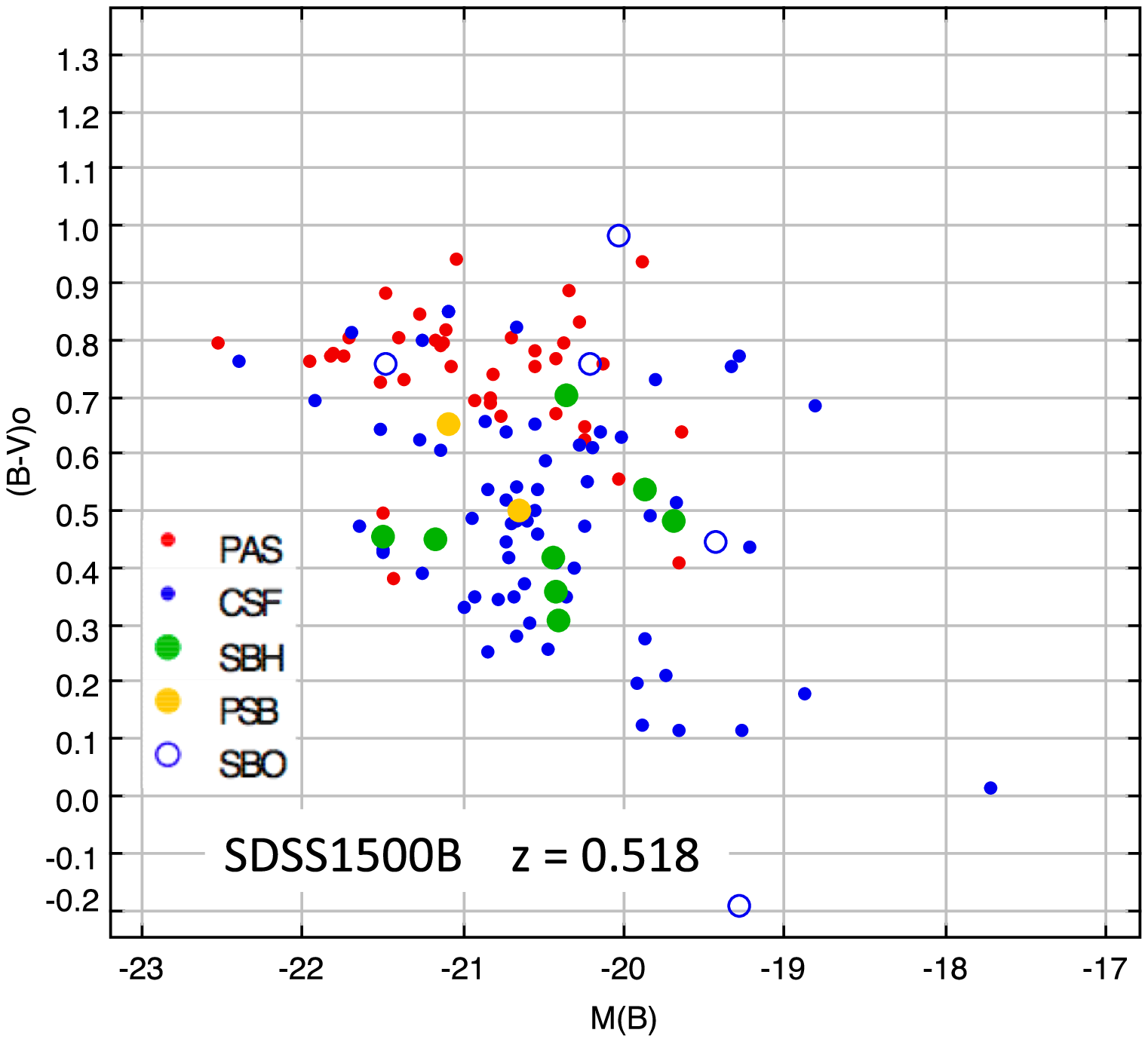}
\includegraphics[width=2.75in, angle=90]{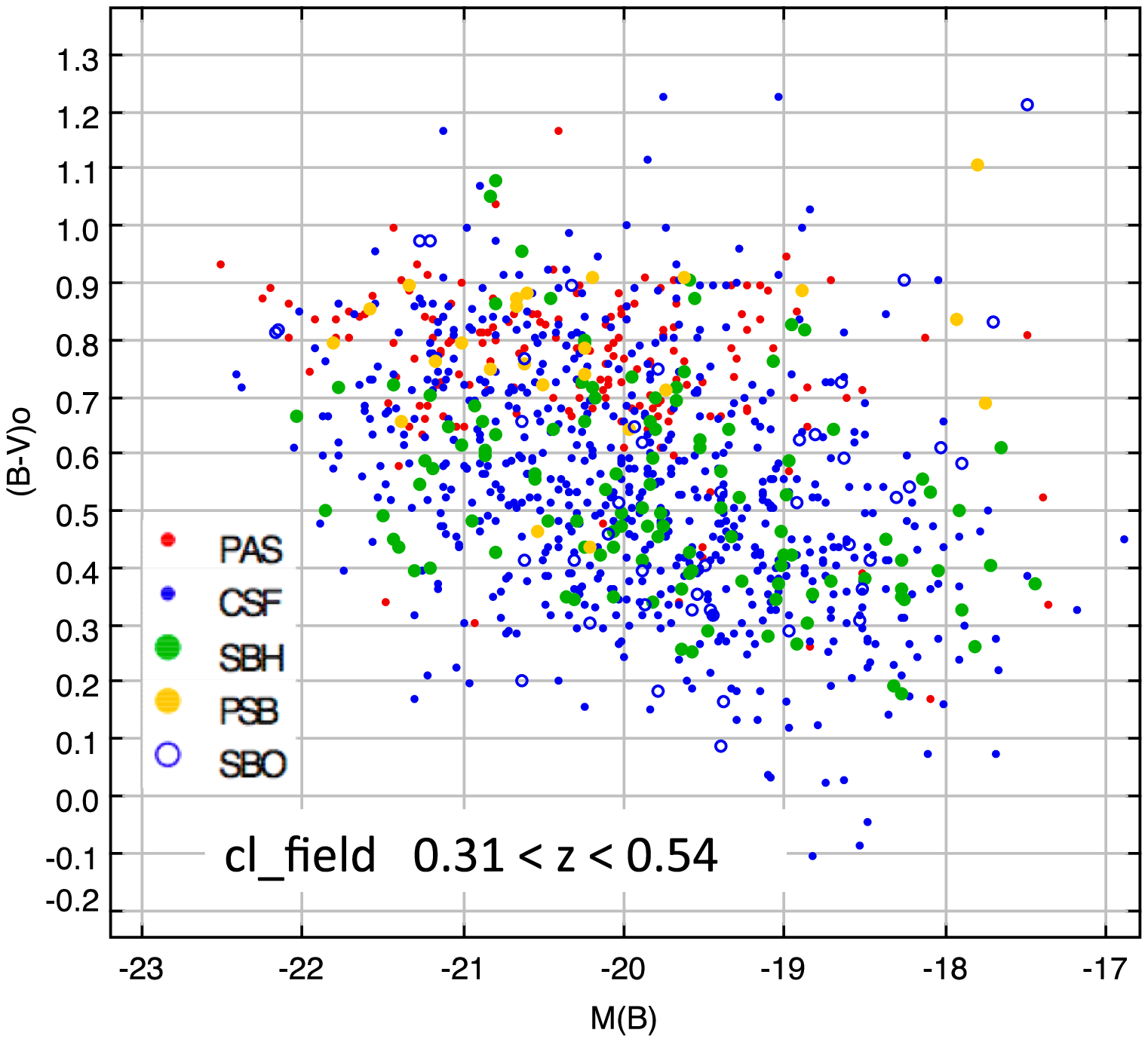}
}
\caption{Rest-frame color-magnitude diagrams for the RCS0221A, RCS1102B, SDSS0845A, SDS1500A, SDSS1500B metaclusters 
and the \emph{cl\_field}, $0.31<z<0.53$.  PAS galaxies define the customary ``red sequence'' and the CSF galaxies cover a wide 
range of (B-V)$_0$ (rest frame) colors, albeit with a larger proportion of bluer galaxies than for the the extended fields of 
intermediate-redshift clusters.  SBH and SBO starburst galaxies have magnitudes and colors similar to CSF galaxies, as might 
be expected, while PSB galaxies cover the region that includes the red sequence and the reddest CSF galaxies, also as expected. 
\label{fig:CM_diagrams}}

\end{figure*}

Figures \ref{fig:CM_diagrams}a-f show color-magnitude (CM) diagrams, M(B) vs $(B-V)_0$, for the metacluster samples and 
for the cl\_field.  These CM diagrams are unremarkable compared to those of the cores of other intermediate-redshift clusters, 
although it is obvious that the fraction of star forming galaxies is higher than for cluster cores alone.  Nevertheless, the PAS galaxies 
define a well-defined red sequence in each cluster and the field, although it is noteworthy that are starforming galaxies of similar 
or slightly redder color that are comparably bright to the most luminous PAS galaxies.   Low-redshift cluster populations, 
with only a small percentage of starforming galaxies to begin with, have negligible numbers of galaxies that are this luminous.  
Presumably these cases have faded by $\sim$1 mag by the present epoch, and of course are candidates to have joined the 
PAS population.

The distributions in M(B) and $(B-V)_0$ are shown in histogram form for all metaclusters combined, and the cl\_field, in 
Figures \ref{fig:sptype_MB_Mass}a and Figures \ref{fig:sptype_BVo_SFR}a, respectively.  The distributions are for galaxies in 
four of the five metaclusters in the sample (solid histogram) and the cl\_field (open historgram).  The metacluster SDSS1500B  is not 
included in these distributions because its higher redshift results in significant incompleteness for lower-luminosity galaxies.  From 
the histograms, it is clear that the luminosity distributions of the five types have large range and overlap substantially.  The
median luminosities of PAS, CSF, and PSB galaxies are similar, with the SBH and SBO distributions shifted to lower luminosities
by a factor of $\sim$2-3, but generally the picture is a lack of clear distinction between classes by absolute brightness.

Figure \ref{fig:sptype_MB_Mass}b shows the stellar mass distribution for the 5 spectral types.\footnote{As described in detail in
Paper 1, masses were calculated with a variant of the Bell \& de Jong (2001) prescriptions, using Bruzual \& Charlot population
models and a Salpeter IMF.  An advantage of this approach is that derived masses are insensitive to internal extinction.}  Despite 
their similar M$_B$ distributions, the mass distributions for the 5 spectral classes show significant variation. PAS and PSB galaxies
are, on average, more massive than CSF galaxies, and the $r=22.5$ completeness limit of the sample now becomes clearly defined 
in the steeper low-mass cutoff of the PAS distribution (due to the small spread in mass-to-light ratio dictated by the small range in 
color).   SBH galaxies have a mass distribution like that of the CSF, with SBO galaxies showing a factor of $\sim$2-3 shift to 
even lower masses.  But, again the basic picture is of a wide-range in masses and a considerable overlap for all classes.  The most
massive galaxies, few in number, are mostly PAS.

The SFRs of spectral types CSF, SBH, and SBO are shown in Figure \ref{fig:sptype_BVo_SFR}b.  It is important to explain that, 
although the latter two types are clearly starbursts, their 

%Figure 1: color-magnitude diagrams

\noindent{SFRs largely overlap with the SFRs of CSF galaxies.  This is because 
the definition of a starburst is related to the rise in the SFR over $\sim10^8$ years compared to the past average over the 
preceding 1-2 Gyr, a factor of 3-10 for the starburst galaxies in the ICBS sample, for galaxies which range in mass by a factor of 
100.  Evidence for a higher SFR for starbursts improves when normalized by the mass --- the specific star formation rate (sSFR) 
shown in the bottom half of Figure \ref{fig:sptype_BVo_SFR}b.  The distribution of sSFR for SBO galaxies is very broad, but the 
mean sSFR of SBO galaxies is a factor-of-4 higher than the CSFs, consistent with (though not sufficient for) their identification as 
starbursts.}  

However, the starburst nature of SBH galaxies is not evident even in Figure \ref{fig:sptype_BVo_SFR}b, partly because the
a large fraction of the SFH population is past the peak of the burst.  But, the principal reasons is that sSFR is not a reliable
criterion for the comparatively low-amplitude starbursts identified with spectral features that sample the composite stellar
population.  Such measurements are sensitive to a factor-of-several increase in the SFR compared to the average of the
priori few Gyr.  In contrast, measurements of sSFR, by construction, compare the present SFR to the past average over the 
full history of the galaxy.  The SFR of a typical galaxy has declined substantially over its lifetime --- by $z=0.5$, a 
higher-ampltidue starburst, like a LIRG or ULIRG, is required to make a notable change in the sSFR.  The converse is 
also true:  a galaxy can have a high sSFR \emph{without a burst} if it has had a rising SFR over a several Gyr period 
of its recent past (see Paper 3).  The sSFR criterion is a blunt instrument for detecting starbursts, because the burst must
reach a sufficiently high amplitude to overcome both of these ambiguities.  Our use of spectral features to quantify 
the time scale of the SBH and SBO phases is uniquely able to localize the burst with respect to the average SFR of the
prior few Gyr.  It is this kind of galaxy, where an increase in starforming ``efficiency" is a more recent and temporary 
condition --- an ``event," that is the moderate starburst we focus on in this study.

Finally, it is interesting to note a clear difference, in the $(B-V)_0$ and star formation rates of the CSF cluster sample compared 
to the CSF cl\_field sample, shown in  Figure \ref{fig:sptype_BVo_SFR} by the open black histogram.  There is a clear excess 
of bluer galaxies, with high star formation rates, for the cl\_field sample.   Although the `cluster' sample is dominated by 
supercluster galaxies outside the dense cluster core, it appears that some suppression of star formation rates has been 
accomplished in this environment, that is, the supercluster field is distinguishable from the purefield population at this redshift.  
Evidence of this effect has, in particular, been documented in studies of the CNOC clusters (Balogh \etal\ 1997, 1999).

As mentioned above, the completeness of detected star formation falls rapidly below an SFR of 1 \Msun\ yr$^{-1}$, or a sSFR
of $10^{-11}$ yr$^{-1}$, for galaxies in the ICBS sample.  The inspection procedure we use attempts to include SFRs that are 
lower than these limits but are nevertheless reliable detections --- cases of unusually good spectra or two or more marginal 
detections that exceed the typical detection limits.

\vspace{1.0cm}

\subsection{Completeness of spectroscopic samples compared to magnitude-limited photometric catalogs}

Basically, our spectroscopic sample is half of the  photometric magnitude-limited ($r \ls 22.5$) sample, that is
for every two galaxies in the photometric sample we have a good quality spectrum of one.  There are, of course,
somewhat different magnitude distributions of the five spectral types, and there is some effect of less-thorough 
sampling in dense regions due to the difficulty of sampling with the spectroscopic multislits, compared to regions
of low spatial density, however, none of these effects differ by much among the different spectral samples ---
PAS, CSF, SBH, PSB, and SBO --- for both metacluster and field.  Table \ref{incompleteness_table} gives these various
levels of incompleteness, in terms of magnitude-incompleteness, spatial-incompleteness, and a combination of both.
These are expressed as weights (averaged from the weights from each galaxy in that sample) that can be applied 
to the spectroscopic samples to fairly represent the photometric catalog.  The samples for all cluster members, and 
all cl\_field galaxies, for the different spectral types are tabulated for each of the incompleteness effects, normalized 
by the incompleteness for the entire sample of cluster or cl\_field galaxies in the spectroscopic catalog.

Table \ref{incompleteness_table} shows that the total weight WTtot that should be applied to compare one spectroscopic
sample to another are all $\sim1.0$, with excursions of typically $\pm$5\% and only one larger than 10\% (+17\% for the 
small sample of cluster SBOs).  These corrections are sufficiently small that, despite the different sampling with magnitude
and spatial location, the spectral samples can be cross-compared without significant correction.

Table \ref{incompleteness_table} also gives these WTtot values for the metacluster group and non-group samples we will describe
below, and for the comparable cl\_field group sample.  Again, the incompleteness corrections are all $\ls10$\%, and unimportant
for the analysis that follows.

\subsection{The utility of mass-limited samples of the 5 spectral types}

The ability to estimate galaxy masses from spectrophotometric measurements has improved in recent years due to improved 
modeling of stellar populations in the integrated light of galaxies and general availability of near-IR fluxes that better constrain 
the higher mass-to-light stellar component.  Many studies of galaxy evolution have taken advantage of this to study 
mass-limited samples preferentially over magnitude-limited samples that have been the norm for extragalactic studies, in 
the field and in clusters (see, e.g., Patel \etal\ 2009a,b).  Since a galaxy's stellar mass is constant or slowly growing over the 
epoch $0<z<1$, while its luminosity may not be, a mass-limited sample is the only reliable way to answer questions about 
quantitative relationships between galaxy samples. In our case, for example, a mass-limited sample is required to decide 
whether a collection of starburst galaxies are one-to-one related to a collection of poststarburst galaxies.  However, many 
interesting questions regarding galaxy populations revolve around issues that are not mass-related, for example, the 
presence or absence of an AGN.  Assembling a sample of AGN galaxies based on the presence of light coming from 
non-thermal sources does not benefit by excluding galaxies below a 

%Figure 2: histograms of sptype -- MB and sptype -- Mass 

\begin{figure*}[t]

\centerline{
%4clusters_sptype_MB
\includegraphics[width=5.5in, angle=90]{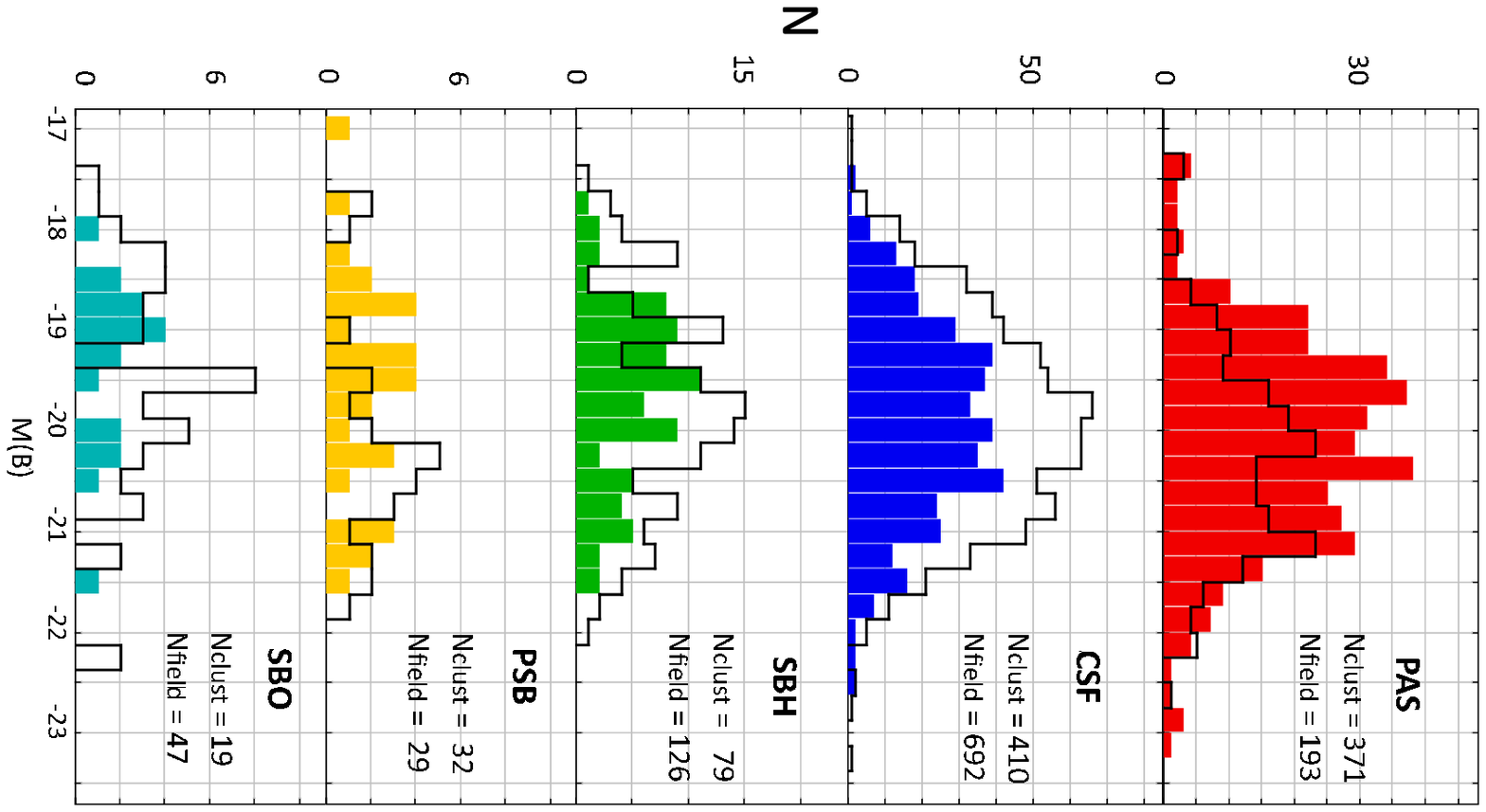}
\hspace{0.35in}
%4clusters_sptype_Mass
\includegraphics[width=5.5in, angle=90]{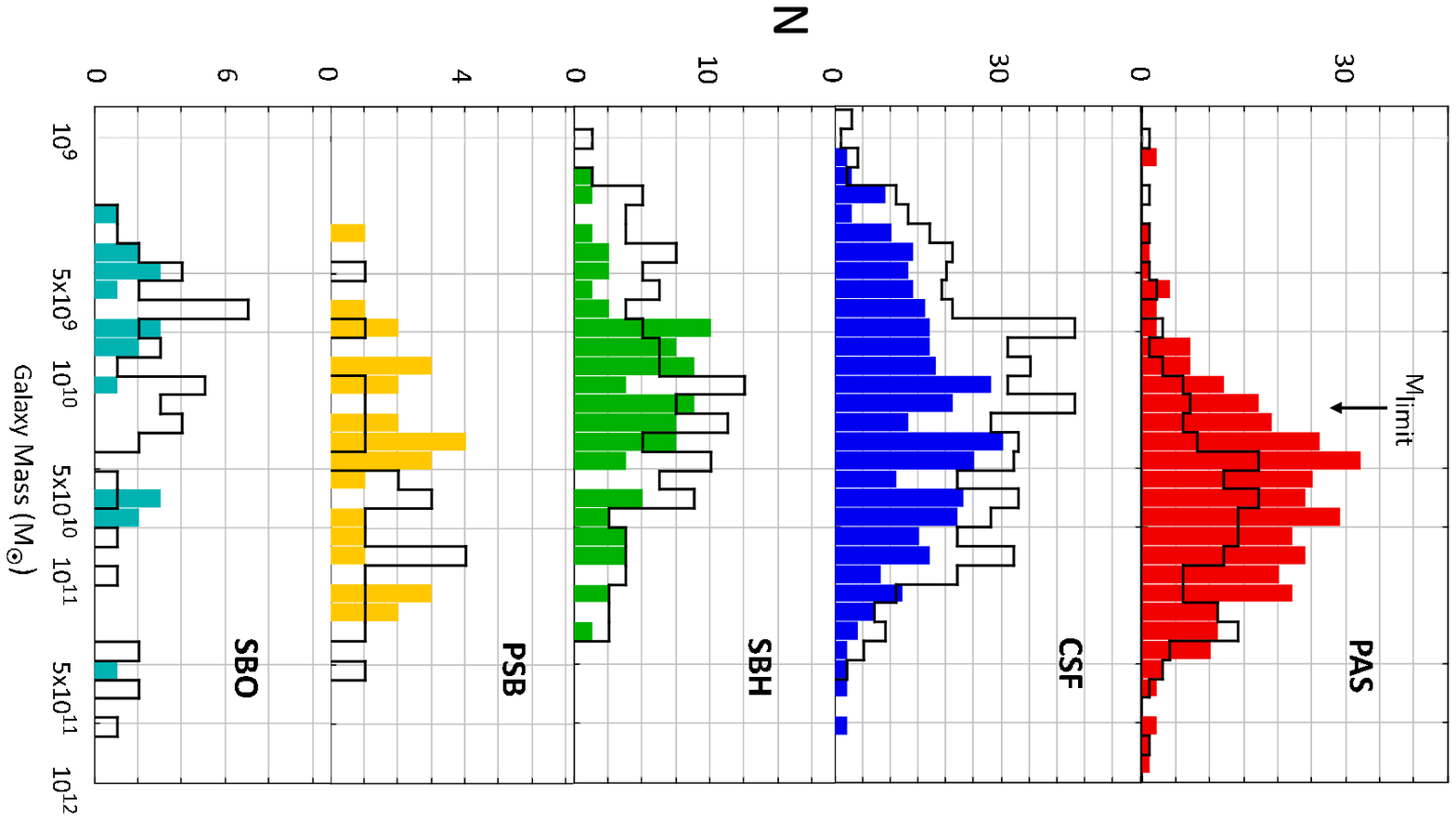}
}

\caption{Solid histograms: Distributions of the five spectral types in (a) $M_B$ luminosity, and (b) galaxy mass, for 915 members 
of the metaclusters of RCS0221A, RCS1102B, SDSS0845A, \& SDSS1500A.  (SDSS1500B has excluded because its higher redshift 
results in significant faint-end incompleteness.)  Open histograms: the same quantities for 1090 cl\_field galaxies,  $0.31< z<0.54$.
The luminosity distributions are broad and overlapping for all 5 types, with only modest variation in the mean.  Mass distributions 
are also broad and overlapping, but PAS and PSB galaxies are clearly shifted to higher mass than the starforming types, CSF, SBH, 
and SBO, which includes galaxies that are among the most massive, and least massive, in the sample.   The PAS mass distribution 
shows that the samples are only complete to a mass limit M$_{limit} = $ 2.5 $\times10^{10}$\,\Msun\ --- marked by the arrow. 
The relatively sharp cutoff is due to the small range of mass-to-light ratios of passive galaxies.  The mass-to-light ratios of the star 
forming systems, CSF, SBH, and SBO, are substantially lower and have a greater range, so the mass distributions of these 
magnitude-limited sample extends well below M$_{limit}$. The poststarbursts, PSB, represent an intermediate population. 
These distributions can be used to convert between results of the magnitude-limited sample and those a mass-limited sample, 
as explained in \S2.2. 
\label{fig:sptype_MB_Mass}}

\end{figure*}

%Figure 3: histograms of sptype -- BVo and sptype -- SFR & sSFR 

\begin{figure*}[t]

\centerline{
%4clusters_sptype_BVo.pdf
\includegraphics[width=5.5in, angle=90]{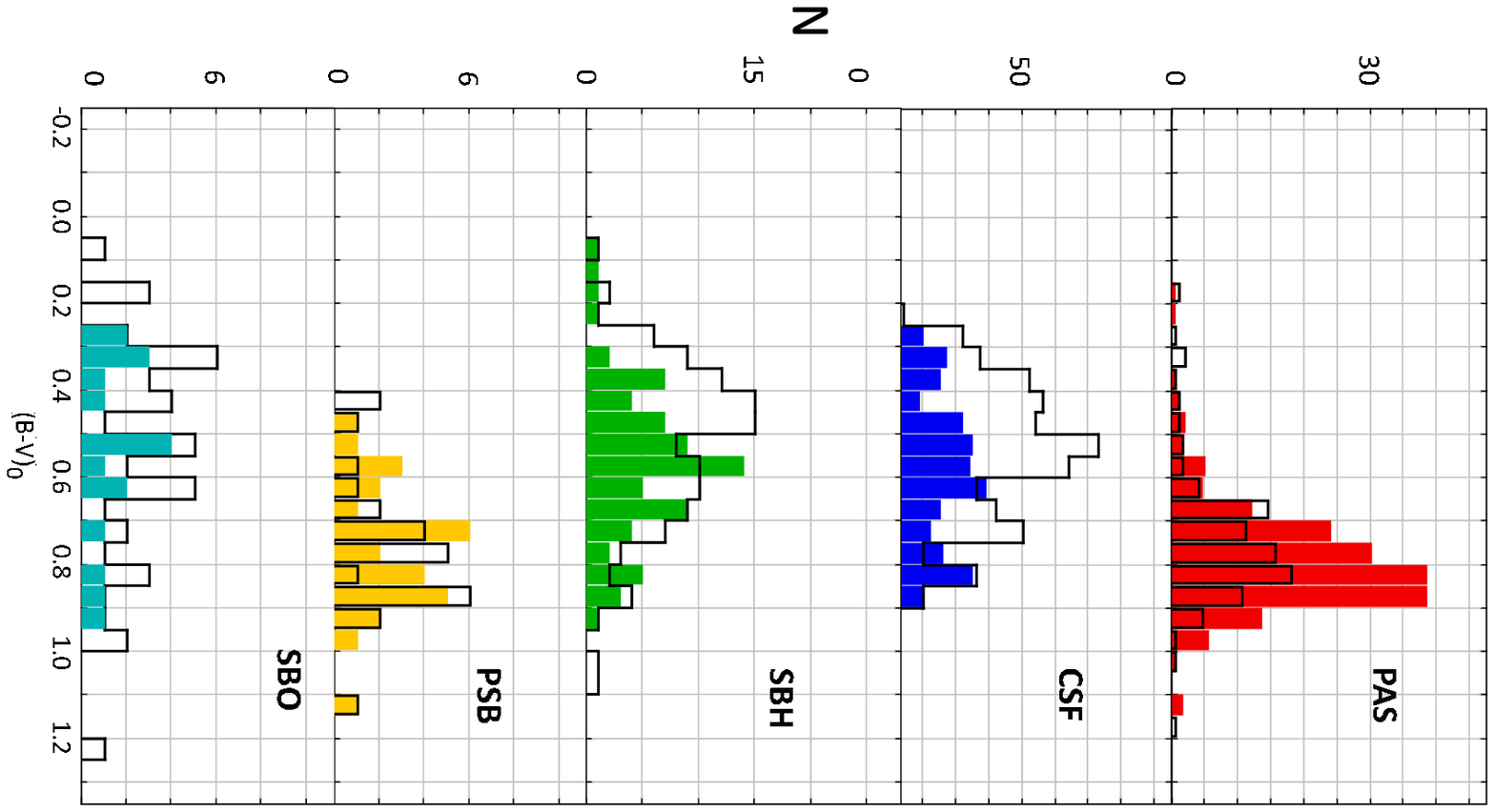}
\hspace{0.35in}
%\4clusters_sptype_SFR_sSFR.pdf
\includegraphics[width=5.5in, angle=90]{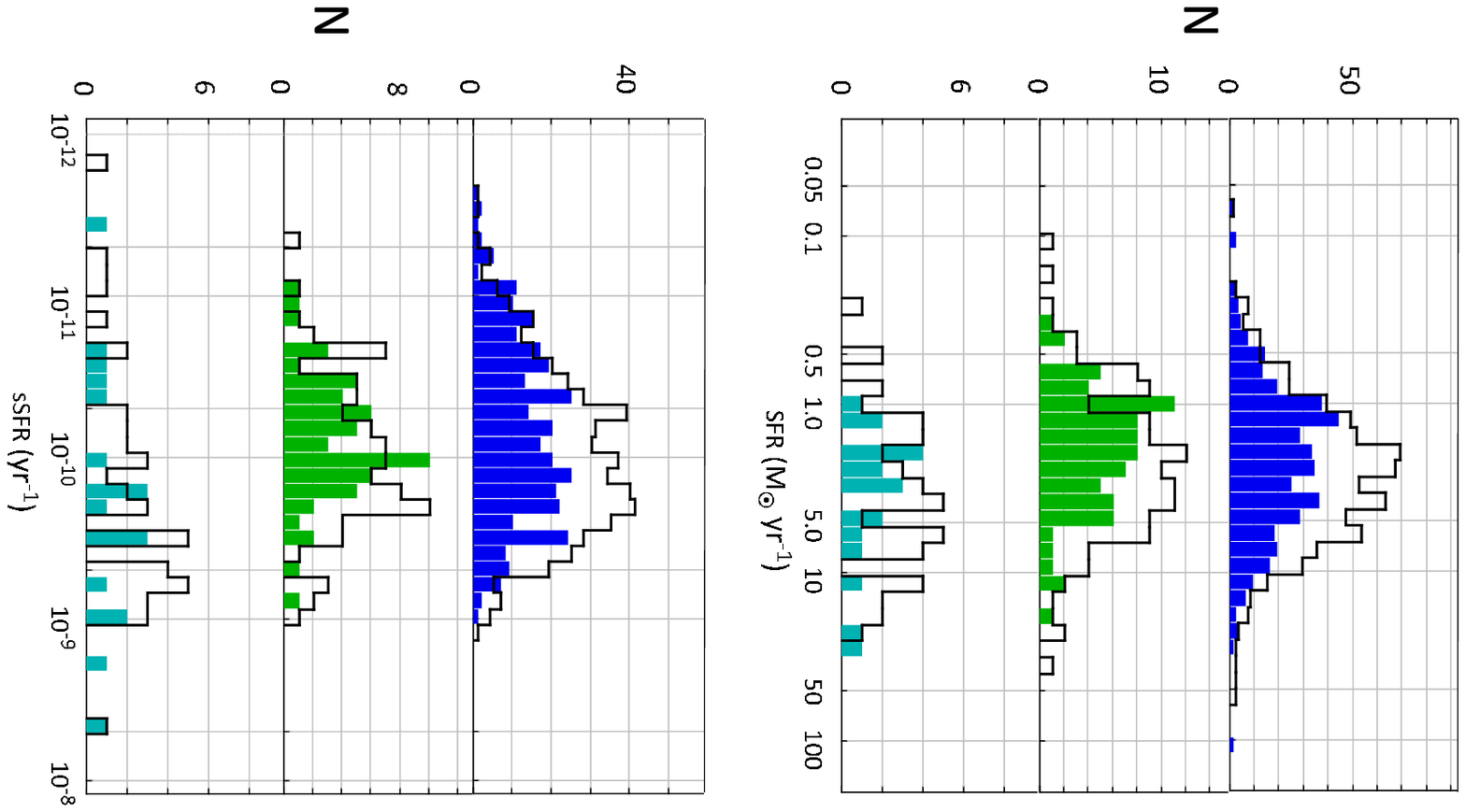}
}
\caption{Solid histograms: Distributions of the five spectral types in (a) (B-V)$_0$ (rest frame) color and (b) star formation rates, 
for 915 members of the metaclusters of RCS0221A, RCS1102B, SDSS0845A, \& SDSS1500A. (SDSS1500B is excluded, 
as in Figure \ref{fig:sptype_MB_Mass}.)   Open histograms: the same quantities for 1090 field galaxies in the redshift range 
$0.31<z<0.54$ covered by the clusters.  CSF and SBH galaxies have very similar distributions of (B-V)$_0$ color, SFR and 
sSFR, but the distribution of SBO galaxies is clearly broader in color and shifted to higher star formation rates.  This 
suggests that sSFR, in particular, might be a good way to distinguish starbursts from general starforming galaxies, but for the 
analysis we do here, with relatively modest amplitude starbursts, this is not the case, as described in \S2.2. The 
most striking feature of these diagrams is the displacement to bluer color and higher star formation rates of cl\_field galaxies
compare to the metacluster sample.  Since PAS galaxies are not included here, this is not likely to be due to the population
in the virialized rich clusters, but is rather a sign that starforming galaxies are in fact influenced by the cluster environment
in a way that suppresses but does not shut off star formation. 
\label{fig:sptype_BVo_SFR}}
\end{figure*}

\noindent{fixed mass limit unless the purpose of the exercise, for example, is to compare the properties of AGN host 
galaxies to those without an AGN.} 

Our study involves both kinds of questions.  In comparing the prevalence of SB (starburst, SBH+SBO) or PSB 
(poststarburst) galaxies in different populations, for example, clusters versus the field, a magnitude-limited sample 
allows us to make maximal use of our data by including objects for which the key feature comes from a component of 
negligible mass, which would be true of either cluster or field galaxies.  However, if we ask to what degree the 
PSB galaxies in either sample will add to the PAS population when the burst has faded, it is necessary to compare the 
numbers (or fraction) of galaxies above a common mass limit.

From Figure \ref{fig:sptype_MB_Mass}b, the mass limit in our ICBS sample can best be seen in the mass distribution 
of PAS galaxies, since the small range in mass-to-light ratio of this spectral type results in a sharp mapping of the
magnitude limit of $r \sim 22.5$ (Figure \ref{fig:sptype_MB_Mass}a) into a well defined mass limit of   
$M_{lim} \approx 2.5 \times 10^{10}$\,\Msun, marked by the arrow.

Table \ref{mass limited samples} compares, for each spectral type in the cluster and cl\_field samples, the total number
of galaxies, the number with measured masses ($M > 10^9$ \Msun), and the number with masses greater than $ M > M_{lim} $.
For example, $\sim$83\% of PAS galaxies in the spectroscopic sample are above $M_{lim}$, but only about 50\% of CSF or
SBH are --- this is expected because star formation yields a lower mass-to-light ratio for these types.  Comparing the fractions
of galaxies above $M_{lim}$ gives us an estimate of how much we need to correct the relative numbers of different spectral
types in going from our magnitude-limited sample to a mass-limited sample.  For example, when comparing the population
of PAS and PSB galaxies, the number of PAS need to be reduced by 0.83 (average of metacluster and field values from 
Table \ref{mass limited samples}), while the PSB population needs to be reduced by 0.679.  The ratio of these numbers are
used to determine, for example, how much a fading population of PSB galaxies will add to the PAS population.  Our
conclusions will not turn on the application of such corrections,  but we will nonetheless bring them in as needed in the discussion 
that follows.

\subsection{The spatial distribution of galaxies of the five spectral types}

\begin{figure*}[h]

\centerline{
%RCS0221A_spmap
\includegraphics[width=2.9in, angle=90]{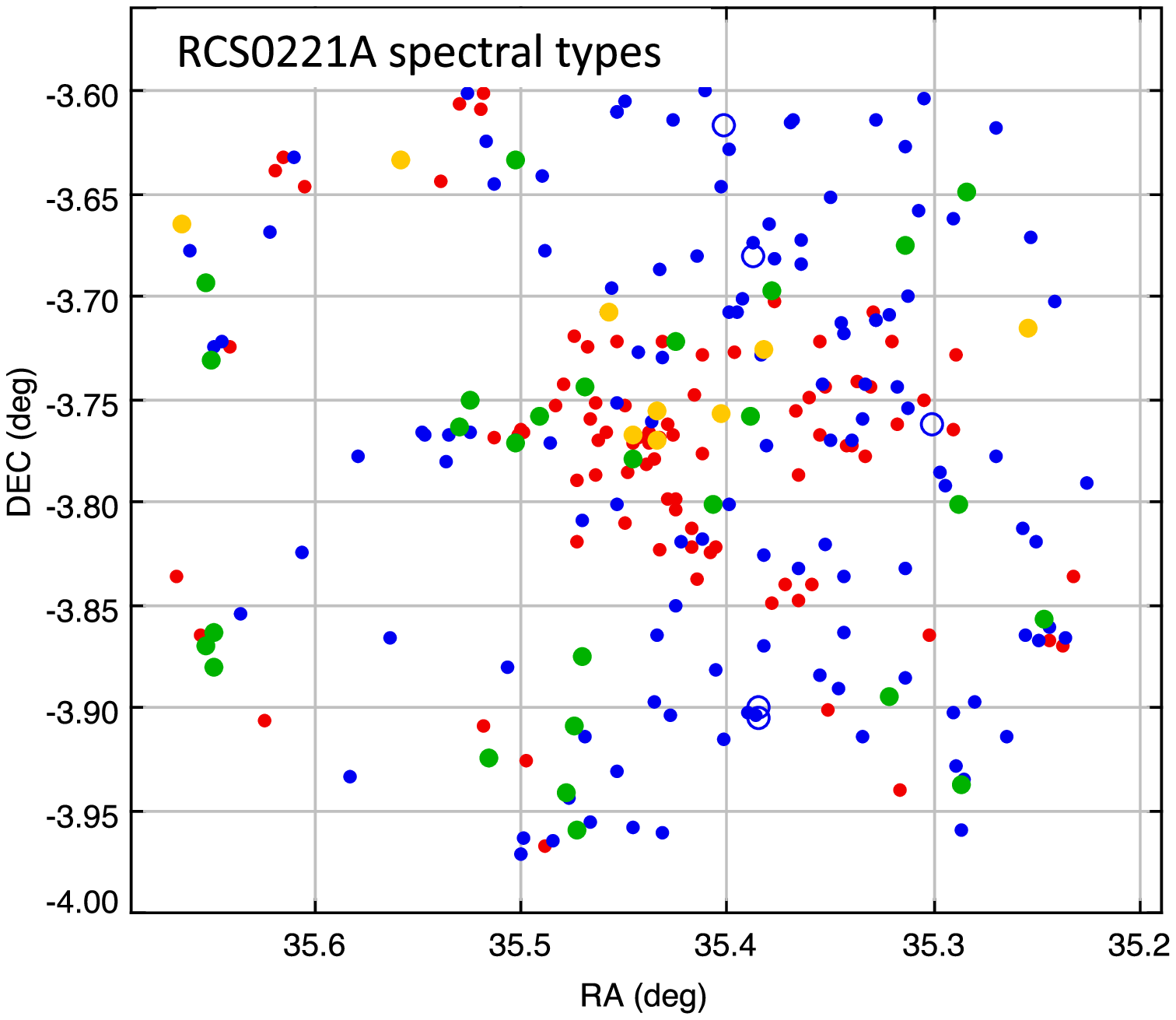}
\qquad
\hspace{-0.50in}
%RCS1102B_spmap
\includegraphics[width=2.9in, angle=90]{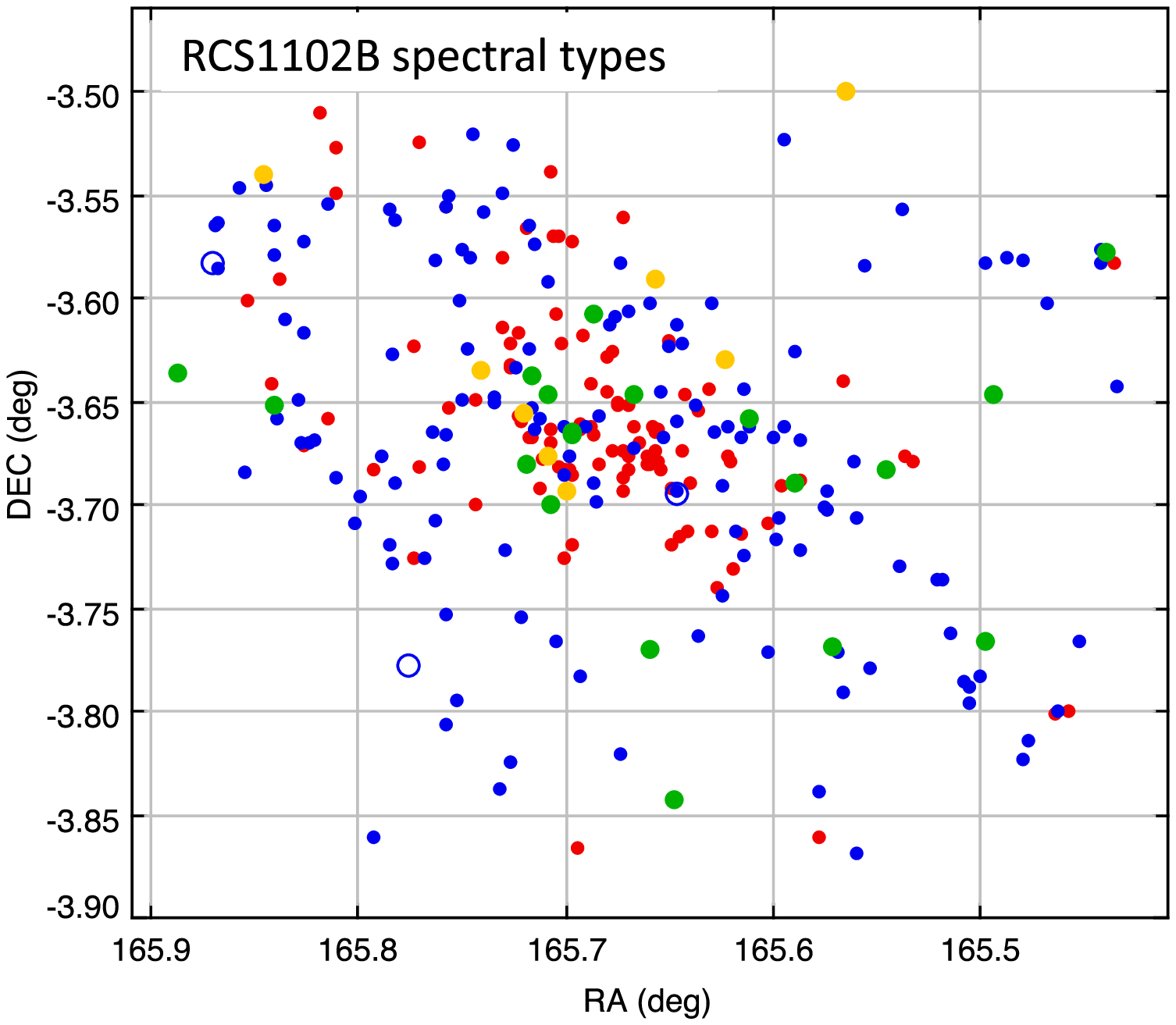}
}
\centerline{
%SDSS0845A_spmap
\includegraphics[width=2.9in, angle=90]{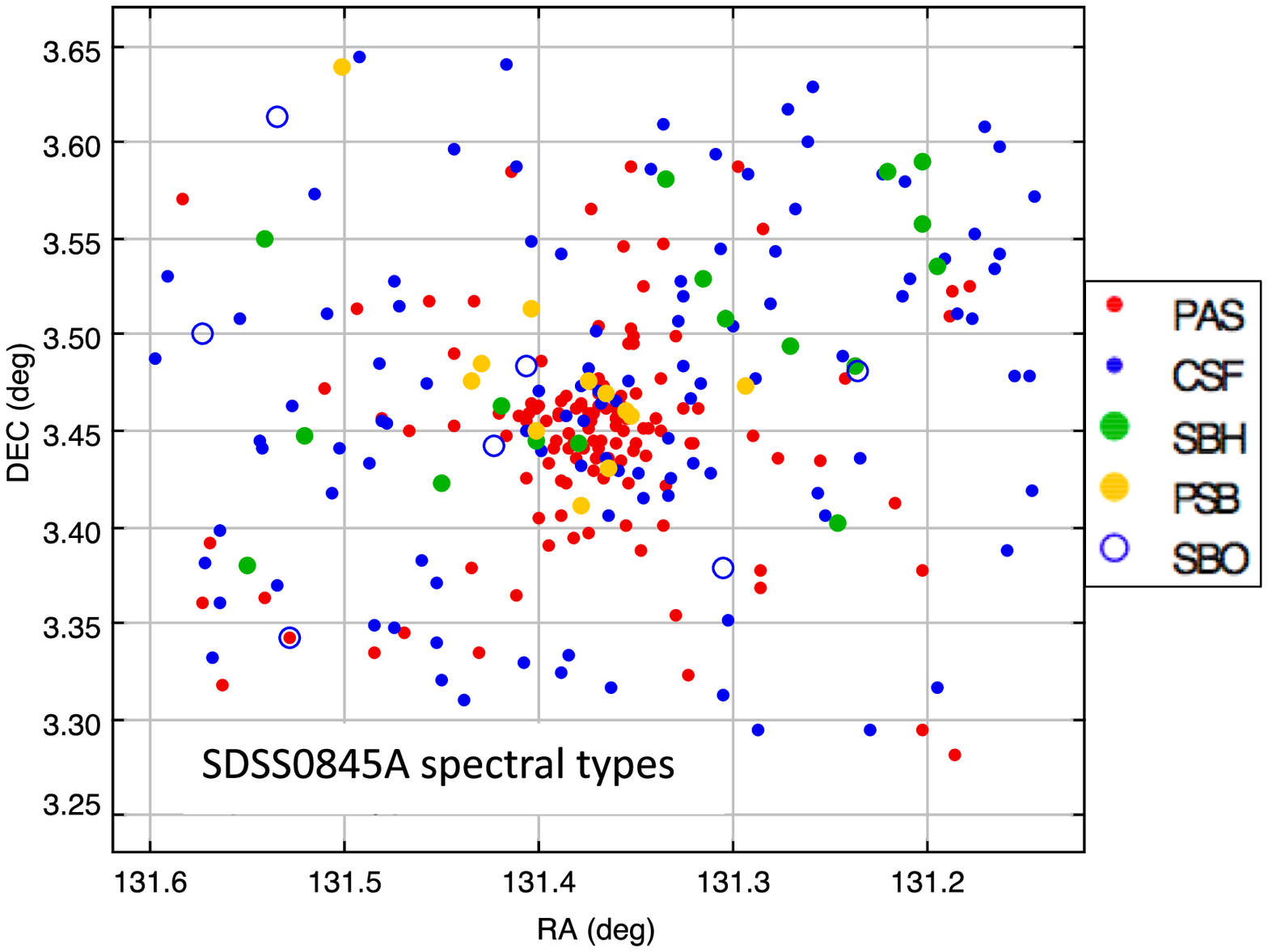}
}
\centerline{
%SDSS1500A_spmap
\includegraphics[width=2.9in, angle=90]{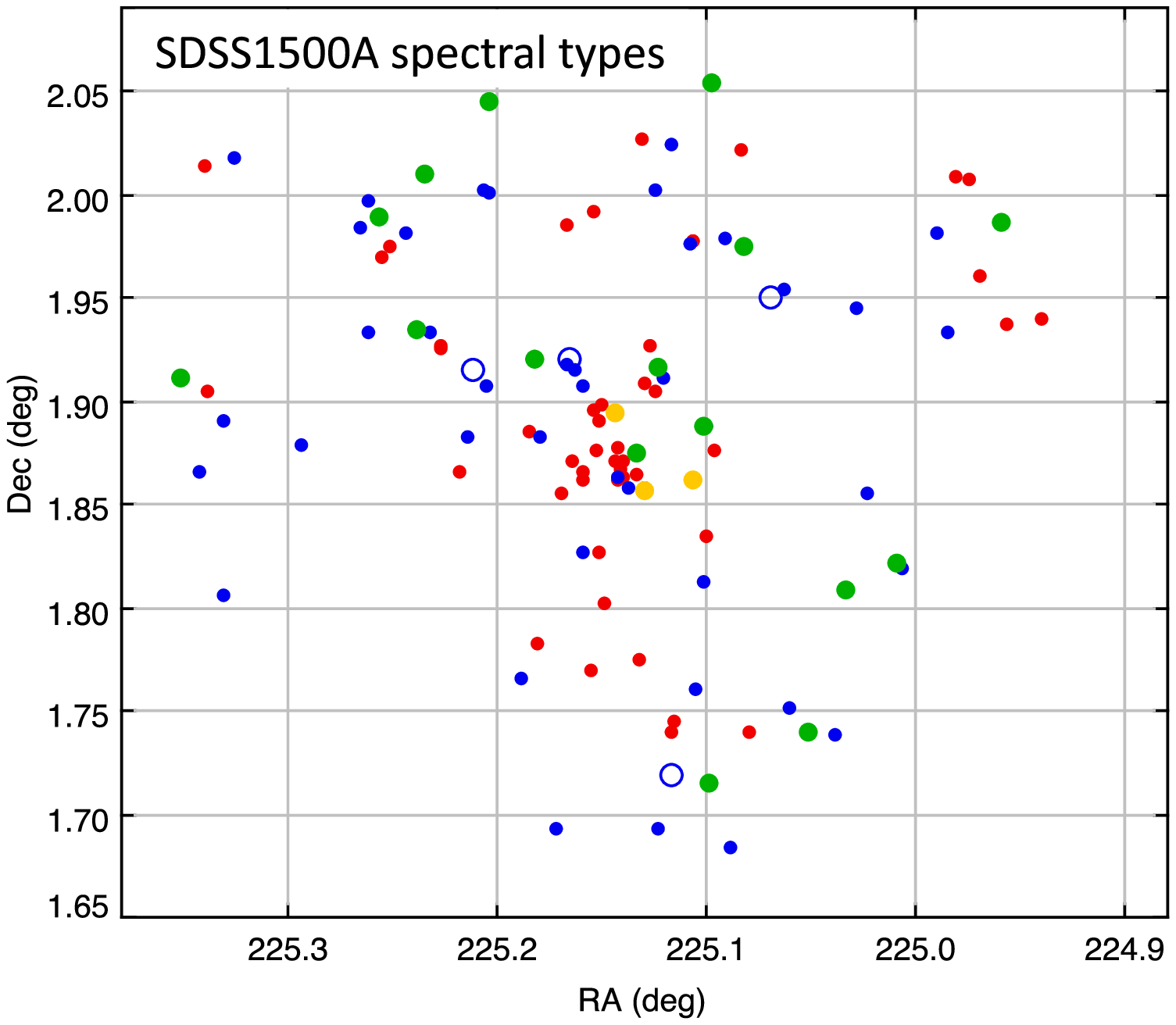}
\qquad
%SDSS1500B_spmap
\includegraphics[width=2.9in, angle=90]{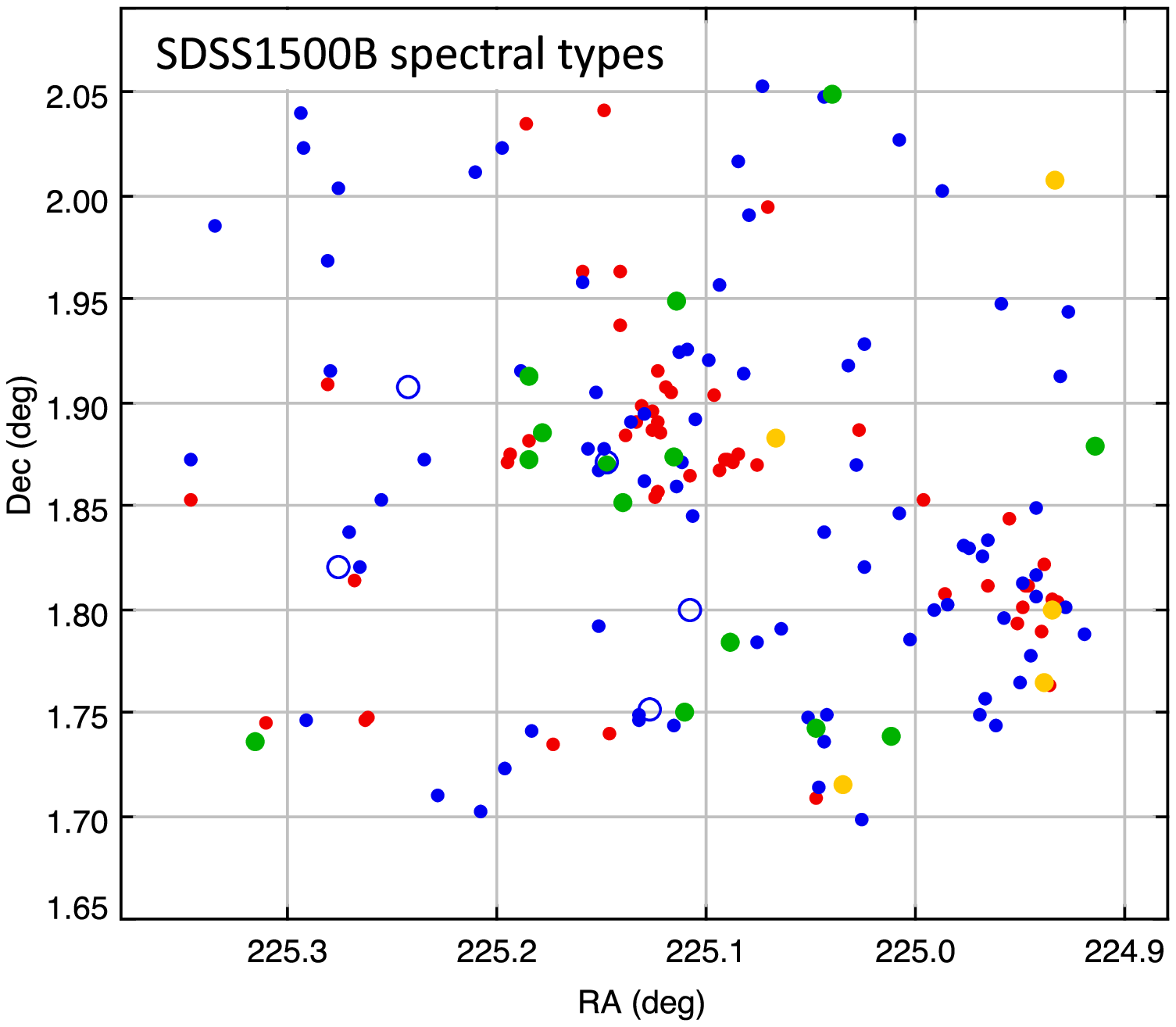}
}
\caption{Sky maps of ICBS metaclusters showing the distribution of spectral types.  Passive galaxies (PAS) are 
strongly concentrated to the cluster center or dense outer groups.  poststarburst galaxies (PSB) trace the PAS 
population.  Continuously-starforming galaxies are more uniformly distributed, as are the starbursts, SBH (strong Balmer 
absorption lines) and SBO (strong \OII emission lines).
\label{fig:sptype_maps}}

\end{figure*}

Figure \ref{fig:sptype_maps} shows the distribution of spectral types in the 5 metaclusters in the 4 fields.  We again recall that 
these are more extensive fields, about 10 Mpc in diameter, than are typically studied around distant clusters of galaxies.  
The typical volume studied in an intermediate-redshift rich cluster is $\sim$2 core radii, and is contained in a sphere of radius 
$R \sim 0.5$ Mpc, about 0.03\deg\ for the clusters in this sample.  In comparison with  the virial radius (which we calculate
in the conventional way as \R200), we see from Table \ref{Clusters} that typically  $R_{vir} \approx 1.3$ Mpc, about 0.07\deg, 
still a small fraction of the R$=$0.45\deg field of \imacs.

With this in mind, several things are nevertheless readily apparent from the plots of Figure \ref{fig:sptype_maps}.   All five 
fields show large numbers of galaxies within the redshift range of the cluster (defined here as $\Delta V < \pm$3000 \kms\ 
in the rest frame of the cluster systemic velocity), that is, there is not a sharp falloff in supercluster members but more of 
a shelf-like distribution.  Furthermore, there appears to be a substantial clumping of these galaxies in four of the five fields.
Only SDSS0845A, shown in the center of Figure \ref{fig:sptype_maps}, appears to show the relatively

%Figure 4: Sptype maps for 5 metaclusters

%Figure 5: Sptype vs surface density and clustocentric distance

\begin{figure*}[h]

\centerline{
\includegraphics[width=2.7in]{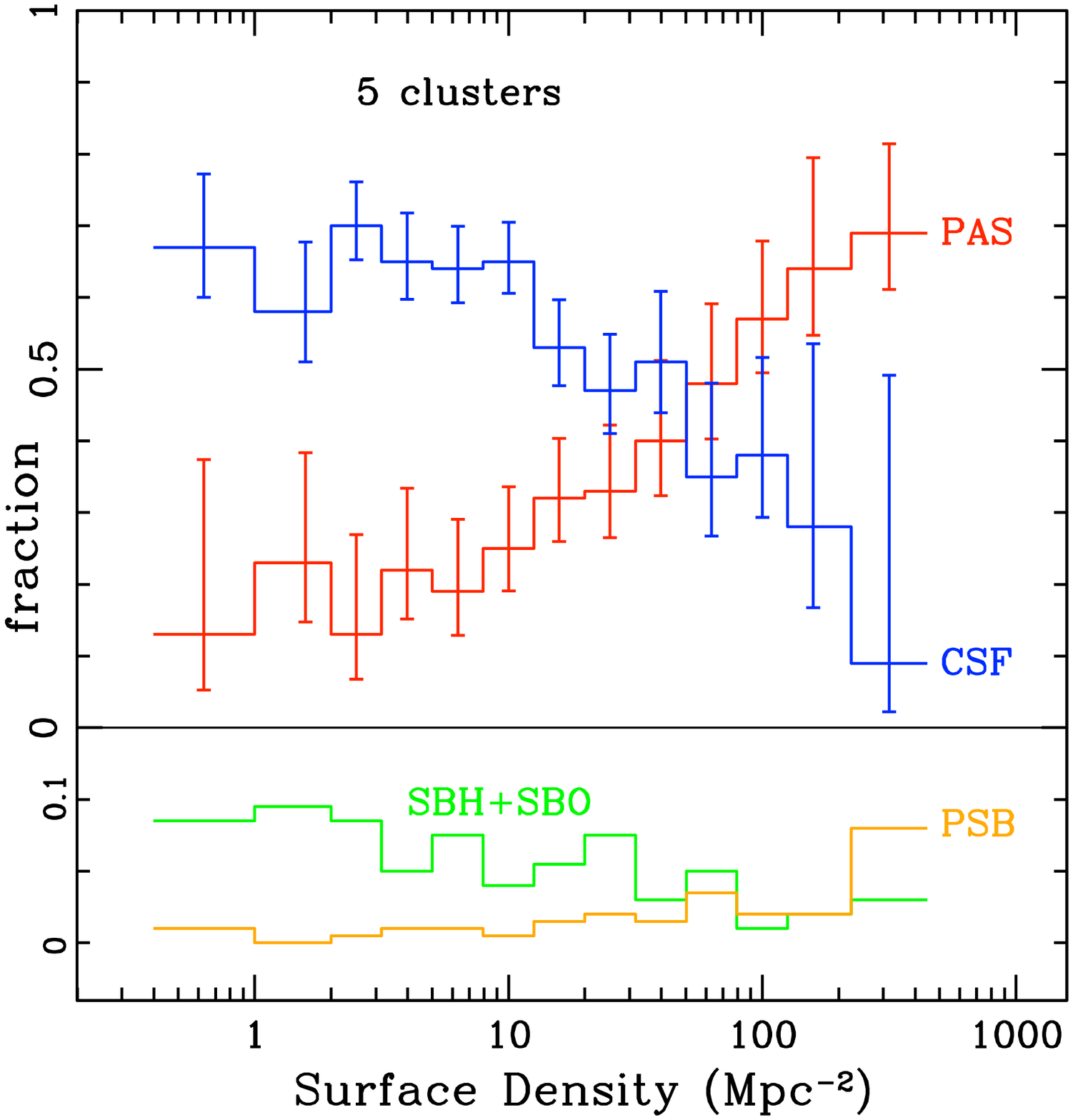}
}
\centerline{
\includegraphics[width=2.7in]{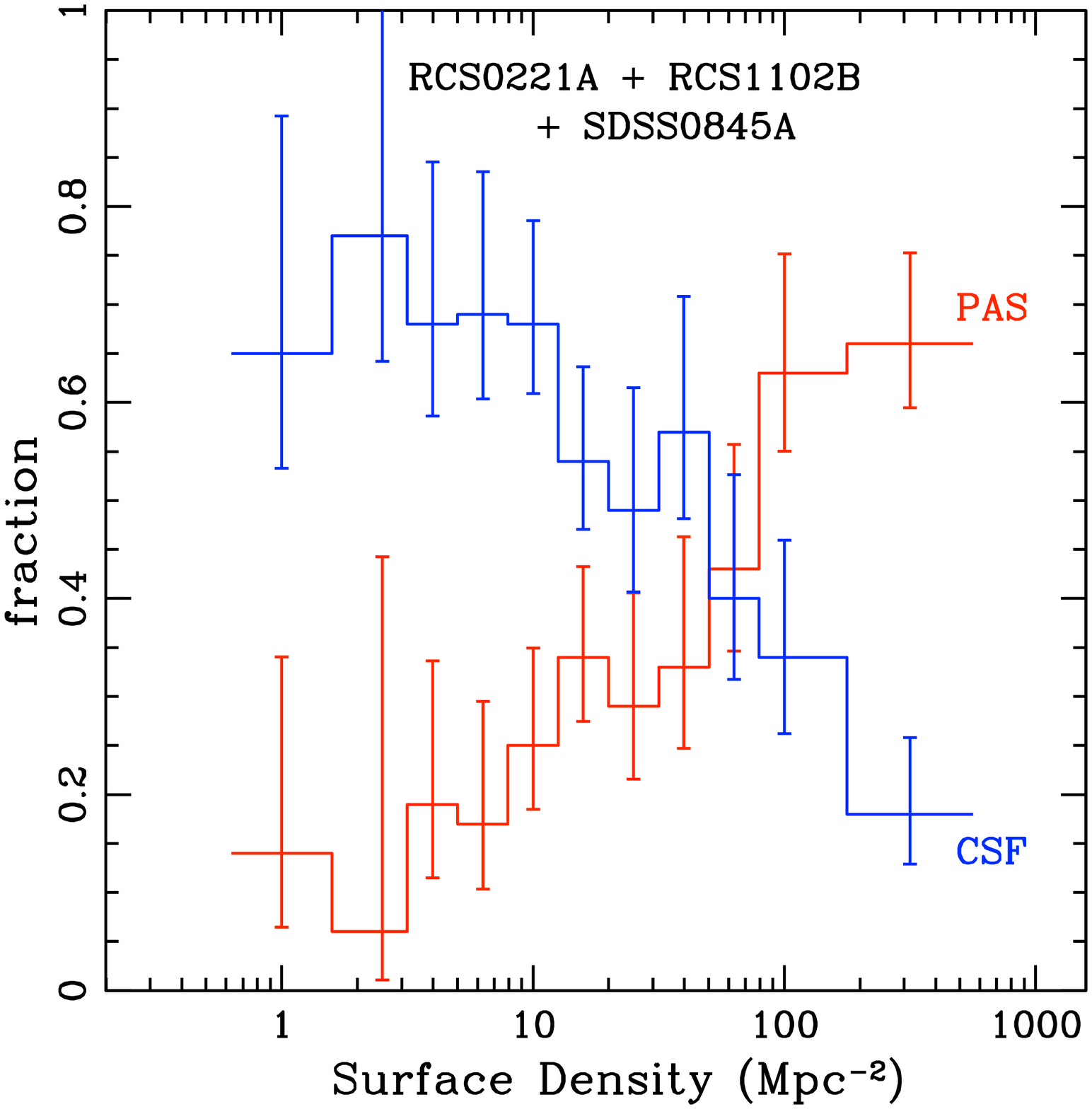}
\includegraphics[width=2.7in]{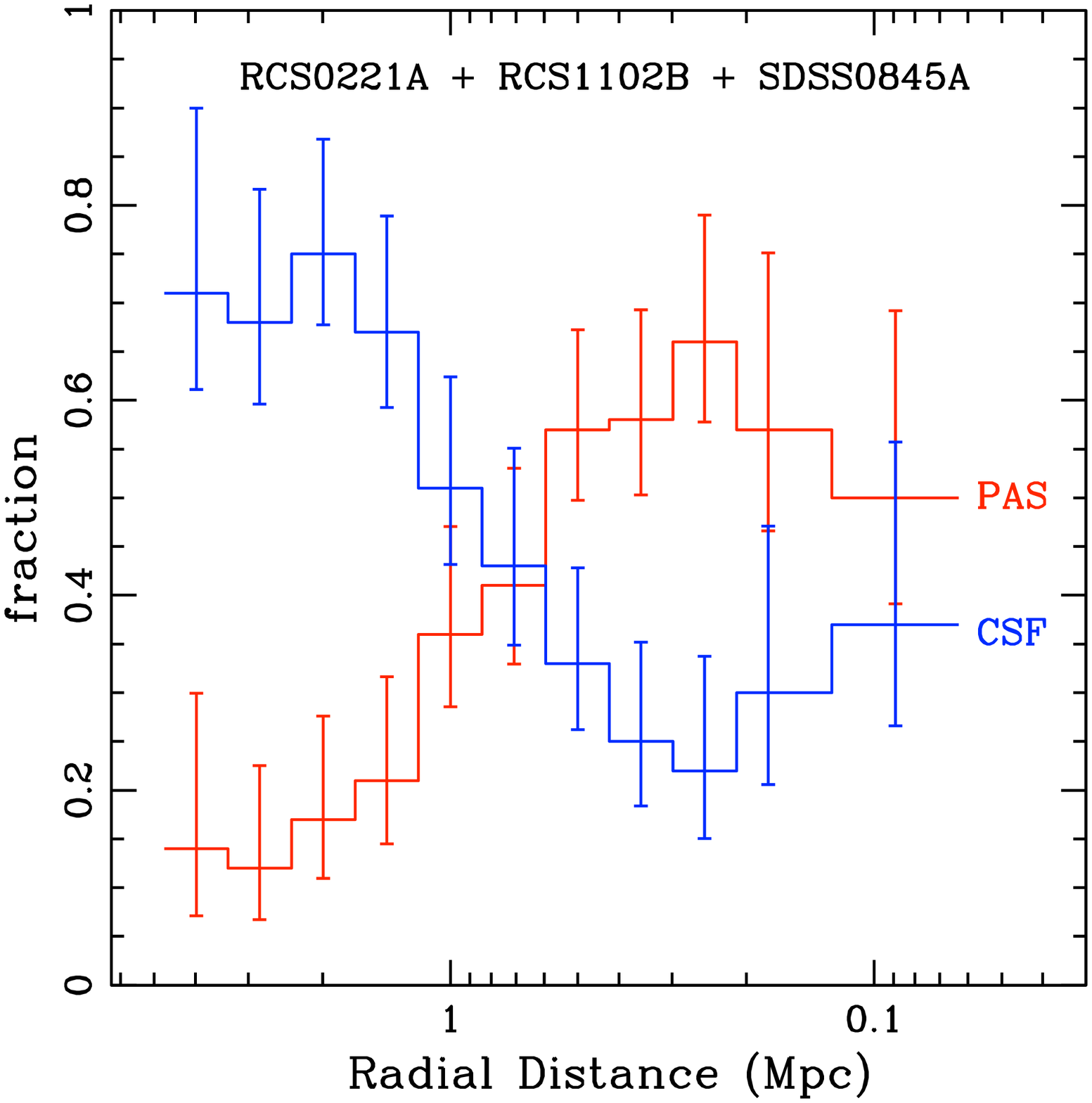}
}
\centerline{
%2clusters_sptype_surfden_fract
\includegraphics[width=2.7in]{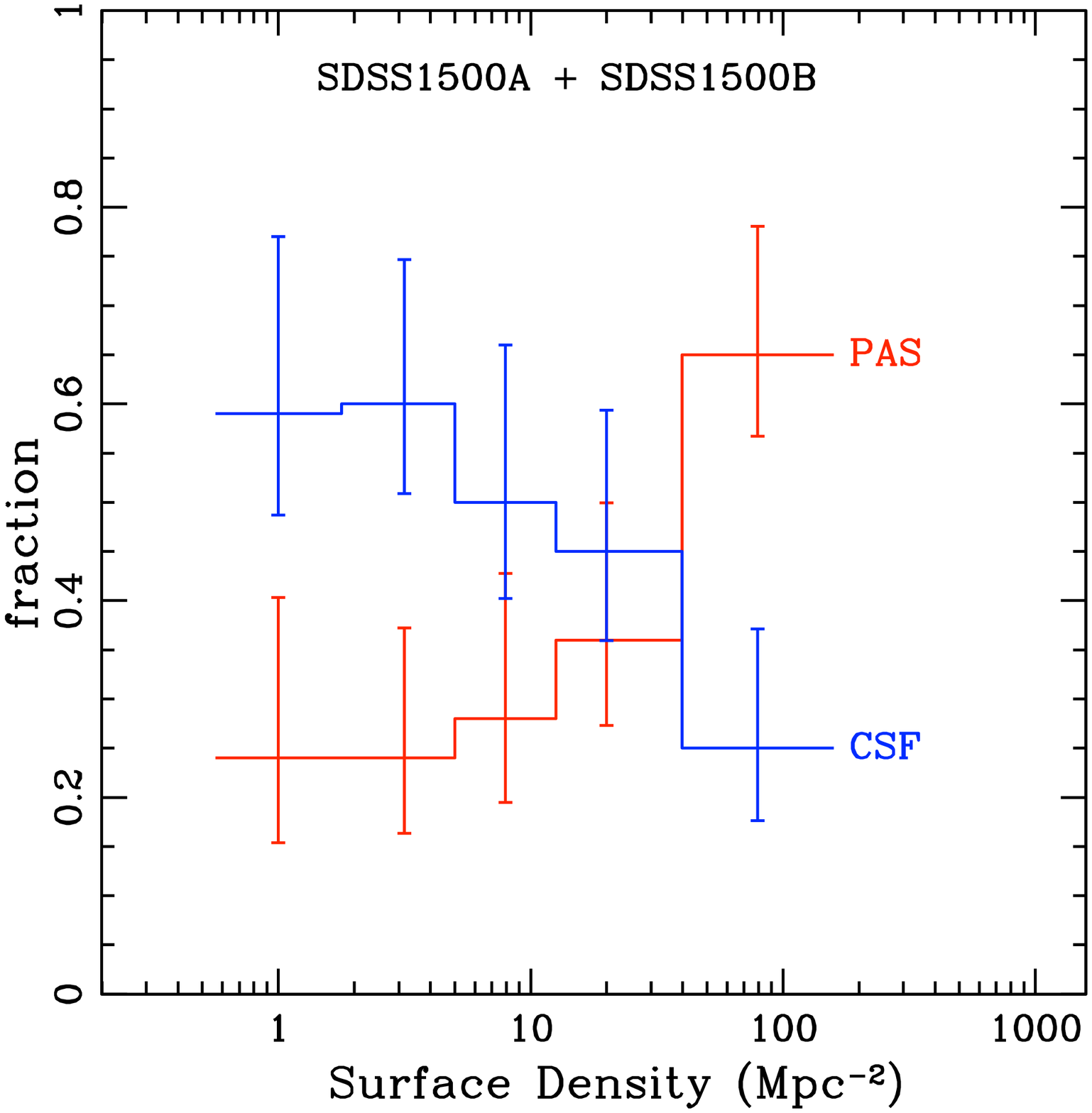}
%2clusters_sptype_log_rcl_fract
\includegraphics[width=2.7in]{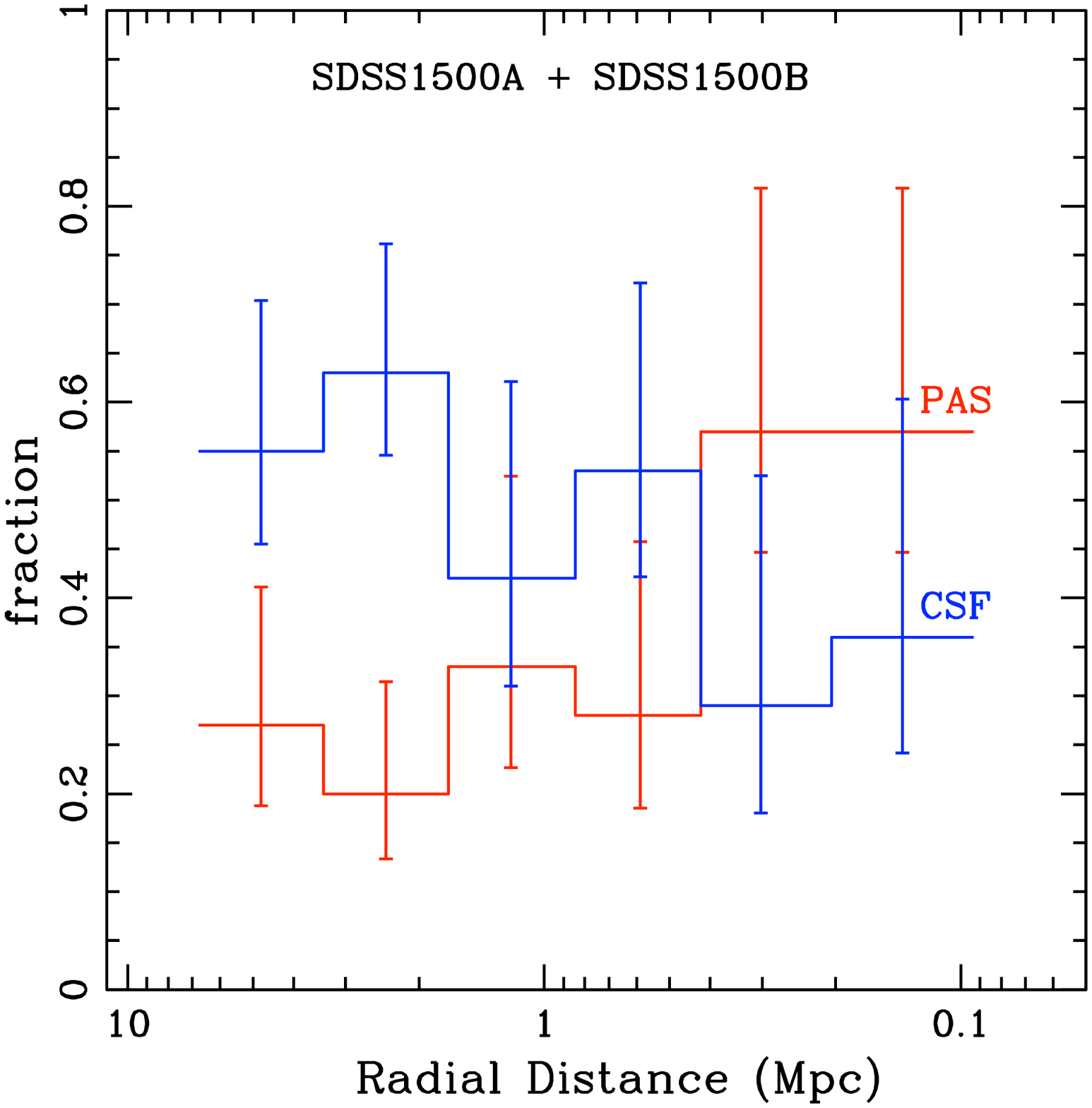}
}
\caption{a) Spectral-type fractions vs. surface density for the full 5-cluster sample.   The upper panel shows
the strong trends for PAS (passive) and CSF (continuously starforming) galaxies, which closely resemble
morphology-density relations (Dressler 1980; Dressler \etal\ 1997).  The bottom panel shows that the fraction of 
SBO + SBH starbursts declines in proportion with the CSF galaxies, while the PSB fraction rises in proportion to the 
PAS galaxies, a feature that suggests a pairing of PAS to PSB and (SBH+SBO) to CSF spectral types.  b) Spectral-type 
fractions relation for 3 concentrated, regular clusters. c) Spectral-type fractions vs. clustocentric radius for 3 concentrated 
clusters. d) Spectral-type fractions vs. surface density relation for 2 irregular clusters composed mainly of rich groups. 
e) Same as (d) for spectral-type fractions vs. clustocentric radius, showing a weaker relation for this compared to both 
(c) and (d).
\label{fig:sptype_density_radius}}

\end{figure*}

%Figure 6: Correlation functions

\begin{figure*}[t]

\centerline{
%RCS0221,RCS1102B,SDSS0845
\includegraphics[width=3.2in]{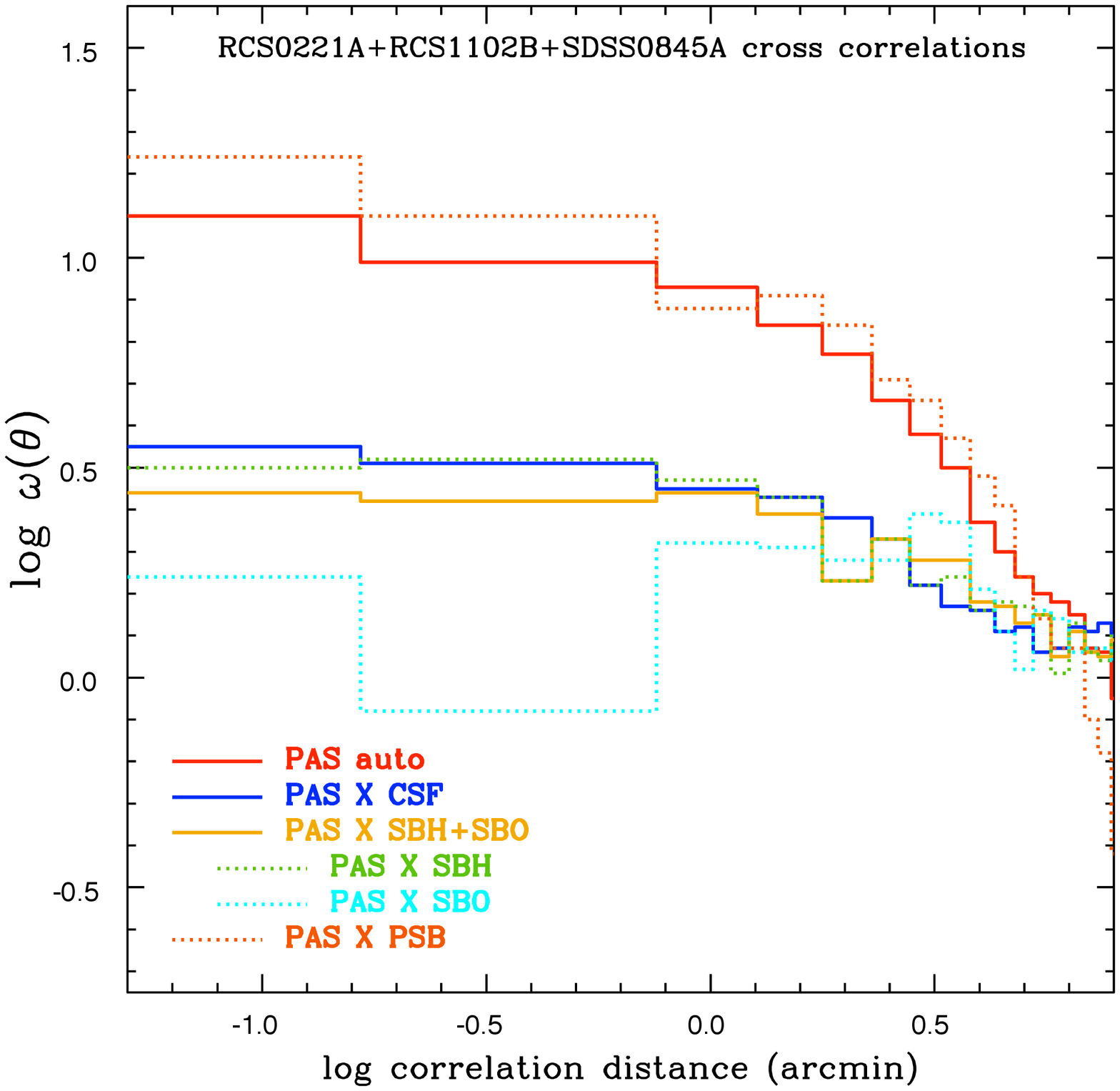}
%SDSS1500A & B
\hspace{0.2in}
\includegraphics[width=3.2in]{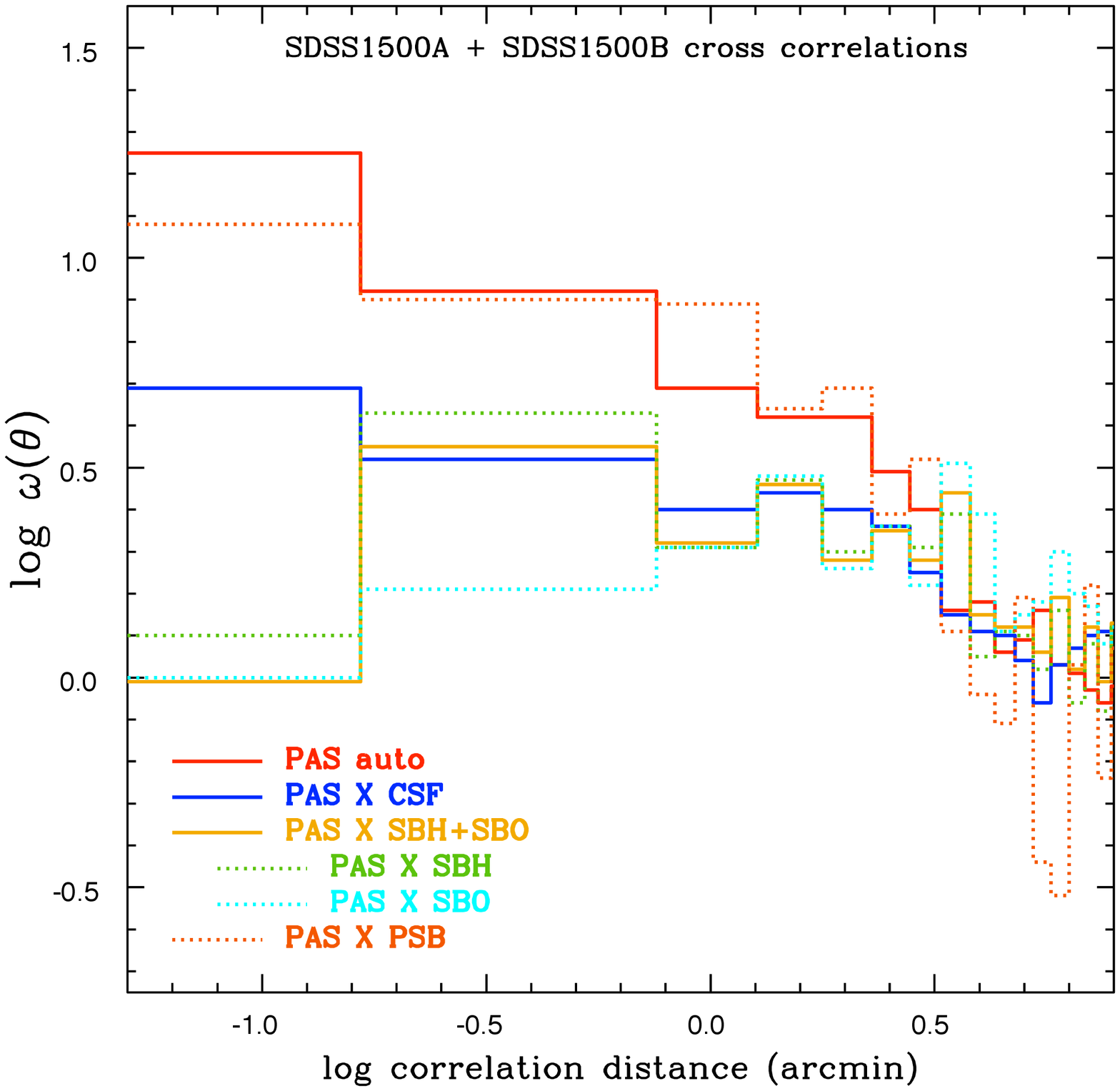}
}

\caption{Angular cross-correlation of different spectral types.  a) (left) The clustering strength of SBH + SBO starbursts
matches the clustering of continuously star forming systems in these three ``regular" clusters, RCS0221A, RCS1102B, and 
SDSS0845A.  Similarly, poststarburst (PSB) galaxies are as strongly clustered as the passive (PAS) members of the population, 
which could indicate that decaying starbursts are adding to the passive galaxies in the clusters.  b) (right) The same 
spatial distribution of PSB tracing PAS, and SBH+SBO tracing CSF as seen in (a), for two less concentrated, more
irregular clusters, SDSS1500A and SDSS1500B. 
\label{fig:cross_correlation}}

\end{figure*}

\noindent{uniform halo of galaxies typical of rich, present-epoch clusters like Abell\,1656 (Coma), Abell\,2199, A2670, and Abell\,2029.  
Indeed, there is very high degree of real substructure in the form of infalling groups in these fields, which we will 
investigate in \S3.}

We begin by considering the distributions of the two most populous spectral types, PAS and CSF.  The well known 
prevalence of passive galaxies in the high-density cluster cores is obvious, but there is clearly a tendency of PAS galaxies to 
follow the higher density structures in the surrounding supercluster.  Echoing the morphology/local-density relation, we 
expect PAS galaxies to be morphological types E \& S0 galaxies (Dressler \etal\ 1999; Postman \etal\ 2005).  Similar 
to the distribution of morphological spiral galaxies in the modest number of intermediate-redshift clusters that have 
been imaged with large mosaics of HST images, the spectral-type CSF galaxies dominate the lower density environs of the
supercluster, avoiding for the most part the dense cores entirely.  

We derived a spectral-type/surface-density relation for these data that quantifies the expected association of spectral
type and morphology.  For this exercise, we defined local-surface-density following Dressler (1980), this time using the 
spectroscopic sample of cluster members (requiring no correction for field galaxies), and calculating a surface density for 
each galaxy as 6 divided by the rectangular area in Mpc$^2$ staked-out by the 5 nearest neighbor galaxies. The different 
clusters were adjusted for the different depths (due to the redshift range) using a Schechter function and normalizing 
to SDSS0845A. 

We show results for the sptype/local density relation, and for the sptype/radial-distance relation, in 
Figure \ref{fig:sptype_density_radius}.  The qualitative, even quantitive resemblance of the top diagram 
with the morphology-density relation reinforces the idea that these spectral types correspond well with the galaxy 
morphologies found for such spectra at the present epoch.  Figure \ref{fig:sptype_density_radius} also shows that 
the spectra-type/radial-distance relation is quite similar to the spectral-type /density relation for the three 
more-or-less regular clusters RCS0221A, RCS1102B, and SDSS0845A. However, the sptype/density relation 
appears to be the stronger for the two more irregular clusters, SDSS1500A \& B, where the spectral-type/radial-distance 
relation is absent outside the two points representing the cluster cores.  This 
issue is relevant for our discussion of the descent of passive galaxies from starforming galaxies, because a preference 
for the density rather than radius as the independent parameter results suggests that the processes at work are more 
likely related to local density (over the lifetime of galaxies), rather than processes that are ``global'' properties of the 
cluster itself, for example, the tidal field or hot intercluster gas.  We will return to this long-standing 
question in \S4.

Returning to Figure \ref{fig:sptype_maps} to consider the distribution of spectral types associated with starbursts, we see that 
the PSB galaxies are in fact distributed like the PAS galaxies, that is, favoring dense cluster cores or density 
enhancements of groups, for example, the lower right corner of SDSS1500B.  However, the active starbursts, both SBH and 
SBO, are spread throughout the metaclusters with no obvious affinity for denser regions, indeed, they seem to share the 
distribution of the CSF galaxies, as if they derive from the same population.  This effect also shows up in the 
spectral-type/surface-density relation of Figure \ref{fig:sptype_density_radius} (bottom), where the active starburst fraction 
rises and the poststarburst fraction falls with increasing galaxy local-surface-density.

We quantify these effects further in Figure \ref{fig:cross_correlation}, where we show the angular auto-correlation functions of PAS 
galaxies and cross-correlation functions of the other spectral types with the PAS galaxies.  Again we have divided the sample between 
(a) the three more regular clusters and (b) the two less regular ones.  The PAS autocorrelation function is strongest, reflecting their 
concentration to smaller, dense regions, as expected.  Less expected, perhaps, is the strength of the PSB cross-correlation: 
these galaxies are as strongly clustered as the PAS galaxies, which means that the PSB must in fact share the spatial distribution to 
high fidelity.  This correspondence of PAS and PSB distributions is clearly seen in both the regular and less-regular clusters -- in 
Figures \ref{fig:cross_correlation}-a and \ref{fig:cross_correlation}-b.  

When we compare the spatial distribution of CSF and active starbursts, SBH+SBO, hereafter SB galaxies, our visual 
impression from Figure \ref{fig:sptype_maps} is confirmed: the close correspondence of the cross-correlation functions 
for the active starburst galaxies with the CSF strongly suggest that the former are a ``random'' draw from the latter, 
in other words, any of the CSF galaxies appear to be candidates for a starburst.

The remarkable way in which the PSB spatial distribution traces the PAS distribution, and the SB spatial distribution
traces the CSF distribution, while the PAS and CSF spatial distributions are so different, suggest that there is more than
a casual connection between PSB/PAS and SB/CSF spectral types.  In \S4 we provide other evidence that 
the PSB-to-PAS and SB-to-CSF connections appear to hold across a wide range of environment, a clue that the 
conventional evolutionary path 

\vspace{0.03in}
\hspace{0.7in}CSF $\Rightarrow$ SB $\Rightarrow$ PSB $\Rightarrow$ PAS
\vspace{0.03in}

\noindent{is not a complete description of the relationships between these spectral types.}

\section{The Structural Evolution of the ICBS Clusters}

N-body simulations of structure formation through hierarchical clustering conventionally locate the building of rich 
clusters at the intersection of dark-matter-and-galaxy filaments and sheets, along which galaxies are channeled 
into regions of very high density where virialization occurs.  One of the motivations of the ICBS was to search for 
evidence of this effect, which the simulations indicate is strong at intermediate redshift.  

\subsection{The identification of infalling groups}

It is clear from inspecting the maps in Figure \ref{fig:sptype_maps} that filamentary structures are not sufficiently obvious in 
projection to allow a simple spatial selection of structures that might be contributing in the building of the ICBS clusters.  
Because of this, we chose to use the Dressler \& Shectman (1988) subclustering test (DS-test) as a tool for identifying 
structures that are kinematically distinct from the high-velocity dispersion environment that characterizes the cluster as 
a whole.  As used here, the test identifies dynamically cold structures by finding the 10 nearest galaxies (in the 
spectroscopic sample) and comparing the velocity dispersion and systemic velocity for each such subset to the velocity 
dispersion and systemic velocity for the metacluster as a whole.  A ``sum-of-squares'' deviation $\delta$ is calculated for each 
galaxy; these values do not identify groups uniquely, but the point clearly to regions where physical groups can be found.  

In the Dressler-Shectman study, the test was used only to demonstrate the statistical significance of subclustering.  A 
`$\Delta$' parameter was defined as the root-mean-square of the individual $\delta$ values, and this was compared 
to the results of a large number of simulated clusters made by randomly shuffling the velocities between galaxies, in 
order to estimate the significance of that $\Delta$ value for that particular sample of cluster galaxies.

Because the DS-test does not find groups per se, and the individual $\delta$ deviations are not at all independent,  
the DS-test is by itself insufficient for the purpose here.  However, we found that, by calculating and plotting the $\delta$
deviations for each galaxy in the field, the test very reliably found genuine physical groupings of galaxies. It was then 
straightforward to investigate galaxy-by-galaxy whether discrete groups --- based on association of their redshifts --- could 
be isolated. In practice, this turned out to be surprisingly easy to accomplish.

\begin{figure*}[t]
\centerline{
\hspace{0.0in}
%RCS0221A_deltamap
\includegraphics[width=2.47in]{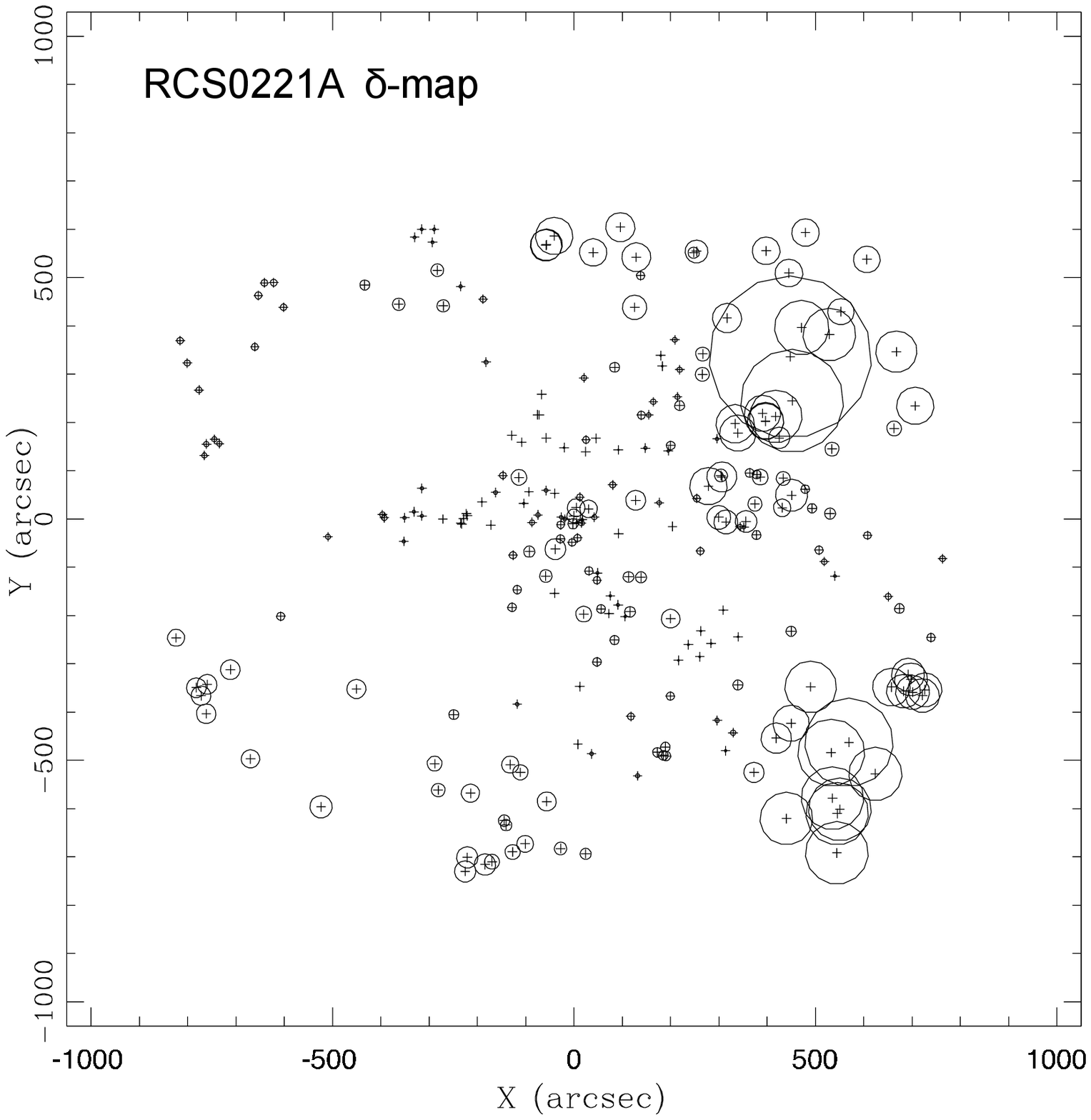}
\qquad
\hspace{0.3in}
%RCS1102B_deltamap
\includegraphics[width=2.53in, angle=90]{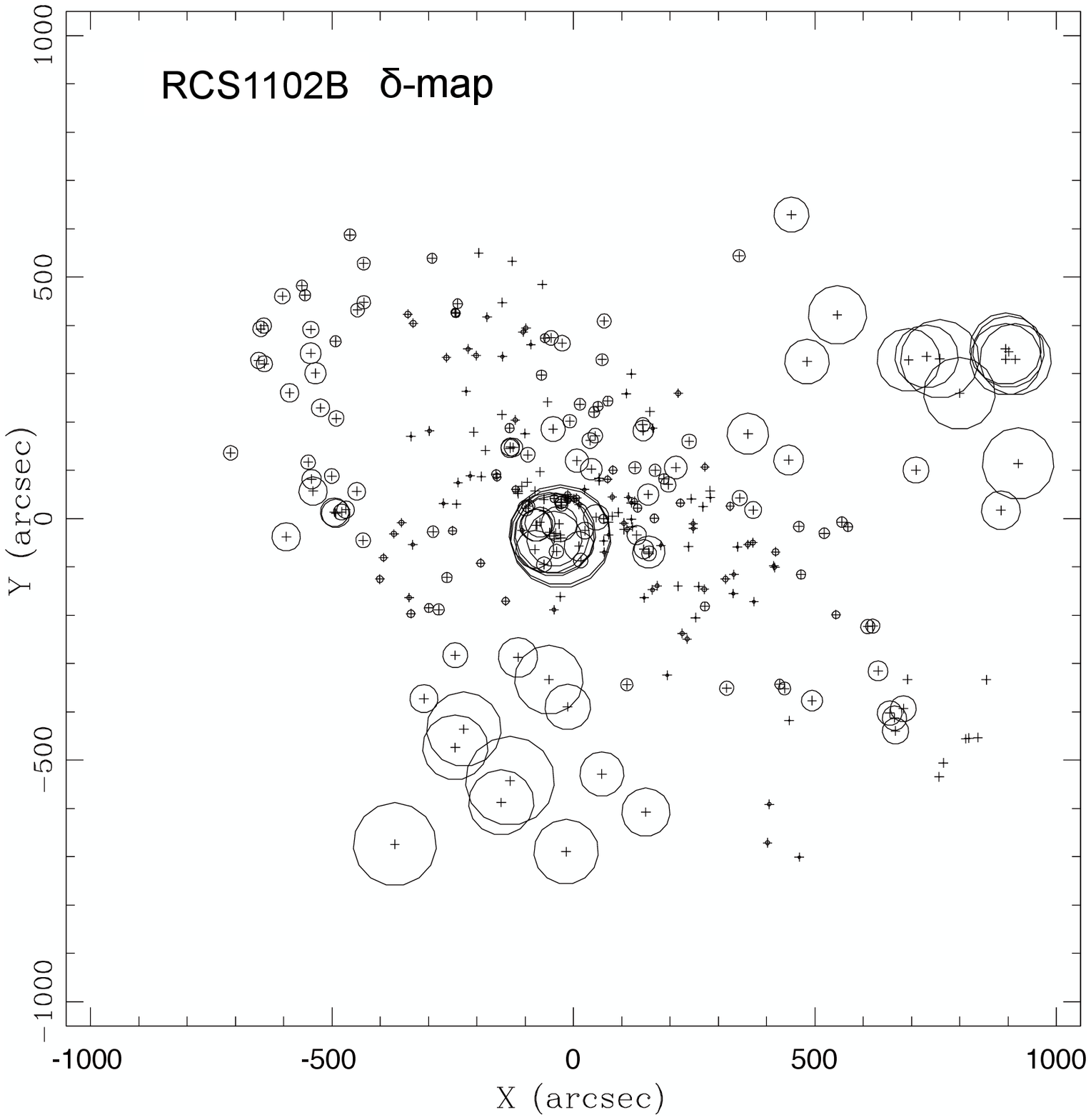}
}

\vspace{0.10in}
\centerline{
\hspace{0.35in}
%RCS0221A_groups_map
\includegraphics[width=2.4in, angle=90]{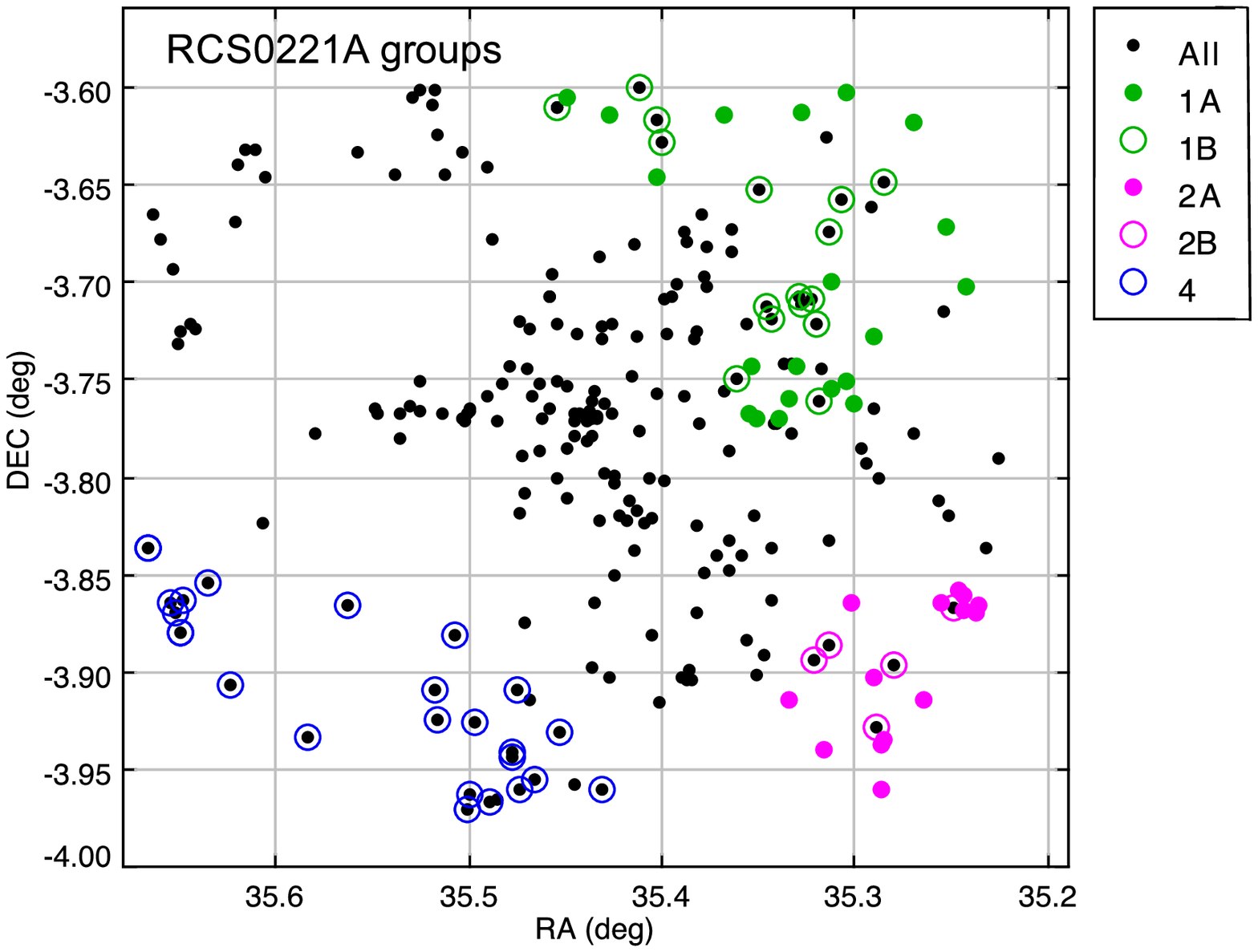}
\qquad
\hspace{-0.30in}
%RCS1102B_groups_map
\includegraphics[width=2.5in, angle=90]{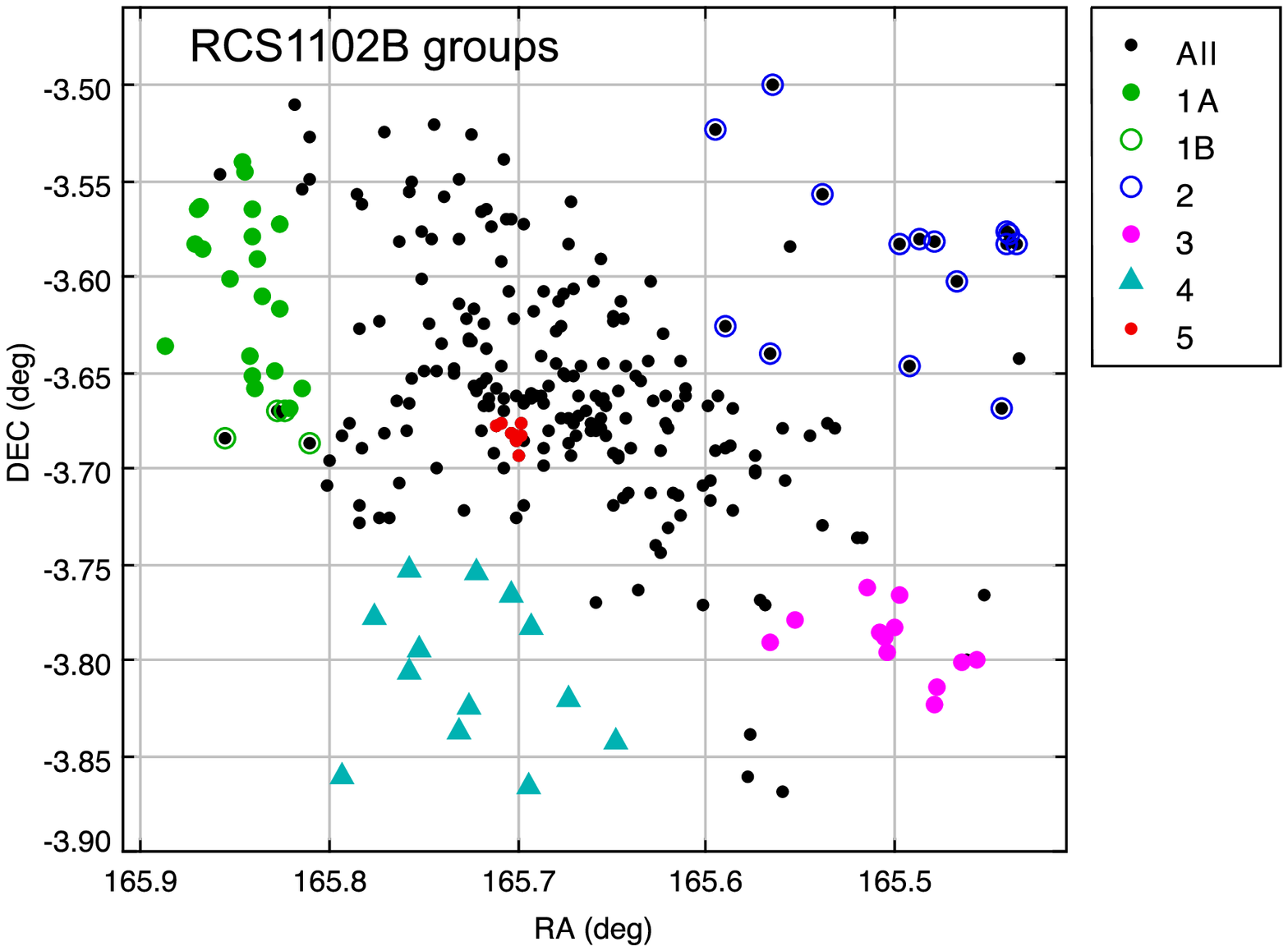}
}

\vspace{0.10in}
\centerline{
\hspace{0.15in}
%RCS0221A_groups_velhist
\includegraphics[width=4.0in, angle=90]{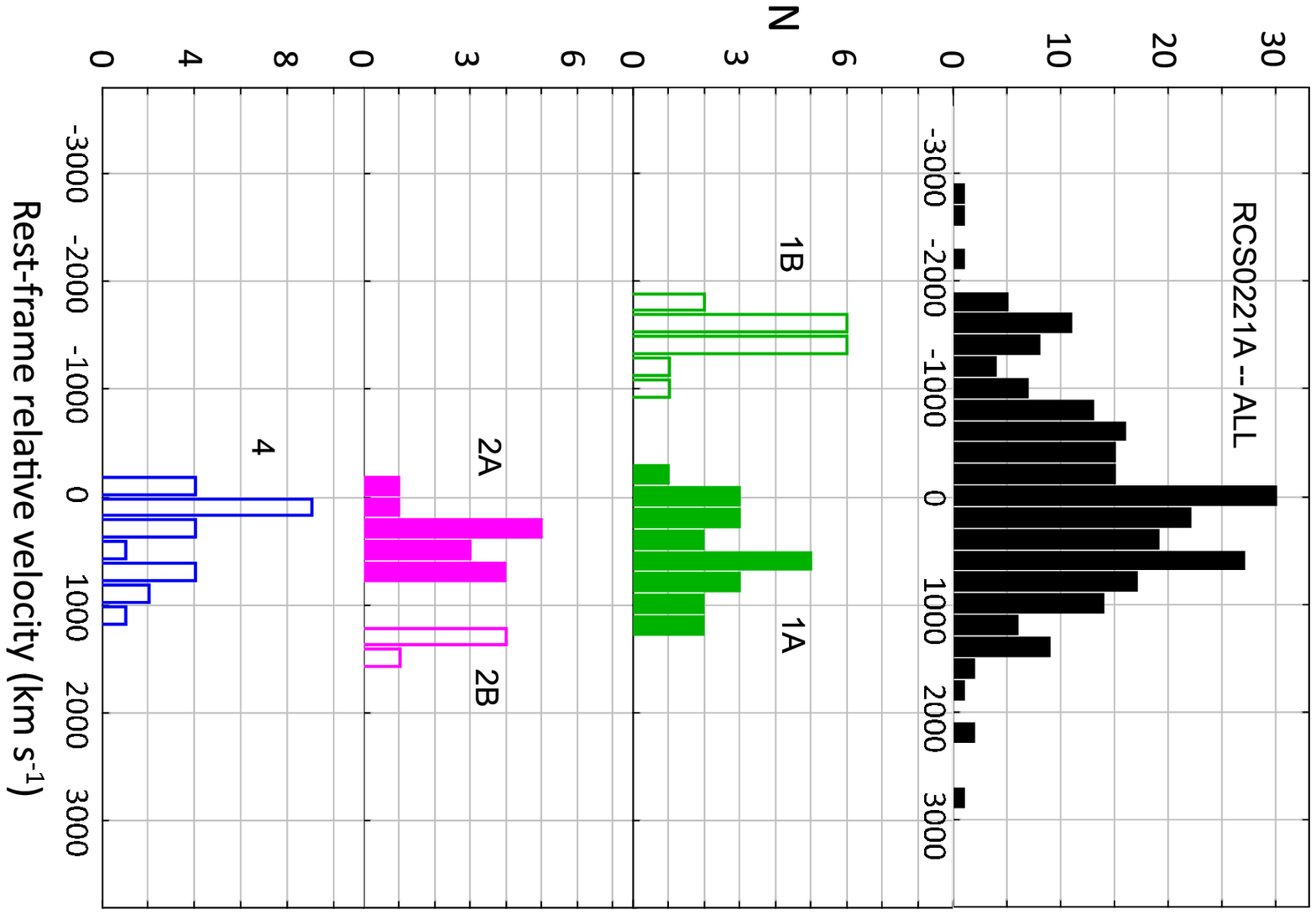}
\qquad
\hspace{0.20in}
%RCS1102B_groups_velhist
\includegraphics[width=4.0in, angle=90]{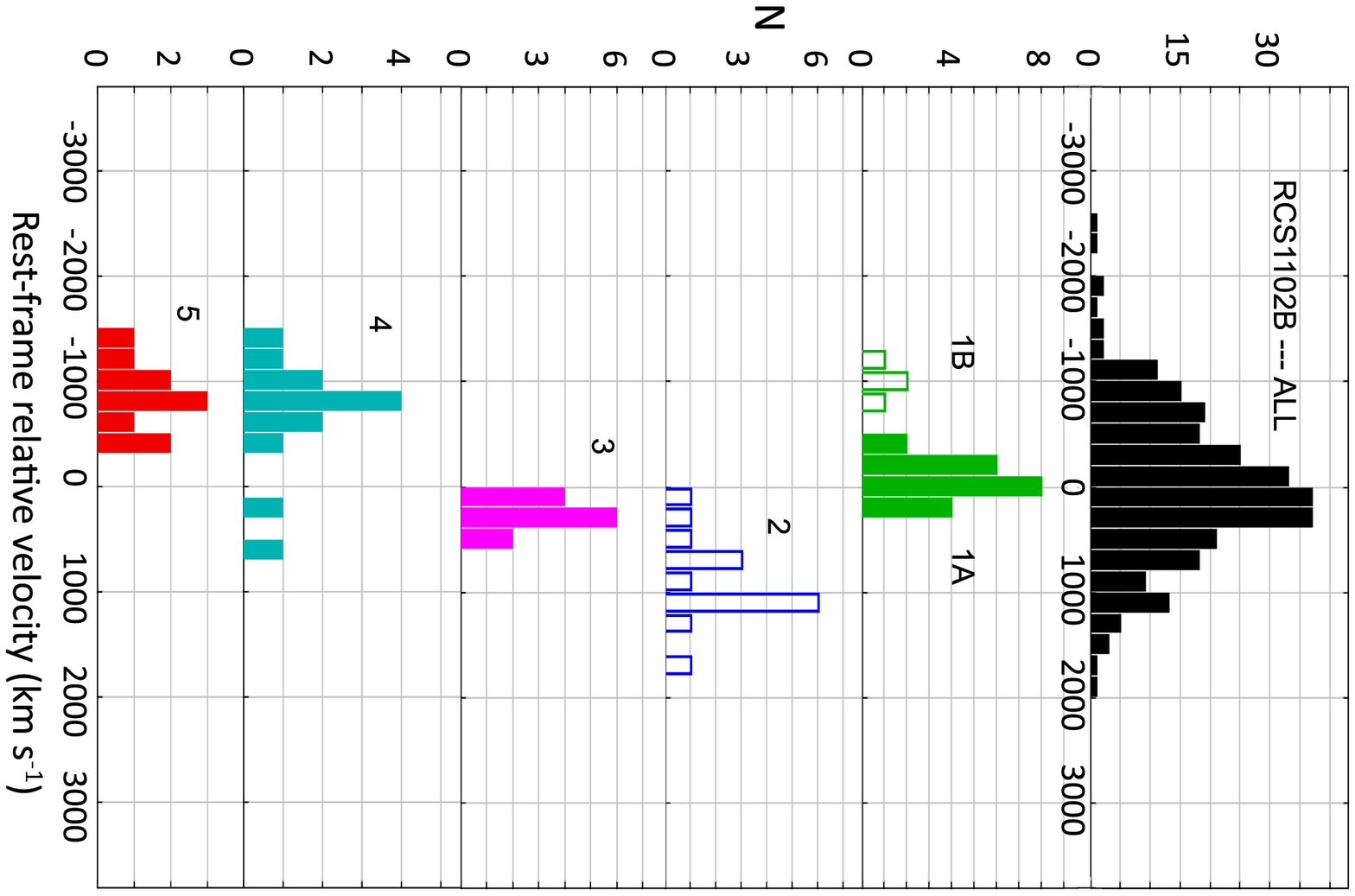}
\qquad
}

\caption{Groups in the RCS0221A and RCS1102B clusters. (left)  RCS0221A: (a) ``delta plot'' (top), 
(b) map (middle), (c) velocity histograms.
(right) RCS1102B: (d) ``delta plot'' (top), (e) map (middle), (f) member velocity histograms (bottom). 
\label{fig:RCS0221A_RCS1102B_groups}}

\end{figure*}

\begin{figure*}[t]

\vspace{0.3in}
\centerline{
%SDSS0845A_deltamap
\includegraphics[width=3.0in]{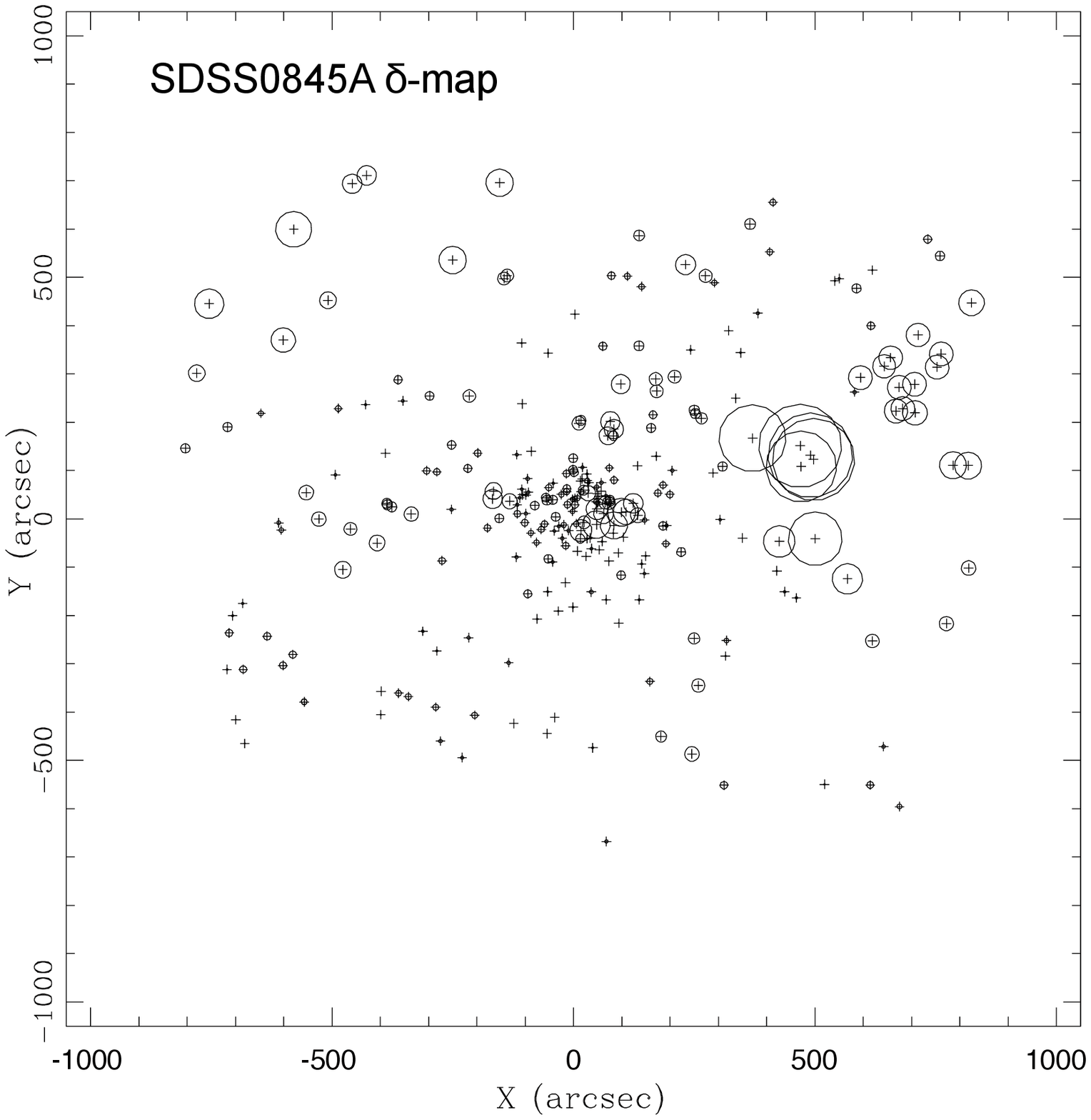}
\qquad
%SDSS0845A_groups_velhist
\includegraphics[width=3.15in, angle=90]{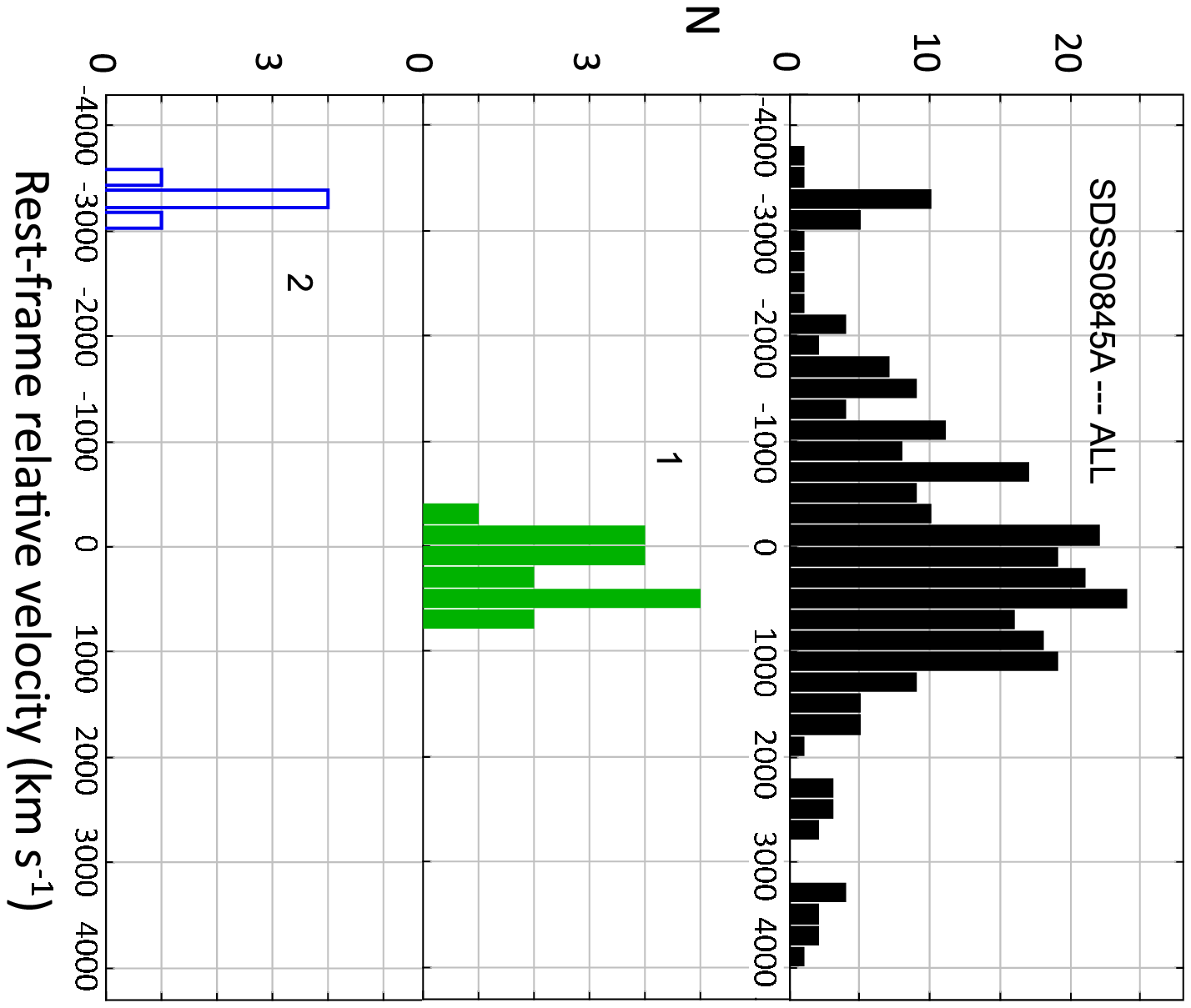}
}

\vspace{0.1in}
\centerline{
%SDSS0845A_groups_map
\includegraphics[width=3.4in, angle=90]{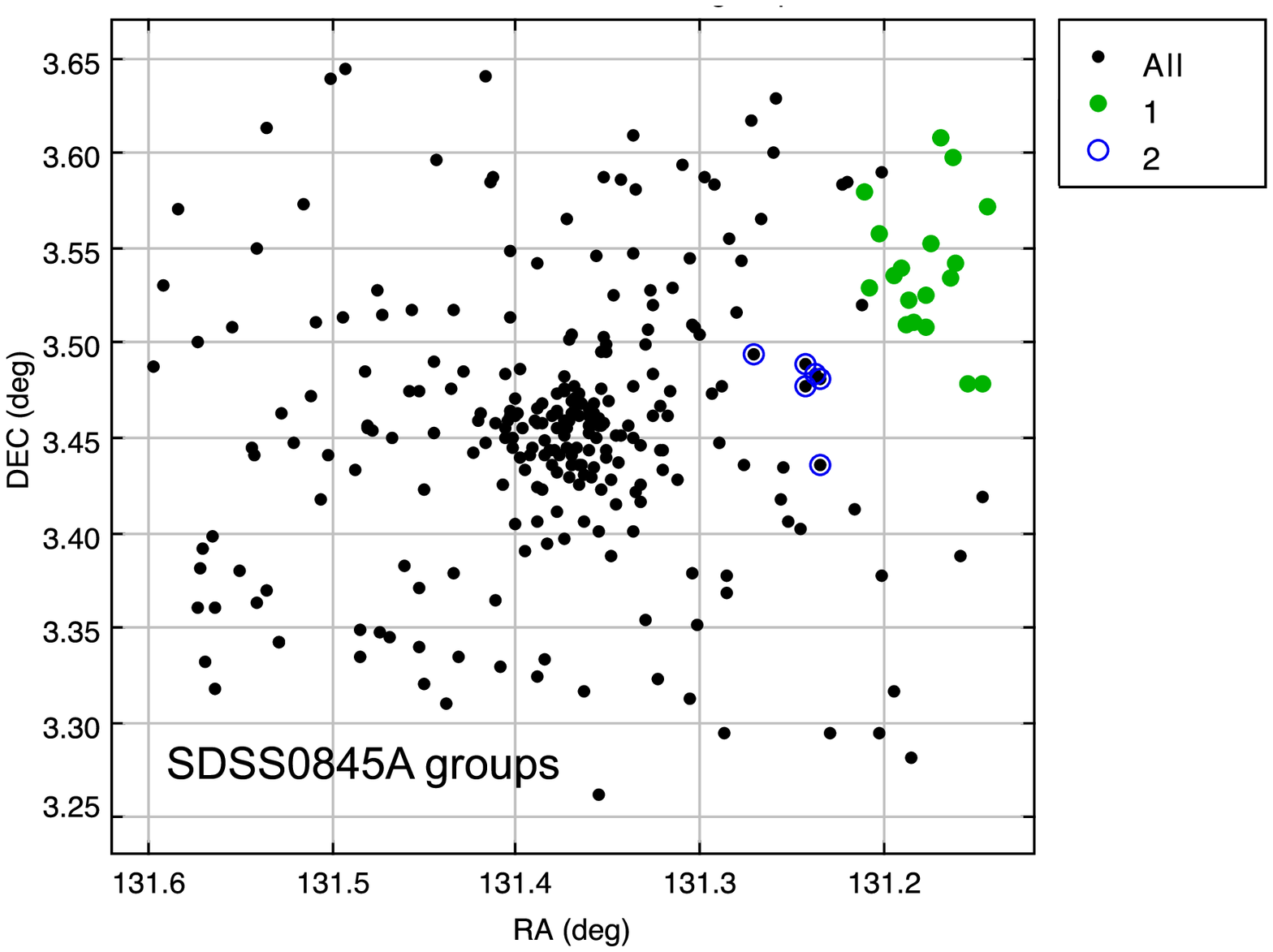}
}
\caption{Groups in the SDSS0845A cluster. (a) ``delta plot'' (top-left), (b) map (bottom), (c) member velocity histograms (top-right)
\label{fig:SDSS0845A_groups}}

\end{figure*}

\begin{figure*}[t]
\centerline{
\hspace{-0.75in}
%SDSS1500A_deltamap
\includegraphics[width=2.50in]{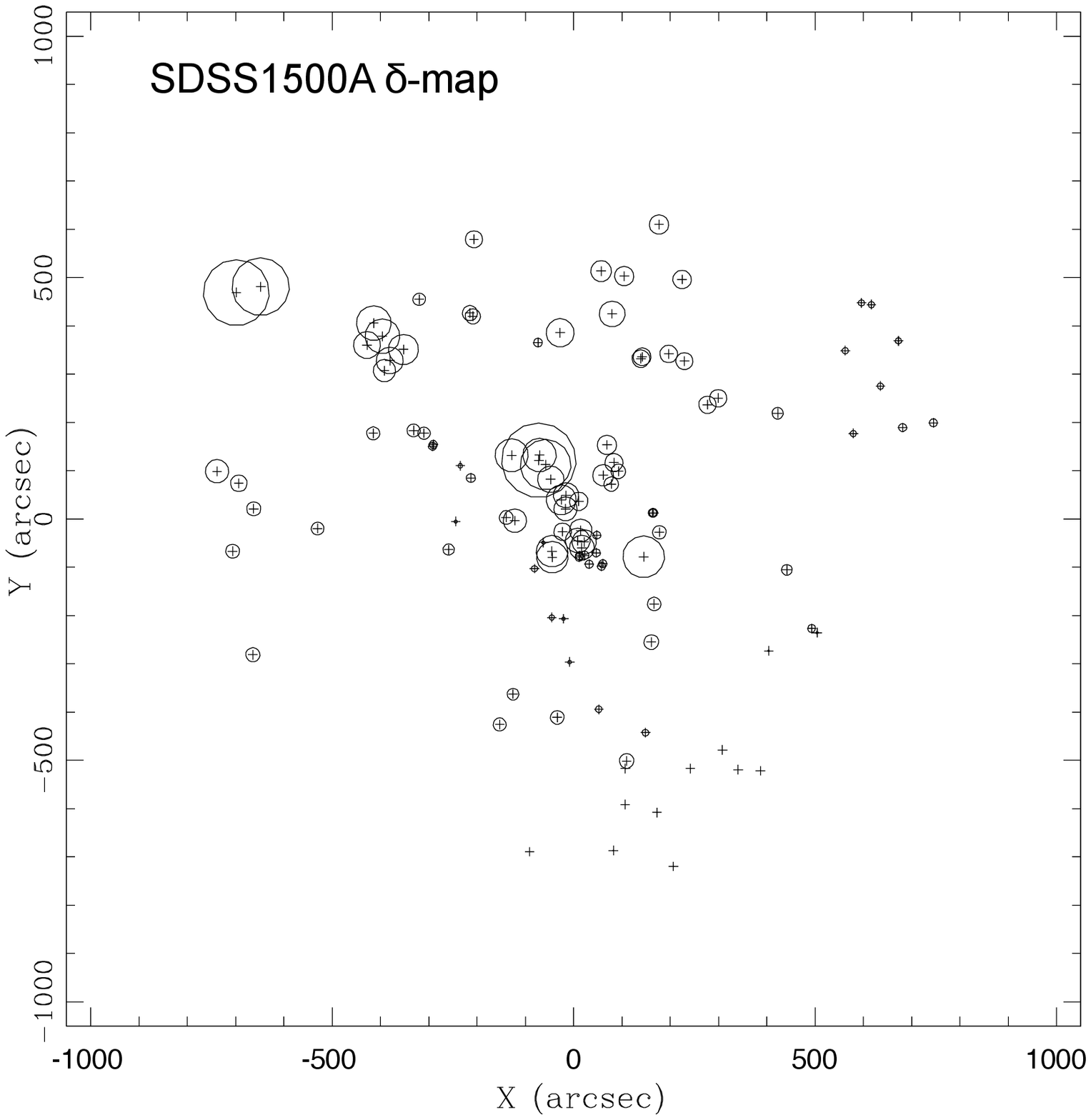}
\hspace{0.70in}
%SDSS1500B_deltamap
\includegraphics[width=2.50in]{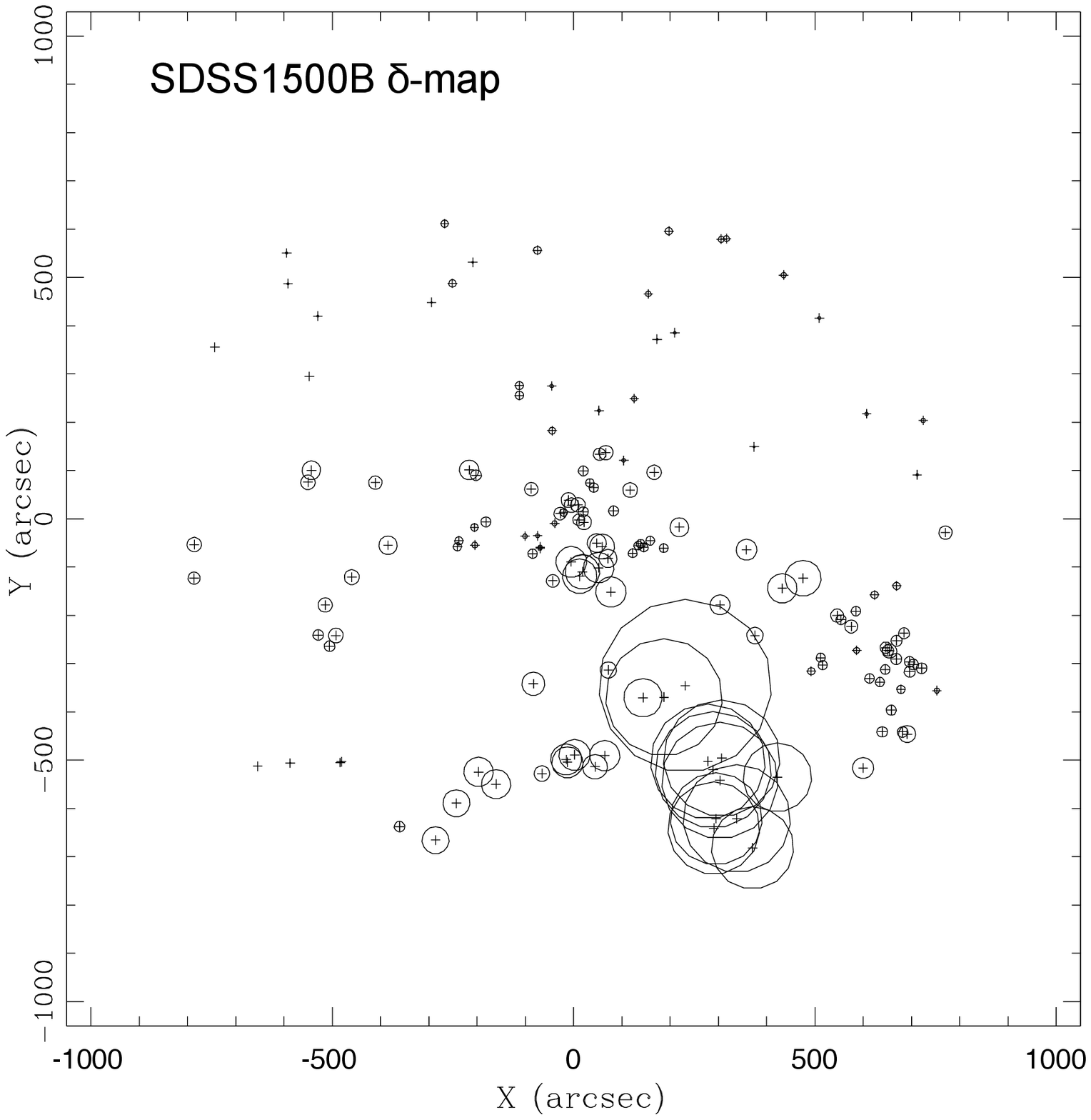}
}

\vspace{0.05in}
\centerline{
\hspace{-0.25in}
%SDSS1500A_groups_map
\includegraphics[width=2.5in, angle=90]{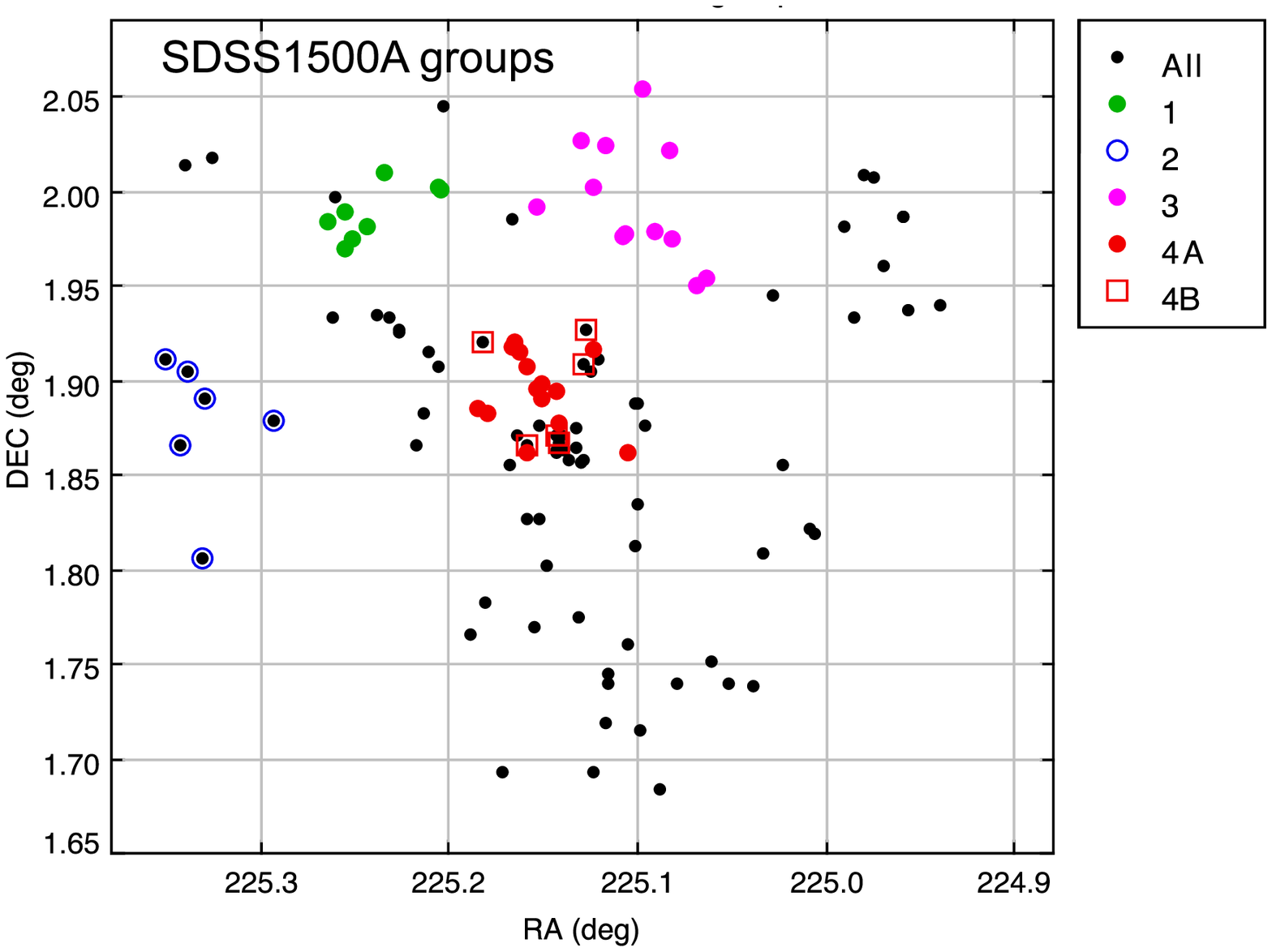}
\qquad
\hspace{-0.45in}
%SDSS1500B_groups_map
\includegraphics[width=2.5in, angle=90]{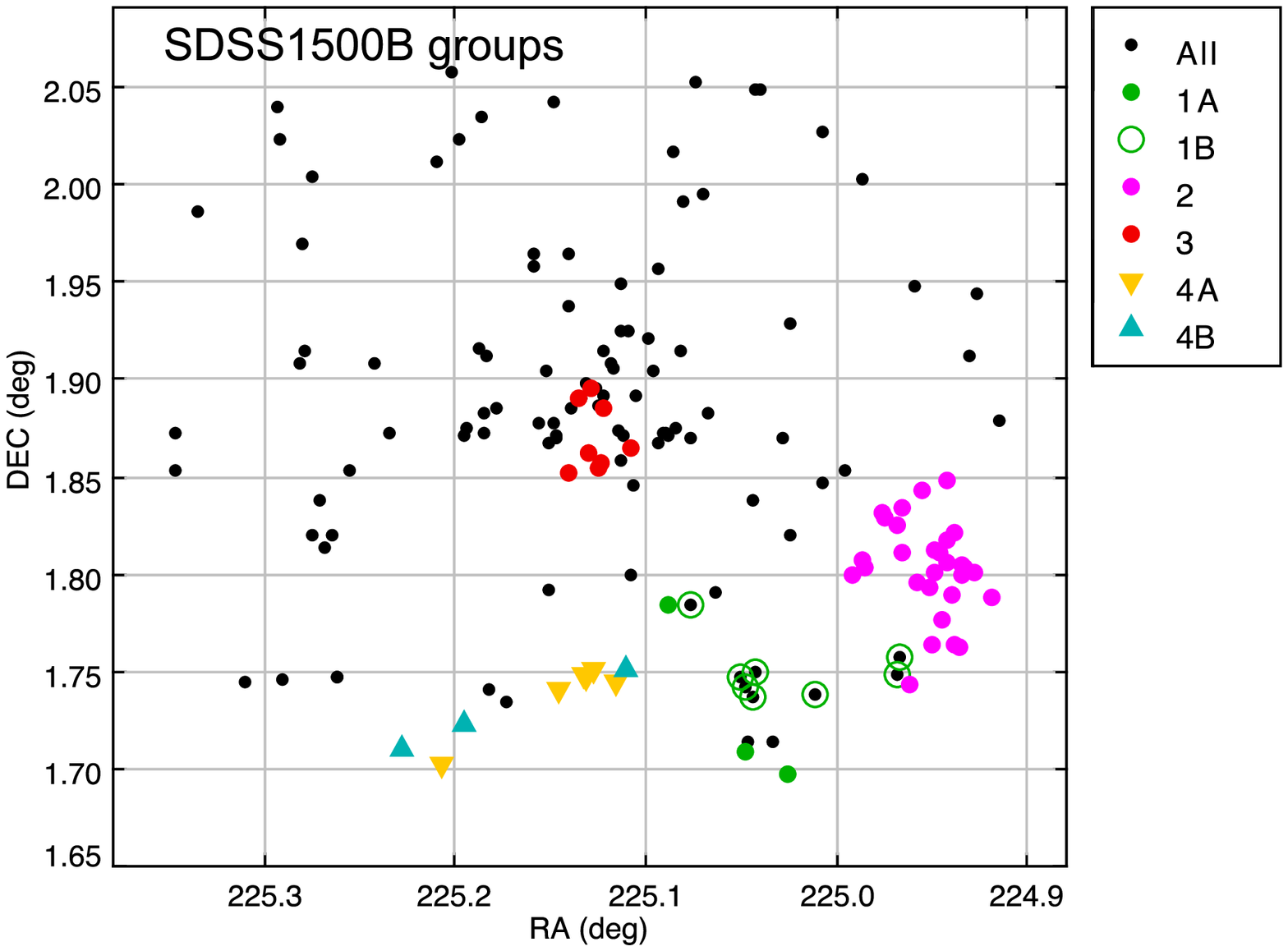}
}

\vspace{0.05in}
\centerline{
\hspace{-0.60in}
%SDSS1500A_groups_velhist
\includegraphics[width=4.0in, angle=90]{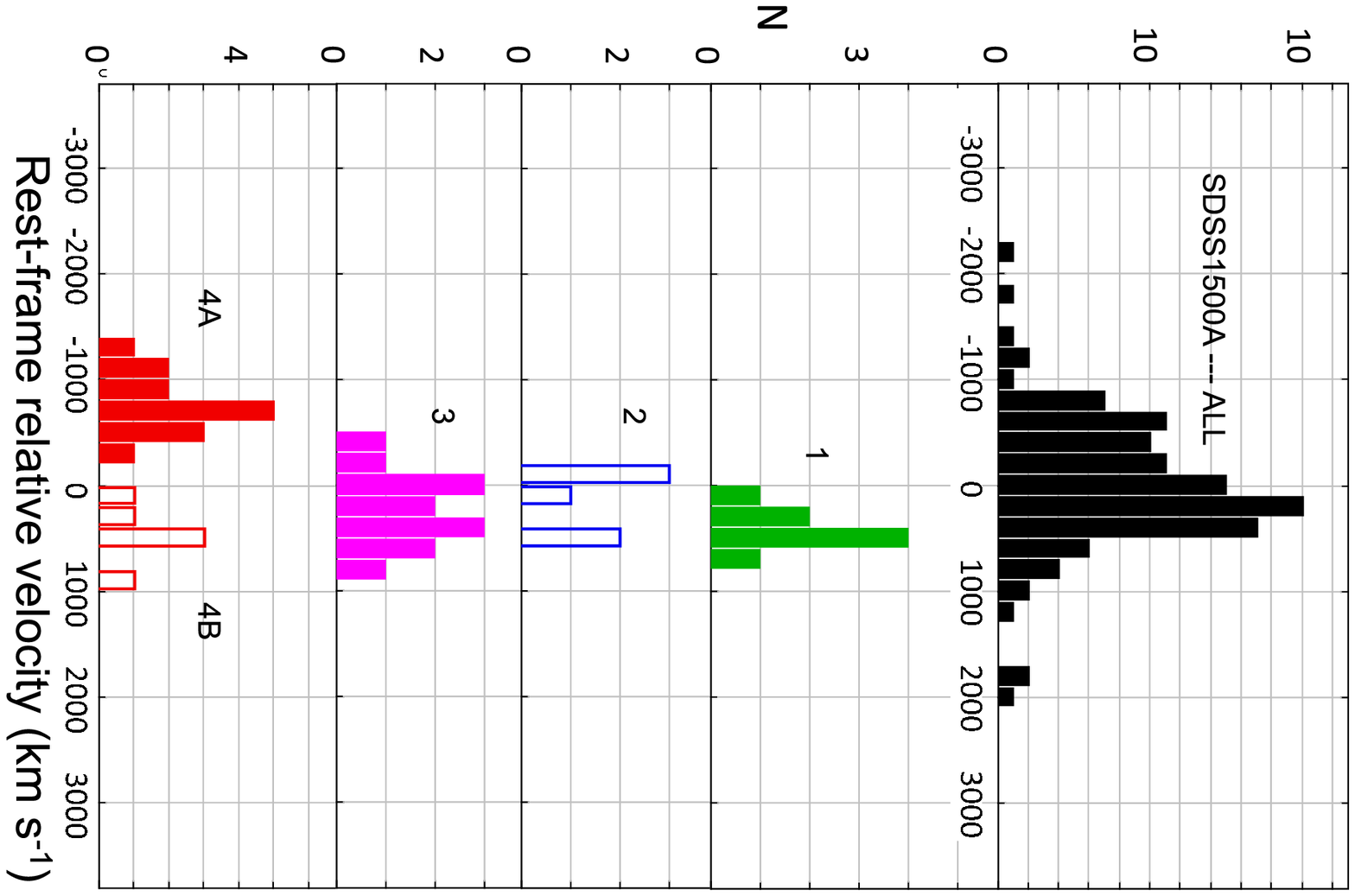}
\hspace{0.50in}
\vspace{0.50in}
%SDSS1500B_groups_velhist
\includegraphics[width=4.0in, angle=90]{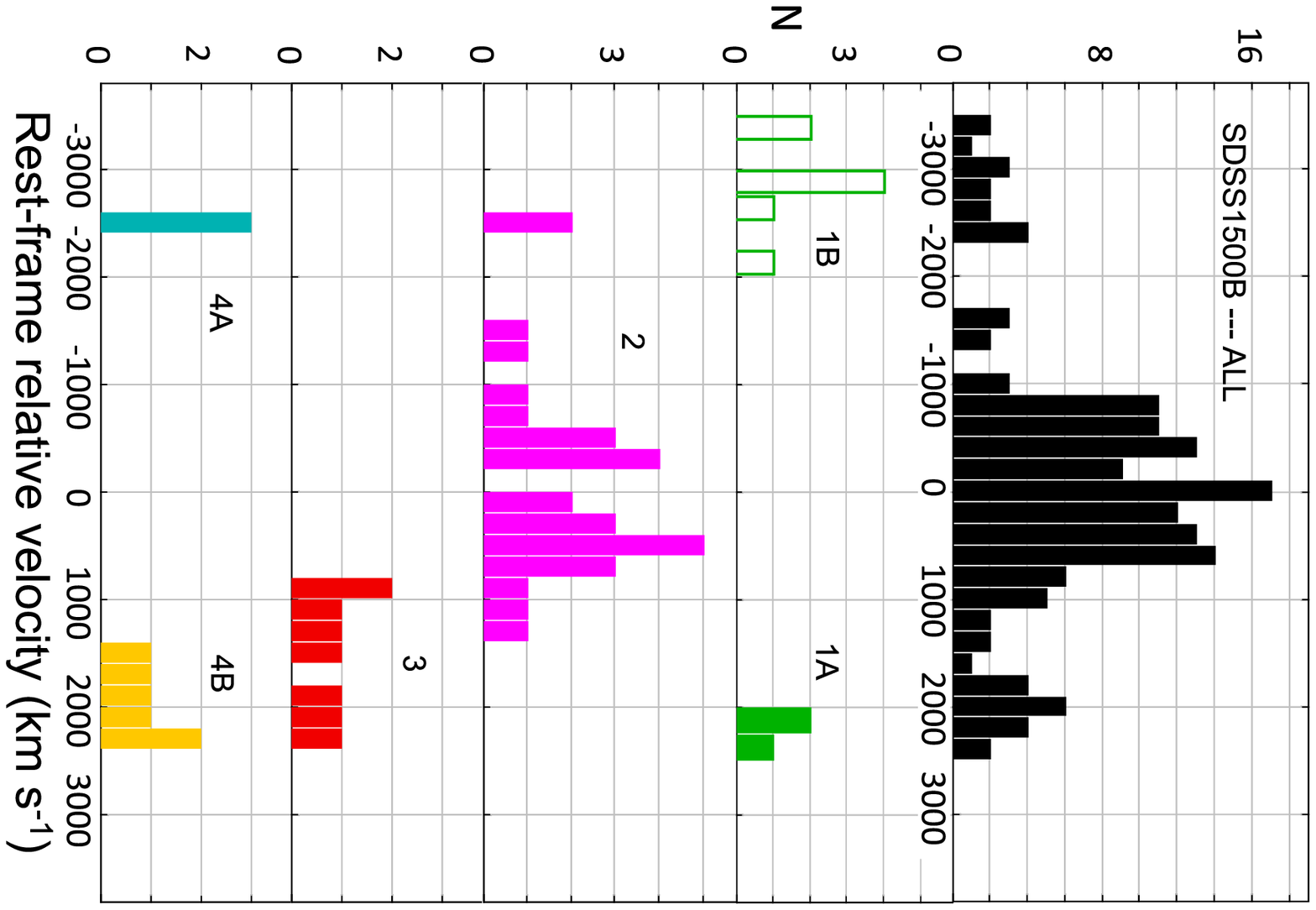}
}

\vspace{-0.4in}
\caption{Groups in the SDSS1500A and SDSS1500B clusters. left,  SDSS1500A: 
(a) ``delta plot'' (top), (b) map (middle), (c) velocity histograms. right, SDSS1500B: 
(d) ``delta plot'' (top), (e) map (middle), (f) member velocity histograms (bottom). 
\label{fig:SDSS1500A_1500B_groups}}

\end{figure*}

In Figures \ref{fig:RCS0221A_RCS1102B_groups} -- \ref{fig:SDSS1500A_1500B_groups} we present the elements of the 
procedure we used to identify and quantify the properties of the groups. For each metacluster, we ran the DS-test and found the 
areas where deviations from the global metacluster values of velocity and velocity dispersion are large.  These are shown at the 
top of the panel for each metacluster.  An open circle whose size is scaled by the $\delta$ deviation represents each 
galaxy with its 10 neighbors, for example, big circles indicate large deviations.  Using this as a map, we selected all 
objects within the area bounded by the big circles, and plotted their velocities relative to the metacluster mean.  
In almost every case a single or double peak of low velocity dispersion ($\sigma\ls350$ \kms) was found; the number of 
galaxies outside of the velocity bounds of these relatively cold structures was always much smaller than those inside the 
investigated area.  This made it unambiguous to eliminate them from the trial groups.  A second pass was made around the 
perimeter of each group to see if the group extended further in any direction (the sensitivity of the DS-test falls as more 
non-deviant objects are among the 10 neighbors), but usually there were at most a few additional objects that fit well into 
the groups.  In practice, the number of objects added to the groups by exploring the perimeter was $<$20\% of those 
originally identified.  Because of this, the process converged rapidly --- no group had to be redefined after this step.  

The groups identified in this manner are shown in the middle map of each panel, with the groups identified by symbols 
and color.  Velocity histograms for each identified group are shown in the bottom plot of each panel.  The group in the upper
right corner of  RCS1102B, Group 2, is an example of one where there is almost no contamination by non-group members 
--- compared to the 15 group members found, only 2 galaxies in the area lay outside the well defined velocity histogram 
Figures \ref{fig:RCS0221A_RCS1102B_groups}-f.   RCS0221A -- 1A and 1B are not well separated from the main body of the 
central cluster, yet here again, only 7 galaxies had to be excluded to form these two groups of 20 and 16 members respectively, 
which separate distinctly in the velocity histograms, Figure \ref{fig:RCS0221A_RCS1102B_groups}-c. 

\subsection{Properties of the groups}

The basic parameters of each of the groups are given in Table \ref{Cluster_Groups}.  Groups were divided into A \& B if two 
different velocity structures were found co-located in projected space.  There are 5, 6, 5, and 6 groups identified for RCS0221A, 
RCS1102B, SDSS1500A, and SDSS1500B, respectively.  Only 2 groups are found for the rich, regular cluster SDSS0845A, and 
one of these is well beyond the 3000\,\kms\ (rest-frame) velocity limit of a 

%Figure 7: RCS0221A  & RCS1102 group finding

%Figure 8: SDSS0845 A group finding

%Figure 9: SDSS1500 A&B group finding

%Figure 10: Velocity histograms comparing group members and non-group 

\begin{figure*}[t]

\vspace{0.3in}
\centerline{
\includegraphics[width=2.0in, angle=90]{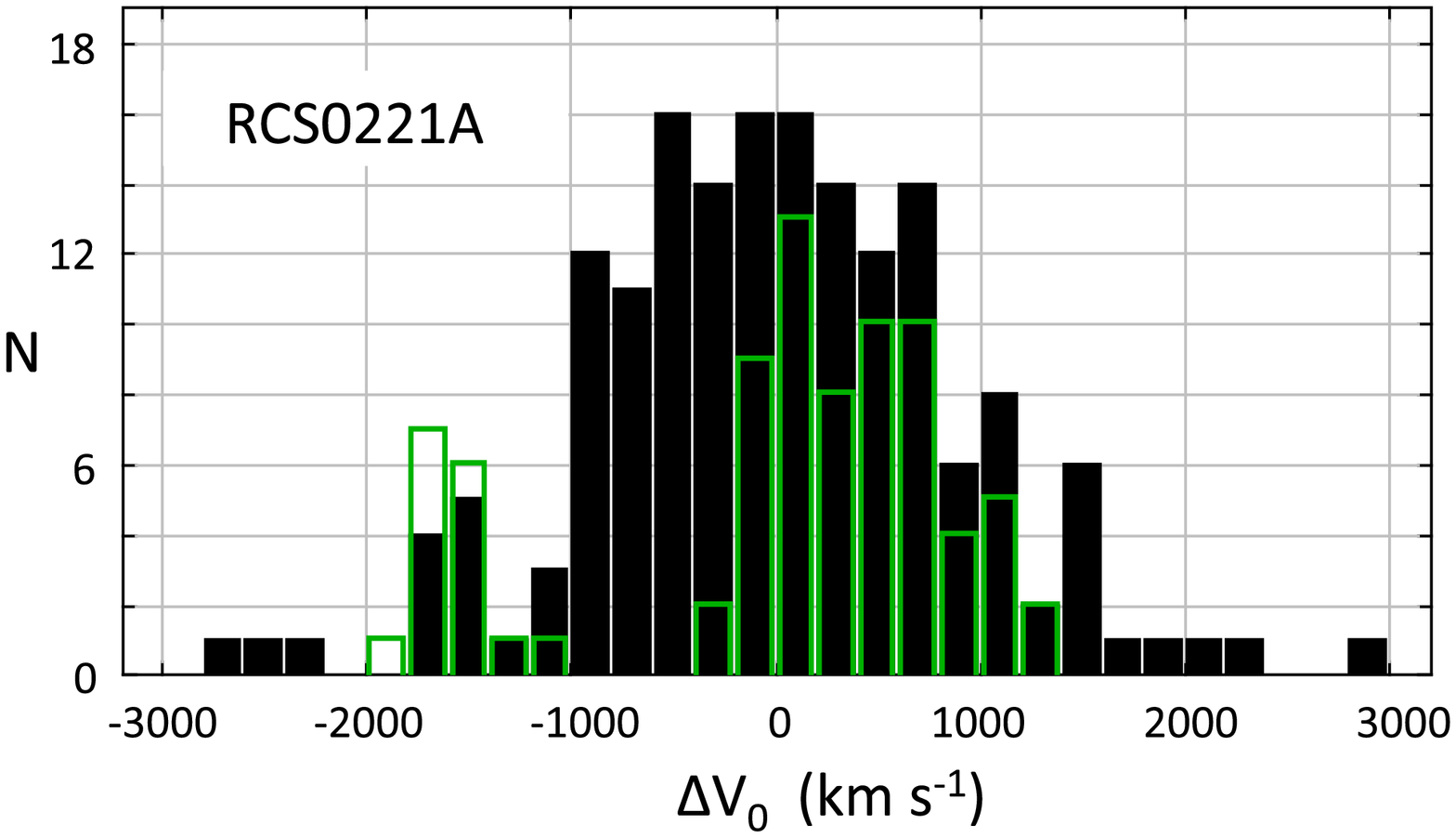}
\qquad
\hspace{-0.20in}
\includegraphics[width=2.0in, angle=90]{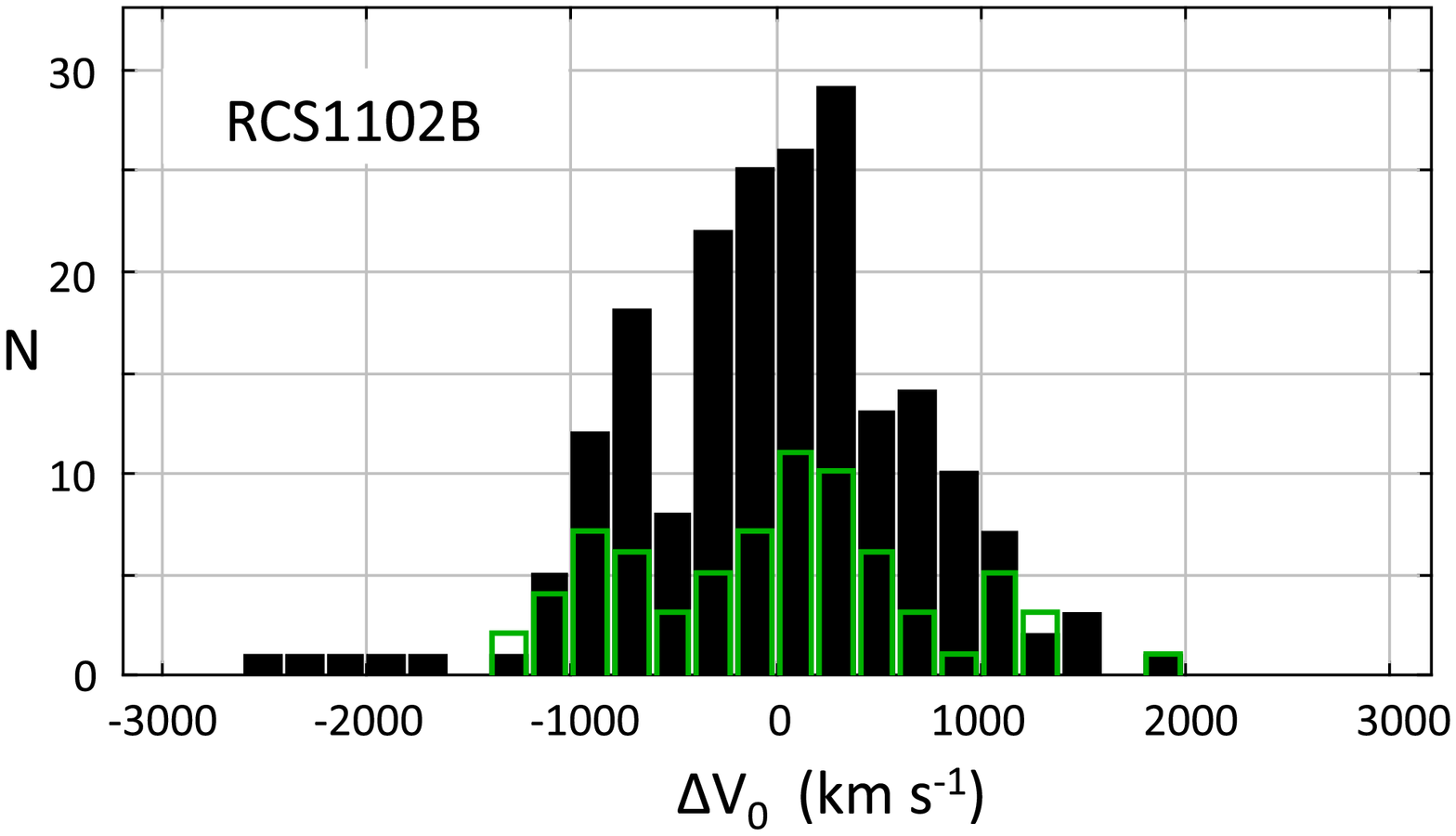}
}

\centerline{
\includegraphics[width=2.0in, angle=90]{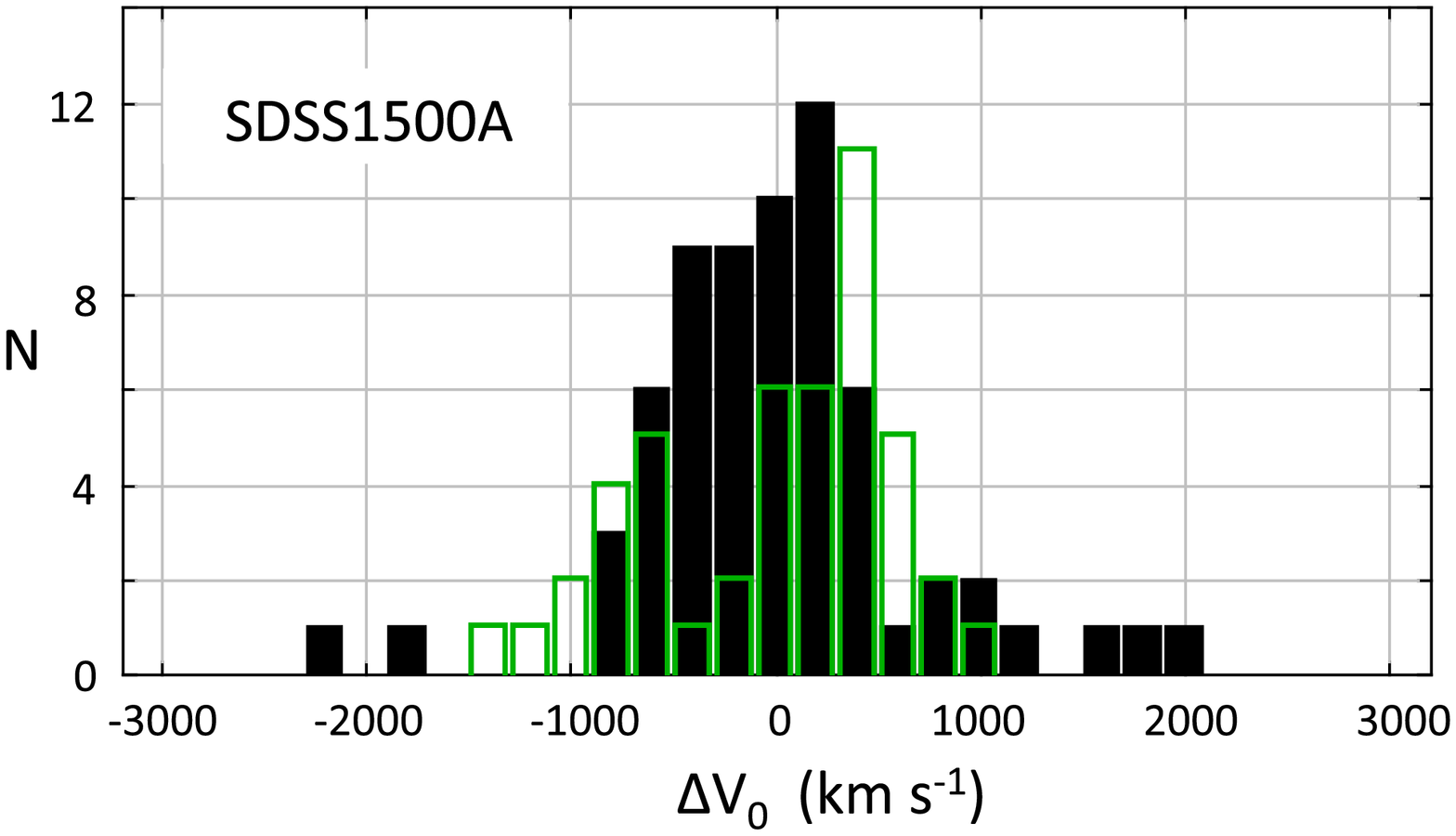}
\qquad
\hspace{-0.20in}
\includegraphics[width=2.0in, angle=90]{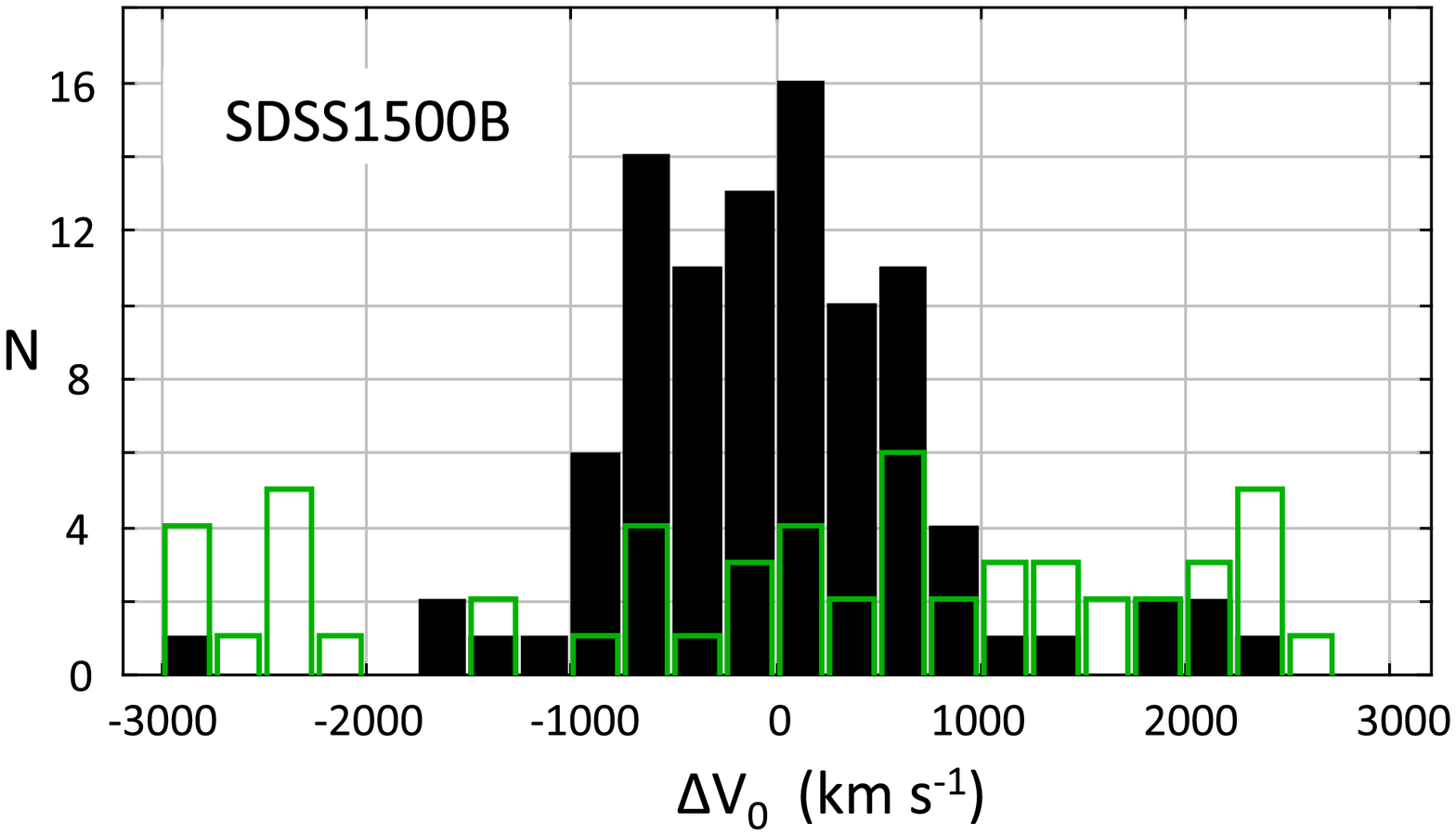}
}

\caption{Comparison of the velocity histograms for the group members (green) compared to the remaining cluster members 
(black) for RCS0221A, RCS1102B, SDSS1500A, \& SDSS1500B.  The groups share the dynamical properties of the 
previously assembled cluster, demonstrating that they are both sampling the same gravitational potential and that individual
groups can have very high infall velocities, even in projection.
\label{fig:cluster_group_velocities}}

\end{figure*}

\noindent{candidate for infall --- this and group 1B in 
SDSS1500B are assumed to be projections of groups that are at least $\sim$40 Mpc in front of the cluster (discounting the
possibility of non-Hubble velocities that are more than 3000\kms).  However, the remaining 23 groups are all candidates for 
delivering future cluster members, even though the infall velocities of a few are as high as $\sim2500$\,\kms\ in projection.  
This is shown in Figure \ref{fig:cluster_group_velocities}, where the velocities of all galaxies in the groups are compared to the 
remaining  metacluster members.  It is clear that the group members trace the same velocity distribution as the cluster 
members, that is, they are sampling the same gravitational potential.  Without a doubt, these groups are delivering the 
next wave of cluster members.}

To assess the statistical significance of these groups, we made Monte Carlo tests based on the overall velocity distribution in
the field, that is, we asked, for a group of N members, how often N random draws from the global velocity distribution yield
a velocity dispersion as small, or smaller, than the measured velocity dispersion of that group.  To be faithful to the procedure 
used in picking the groups, we had the program select N + Nex members (where Nex is the number of by-hand 
excluded galaxies within the bounded region of the group), and then to form the lowest velocity-dispersion group of N 
members from that sample (like making the best 5-card poker hand from 7 dealt cards).  The test is very conservative since
we put no spatial constraint on the selected galaxies --- they could come from anywhere in the cluster, whereas galaxy in the 
actual groups are (of course) from the same region. 

Table \ref{Cluster_Groups} contains the Nex values for each group and the derived probability of randomly concocting such 
a group from galaxies at any location in the metacluster.  The groups we had identified were found to be highly significant, 
with a typical probability of  $P \sim 10^{-3}$ --- 22 of the 24 groups have probabilities $P \ls 1$\%.  SDSS1500A group 2 
has a near-zero difference from the systemic velocity of the metacluster, and with only 6 members its 200 \kms\ velocity 
dispersion could be a random draw 16\% of the time, according to the Monte Carlo test.  It is, nevertheless compact, and 
isolated, so it is likely to be a real subgroup.  SDSS1500B group 2  is more interesting: it has a high probability of being 
a random selection from the cluster velocity distribution, 23\%, but this is because it is as hot as the cluster 
($\sigma_0 = 915$\kms) at essentially the cluster systemic velocity ($\Delta V_0 = -100$\,\kms).  There is no doubt that 
this is a dynamical group, however, as it has 29  members with an effective radius of 0.7 Mpc, a remarkable structure 
that is denser than the SDSS1500B cluster core.  It might be reasonable to suggest that this is the core of another
rich cluster which is merging with the main body of SDSS1500B, but there is no evidence of a surrounding population
attached to SDSS1500B group 2 over the semicircular area within the IMACS field. We discuss this rich group further 
in the next section.

All of the groups share basic morphological features --- they all are roughly round in shape rather than obviously filamentary.  
Although we had expected to see filaments like those in the N-body simulations, it could be that our fields, though large, still 
do not extend far enough to reach these filaments.  Regardless, our finding of many infalling groups that are well bounded 
and more round than flat suggests that, if filaments are feeding the growth of these clusters, the formation of groups 
would have to come from these further-out filaments. The typical group has 10-20 spectroscopic members (implying 20-40 
photometric members), a effective radius of 1 Mpc, and a velocity dispersion of $\sim$250\,\kms.  In addition to the 
homogeneity of the cluster groups, however, there a few interesting cases that we now describe.

\begin{enumerate}

\item{RCS0221A group 1A and 1B appear to cover the same kidney-bean-shaped region on the sky, yet they appear 
kinematically distinct with a relative velocity difference of $\sim$2000\,\kms.  The same appears to be true for RCS0221A 
groups 2A and 2B, but in this case it is also possible that they are part of a single velocity distribution. RCS1102B group 
1A and 1B may similarly be from a single, though very asymmetric, velocity distribution.}

\item{RCS1102B group\,5 is extremely compact ($R_{pair}=0.14$ Mpc!), relatively cold ($\sigma_0=335$\kms) 
and has a high relative velocity of $\Delta V_0\approx843$\,\kms\ (rest-frame) with respect to the cluster mean.

%Figure 11: field group and filament maps

\begin{figure*}[t]

\centerline{
%RCS0221_field_groups
\includegraphics[width=2.45in, angle=90]{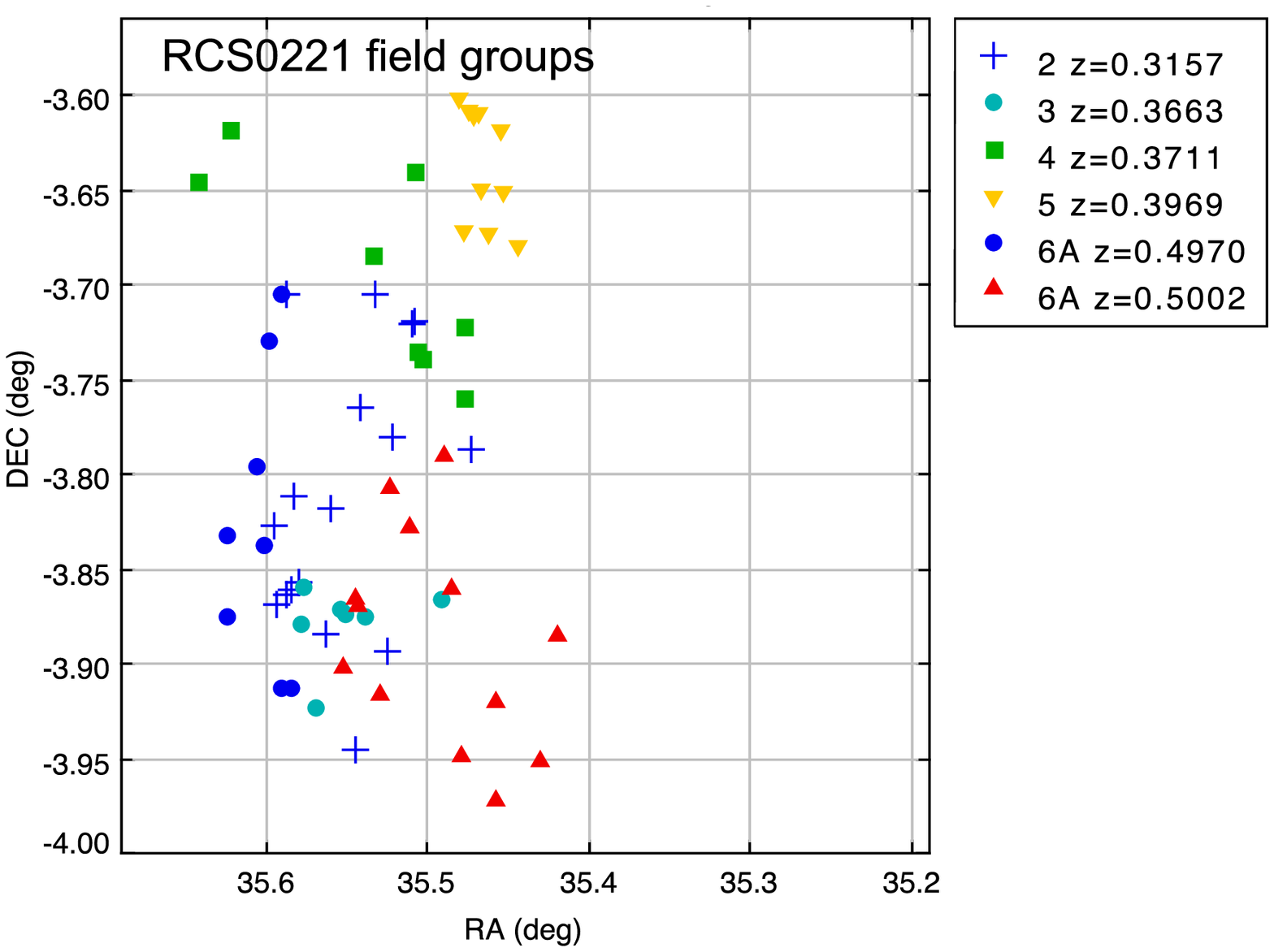}
\qquad
\hspace{-0.40in}
%RCS0221_field_filaments
\includegraphics[width=2.45in, angle=90]{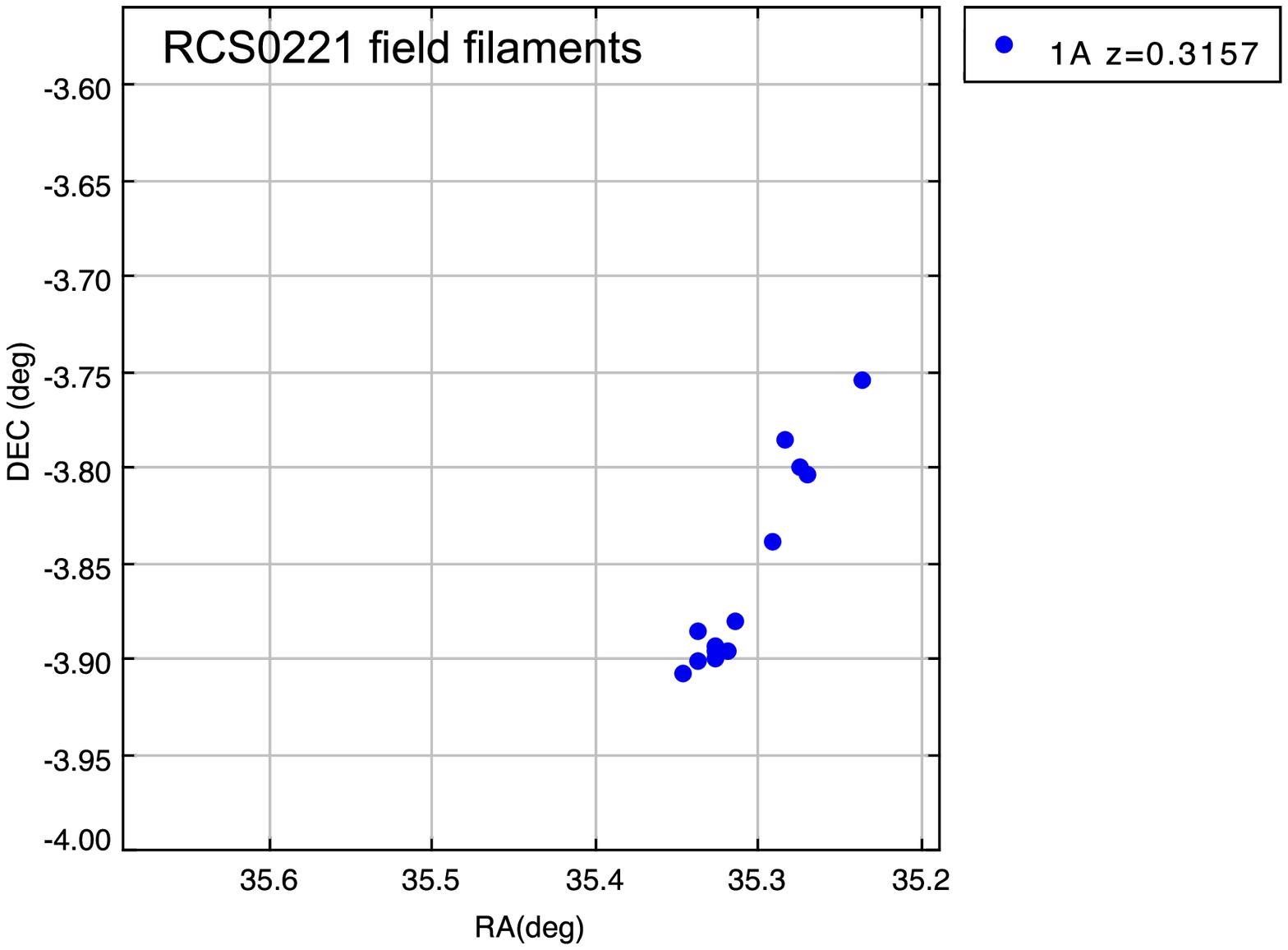}
}

\centerline{
\hspace{-0.1in}
%RCS1102_field_groups
\includegraphics[width=2.3in, angle=90]{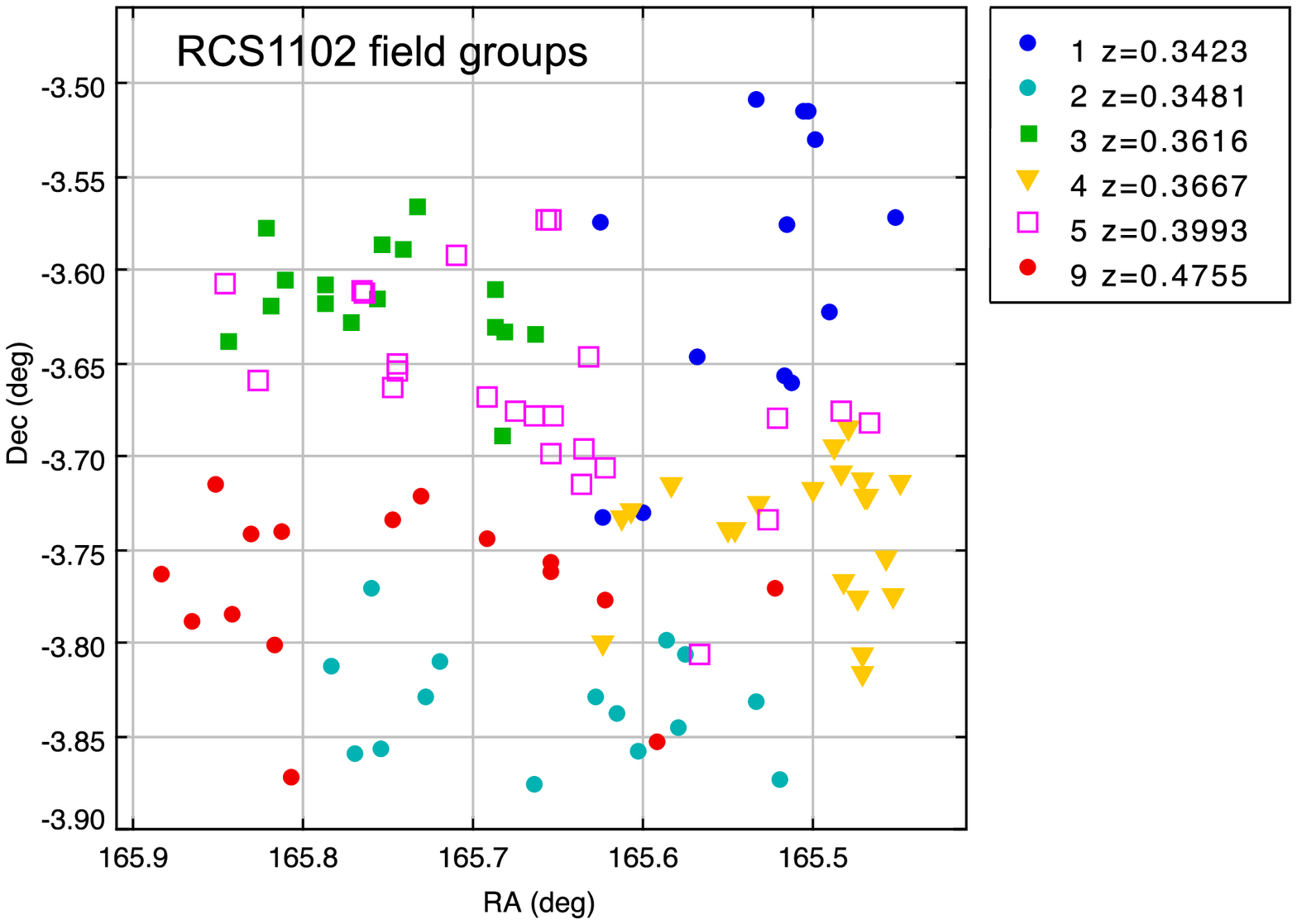}
\qquad
\hspace{-0.40in}
%RCS1102_field_filaments
\includegraphics[width=2.3in, angle=90]{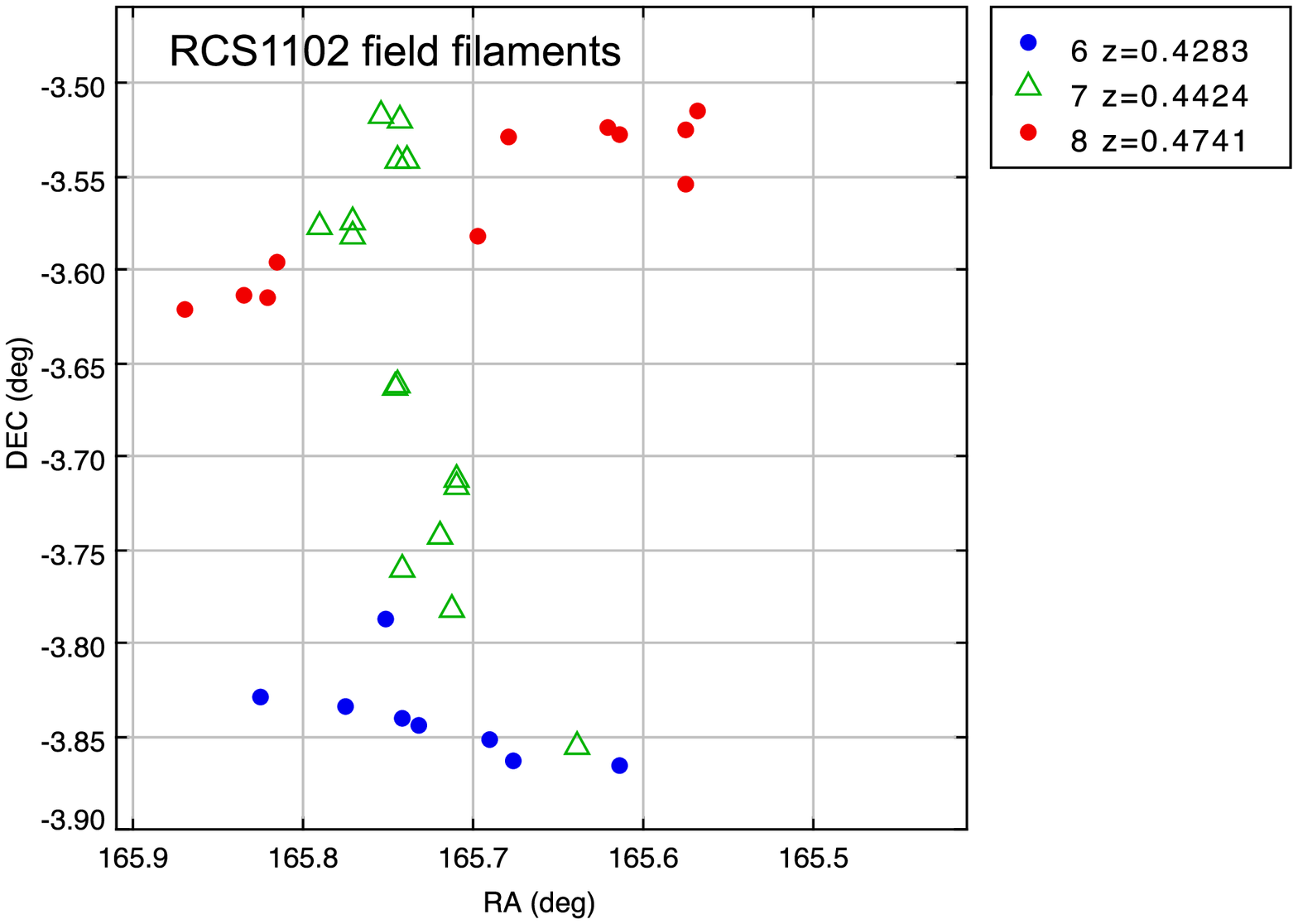}
}

\centerline{
\hspace{-0.12in}
%SDSS0845_field_groups
\includegraphics[width=2.3in, angle=90]{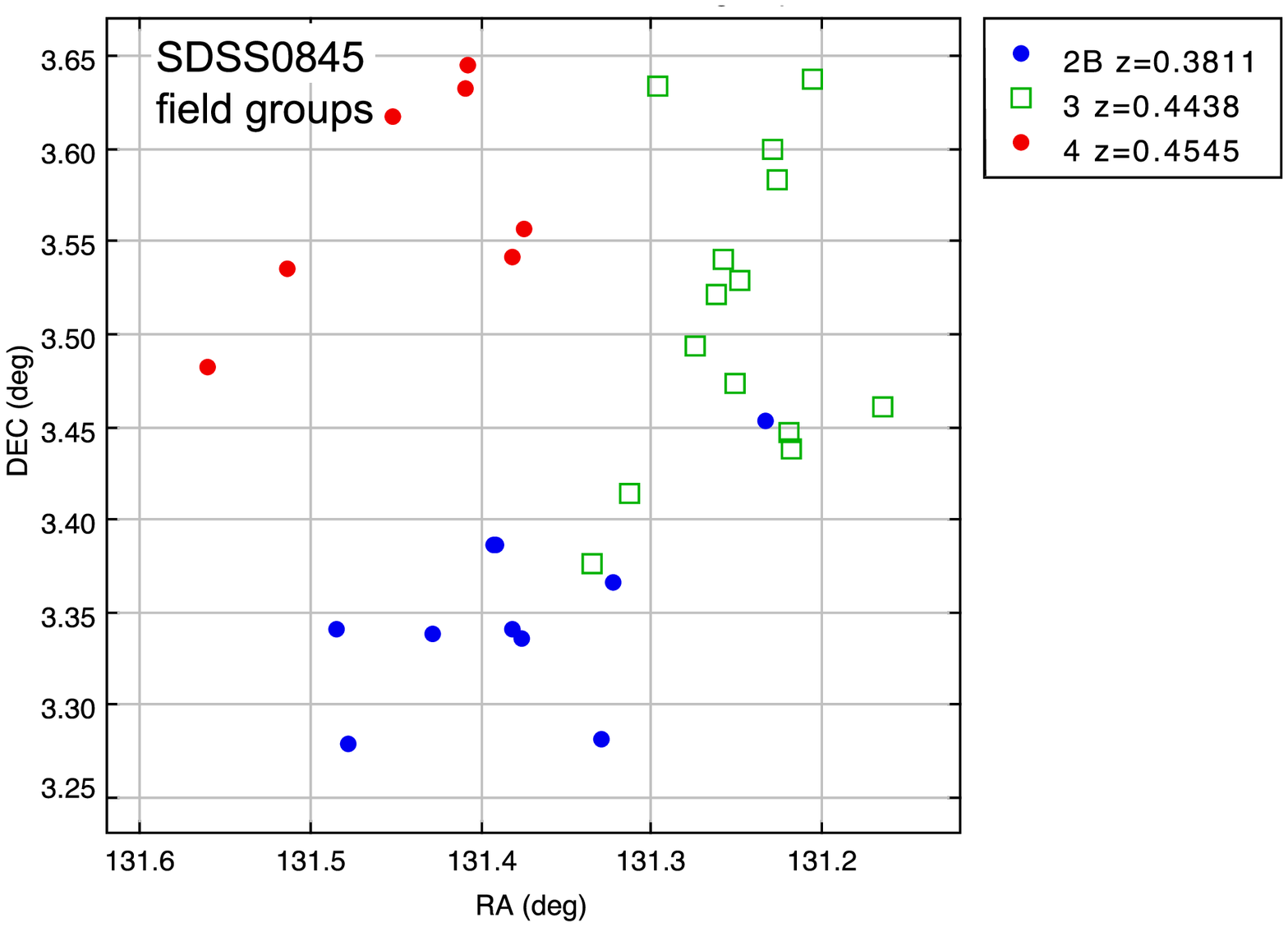}
\qquad
\hspace{-0.40in}
%SDSS0845_field_filaments
\includegraphics[width=2.3in, angle=90]{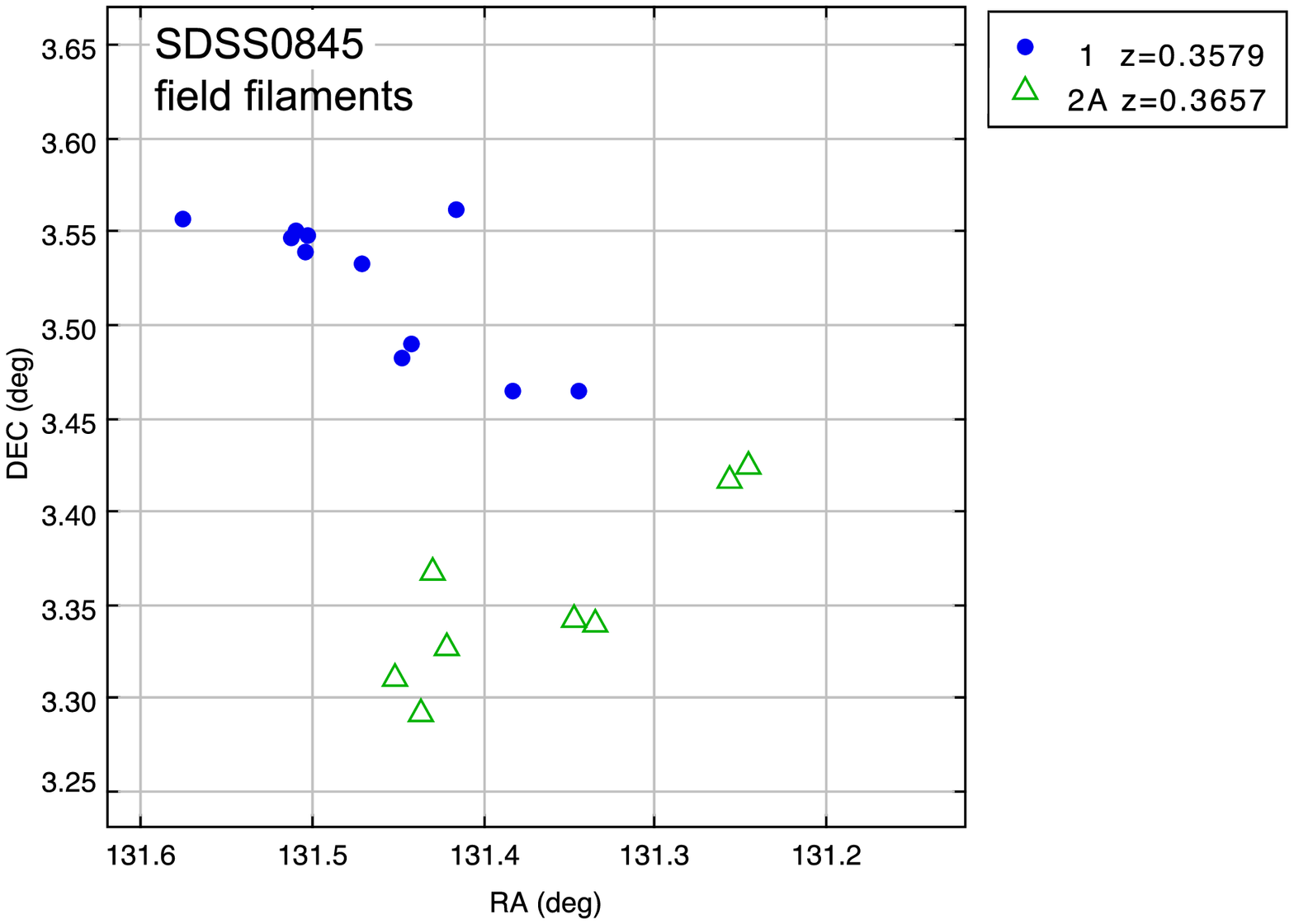}
}

\centerline{
%SDSS1500_field_groups
\includegraphics[width=2.3in, angle=90]{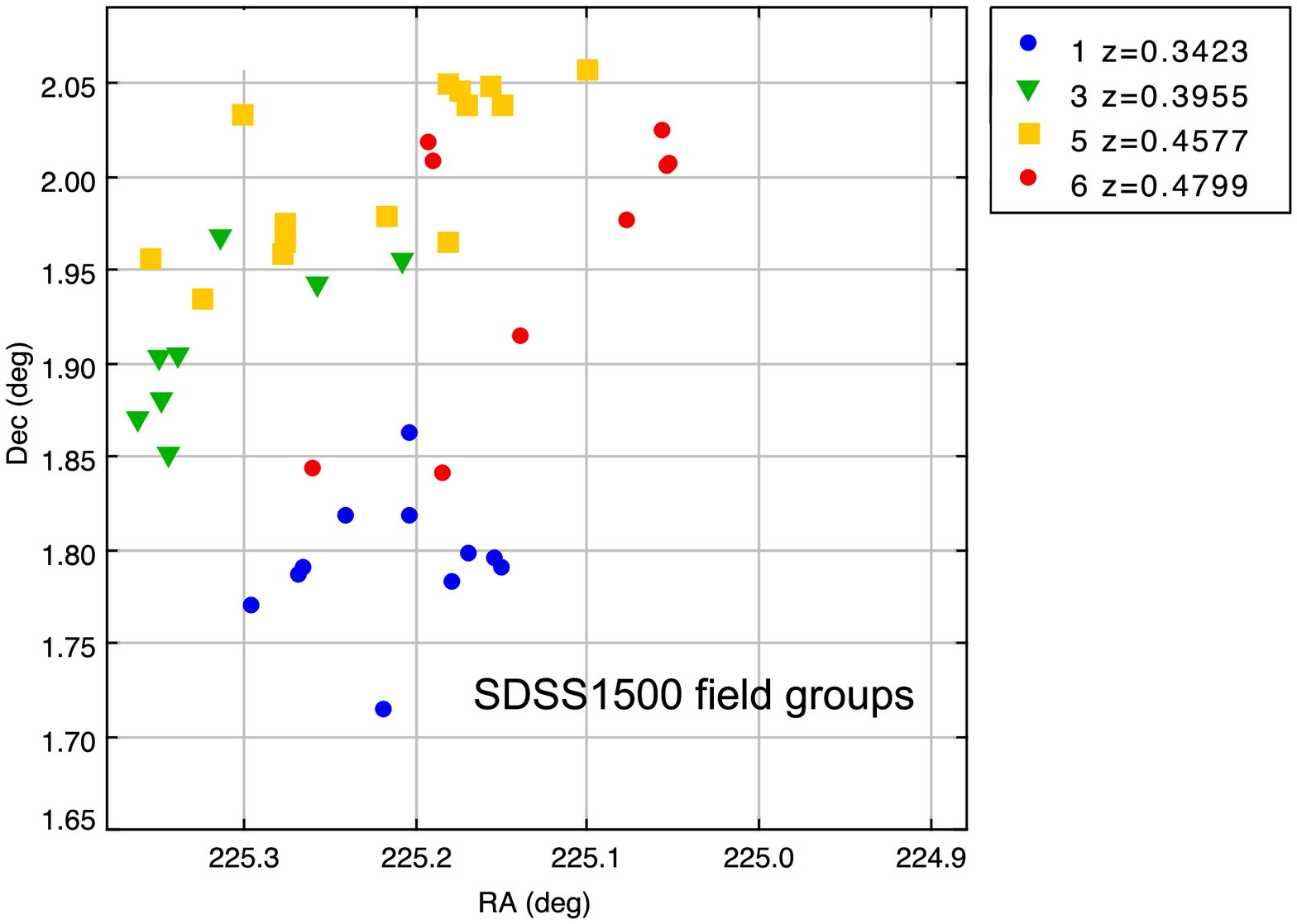}
\qquad
\hspace{-0.40in}
%SDSS1500_field_filaments
\includegraphics[width=2.3in, angle=90]{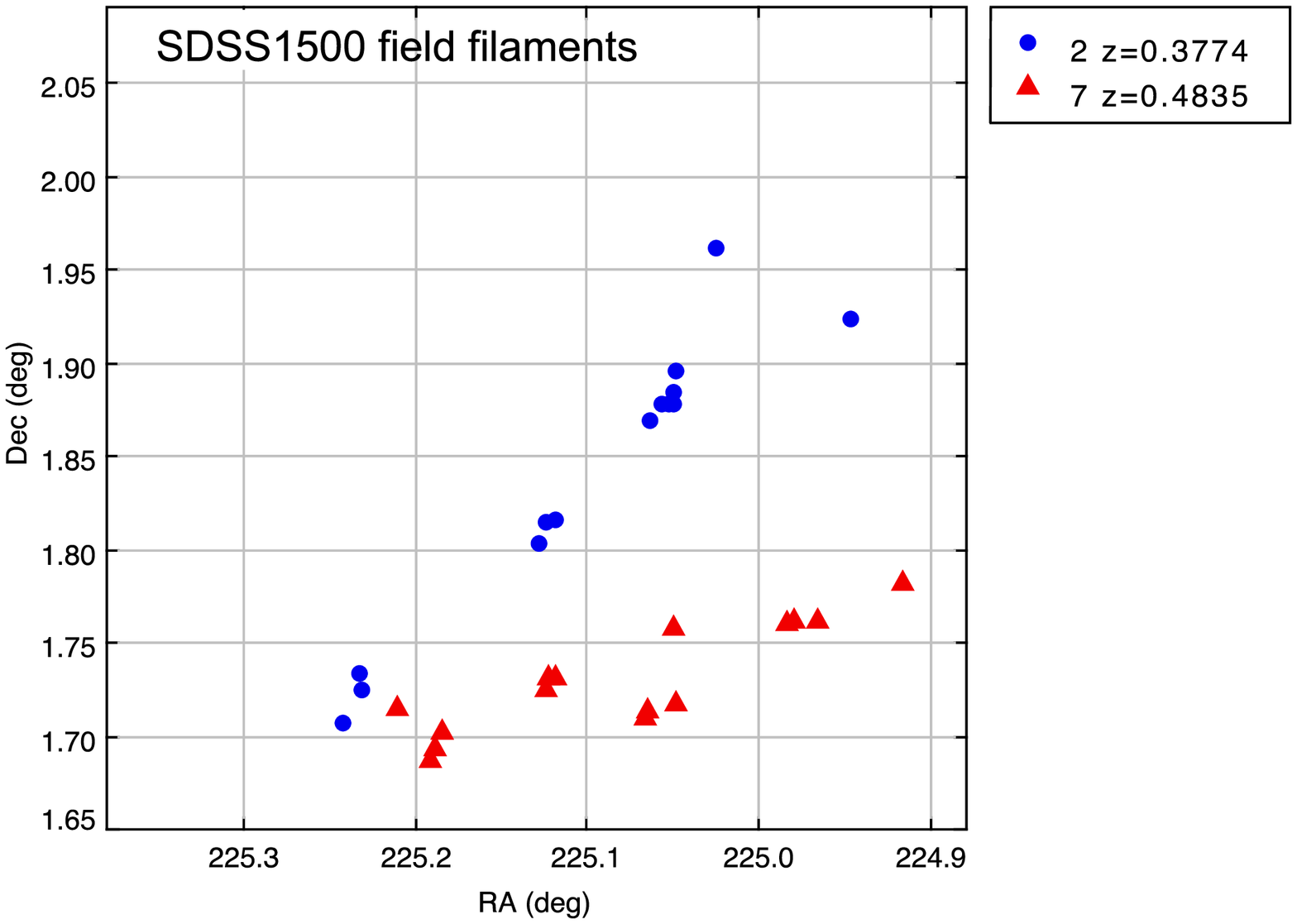}
}

\caption{Structures in the field $0.31<z<0.54$ that resemble --- in number and spatial extent -- the infalling groups 
found in the 5 rich clusters.  The left panels show the analogous groups while the right panels show filamentary structures
found in the search, chosen by the same criteria but judged as filamentary based on purely on their shape. No such 
filamentary structures were found among the groups of the ICBS metacluster sample.
\label{fig:field_group_maps}}

\end{figure*}

This would seem to be a small clump or filament that is falling in from the backside or has fallen through the core.  
The spectral type distribution is 60\% PAS, 20\% CSF, and 20\% PSB, an unusually high fraction of PAS and PSB 
galaxies for a relatively low-mass group.  It is tempting to consider this a case of a strong environmental influence from 
the cluster core on an infalling group or projected filamentary structure.}

\item{SDSS1500A groups 4A \& 4B are compact central concentrations or filaments projected directly on the cluster core.  
The members appear to divide into two cold groups at $+500$\,\kms\ and $-800$\,\kms\ with respect to the cluster systemic 
velocity.  Of its combined 20 members, 10 are PAS, with 3 SB and 2 PSB as well.  As in the case of RCS1102B, a cluster-core 
influence could be inferred.}

\item{SDSS1500B group 3 is another very compact structure projected against the cluster core.  It has a high infall velocity 
($V_0 \approx 1500$\,\kms) from the frontside and is relatively hot, $\sigma_0 \approx 500\,\kms$, and 5 out of 8 members
are PAS.} 

\end{enumerate}

\subsection{The identification of comparable groups in the field}

The groups identified in the previous section are clearly providing a major, if not the dominant component in the
building of these clusters.  One of the goals of the ICBS is to look for evidence of spectral evolution in the infalling
population to help understand what role, if any, is played by the cluster environment as distinct from that of the groups 
that have brought the cluster to this state of assembly.  In order to address this issue, it is important to compare the
properties of these supercluster groups to those of the general field.  To accomplish this, we searched for and 
identified groups in the cl\_field sample (the field over the same redshift range as the cluster observations) with 
the goal of finding groups whose basic parameters --- size, richness, and velocity dispersion --- were similar to the
cluster groups. 

The results of an automated ``friends-of-friends" search (Paper 3) had already provided a catalog of 
groups, but these were mainly poor and small, and not a good match to the cluster sample.  Because there is not a 
comparatively narrow redshift interval in this field sample as there is for each cluster, it is not straightforward to use 
the Dressler-Shectman test, which is based on the fact that all objects in the sample are members of the cluster and 
relies on global values of velocity dispersion and systemic velocity.  We therefore decided to look for `spikes' in the 
redshift distribution and look for spatial segregation for galaxies in these spikes. In practice, this was accomplished by 
investigating $\Delta z=0.02$ slices, stepped in $\Delta z=0.01$ increments through the full depth.  In order to match 
the metacluster sample, where the dense cluster environment dominates over the central few megaparsecs of the field 
(leaving only the region beyond available for a group search), we concentrated on groups that were a few Mpc or smaller 
in projection on the sky, but our search turned up larger systems that had not been found in the metacluster sample.  A 
couple of field groups stretch across most of the \imacs\ field.  Possibly, such large groups cannot survive in close 
proximity to a rich cluster because of tidal disruption.  

The search yielded 30 groups, whose properties are listed in Table \ref{Field_Groups}.   Again, we found that this
process was quite unambiguous: we consistently found well-defined structures with little confusion as to what was or was
not a likely member.  This is demonstrated by the size of the groups and their low velocity dispersions --- 29 of the 30
groups have velocity dispersions $\sigma<350$\,\kms and 18 of 30 have $\sigma<250$\,\kms.   Members of these
groups comprise $\sim$40\% of the cl\_field sample. 

The morphology of most of these field groups overlapped that of the cluster groups, but a sizable minority have a narrow 
filamentary shape that was not found in the cluster sample.  In Figure \ref{fig:field_group_maps} we show these groups and 
filaments separately for the 4 fields.  Histograms of basic group parameters are shown in Figure \ref{fig:group_properties} and 
discussed below.

\section{Discussion}

\subsection{Galaxy clusters under construction}

It has been well recognized for the last two decades that clusters have grown through the accretion of systems 
of all scales, from single galaxies and moderate-sized groups to cluster-cluster mergers.  Accordingly, finding  
substructure is an unsurprising result of any study of rich clusters.  However, this point of view developed slowly 
in the 1980's as substructure in clusters was recognized as the consequence of hierarchical structure growth, as first 
suggested by White (1976).  Up to this time the prevailing view was that clusters, typified by the only very well-studied 
cluster --- Abell 1656 (Coma), have smooth, axially symmetric distributions of galaxies.  This picture suggested 
a process of cluster formation that either did not involve merging or accretion of smaller structures, or a
process that actually destroyed them, in particular, the Lynden-Bell (1967) ``violent relaxation'' model 
that described the gravitational collapse of a volume of roughly uniform density --- an uncommon occurrence
in a hierarchical universe.
 
Dressler's (1980) discovery of a correlation between galaxy morphology and local projected density was regarded skeptically 
because violent relaxation was the prevailing picture of cluster formation at that time: if apparent subgroups in clusters were 
merely statistical density fluctuations they would be too short-lived to be seriously involved in morphological evolution.  In 
reviewing the observational data on substructure, and making the first quantitative estimate of the prevalence of substructure 
through surface-density contour maps, Geller \& Beers (1982) found statistically significant substructure in approximately 40\% 
of a sample of 65 rich clusters studied by Dressler (1976, 1980), a necessary if not necessarily sufficient degree of subclustering 
to account for the morphology-density relation.  As the number of available redshifts in such clusters grew, more discriminating 
tests became possible.  Employing the test described above with $\sim$1000 cluster redshifts divided among 15 of the same 
clusters, Dressler \& Shectman's (1988) came to a similar conclusion, that ``In 30-40\% of the cases, the subclusters contain 
a large fraction of the galaxies found in the main body of the cluster."

Hierarchical clustering suggests that the clusters of the relatively recent past, $0.3<z<1.0$ should exhibit much stronger 
substructure compared to present-epoch rich clusters (Kauffmann 1995), and indeed, observations have produced some 
striking examples (e.g., De Filippis \& Schindler 2003; Kodama \etal\ 2005; Oemler \etal\ 2009).  However, the selection of 
intermediate-redshift rich clusters for study has been substantially biased to clusters with strong X-ray -emission: at any 
epoch, these are the most dynamically evolved and accordingly exhibit the least amount of substructure.  The ICBS 
includes one such cluster, SDSS0845A, which is populous and has a  smooth symmetric distribution: significantly, 
it includes only one small infalling group in the field (Figure \ref{fig:SDSS0845A_groups}-b), in  contrast with the many 
infalling groups of each of the other 4 clusters.  Furthermore, most studies of distant clusters, particularly those making 
use of HST imaging, cover relatively small volumes of space around intermediate-redshift clusters, $R\ls1$ Mpc, 
approximately the virial  radius of clusters of this richness and essentially the inner regions of the cluster where 
substructure is more likely to have been erased.

For these reasons, we believe that the ICBS program may be the first to investigate this question of cluster growth 
for \emph{typical} rich clusters over the volume needed to see the infalling population that will be incorporated into 
the cluster between the redshift of observation and the present epoch.  Our finding of a robust population of infalling 
groups of 10-20 spectroscopic (20-40 photometric) members in 4 of the 5 clusters of our study, shown in Figures 
\ref{fig:RCS0221A_RCS1102B_groups} and \ref{fig:SDSS1500A_1500B_groups}, may be in fact the most  representative 
view to-date of how a typical rich cluster of today was assembled.

%Figure 12: group member, and non-member maps 

\begin{figure*}[t]

\centerline{
\includegraphics[width=3.5in, angle=90]{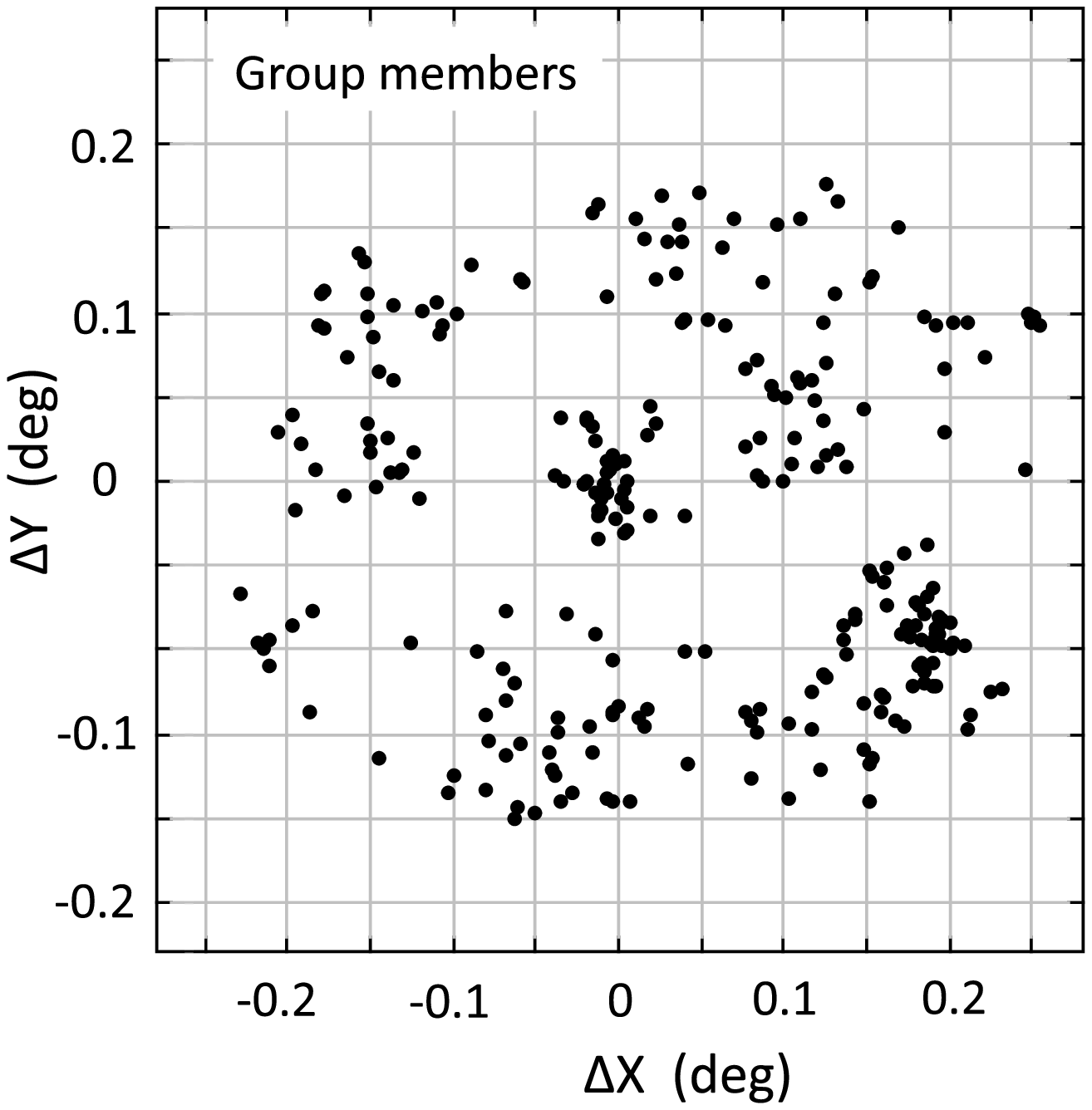}
\includegraphics[width=3.5in, angle=90]{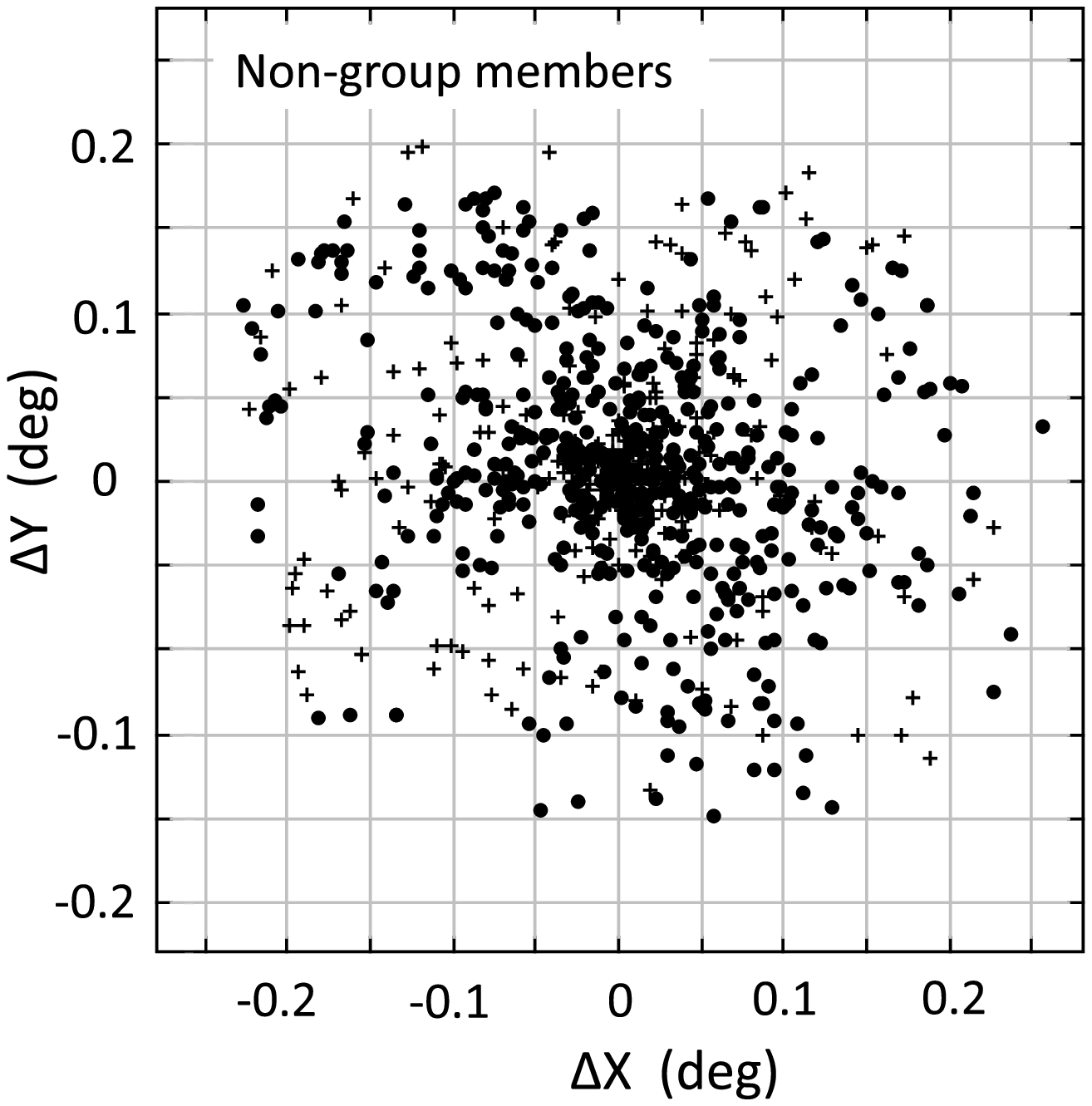}
}

\caption{Composite `sky maps' of metacluster members.  a) (left) closed dots -- members in groups in metaclusters RCS0221A, 
RCS1102B, SDSS1500A \& B;   b) (right) closed dots -- non-members of groups for all 5 metaclusters; plus signs -- SDSS0845A. 
\label{fig:member_nonmember_skymaps}}

\end{figure*}

\subsection{Building Clusters Through Group or Galaxy Accretion?}

The identification of kinematically distinct groups in the ICBS clusters offers the possibility of a quantitative test of the
paradigm $\Lambda$CDM model (Springel \etal\ 2005).  There are, of course, many subtleties involved in comparing easily 
identifiable galaxies in the sky to the dark matter halos traced by N-body simulations.  The ICBS directly samples only about 
1.6 Gyr of cosmic time: although we have argued that the ICBS clusters are typical clusters at this epoch in terms of the 
maturity of their dynamical evolution, the infall we measure is limited to a few billion years of cluster history. For this
reason, it is not straightforward to compare our results with apparently suitable theoretical studies on galaxy infall 
into clusters, for example, the $\Lambda$CDM N-body simulations by McGee \etal\ (2009), Berrier \etal\ (2009), and 
De Lucia \etal\ (2012).  A principal motivation of these studies was to investigate whether so-called ``preprocessing'' in 
groups of galaxies --- outside the rich cluster environment --- could partially or fully achieve the high fraction of 
passive galaxies in rich clusters, before cluster-specific processes such as ram-pressure or tidal stripping `kick in.' 

Berrier \etal\ conclude that such preprocessing is not important, based on their simulation which showed that:

\begin{quotation}
\small
\noindent{On average, 70\% of cluster galaxies fall into the cluster potential directly from the field, with no luminous 
companions in their host halos at the time of accretion; less than 12\% are accreted as members of groups with five 
or more galaxies.}
\normalsize
\end{quotation}

McGee \etal\ find essentially the opposite:

\begin{quotation}
\small
\noindent{We find that clusters at all examined redshifts have accreted a significant fraction of their final galaxy populations
 through galaxy groups. A 10$^{14.5} h^{-1}$ \Msun\ cluster at $z=0$ has, on average, accreted $\sim$40\%
of its galaxies (M$_{stellar} > 10^9 h^{-1}$ \Msun) from halos with masses greater than 10$^{13} h^{-1}$ \Msun.}
\normalsize
\end{quotation}

Confirming the conclusions of McGee et al., De Lucia \etal\ (2012) attribute the importance of distinguishing
between different timescales of accretion into a group --- as distinct from accretion into the final cluster --- as
important to reconciling the apparently conflicting result of Berrier et al. 

It is not obvious how to decide if these criteria are met by the ICBS infalling groups, or whether the conclusions of these 
theoretical studies refer only to virialized halos, which likely describes only some of the ICBS groups.  Furthermore, the 
percentages given by these studies are averaged over some longer history of the cluster, while the ICBS samples a narrower 
epoch, albeit one of significant growth for the cluster:  the infall we are observing at $z\sim0.5$ will substantially increase the 
cluster's mass in the several gigayears required to incorporate the groups into the cluster. 

%Figure 13: Rcl/R_200 vs delta-v0  

\begin{figure*}[t]

\centerline{
\includegraphics[width=3.2in, angle=90.]{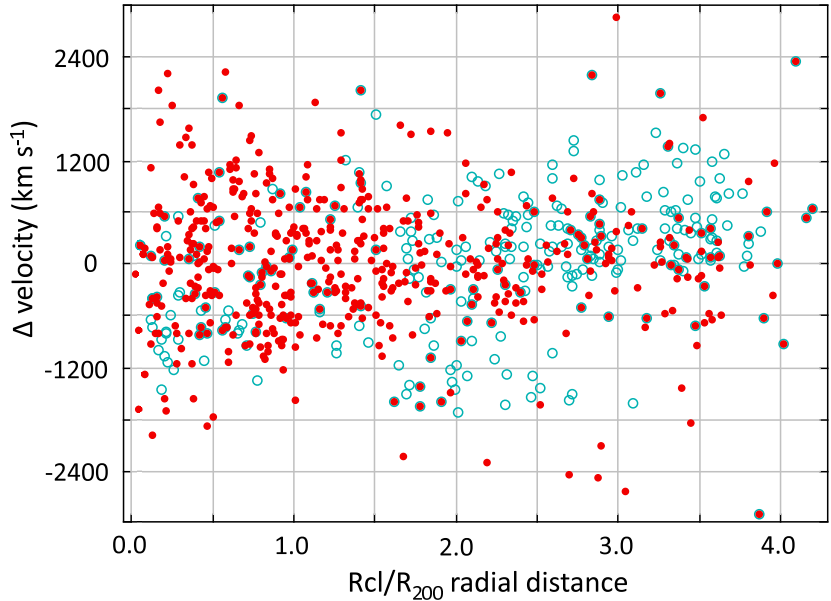}
}

\caption{Delta redshift from the mean cluster velocity as a function of Rcl, for 4 combined metaclusters --- 
RCS0221A + RCS1102B + SDSS1500A + SDSS1500B -- for members that are not identified as in groups 
(red dots) and for those in groups with 5 or more members (blue open circles).
\label{fig:Rnorm_vs_delVo}}

\end{figure*}

Modulo these uncertainties, an estimate of the fraction of all infalling galaxies --- in and out of groups --- is necessary to 
compare with model simulations.  For the four ICBS metaclusters RCS0221A, RCS1102B, SDSS1500A, and 
SDSS1500B, 257 galaxies have been identified as members of 24 groups, compared to 532 clusters members 
that are not members of groups.  This alone indicates that the mass of the virialized clusters will grow by at least $\sim$50\% 
by the present epoch.  This contests with $\ls$7\% for the minimum growth of SDSS0845A, an already relaxed, concentrated 
cluster, based on the single infalling group we identified.  Figure \ref{fig:member_nonmember_skymaps} shows these 
populations in graphical form, as a composite of the groups in the four metaclusters and a composite of the non-group 
population (with members of the relaxed cluster SDSS0845A also shown, as distinct symbols).  The group composite is 
a thick shell, possibly because of tidal destruction of groups within the inner radius, or perhaps just because of the difficulty 
identifying groups further in.  Like the dynamically evolved cluster SDSS0845A, the composite of the non-group members 
in the four other metaclusters shows a smooth distribution indicative of a spherically-symmetric potential well.

The 257 infalling galaxies in groups is of course a lower-limit, because it is much of the the infall will be in the groups in 
smaller (N$<$5) groups and single galaxies.  An estimate of the total infalling population would 
correct this deficiency.  It is, of course, impossible to distinguish individual galaxies as members of the viralized versus 
infalling populations, but numerical studies by Balogh \etal\ (2000) and Moore \etal\ (2004) can be used to estimate the 
two populations as a function of $R/R_{virial}$.  These studies have used N-body simulations to follow the ``backsplash,"
or ``overshoot'' of galaxies that have passed through the cluster and joined the virialized system.  Both works identify the 
zone $1<R/R_{virial}<2$ as the overlap region where roughly half the galaxies are members of the cluster and half are 
infalling, and find further that the fraction of cluster-members to infalling-galaxies falls rapidly beyond this zone.  

For our sample we adopt \R200\ as a proxy for $R_{virial}$ and for each of the four metaclusters clusters divide the 
measured radial distances from the cluster center by the \R200\ value listed in Table \ref{Clusters}.  Figure 
\ref{fig:Rnorm_vs_delVo} shows this normalized radial distance from the cluster center, \Rnorm, plotted against 
$\delta V_0$ for the galaxies at the cluster redshift that are are, and not members, of the groups described about.  
There is a suggestion here that $\sim2$\,\Rnorm\ is a transition from the virialized cluster to the infalling population.  The 
velocity dispersion $\sigma$ for  three roughly equally populated inner zones, \Rnorm\ = 0.0-0.5, 0.5-1.0, and 1.0-2.0,  is 
slowly deceasing --- 828$\pm83$\,\kms, 772$\pm70$\,\kms, 702$\pm56$\,\kms, respectively, but beyond 2.0\,\Rnorm\ to 
the limit of the sample at 4.3\,\Rnorm, $\sigma$ increases to 826$\pm67$\,\kms.  The galaxies in infalling groups beyond 
2.0\,\Rnorm\, also seen in Figure \ref{fig:Rnorm_vs_delVo}, show an asymmetric distribution and even higher dispersion 
of $\sigma = 1111\pm82$\,\kms.  

Although insufficient to confirm a transition from virialized to infalling populations at about 2\,\R200, these kinematic signatures 
are at least consistent with the predictions of Balogh \etal\ and Moore \etal.  We use this information, then, to estimate the relative 
sizes of the virialized to infalling populations for this four metacluster cluster sample of 789 galaxies.  We assign all 258 galaxies 
with \Rnorm$<1$ to the virialized cluster, split equally the 195 galaxies $1<$\Rnorm$<2$ between cluster and infalling,  and 
identify of 80\% $2<$\Rnorm$<3$ and 100\% of \Rnorm$>3$ as infalling.  The result of this simple estimate is 393 members of the 
virialized clusters, 396 infalling galaxies.  This equal split between infalling and virialized galaxies is approximate --- 60/40 or 
40/60 is just as likely --- but it is good enough to indicate that the mass of each of these clusters will approximately double over the 
next $\sim$4\,Gyr as a result of the incorporation of infalling galaxies.  Even without an accounting of the groups that might be 
infalling from $R\sim5$\,Mpc and beyond, it is reasonable to conclude that $z\sim0.4$ is the major epoch of growth for these 
systems.  

Two more issues are worth discussion.  First, if $\gs$\,90\% of the 336 galaxies in these four metaclusters with \Rnorm$>$2 are 
identified as infalling, and the identified groups --- which are certainly infalling --- account for 183 of them, this means that the 
$\sim$120 other infalling galaxies are either isolated galaxies or members of small groups with 4 or less members.  Identification 
of individual poor groups is very difficult in this supercluster environment, but such groups may still be massive enough to support 
some kind of preprocessing, which we discuss below.

Second, the Balogh \etal\ and Moore \etal\ identification of 1-2\R200 as the overlap zone of infalling and backsplash galaxies provides a way to roughly divide our metacluster sample into ``supercluster" (infalling) and ``cluster" (virialized), by splitting the galaxies at \Rnorm\ = 1.5 ($\sim$2\,Mpc).  This will produce two different samples that have only modest cross-contamination.  We note here that, making this division, the cluster sample has a 52\% PAS and 5.6\% PSB fraction, while percentages for the supercluster are 25\% PAS and 1.9\% PSB (which 
is similar to fractions for the cl\_field).  This substantial difference is further evidence that the 1.5\,\R200\ division has physical significance.  
The parameters of the clusters and superclusters defined by the \Rnorm\ = 1.5 split can be found in Table \ref{Other_Samples}.

In conclusion, while there is a range of mass growth represented in these clusters, an increase by a factor of two by the present epoch seems typical.  From our observations of four ICBS clusters, the fraction of infalling galaxies that are in groups where preprocessing might occur is substantial, of order 50\% or greater.  This appears to be consistent with with the predictions by McGee \etal\ and De Lucia et al. but inconsistent with the prediction by Berrier et al., which was specifically addressing the issue of preprocessing.  Again, quantifying the degree of agreement or contradiction requires a reliable correspondence to be drawn between the ICBS cluster groups --- dynamically cold, discrete groups of about 10-50 L$^*$ galaxies --- with the dark halo groups identified in the simulations.  Many of the ICBS groups may be young, even unvirialized, but their galaxies are already experiencing the group environment.  Regardless of the outcome of this comparison between theory and observation, the ICBS results are by themselves unambiguous: many, perhaps most galaxies are members of groups where some sort 
of preprocessing of star forming galaxies into passive galaxies could occur, well before these galaxies enter the more extreme cluster environment.

\subsection{Evidence for preprocessing from the spectral types of group galaxies}

The PAS galaxies, which are non-starforming at the level sSFR $<10^{-11}$ yr$^{-1}$, and the PSB (poststarburst) galaxies 
that are in the process of joining or rejoining the PAS population, are systems where star formation has been effectively 
ended, either by internal processes or external agency.  With our sample of groups in clusters and the field we can look to
see if the PAS+PSB fraction is correlated with any properties of the groups themselves.  We exclude six groups
for this exercise, three cluster groups with $N<5$ members (too small for a statistical result) and four field groups
with $\tauenc > 6$\,Gyr (described below).

\begin{figure*}[t]
\vspace{0.3in}

\centerline{
\includegraphics[width=1.52in, angle=90]{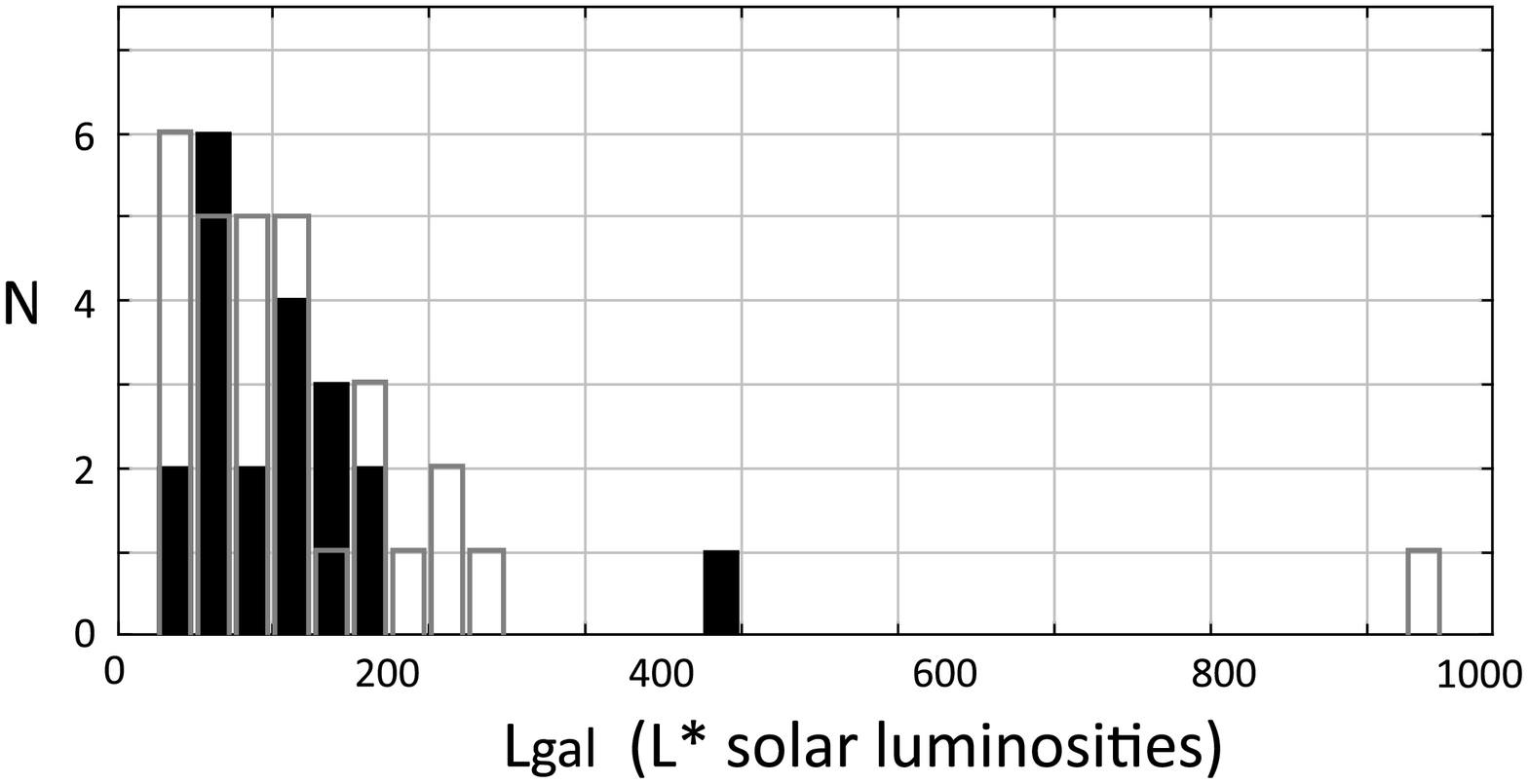}
%\hspace{-0.2in}
\qquad
\includegraphics[width=1.5in, angle=90]{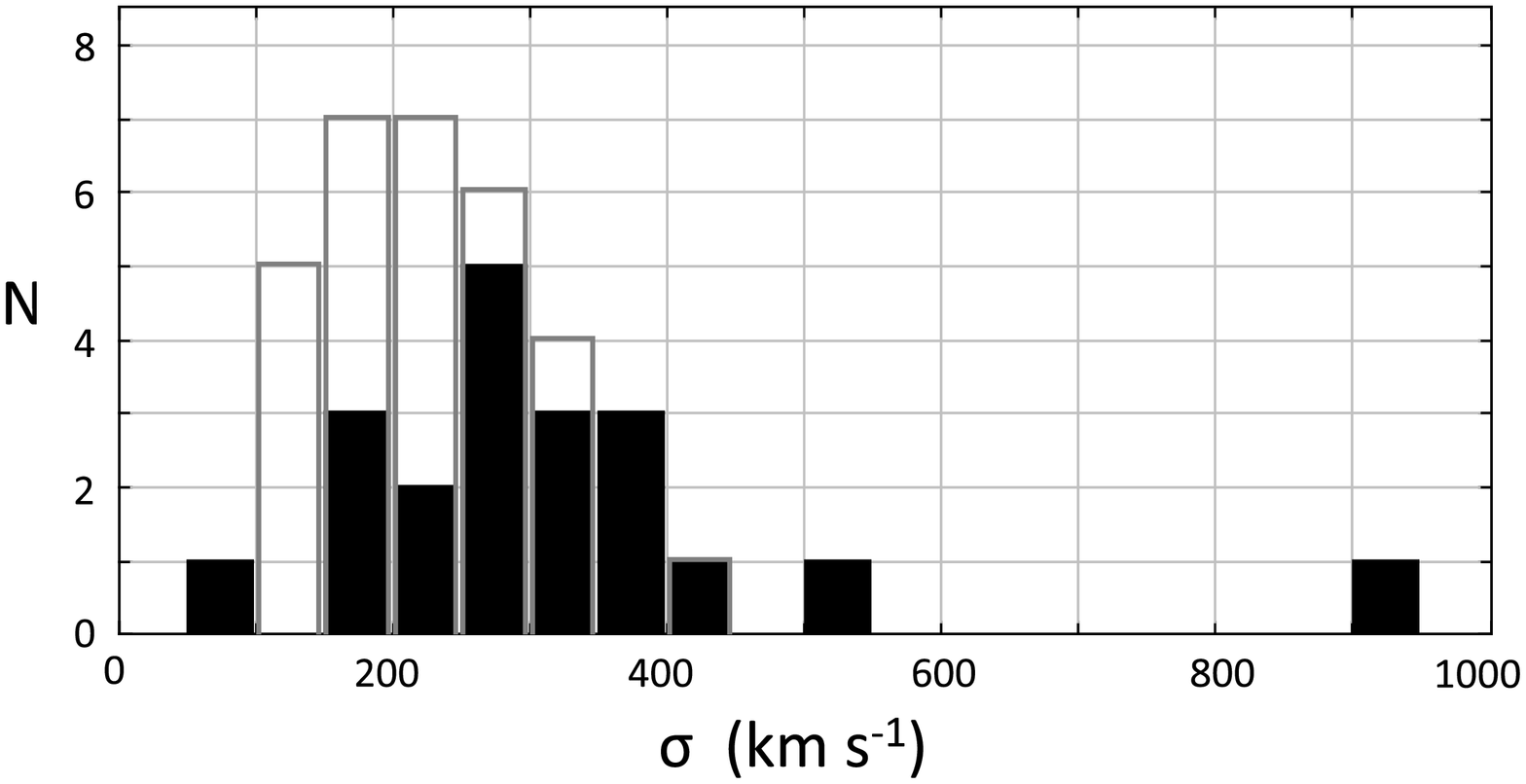}
}
\centerline{
\includegraphics[width=1.5in, angle=90]{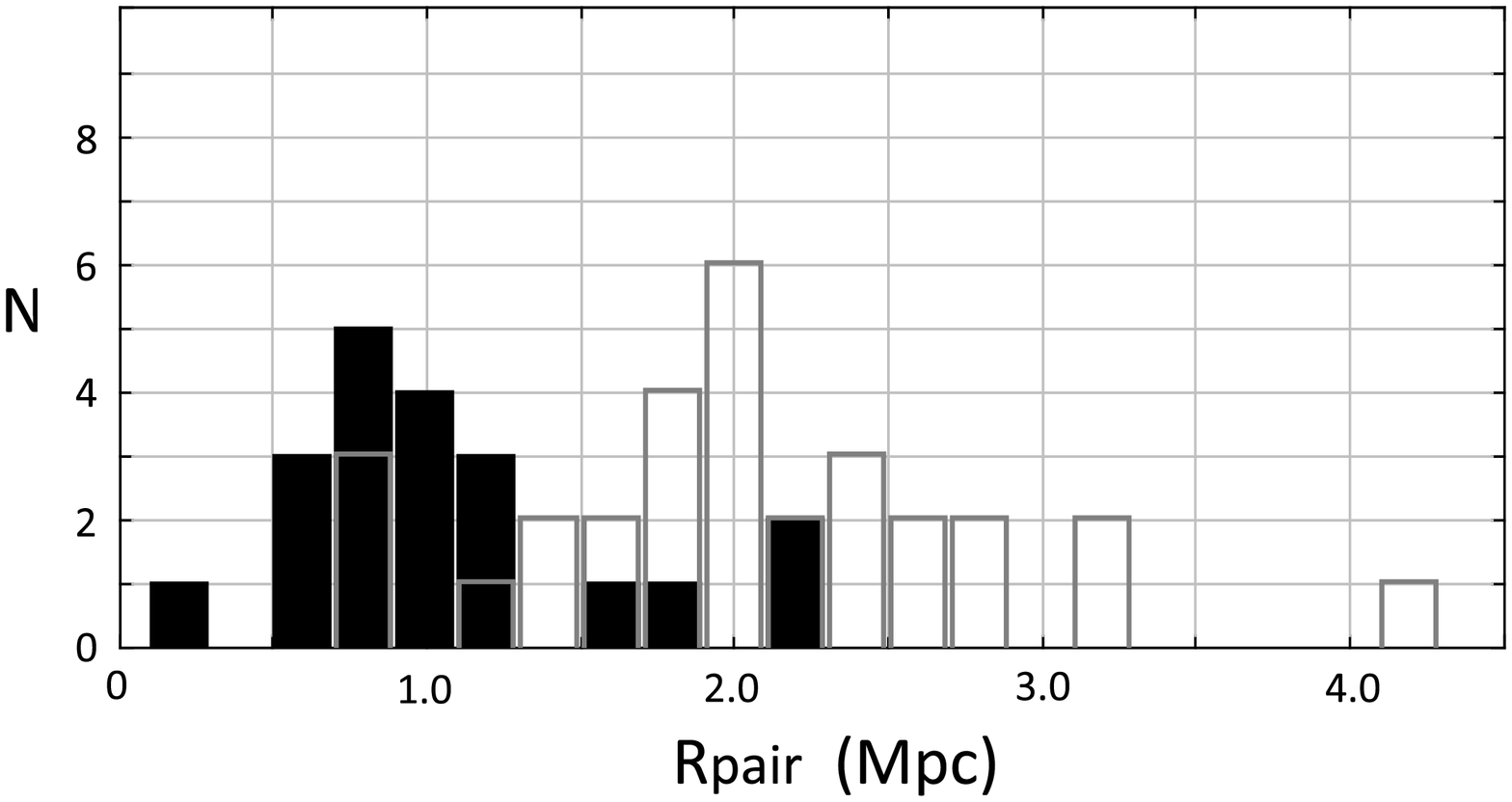}
%\hspace{-0.2in}
\qquad
\includegraphics[width=1.5in, angle=90]{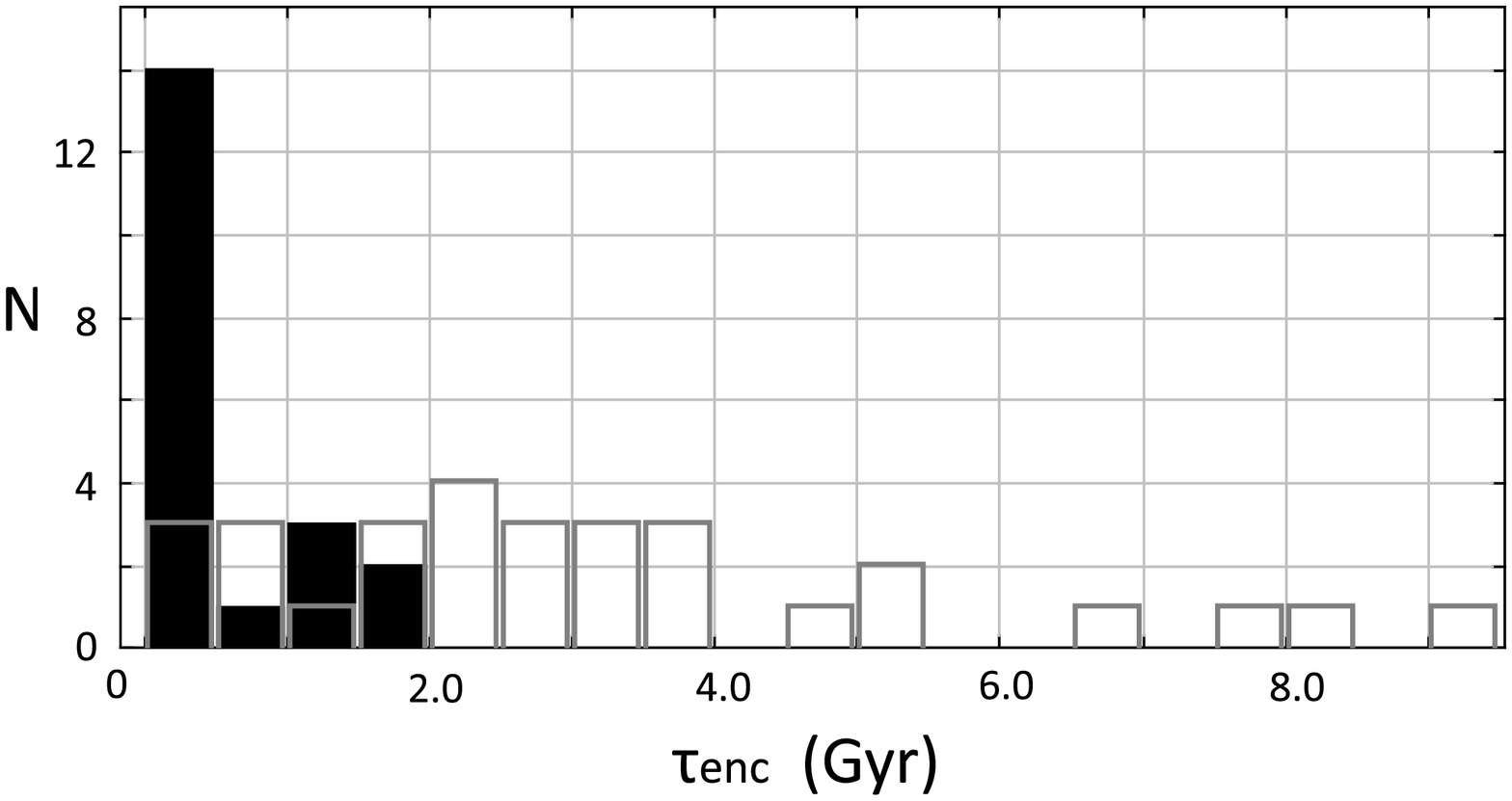}
}

\caption{Properties of the metacluster groups, and field groups and filaments.  Metacluster groups -- solid histogram; field 
groups and filaments --- open histogram.  a) (upper-left) Group luminosity \Lgal (see \S4.3); b) (upper-right) group velocity 
dispersion; c) (lower-left) Rpair (size) in Mpc; d); d) (lower-right) group encounter time.    The distribution of number of 
members (not shown) is very similar for the cluster and field (see Tables \ref{Cluster_Groups}, \ref{Field_Groups}, 
and \ref{Field_Filaments}) , as is mirrored in the distribution of Lgal upper left.  The velocity dispersions of the 
metacluster and field groups are also very similar, though the metacluster sample contains a significant number of 
higher-dispersion systems.  More distinct, however, is the difference in size, Rpair: field groups extend to substantially 
larger sizes, which accounts for the larger ``encounter times" --- the characteristic time in Gyr for a group member to 
encounter a neighbor.
\label{fig:group_properties}}

\end{figure*}

In Figure \ref{fig:group_properties} we show distributions of some basic properties for metacluster groups and cl\_field 
groups and  filaments.  The number distribution of group members is essentially the same for these two samples (see 
Tables \ref{Cluster_Groups}, \ref{Field_Groups}, and \ref{Field_Filaments}), but a more useful parameter is \Lgal\ --- the 
``total'' luminosity of the in units of L$^*$.   which is calculated from an extrapolation of a Schechter (1976) function to 
bring all the groups (sampled at different redshifts and luminosities) to the same richness scale.  \Lgal\ is a luminosity, 
but it is related to group \emph{stellar mass} by a stretched scale that reflects steadily increasing mass-to-light ratio of 
the growing PAS+PSB fraction, and a modest scatter of 20-30\% generated by the specific mix for each group of 
starforming and passive galaxies.   

%Figure 14:  Properties of groups and filaments

%Figure 15:  PAS+PSB fraction vs various group parameters 

\begin{figure*}[t]

\centerline{
%Rpair
\includegraphics[width=1.56in, angle=90]{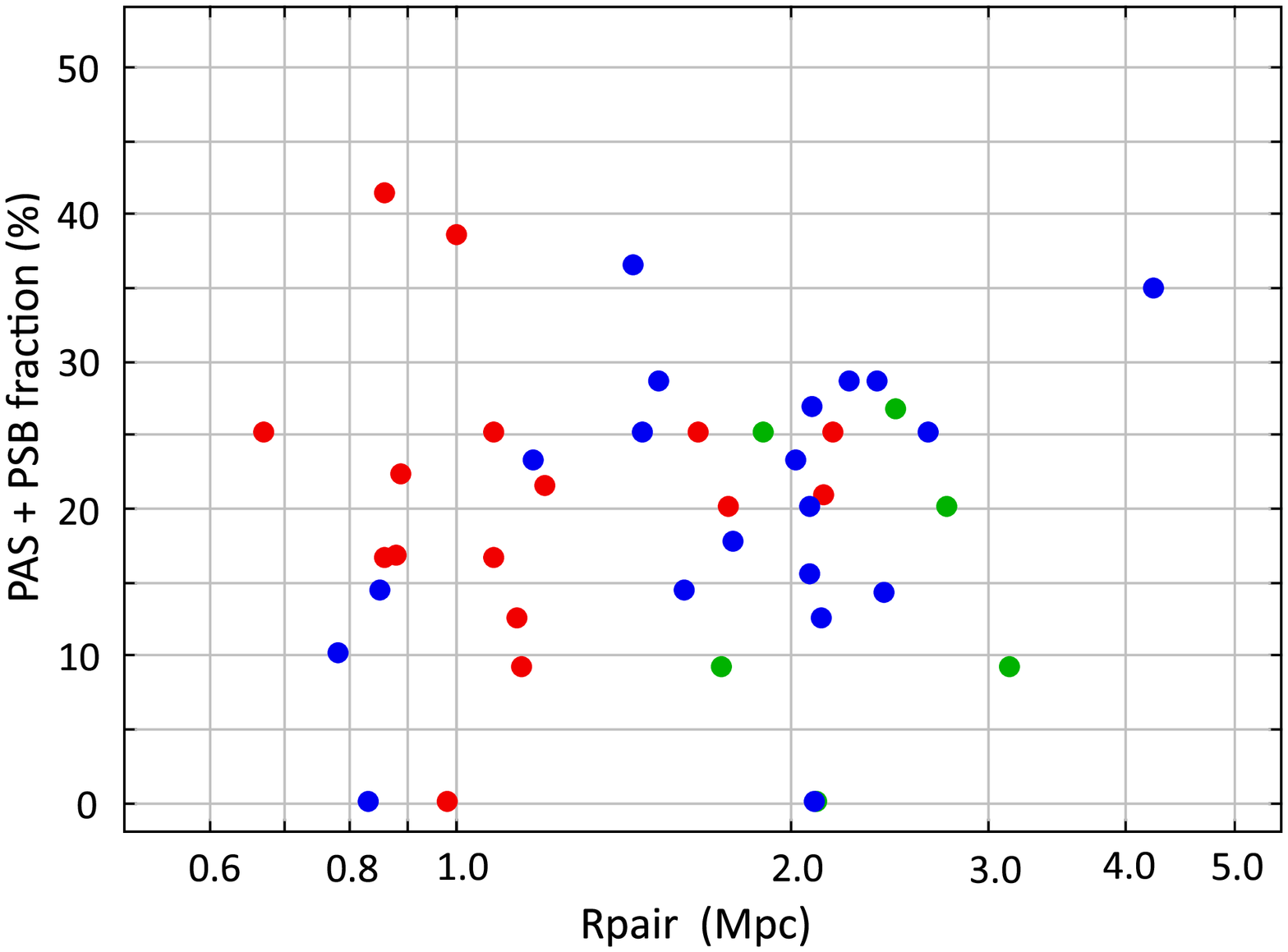}
\qquad
%sigma
\includegraphics[width=1.65in, angle=90]{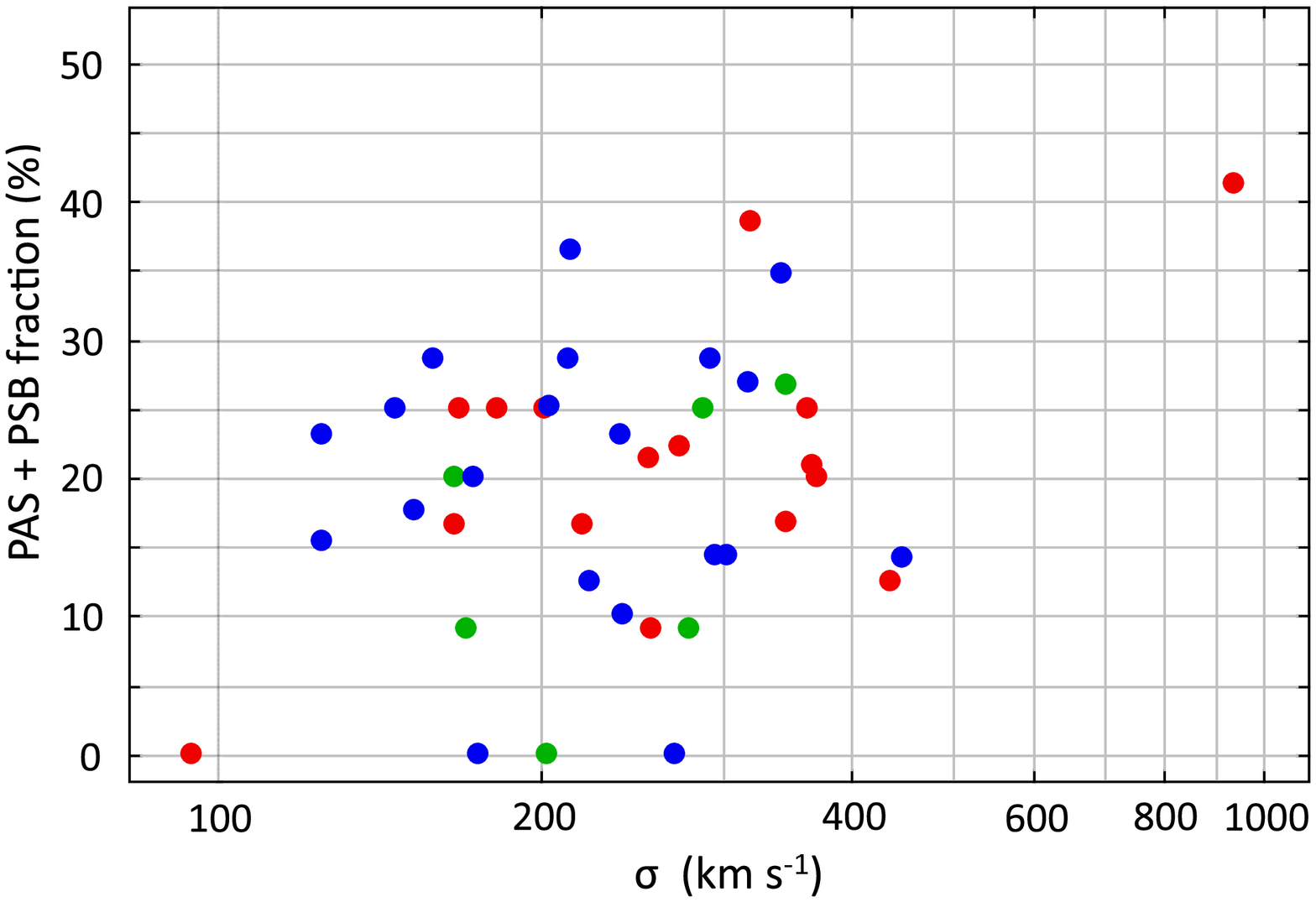}
%tau_encounter
\qquad
\includegraphics[width=1.65in, angle=90]{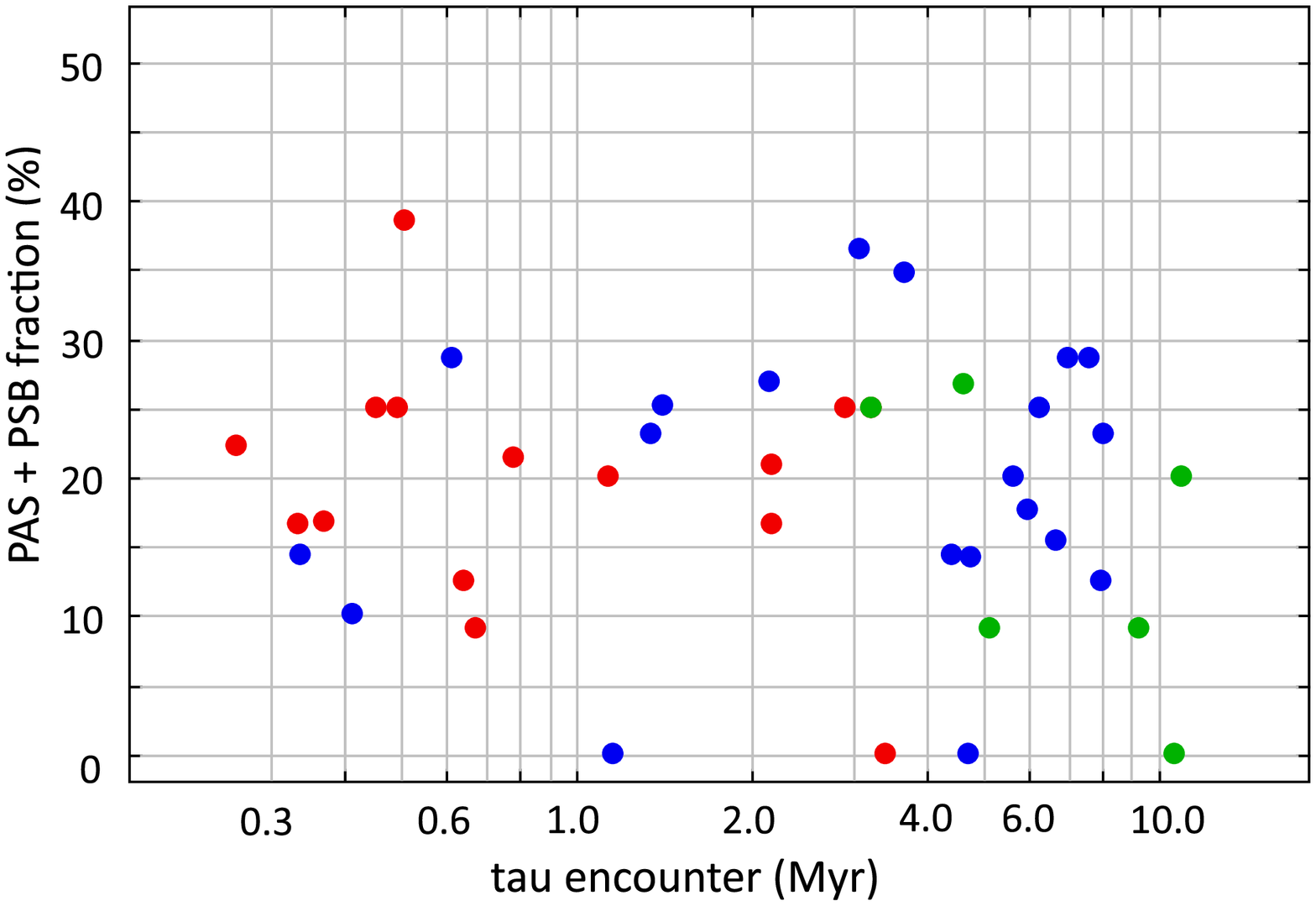}
}

\centerline{
%Rpair*sigsq
\includegraphics[width=1.60in, angle=90]{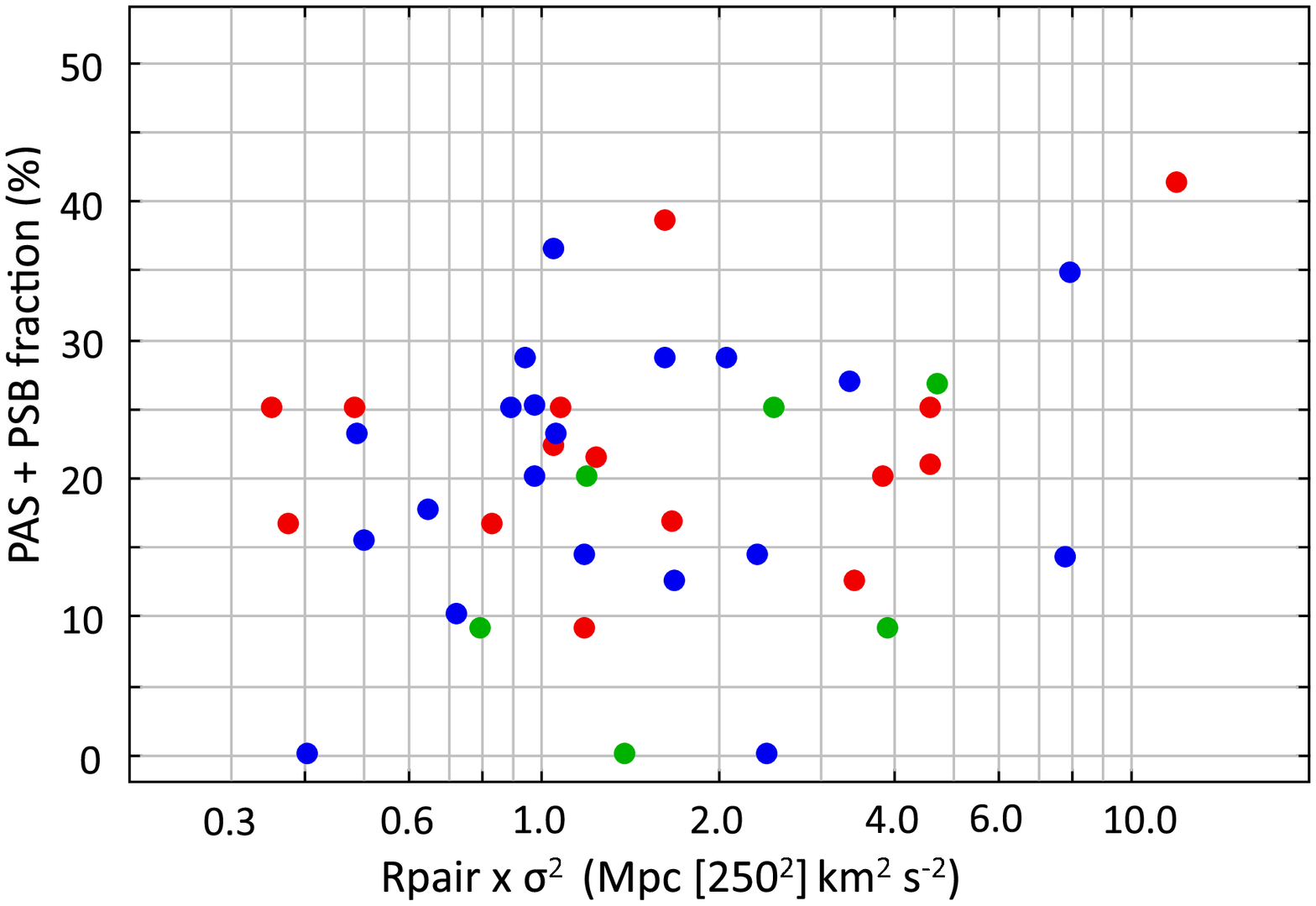}
\qquad
%Ntot
\includegraphics[width=1.65in, angle=90]{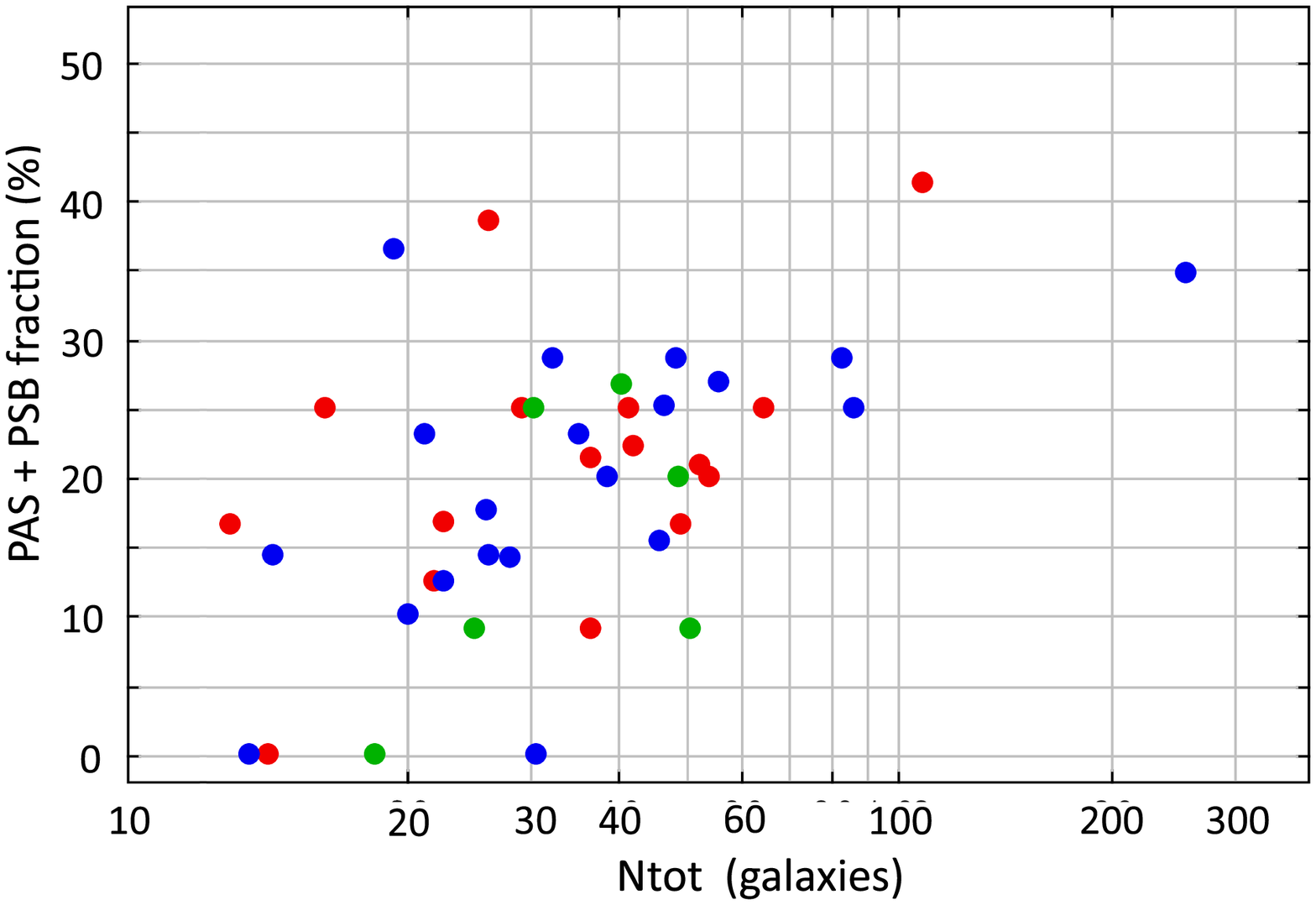}
\qquad
%Lgal
\includegraphics[width=1.65in, angle=90]{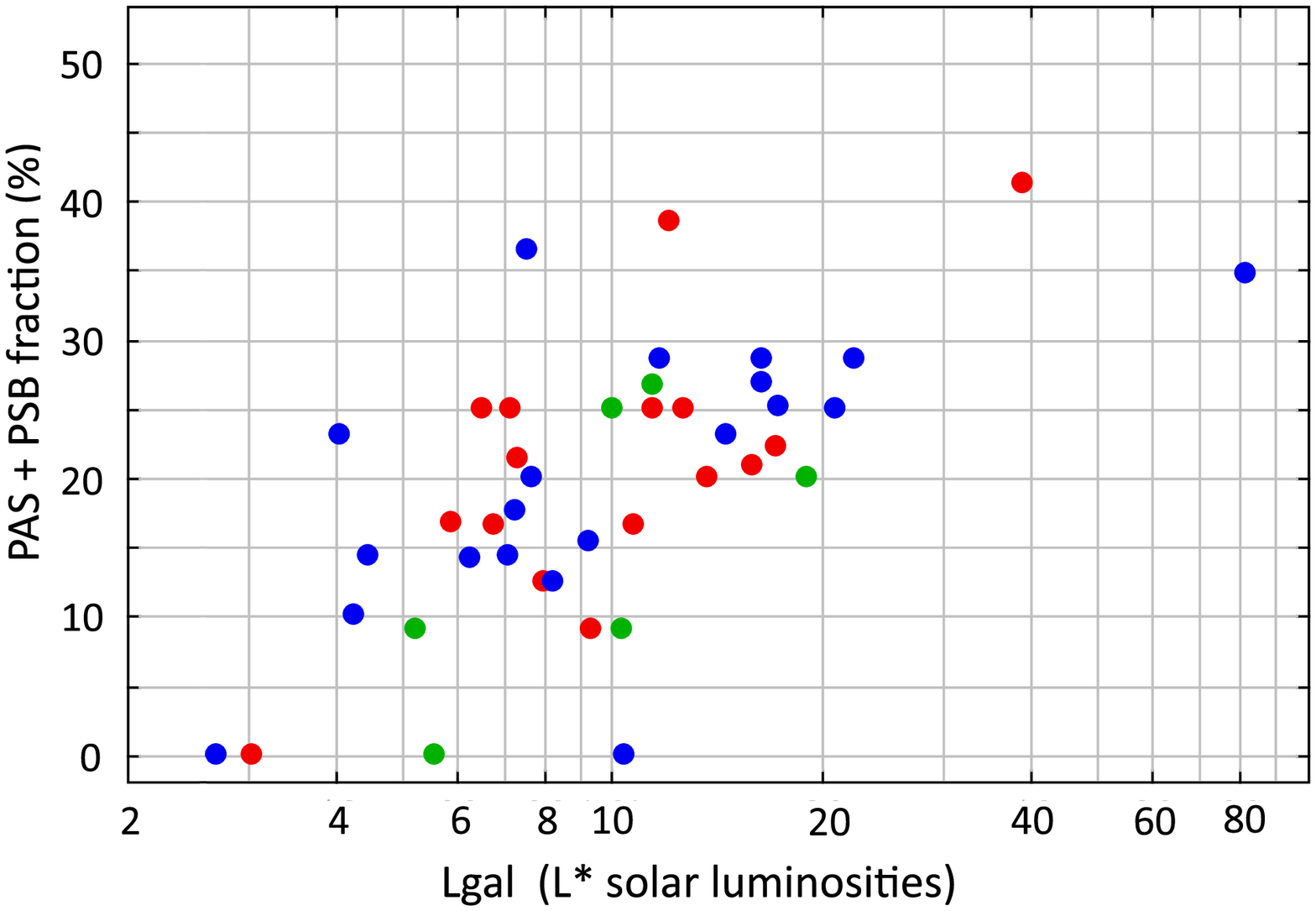}
}

\caption{Dependence of fraction of passive galaxies, PAS+PSB on various properties of the metacluster groups 
(red points), field groups (blue points), and field filaments (green points).  (The 4 groups projected onto the 
cluster cores (see \S3.2) have been excluded from this exercise.) The figures have been roughly ordered 
by the strength of the correlation.  (a-c) There is no significant correlation of the passive fraction with group size 
$R_{pair}$ (top-left), velocity dispersion $\sigma$ (top-middle) or $\tau_{enc}$ --- a typical time for any galaxy to 
encounter another group member (top-right). (d-f)  Correlations \emph{are} found with parameters describing the 
``richness" or ``scale'' of the group. There seems to be a weak correlation of $R_{pair} \times \sigma^2$ --- a 
measure of group dynamical mass (bottom left)  and a better correlation with the parameter \Ntot that is the 
number of galaxies Ntot, which is corrected for the different depths to which the groups are probed (bottom-middle).  
The best correlation (bottom-right) is with group luminosity, \Lgal, related to group mass by a stretched scale 
with modest scatter reflecting the increasing mass-to-light ratio of increasing PAS+PSB fractions.  Both \Ntot and 
\Lgal refer to the spectroscopic sample and should doubled to represent the richness and luminosity of the full photometric 
sample.  The \Lgal vs. PAS+PSB relationship is explored in more detail in Figure \ref{fig:Lgal_vs_PAS+PSB_fraction}.
\label{fig:group_correlations}}

\end{figure*}

Figure \ref{fig:group_properties}-a shows that the distribution \Lgal\ for the two samples --- metacluster groups and 
cl\_field groups --- is very similar, as is the distribution with velocity dispersion, $\sigma$ (Figure \ref{fig:group_properties}-b).  
However, the distribution of sizes, R$_{pair}$ (the mean of all pair separations), is clearly different for the two samples 
(Figure \ref{fig:group_properties}-c).  The field distribution overlaps the metacluster distribution but includes much larger 
systems. This may be a selection effect in that groups R$_{pair}>2$ Mpc are more difficult to pick out in fields dominated 
by a rich cluster, or it may be that such large, loose groups have been tidally dispersed, or their formation suppressed, 
in the supercluster environment.

We also calculate \tauenc\ --- a typical `interaction time' for a group member, moving at the speed of the velocity dispersion, 
to encounter another galaxy within a fairly large impact parameter, R$\sim$0.5 Mpc.  (The full photometric sample in these 
groups is a factor-of-two larger than the spectroscopic sample, so \tauenc\ has been divided by two.) Since this encounter 
time  depends linearly on the size of the group, there are some field groups with significantly longer times than those of the 
metacluster groups, all of which have \tauenc $\ls$2\,Gyr (Figure \ref{fig:group_properties}-d).  Three of the field groups 
have $\tauenc> 6$\,Gyr, a significant fraction of a Hubble time, long enough to doubt the reality of the group as a physical 
association.  These were dropped from the sample.

In Figure \ref{fig:group_correlations} we show the correlations of these various properties with the PAS+PSB fraction.  There 
is no significant correlation of PAS+PSB with Rpair, the characteristic group size (Figure \ref{fig:group_correlations}-a), 
or, perhaps more surprisingly, with velocity dispersion $\sigma$ (Figure \ref{fig:group_correlations}-b), or with
\tauenc, the characteristic interaction time (Figure \ref{fig:group_correlations}-c) --- also a scatter diagram. 

Poggianti \etal\ (2006 --- see Fig.\,10) have explored a relation like Figure \ref{fig:group_correlations}-b), but using
the fraction of starforming galaxies (the inverse of what we plot here).  Poggianti \etal\ find a correlation between $\sigma$ 
and the fraction of [O II]-emitting galaxies for clusters with $\sigma > 500$\,\kms, in the sense that this fraction is bounded 
at progressively higher  values as $\sigma$ decreases.  There is some evidence that this trend continues for poor clusters 
and groups, $\sigma < 500$ \kms, the range covered by the ICBS groups.  However, the dominant feature of this 
low-$\sigma$ part of the diagram is the wide scatter in the starforming fraction, with values ranging from 0\% to 100\%, with 
a median of about 50\%.  With such scatter and only 10 groups, it is hard to demonstrate a correlation between $\sigma$ and 
starforming fraction over this range.  This is at least consistent with the lack of correlation for the ICBS groups between the 
non-starforming fraction and the $\sigma$, but it is perhaps interesting that the ICBS sample does not have such a wide 
scatter: 37 out of 42 values range in starforming fraction 70-100\%, and the median value is 80\%.  Given the small 
Poggianti \etal\ sample for $\sigma<500$\,km s$^{-1}$, these differences may not be statistically significant.  Even so, the lack 
of a trend in the ICBS data for  these relatively cold groups suggests that $\sigma$ is a less reliable indicator of ``scale'' for poorer, 
less dynamically mature systems compared to the $\sigma>500$\,\kms\ clusters.  It is for this reason that we think simply counting 
up the total luminosity  or mass in galaxies  is the best way to look for a correlation with group/cluster scale, and it could be 
interesting to recast the Poggianti \etal\ plot in this way.

Correlations of the PAS+PSB fraction \emph{are} found with parameters describing the ``richness" or ``scale'' of the group. There seems 
to be a weak correlation of $R_{pair} \times \sigma^2$ --- a measure of group dynamical mass (Figure \ref{fig:group_correlations}-d), and 
a clear correlation with the parameter \Ntot, which is the observed N members corrected (like \Lgal) for sampling depth.  The best correlation 
 --- a very good one --- is with group luminosity, \Lgal (bottom-right). \Ntot\ and \Lgal\ (see Tables4, 5, \& 6) are normalized galaxy counts and 
luminosities for each group that were calculated by first correcting the observed galaxy population for incompleteness above the limiting 
magnitude of $r_{lim}=22.50$, then normalizing the luminosity and counts to the limiting absolute magnitude reached for $r_{lim}=22.50$ 
at a fiducial redshift of $z=0.30$. To do this, we used a Schechter function with parameters, as a function of redshift, determined from the 
analysis of the evolution of the field luminosity function described in Paper 3.  Both \Ntot\ and \Lgal\ refer to the spectroscopic sample and 
should be doubled to represent the richness and luminosity of the full photometric sample. 

It is especially interesting that Figure \ref{fig:group_correlations}-f shows a clear correlation of PAS+PSB versus \Lgal, which is 
essentially one with group mass, while the correlation of PAS+PSB with the dynamical mass, $R_{pair} \times \sigma^2$ 
(Figure \ref{fig:group_correlations}-d), is weak at best.  This suggests that total mass inferred from the total luminosity via 
galaxy mass-to-light ratios is more reliable than dynamical mass, probably because many of these systems are not virialized.  
Although it is less than obvious why total group mass should be the independent variable best correlated with the PAS+PSB 
fraction, this certainly seems to be the case for the ICBS sample, so we will investigate next this correlation, and its 
implications, for the group sample and the larger and smaller mass scales also covered by the ICBS data.

\subsection{Growth of the passive population with structure scale}

The good correlation we found in Figure \ref{fig:group_correlations}-f between passive galaxy fraction and the
total luminosity, \Lgal, suggests a process that occurs in the hierarchical assembly of galaxy groups --- 
environmentally-driven --- that converts some starforming galaxies into passive galaxies.  The nomenclature 
``preprocessing" refers to a mechanism that operates before such groups are incorporated into the even denser
environment of rich clusters, where unique mechanisms for suppressing or stopping star formation are expected.

For the groups infalling into four of the ICBS clusters, and the comparable field groups and filaments we have identified, 
group luminosity seems to be well correlated with the PAS+PSB fraction, while \tauenc, a measure of the 
galaxy-galaxy interaction rate, is not.  A possible explanation is that the passive fraction grows in discreet events 
associated with the building of larger and larger groups through hierarchical clustering, rather than a steady 
transformation from starforming to passive galaxies through galaxy-galaxy interactions as these stable groups age.  
This topic is explored further below.

\begin{figure*}[t]

\centerline{
\includegraphics[width=3.2in]{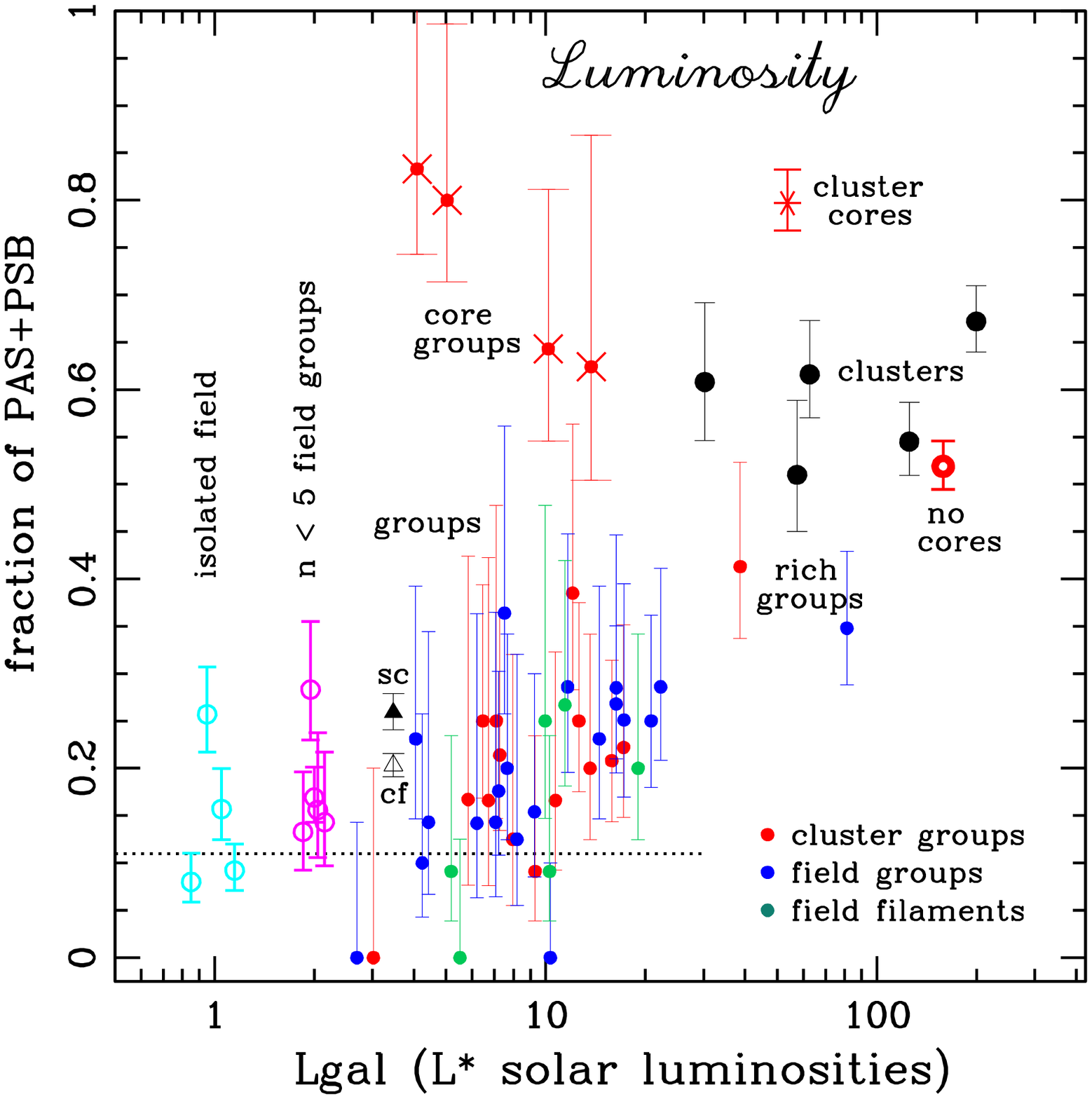}
\qquad
\includegraphics[width=3.2in]{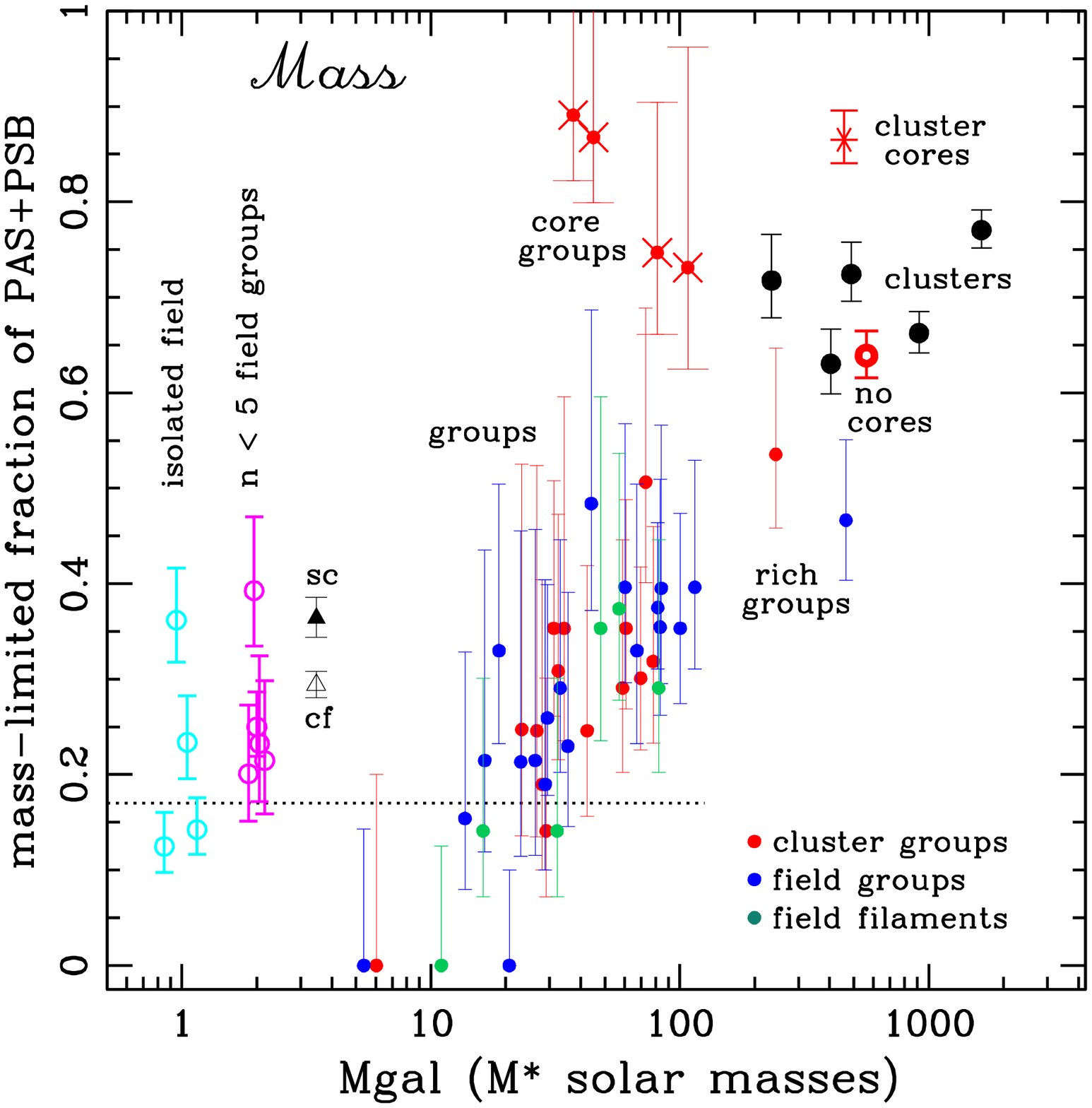}
}

\caption{Fractions of PAS+PSB spectral-type galaxies (a, left) as a function of \Lgal, the group luminosity, and (b, right) as a function 
of \Mgal, the group mass, derived from \Lgal\ through photometry and approximate mass-to-light ratios. The fractions of PAS+PSB 
galaxies in (a) are derived from the full spectroscopic sample, while the fraction in (b) are for a mass-limited sample calculated 
using Table \ref{mass limited samples}.  Cluster groups (red dots), field groups (blue dots), and field filaments (green dots) all show 
a trend of increasing PAS+PSB fraction with increasing \Lgal\ or \Mgal\ that we identify as ``preprocessing'' --- a turning-off of star formation
in some galaxies in the building of such groups.   A linear extrapolation of the trend for typical groups (\Lgal$<$25, \Mgal$<$100) intercepts 
two rich groups --- one in a metacluster and one in the cl\_field --- whose \Lgal\ and \Mgal\ values are within a factor of a few of the five 
clusters (black dots).  The clusters themselves (defined by Rcl/R$_{200}<1.5$) appear to lie somewhat above the linear extrapolation of the 
group trend, but the much higher values of the ``core groups" (red X's -- see \S3.2), and cluster cores (R$<$500\,kpc) alone suggest that 
one or more processes specific to this extreme environment, for example, ram-pressure or strong tidal stripping, has further boosted the 
passive population.  The remaining cluster with the core removed (labeled ``no core"), is perhaps consistent with the `group trend' for 
preprocessing, suggesting that only in the cluster cores is a more potent mechanism for passive production implicated.   Also shown for 
the cl\_field are isolated galaxies (open cyan circles, \Lgal, \Mgal\ $\equiv$ 1.0 ) and less populous groups (N$<$5, \Lgal, \Mgal\ $\equiv$ 2.0) 
identified with a friends-of-friends algorithm (open magenta circles).  The average supercluster populations (Rcl/R$_{200}\ge$1.5) are 
represented by a (single) black solid triangle  (\Lgal, \Mgal\ $\equiv$ 3.5) labeled ``sc,'' and compared to the average cl\_field, the open 
black triangle labeled ``cf.''  The data support a picture in which a ``floor" of $\sim$10\% (17\% by mass) fraction of PAS+PSB  galaxies, 
for isolated galaxies and small groups, grows as more massive groups are assembled, and implicates slow quenching mechanisms involving
galaxy-galaxy interactions, for example, tidal stripping or starvation of star formation through gas removal.   
\label{fig:Lgal_vs_PAS+PSB_fraction}}

\end{figure*}

In Figure \ref{fig:Lgal_vs_PAS+PSB_fraction} we expand this discussion to the other environments explored in the ICBS.
Figure \ref{fig:Lgal_vs_PAS+PSB_fraction}-a casts the relation in the observational parameters of our magnitude-limited 
(or luminosity-limited) sample --- from isolated field galaxies, through group galaxies, to rich clusters and their cores.
We add the ``core" groups, metacluster groups projected on the cluster cores (see \S3.2) that were not included in the
in Figure \ref{fig:group_correlations} of \S4.3.

We now add Poisson error bars for the cluster groups, field groups, and the subset of field groups that are filamentary, 
and again code them by red, blue, and green, respectively.  It is remarkable that the increasing PAS+PSB fraction with 
\Lgal\ appears the same in all three samples, given the different environments of superclusters and the field, and the 
clearly different structure of filaments.  If verified by other, independent samples, this correlation suggests a process that 
is truly generic. 

%Figure 16:  Lgal vs PAS+PSB fraction across environment

In Figure \ref{fig:Lgal_vs_PAS+PSB_fraction}-b we recast Figure \ref{fig:Lgal_vs_PAS+PSB_fraction}-a as mass-limited
passive fractions as a function of mass scale, using simple relations to accomplish this transformation.  The passive
fraction in the luminosity-limited sample is converted to a mass-limited (M $\ge2.5\times10^{10}$\,\Msun) passive 
fraction by using the corrections of Table \ref{mass limited samples} for PAS and PSB galaxies.  Since these types have 
the largest fraction of galaxies above the mass limit, making this conversion raises the passive fraction, by a factor of 
$\sim$1.5 for the group sample.  \Lgal\ values are converted to \Mgal, the total mass  in units of M$^*$ --- the characteristic 
mass scale that corresponds to L$^*$ in the Schechter (1976) parameterization.  \Mgal\ is estimated from \Lgal\ by assigning 
M/L = 10 for the fraction of galaxies that are PAS or PSB, and M/L  = 2 for all starforming types.  Applying these corrections 
stretches the axes of Figure \ref{fig:Lgal_vs_PAS+PSB_fraction}-a in a non-uniform way.  Although there are no additional data 
added in this process, Figure \ref{fig:Lgal_vs_PAS+PSB_fraction}-b is more likely to present a clearer picture of this correlation, 
one we think may be helpful in addressing in particular one interesting question about the possible departure of cluster samples 
from the group samples, as described below.

A relevant check on our measurements of passive fraction can be made by comparison with the investigation by Balogh 
\etal\ (2009) of star formation in field groups culled from the CNOC study (Carlberg \etal\ 2001).  This sample covers the 
redshift range $0.25<z<0.55$ and is probably the one in the literature most comparable in basic parameters to the ICBS groups. 
Based on broad-band photometry, Balogh \etal\ report a passive fraction of 45$\pm$7\% for their faintest group samples, which 
are roughly the same depth of the ICBS survey.  The Balogh \etal\ groups are systematically more massive systems, with a mean 
$\sigma\sim350$ \kms, the upper mass limit of the ICBS group sample.  Since the Balogh \etal\ results are for mass-limited samples, 
albeit one with a lower mass limit  ($10^{10}$\,\Msun) than the ICBS sample, we refer to Figure \ref{fig:Lgal_vs_PAS+PSB_fraction}-b 
and see that the ICBS groups (with the exception of two more massive groups) end at a passive fraction of $\sim$40\%.  This is good 
agreement, but it may be fortuitous: the lower galaxy mass limit of Balogh \etal\ groups should have led to a lower passive fraction, but 
this is likely more than compensated by a SFR limit that appears to be 2-3 time less sensitive than the ICBS.  This difference is due
to the relative insensitivity of broad-band photometry to low-levels of star formation compared to spectroscopic features that can be
readily measured for SFRs of 1 \Msun\ yr$^{-1}$ or less.  

The $\sim$15\% passive fraction Balogh \etal\ find for the field population is in good agreement with the ICBS value that we 
now discuss.  With the full range of environments covered by the ICBS, the dependence of PAS+PSB on \Lgal\ and \Mgal\ 
that we found for the groups in Figure \ref{fig:Lgal_vs_PAS+PSB_fraction} can be widened to include smaller and larger 
systems.  Field galaxies that are not members of the groups and filaments listed in Tables \ref{Field_Groups} \& \ref{Field_Filaments} 
have been subdivided into galaxies that are (1) truly isolated (to the depth of our sample, roughly M$^*+2$), and (2) galaxies 
in smaller groups ($N<5$) as found by a friends-of-friends algorithm (see Paper 3).  Most of the latter are relatively compact 
pairs and triplets, so we have assigned for purposes of display \Lgal\,\,(\Mgal) $\approx1.0$ for the isolated galaxies and 
\Lgal\,\,(\Mgal) $\approx2.0$ for the $N<5$ field groups.  For both these samples there seems to be a floor of the PAS+PSB 
of $\sim$10\% (17\%) for the luminosity-limited (mass-limited) sample.  This is consistent with the smallest systems in the 
$N\ge5$ group sample: these also scatter around 10\% (17\%) --- the small groups that contain no PAS+PSB galaxies are 
merely statistical fluctuations.  In other words, there is a base level of about $\sim$10\% (17\%) PAS+PSB galaxies that is 
found for small groups and isolated galaxies.  It is reasonable to speculate that these have been in place for a relatively long 
time ($z>1$), and that, as is well known for massive galaxies, averaged-sized galaxies can also reach a terminal state of star 
formation, either from very early processes that are properly thought of as early galaxy assembly, $z>2$, or through later 
processes such as major mergers or starvation at $0<z<2$.

%Figure 17:  maps comparing RCS1102B cl_field group 10 and SDSS1500B group 2.

\begin{figure*}[t]

\centerline{
\includegraphics[width=2.5in, angle=90]{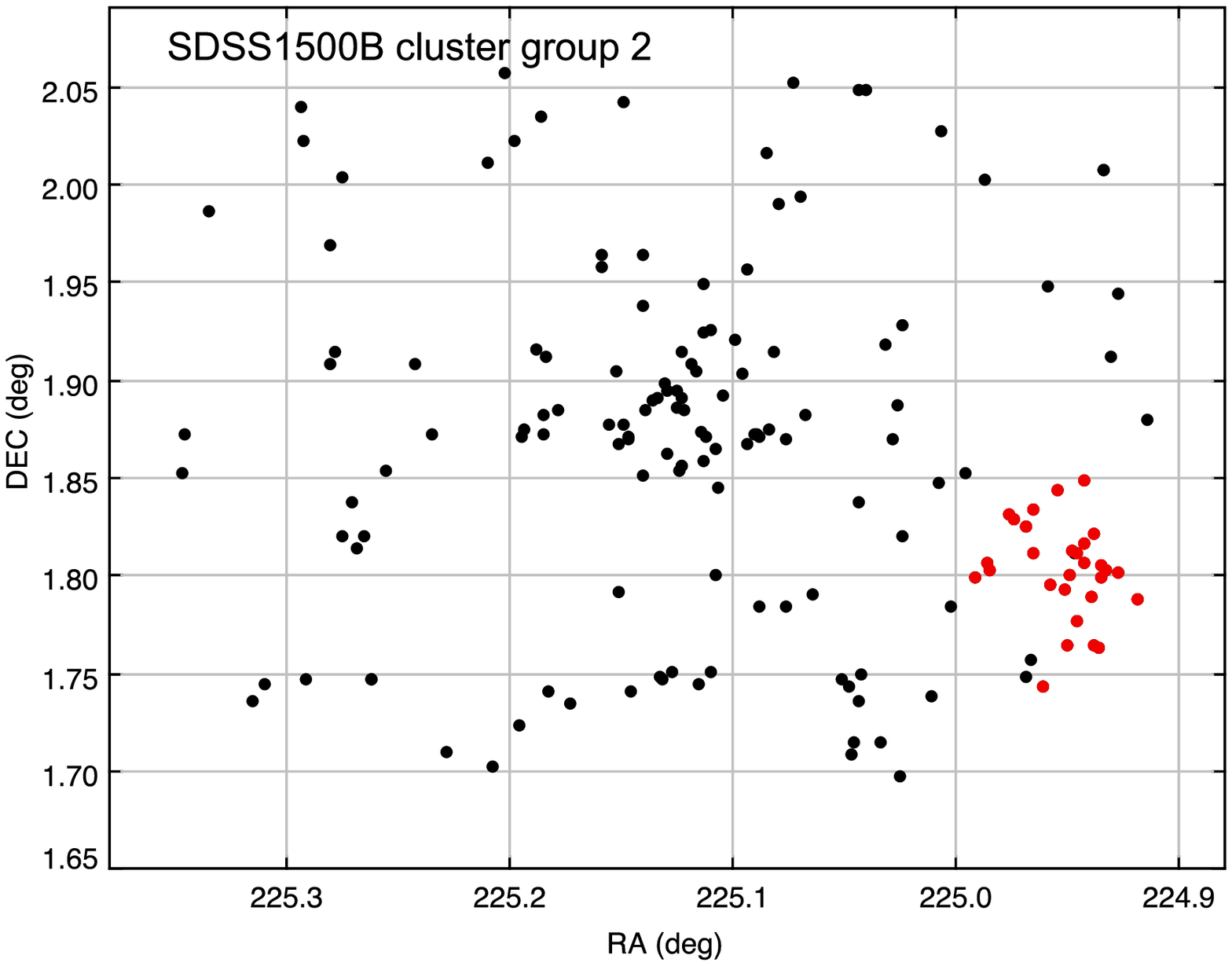}
\qquad
\includegraphics[width=2.5in, angle=90]{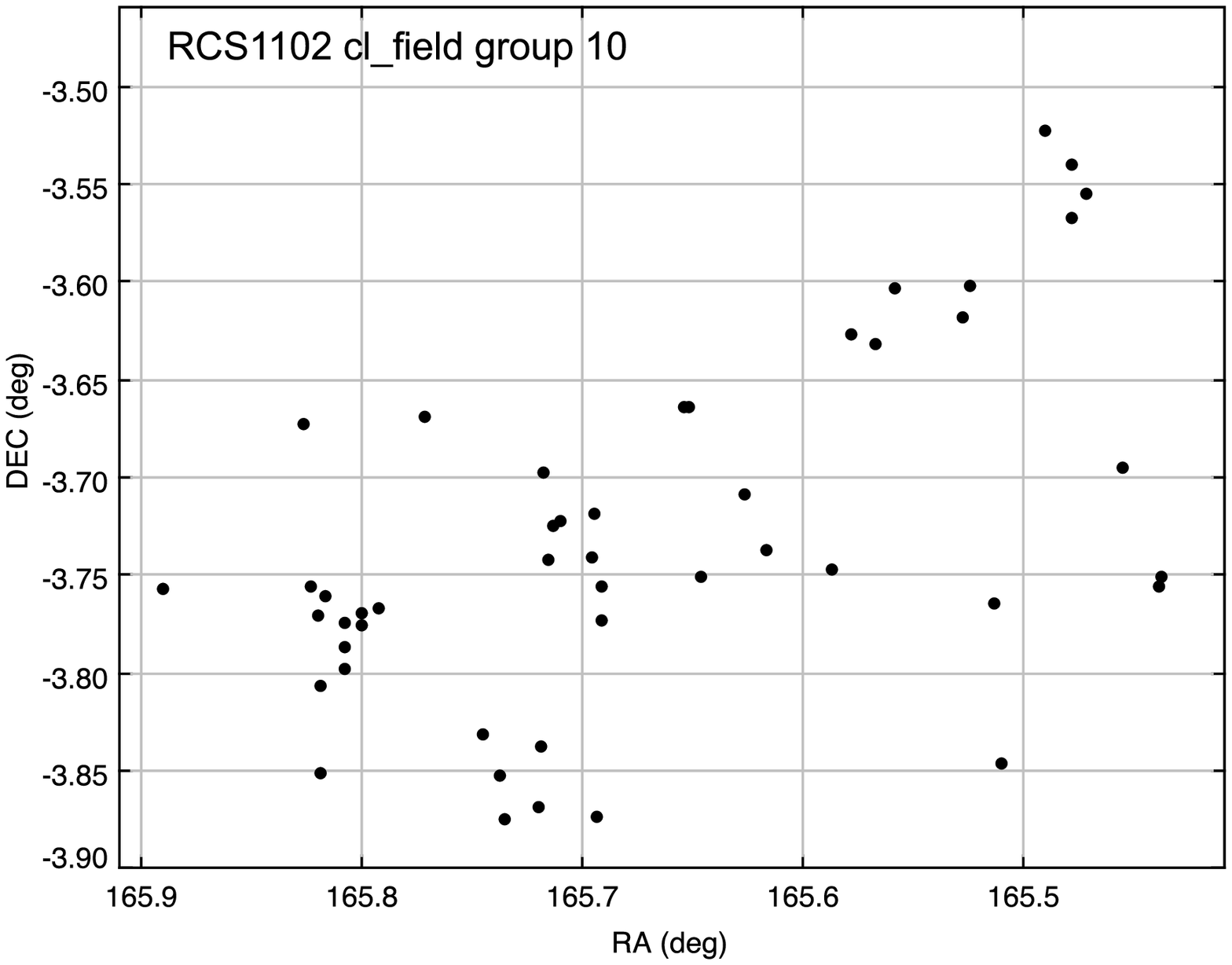}
}

\caption{Two rich groups, Group 2 (z=0.5175) 1in the metacluster SDSS1500B, and Group 10 (z=0.4992) in the 
cl\_fielddistribution of RCS1102.  The two groups have similar richness (Ntot and Lgal) and spectral type 
distribution (see the ``rich groups'' marked on Figure \ref{fig:Lgal_vs_PAS+PSB_fraction}), but are very different 
in structure.  PAS and PSB galaxies are in a very concentrated structure in SDSS1500 group 2 but widely 
distributed in RCS1102 cl\_field group 10.
\label{fig:two_rich_groups}}

\end{figure*}

At the other end of the \Lgal\ \& \Mgal\ scales in Figures \ref{fig:Lgal_vs_PAS+PSB_fraction}-a \& 16-b, we note that the richest of 
the infalling cluster groups, SDSS1500B-2 (\Lgal$\sim$40, \Mgal$\sim$200, PAS+PSB=41\%), and the richest field group, 
RCS1102-10 (\Lgal$\sim$80, \Mgal$\sim$400, PAS+PSB=35\%), lie on an extrapolation of the trend of PAS+PSB versus 
\Lgal\ or \Mgal\ established by the typical cluster and field groups, \Lgal$<$25, \Mgal$<$100.  As Figure \ref{fig:two_rich_groups} 
shows, these rich groups have very different structures: the cluster group SDSS1500B-2 is as concentrated as the cores of the 5 
ICBS clusters; the field group RCS1102-10 is spread over the entire \imacs\ field and may in fact continue to the southeast 
(lower left).  While the PAS and PSB galaxies in SDSS1500B-2 are, of course, limited to a high-density environment, these 
spectral types are also found in RCS1102-10 in similar abundance, within its full range of environments from high-density knots 
to medium-density groups to isolated galaxies.  This may be an expression of the spectral-type/local density relation (Figure \ref{fig:sptype_density_radius}) --- the global environments are quite different, but from a ``local" perspective, there seems a sufficient 
volume of high-density environment in RCS1102-10 to preprocess the $\sim$35\% (47\%) passive fraction. 

For the rich cluster environment we use the samples we extracted from the full metacluster sample sample by splitting at Rcl/R$_{200} = 1.5$, 
as we have discussed earlier.  The PAS+PSB fraction of $\sim$60\% ($\sim$70\%) for each of the five ICBS clusters appears significantly 
above the two rich groups and the extrapolation of the trend for the typical groups, for both the \Lgal\ and \Mgal.  However, we have no
way of knowing the underlying relationship, which in this semi-log diagram may surely depart from linear, so this offset from the extrapolation
for smaller scale systems is only suggestive.  We believe, however, that the very high PAS+PSB fractions of the four metacluster groups 
projected on the cores (the ``core groups'' see \S3.2) offer additional insight into the possibility that the cluster environment is ``special."  
These four groups are all rich in passive galaxies --- 60--80\% (70--90\%), far above the trend for the remaining groups, and they all 
have systemic velocities substantially off the cluster mean.  It seems clear, then, that these groups have been affected by the extreme 
environment of the cluster core, and a rapid conversion of starforming galaxies into passive galaxies has been the probable result.

Encouraged by this observation, we created a ``cluster core" population (see Table \ref{Other_Samples}) where all galaxies within 500 kpc 
of the cluster center in each of the five clusters are gathered together, and a ``no core" sample (a composite of what is left when the 
cores are removed).  The ``cluster core" point rises to the passive fraction of the ``core groups" --- 80\% (88\%), and the ``no core" 
cluster sample falls to a passive fraction of 52\%, (64\%), which is arguably consistent with the group trend that we have identified as 
preprocessing.  We suggest, then, that the process(es) that are occurring in galaxy groups to raise the passive fraction are sufficient to 
account for all but the highest passive fractions, the ones found in cluster cores, and for the fast-moving, high PAS+PSB groups that 
are associated with them.  

The basic conclusion to be drawn from Figure \ref{fig:Lgal_vs_PAS+PSB_fraction} is that there is considerable preprocessing in groups that 
raise the passive fraction substantially above the $\sim$10\% (17\%) level found for isolated field galaxies and those in poor groups.  It
appears that a much of the high fraction of passive galaxies in dense environments, up to a level of at least $\sim$40-50\%, could be  
from processes in modest-sized groups.  Beyond that, there is persuasive evidence that cluster cores `drop the hammer' on what is left 
of star formation in cluster galaxies.

\subsection{Cluster building, preprocessing, and implications for the ``quenching" of star formation in galaxies}

Our results concerning infall into the ICBS clusters show that this is an epoch of substantial growth in the history
of massive clusters at $z\sim0.5$.  Choosing clusters by their richness instead of strong X-ray emission, the ICBS 
shows how more typical rich clusters grew during this epoch.  For RCS0221A, RCS1102B, SDSS1500A, and 
SDSS1500B, we find an easily identifiable infall of groups comprised of 257 galaxies, and estimate that 
100-200 additional galaxies are either isolated or in small groups.  The number of infalling galaxies in these
four fields is roughly equal to the virialized cluster population.

``Quenching'' is a popular shorthand for a process capable of ending star formation in starforming galaxies. 
As described by Peng \etal\ (2010), quenching refers to one or more physical processes, driven internally 
(e.g., secular evolution and starbursts {\it within} a galaxy) or externally (e.g., mergers or ram-pressure 
stripping).  These authors distinguish this from the general decline in the SFRs of starforming galaxies since 
$z\sim1.5$, probably the result of a decline in available gas that is capable of sustaining star formation.  
This separation may not be a clean one, however, since the decline in available gas may itself be a function of 
environment: many of today's passive galaxies, especially massive ones, may have ceased star formation at very 
early times because they quickly processed the accessible gas into stars.

Our results here suggest that some $\sim$10\% of galaxies, even in the lowest density environments, had already 
ceased significant star formation by $z\sim1$.  Presumably this could be a mix of early mergers of individual
galaxies, fossil groups, or the occasional massive galaxy with unusually efficient star formation at $z\sim2$ that 
exhausted the local gas supply.  Measurements of properties of this ``base level" passive population at very early
times,  and studies of mass and luminosity functions, morphology and structure over cosmic time, should be able 
discriminate which paths lead these galaxies to a permanently passive state.

\begin{figure*}[t]

\centerline{
\includegraphics[width=3.05in]{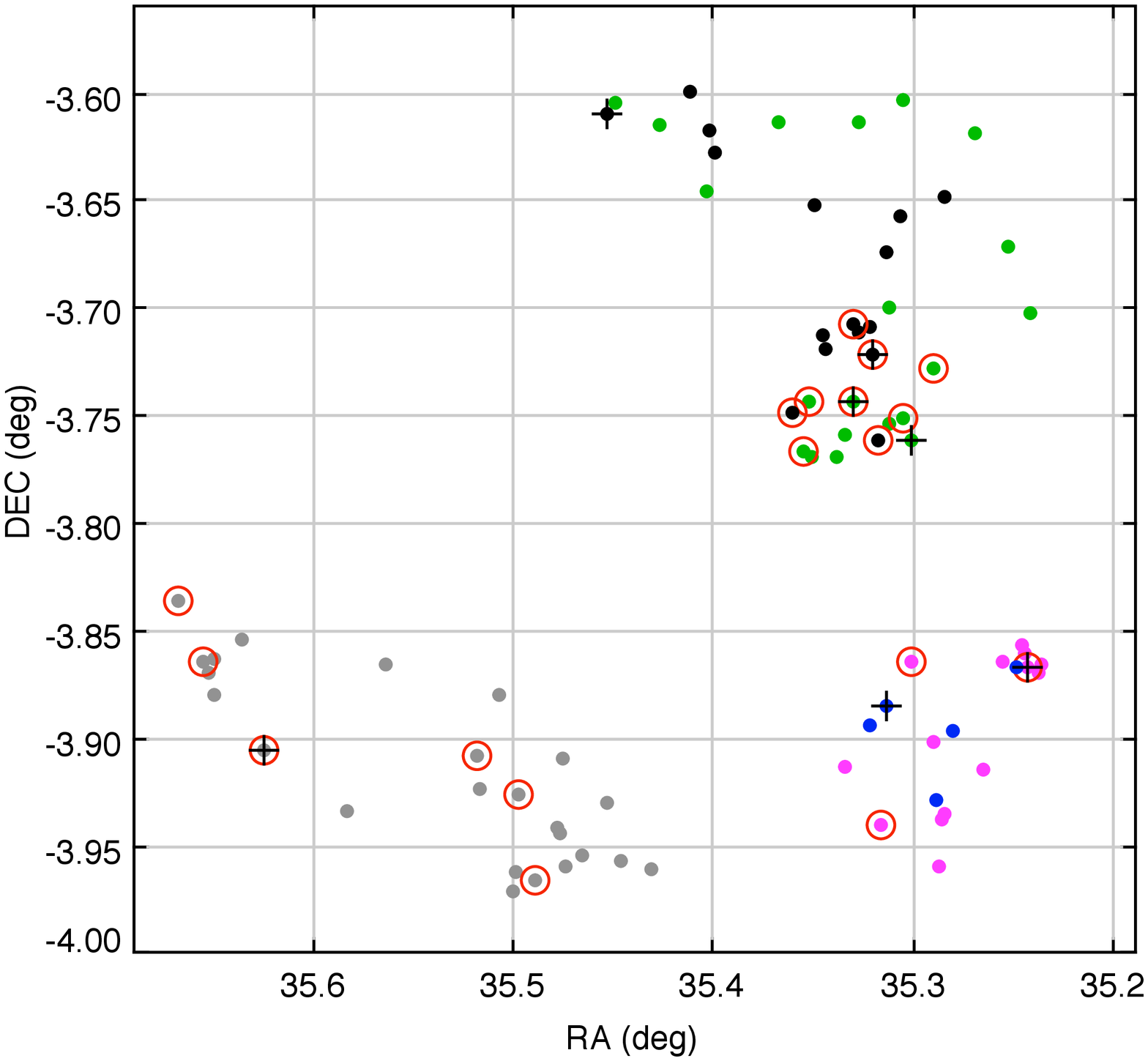}
\qquad
\includegraphics[width=3.05in]{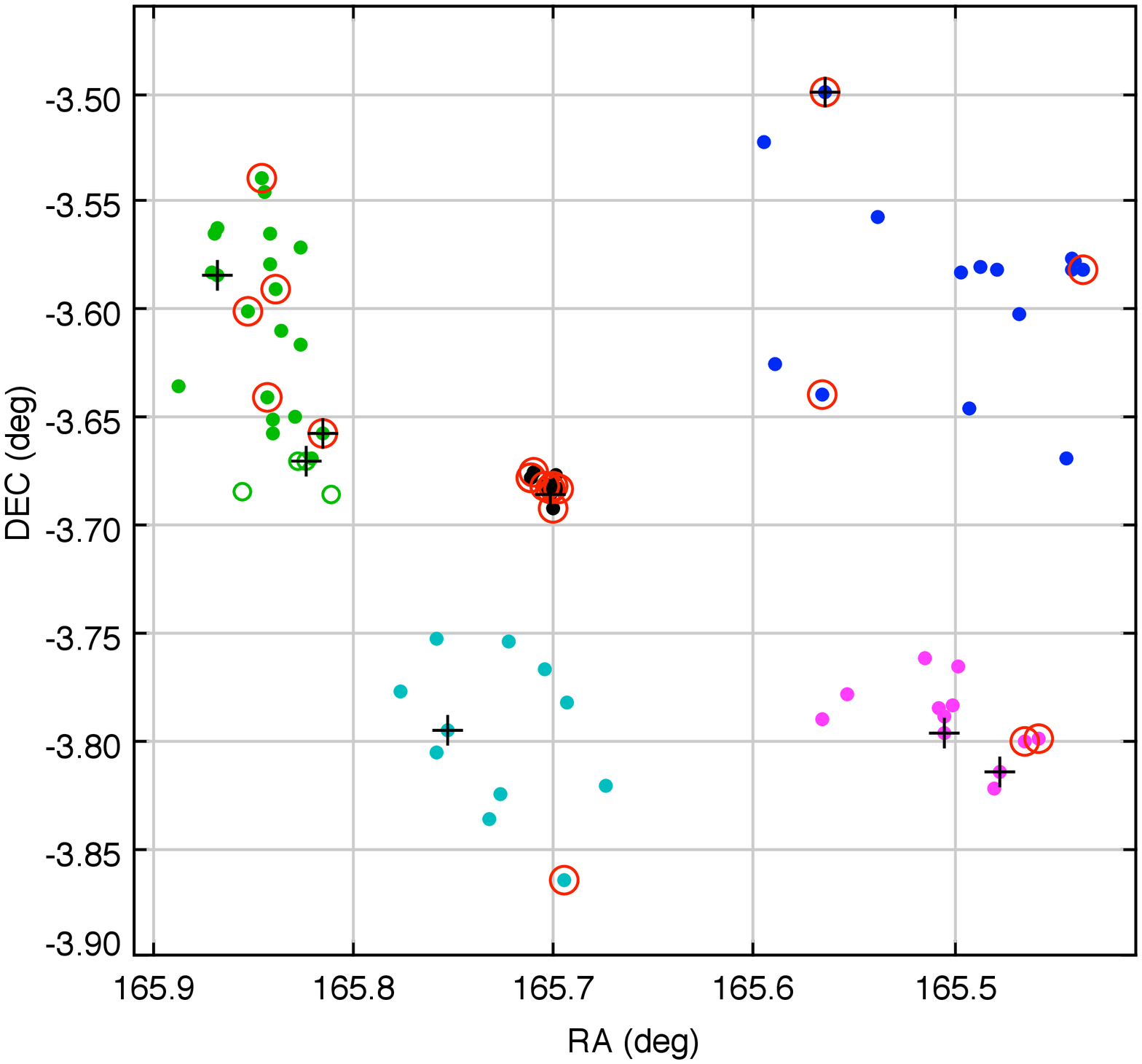}
}

\centerline{
\includegraphics[width=3.00in]{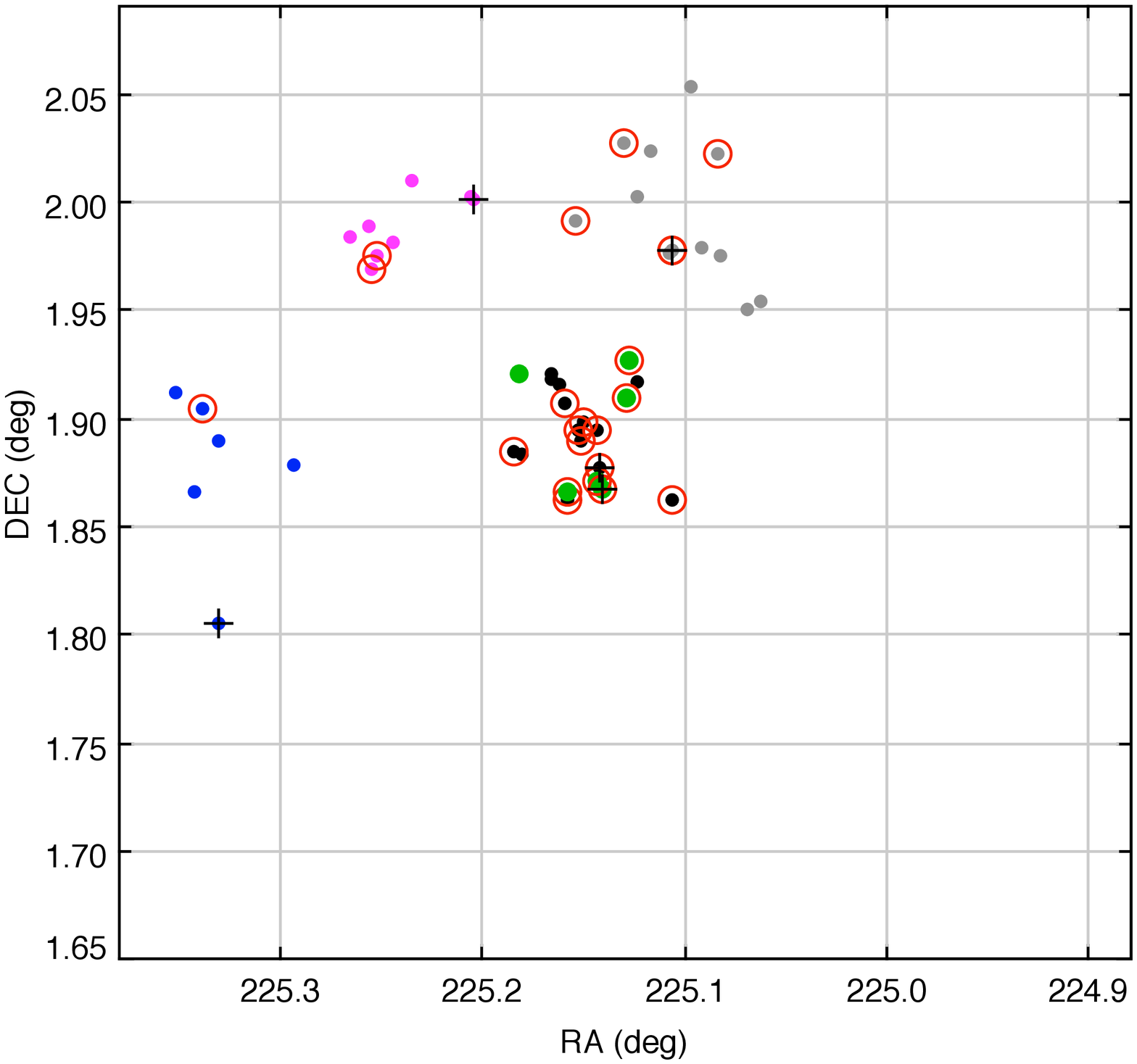}
\qquad
\includegraphics[width=3.00in]{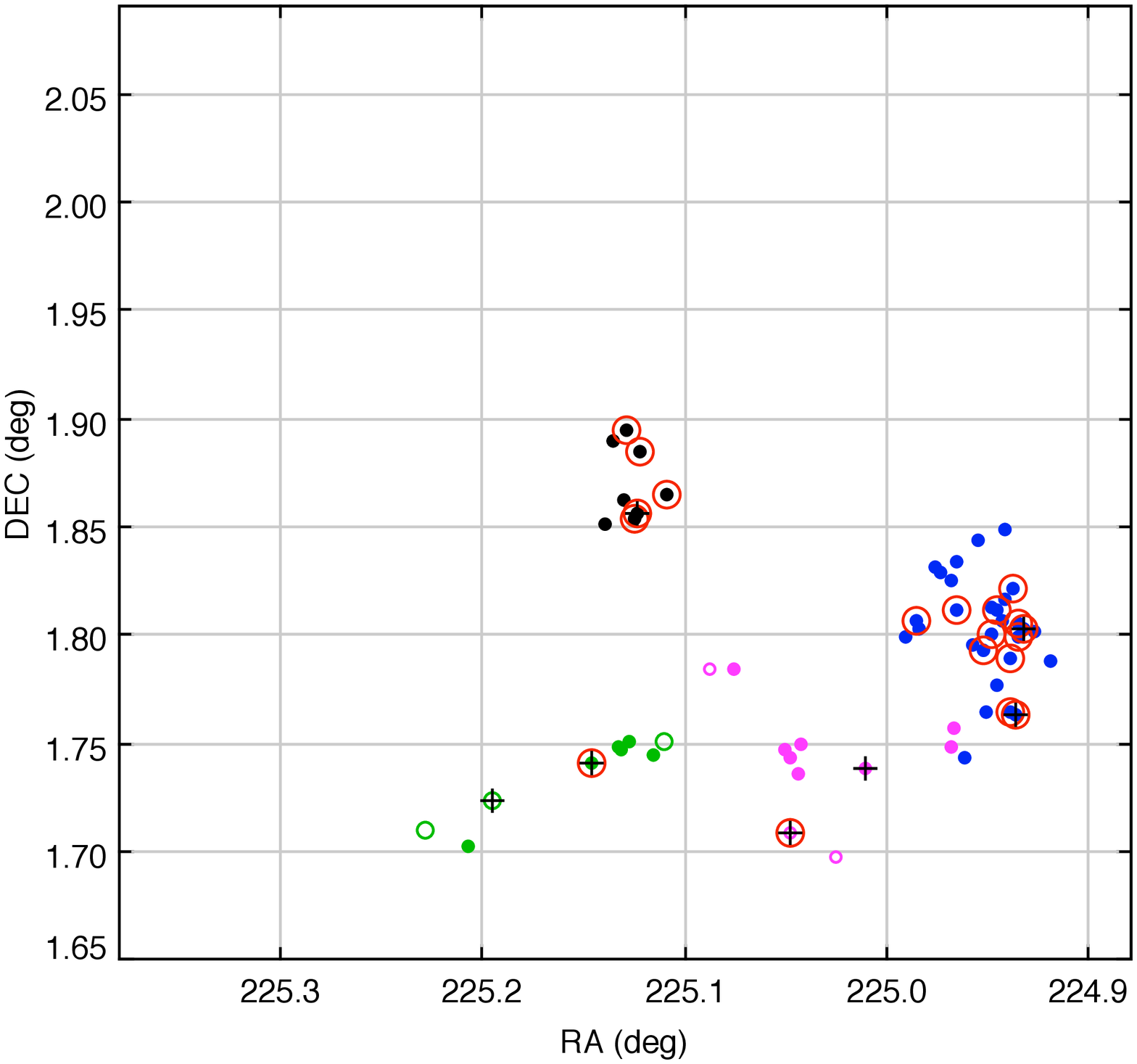}
}

\caption{Identification of PAS and PSB spectral type galaxies in the metacluster groups, marked by open circles.  
These types are found in locally dense environments, consistent with the idea that incorporation into a larger halo
system ends star formation in a ``satellite'' galaxy, but they also fairly common in low-density or even
isolated environments where no such more-massive companion can be responsible.
\label{fig:cluster_groups_PAS_PSB}}

\end{figure*}

From this base level, we see a clear signature of increasing fraction of passive galaxies once the mass scale of a group rises 
above a few L$^*$ (M$^*$).  As we have shown here, this increasing fraction is usually associated with higher density environments, 
but some passive galaxies are also found in relatively low-density parts of these moderate-sized groups.  A prevalent idea that 
has come from $\Lambda$CDM simulations is that passive galaxies are satellites whose halos have  been incorporated into 
more massive ``central'' galaxies, as discussed in van den Bosch \etal\ (2008, see also Weinmann \etal\ 2010).  In this
connection, we show in Figure \ref{fig:cluster_groups_PAS_PSB} the distribution of passive galaxies in the ICBS cluster groups.  
Although our spectroscopic sample is only $\sim$50\% complete, there seems to be a wide distribution of environments --- and no 
clear companions of comparable mass --- for a majority of the passive systems in our study.  While apparently 
inconsistent with the notion that passive galaxies are mostly likely to be satellite galaxies, van den Bosch \etal\ show 
in their study of a large SDSS sample that this phenomenon is a strong function of mass, declining from the dominant 
fraction ($\sim$70\%) of the population for galaxies with stellar masses of $\sim2 \times 10^9$ \Msun\ to virtually nil at 
$\sim2 \times 10^{11}$ \Msun.  The majority of our sample are galaxies intermediate between these two limits, so 
perhaps a sizable minority of galaxies that ceased star formation when they were incorporated into central galaxies 
is --- after all --- consistent with what we find in the ICBS group sample.

The trend of gradually rising PAS+PSB fraction with increasing group mass might be explained by the effects of mergers 
(roughly equal mass systems) and accretions (higher mass ratios), or the increasing loss of gas supply (starvation), all 
processes that should be favored in more populous systems.  However, our data would seem to argue against any 
\emph{steady} transformation process that operates over the lifetime of the group, by the lack of any correlation 
with \tauenc\ (see Figure \ref{fig:group_correlations}-c).  These observations could be reconciled by recognizing 
that, in a hierarchical model of structure growth, such systems are built from the mergers of smaller groups.  If we regard 
the coalescence of groups themselves as the event that bumps up the number of passive galaxies through one or more 
of the ``interaction'' processes, the lack of a correlation with \tauenc, and the correlation with group mass, is 
readily explained.  For example, the rapidly changing gravitational potential of a group merger could deflect one or 
more starforming galaxies on previously ``clear" orbits and lead to tidal encounters with other galaxies (particularly 
large ones) that are sufficiently strong to remove the gas reservoir that maintains star formation, in this way linking the
event of the group merger to an increase in the passive fraction.\footnote{In this picture, preprocessing would be akin to the 
punctuated equilibrium model of biological evolution that appears to explain the paucity of the evolutionary links that
are expected in a more continuous evolutionary model.}   We discuss other evidence bearing on this model in \S4.7 
and \S4.8.

%Figure 18:  locations of PAS and PSB galaxies in metacluster groups

Projecting the relationship to the higher masses of the clusters, we see in Figure \ref{fig:Lgal_vs_PAS+PSB_fraction}
a track that could well be populated with increasingly rich groups not in our sample, reaching the $\sim$50\% fraction
of PAS+PSB galaxies in these more massive systems, analogous we believe to the rich groups identified in the CNOC 
sample of strong x-ray-emitting clusters studied by Li \etal\ (2009).  Indeed, the two rich groups in the ICBS sample seem 
to confirm the idea that this modest process that we associate with interactions in the group environment is capable of building 
this $\sim$50\% fraction of passive galaxies.  On the other hand, our sample presents what appears to be prima facie evidence 
for a cluster specific process (see also Li et al.) --- the 60-80\% fractions of PAS+PBS galaxies in the four `core' groups (marked 
with a red `X' in Figure \ref{fig:Lgal_vs_PAS+PSB_fraction}).  Whether these are groups or filaments, the projection of these 
four cold substructures directly on the highest density regions of the clusters leaves little doubt that they are passing through, 
or have passed through, the cluster center, and the cluster core sample, with its similarly high passive fraction, seems to
confirm a  cluster-core-specific mechanism --- for example, ram-pressure stripping or harassment (more generally, tidal 
stripping) --- that is more rapid and efficacious than any preprocessing that is happening in the other metacluster and field 
groups.  Strong supporting evidence for this idea has been presented by Ma \etal\ (2010), who find an unusually high incidence 
of poststarburst galaxies in a cluster merger at $z=0.586$, where a population of galaxies in a substructure has been thrust 
into the main cluster core.  Ma \etal\ link this PSB population to a 70\% fraction of galaxies of 'transformed,' they suggest, 
into morphological type S0 galaxies by ram pressure and/or tidal forces near the cluster core.

Finally, is there evidence that preprocessing in groups is different in superclusters than in the general
field? One of the goals of the ICBS was to look for similarities or differences between the supercluster environment 
surrounding a rich cluster and the general field population.  In terms of the fraction of PAS+PSB galaxies, the 
values are very similar, 26$\pm$2.5\% for the average of the 5 superclusters defined by \Rnorm\ $>$ 1.5, and 
20$\pm$1\% for the cl\_field.  (These values are plotted in Figure \ref{fig:Lgal_vs_PAS+PSB_fraction} at the 
somewhat arbitrary \Lgal, \Mgal\ value of 3.5 that approximately represents the average environment.)  Another point of 
comparison is the fraction of galaxies found in groups in the superclusters compared to the fraction found for 
the cl\_field. Again, the values are very close, 40$\pm$2\% for the supercluster, 39$\pm$1\% for the cl\_field.
In \S4.3 we examined a number of properties of supercluster groups compared to the same for the cl\_field and
found the only significant difference to be the substantially larger size of the field groups, but acknowledged that
this could be the result of selection effects.  When we compared the relationships of PAS+PAS fraction with 
different properties we found the same correlations with various parameters of scale, and the same lack of 
correlation with other properties such as velocity dispersion or \tauenc.  The most convincing correlation, PAS+PSB 
vs. \Lgal\ or \Mgal\ (Figure \ref{fig:Lgal_vs_PAS+PSB_fraction}) looks the same, within statistical errors, for 
supercluster groups, and field groups or filaments.  

The implication seems to be that there is nothing obvious the differentiates the supercluster environment from the 
general field.  If preprocessing is happening in groups in both environments, perhaps the only distinction is the
way the preprocessed populations of groups in superclusters are further effected by entering the dense environment
of a rich cluster.

\subsection{Do starbursts play a major role in the production of PAS galaxies?
\label{discussion_starburst_role}}

In this section we address the question of the numerous starburst galaxies we have identified in the ICBS program ---
both active- and post-starburst.  As described in Paper I, we have adopted specific criteria for identifying these from 
\Hd and \OII emission that have been validated with well-studied present-epoch samples.  For the ICBS sample $0.31<z<0.54$ 
we find a level of 15$\pm$5\% for all starbursts (SBH+SBO+PSB) in every environment studied in the ICBS, from isolated field, 
to groups, to rich clusters.  Careful application of specific criteria are important, for our definitions of SBO and SBH starbursts 
extends down to galaxies where the SFR at the epoch of observation is only $\sim$3 times that of the (few Gyr) past average.  
The lower end of the range selected with our criteria includes systems with a more moderate starburst than even the lower 
luminosity \emph{LIRG} galaxies that have been discovered through infrared surveys.  As with CSF/PAS galaxies, the 
poststarburst category requires a uniform and well-defined boundary between starburst and poststarburst systems.

Concerning poststarbursts in particular, we have previously presented evidence for the ubiquity of this minority but potentially 
important population at intermediate-redshift in a series of papers (see, e.g., Dressler \& Gunn 1983;  Oemler \etal\ 1997;  
Dressler \etal\ 1999;  P99) as have other studies (e.g. Couch \& Sharples 1987; Barger \etal\ 1996; Tran \etal\ 2003; 
Lemaux \etal\ 2010).  Poggianti \etal\ (2009a) have in particular conducted an extensive study of poststarbursts over a wide 
range of environments at $z = 0.4-0.8$.  Nevertheless, some other studies have either questioned the prevalence of such a 
population (see Balogh \etal\ 1999 -- cf Dressler \etal\ 2004; Kelson \etal\ 2001) or de-emphasized it's importance (e.g., van 
Dokkum \etal\ 2000; Ellingson \etal\ 2001).  Key to the doubt expressed about the importance of the starburst is the suggestion 
that the poststarburst signature is really nothing more than the sharp truncation of star formation in a very active galaxy, 
without a burst.

It is true that an abrupt end ($\tau \ls 200$ Myr) of star formation in a galaxy whose current SFR is close to its past average 
(roughly constant) will indeed develop a spectrum that is hard to distinguish from the termination of a \emph{mild} starburst.  
However, in modeling this affect, (P99, see also Poggianti 2004) concluded that \Hd$\approx$ 5\ang\ is a limit reached by 
truncating such a system (in particular, for an intermediate-redshift galaxy that has been forming stars with a normal 
initial-mass-function and constant SFR since $z\gs2$). Poggianti \etal\ noted that approximately one-third of the galaxies 
identified as PSB in the \emph{Morphs} sample exceed the 5\ang\ limit; we find the same fraction in the PSB sample of the 
ICBS: 19/55 = 35\%.  Taking into account the decline in \Hd strength that these systems \emph{must} experience as they age, 
over a longer time scale, this accounts for about another third of the observed sample, leaving at most one-third to be identified 
as simply truncated systems with $3$\ang$<\Hd<$5\ang.  

The one-third fraction should moreover be an upper limit because most SFRs decline with cosmic time. On the other hand, a 
new determination of the histories of star formation in Paper 3 (see also Paper 4) suggests that  a small but non-negligible fraction 
of galaxies are genuinely younger, in the sense of SFRs that have peaked more recently than $z\sim2$.  In a future paper we 
will use a representative distribution of star formation histories to refine this estimate of the fraction of poststarburst galaxies 
with \Hd$>3$\ang\ that could be the result of simple truncation of star formation, with no prior burst required.

These considerations do not, however, affect the general conclusion that the majority of galaxies classified as PSBs are 
poststarbursts. The abundance of SB galaxies in the ICBS sample, which are unambiguous cases of mild-to-moderate 
\emph{active} starbursts, strongly supports the conclusion that a sizable fraction of the PSB sample must come from starbursts 
rather than truncation.  From \emph{Spitzer} \24m observations of the rich cluster Abell 851 at $z=0.41$, Dressler \etal\ (2009b) 
concluded that even some of the PSB are at some level \emph{active} starbursts, probably nuclear bursts that are more easily 
obscured by dust. 

For the purposes of this discussion, then, we take as a given that starbursts and poststarbursts are a significant 
component of the intermediate-redshift galaxy population, and turn our attention to how the SB and PSB galaxies
relate to the ordinary CSF and PAS galaxies.

\begin{figure*}[t]

\centerline{
\includegraphics[width=3.2in]{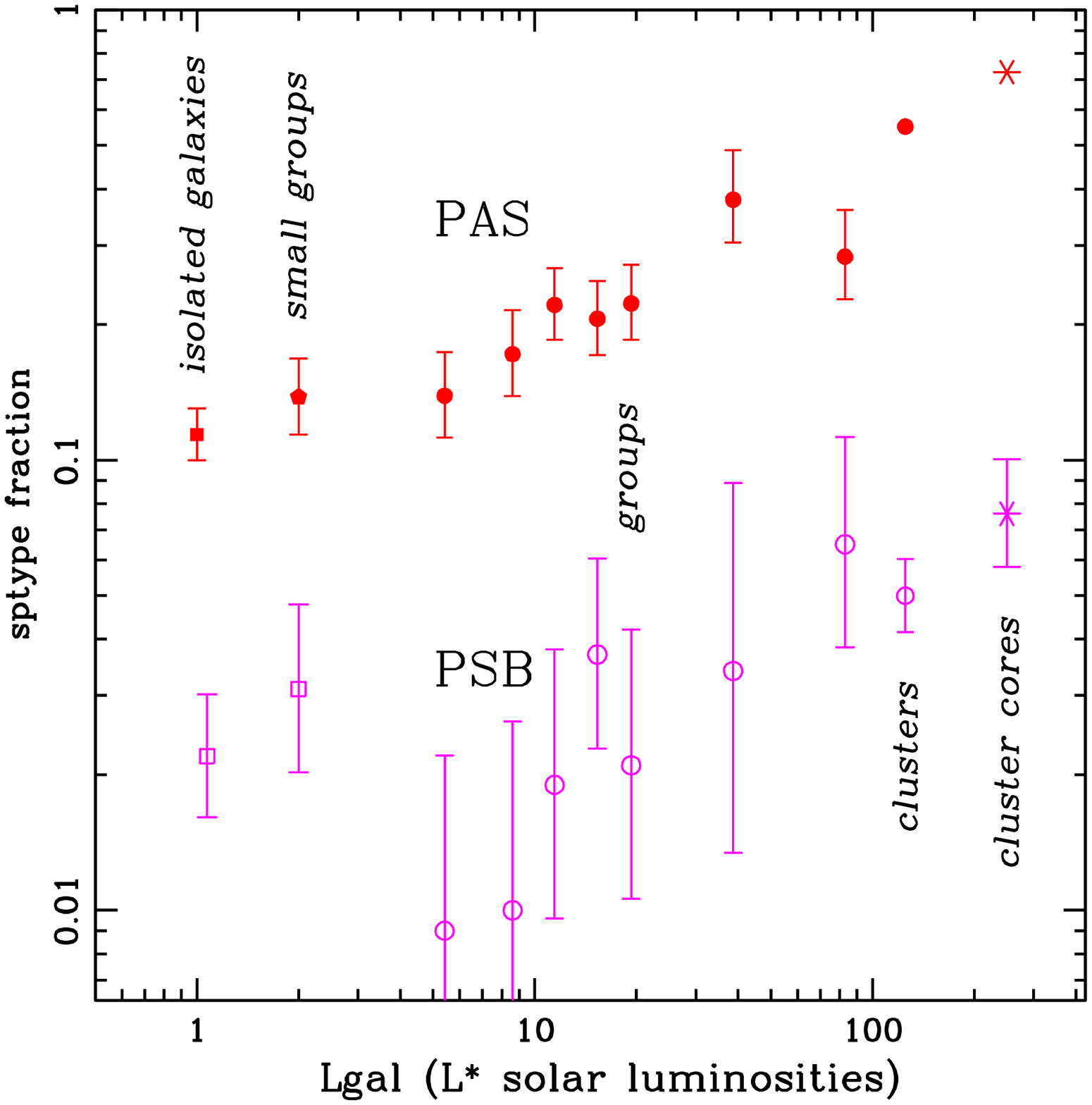}
\includegraphics[width=3.2in]{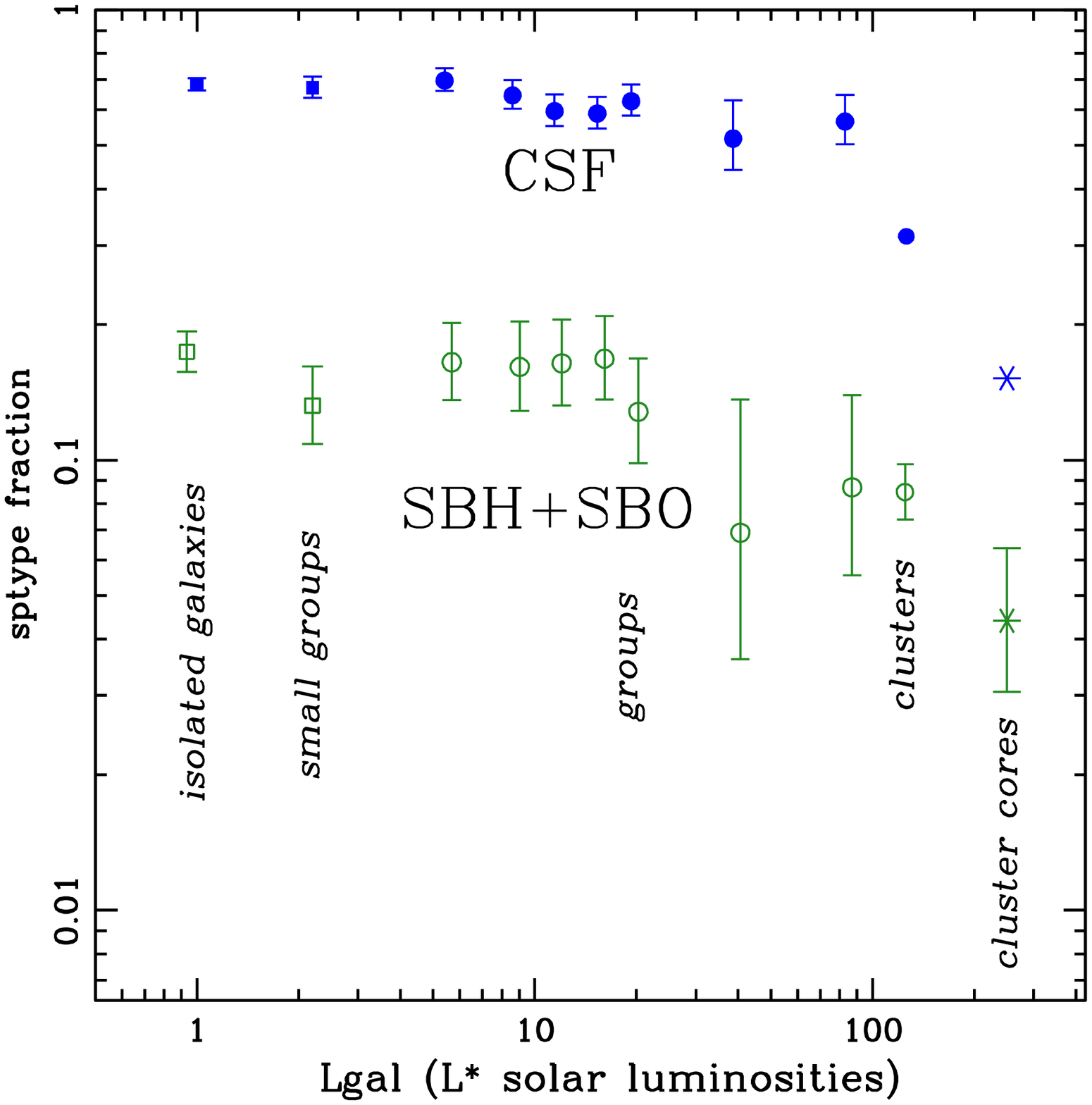}
}

\caption{(a -- left) The fraction of PAS (passive) and PSB (posbstarburst) galaxies across the full range of environments 
sampled in the ICBS, including isolated galaxies, small groups, groups, clusters, and cluster cores.  The data are from 
the samples shown in Figure \ref{fig:Lgal_vs_PAS+PSB_fraction}, but with averaged values for the individual samples for 
isolated galaxies, small groups, clusters, and cluster cores ($R<500$ kpc), and for binned samples of moderate-sized 
groups and filaments.  Across the full range of increasing scale size, represented by total luminosity \Lgal, the PAS fraction 
rises and is tracked by the PSB fraction, displaced lower by a factor of 5--10.  (b -- right) The fraction of CSF (continuously 
star forming) falls slowly until dropping sharply in rich groups and clusters; this behavior is tracked by the fraction of SB 
(SBH+SBO starbursts), displaced lower by a factor of $\sim$4.  The plot shows that that most SB are not on the path to 
becoming passive galaxies: in all but the most luminous (massive) systems they are sufficiently numerous that they would 
overproduce the PAS galaxies (see \S4.6).  Many PSB galaxies could be on the way to becoming PAS galaxies, a conclusion 
that is supported by the spatial concordance of the two types (see Figure \ref{fig:cross_correlation}) in the cluster environment.  
However, as discussed in \S4.7, another and possibly more natural interpretation of these approximately constant ratios of 
PAS/PSB and SB/CSF types, is that most SB galaxies begin in, and return to, CSF galaxies, and PSB galaxies began with 
starbursts in PAS galaxies.  Minor mergers and accretions of gas rich companions --- as the cause of these moderate 
starbursts and poststarbursts --- could provide a natural way to produce the effect.
\label{fig:Lgal_sptype_fractions}}

\end{figure*}

In Figure \ref{fig:Lgal_vs_PAS+PSB_fraction} we showed the fraction of PAS+PSB galaxies over the full range of galaxy 
environments.  Assuming that the PSB galaxies are unlikely to regain future SFRs of even a few tenths of a solar-mass per 
year, the PAS+PSB are the complete population of galaxies with masses M $\gs10^{10}$ \Msun\ that have been ``quenched,'' 
by whatever internal or external means.  If we instead consider the fractions of PAS and PSB  galaxies separately, over the full 
range of environments, we can apply a simple timescale argument to investigate whether starbursts play a significant role in 
increasing the PAS population.  Figure \ref{fig:Lgal_sptype_fractions}-a shows the fraction of PAS and PSB galaxies 
individually over the full range of environments sampled in the ICBS.  The samples are much the same is in Figure 
\ref{fig:Lgal_vs_PAS+PSB_fraction}, but we have combined the results for the different fields for the isolated galaxies, 
small groups, cluster cores, binned the cluster and field groups (including filaments), and omitted the composite populations 
of superclusters (which are mixed rather than unique environments). The four ``core groups" that appear unique in their 
association with processes of the cluster cores are also omitted.  The points with (Poisson) error bars are (1) four-field 
averages of isolated galaxies and the small groups, placed at \Lgal\ $\equiv$ 1.0 \& 2.0, respectively, and at increasingly 
larger \Lgal (2) averages of cluster groups, field groups and field filaments, in bins in which the summed $\Sigma$\Lgal\ 
$\approx$ 80\,L$^*$ (containing between 94 and 129 galaxies in the spectroscopic sample per bin).  The stars representing 
the combined 5 ICBS cluster cores (R $<$ 500 kpc) have error bars that are comparable in size to the symbol, and their 
placements along the \Lgal\ axis represents the effect of a yet-more-extreme manifestation of the cluster environment.  

%Figure 19: log-ratio diagrams, PAS-PSB, and CSF/SB versus sptype fraction 

We note that the PSB and SB trends we observe qualitatively resembles the trends of the post-starburst and starburst 
fractions with environment (from field, to poor-groups, to groups and clusters) found by Poggianti et al. (2009) at $z=0.4-0.8$.
Our measured 2$\pm1$\% PSB fraction for the general field (from isolated galaxies through the $\Lgal<20$ groups) at
$z\approx0.4$ is in good agreement with the PSB fraction measured by Yan \etal\ (2009) for the Deep2 Survey at 
$z\approx0.8$.

In addition to the previously discussed result of the steady increase in the PAS fraction of intermediate-redshift galaxies, 
from $\sim$10\% for the isolated field galaxies, through groups, to $\sim$70\% for the cores of rich clusters, Figure 
\ref{fig:Lgal_sptype_fractions}-a shows what we consider a remarkable result: the PSB track the PAS fraction in the 
sense that the PSB fraction is 15\% $\pm5$\% of the PAS fraction \emph{in all environments}. 

A comparison of the fraction of SBO and SBH starbursts, collectively SB, to the fraction of continuously star forming galaxies, 
CSF, shown in Figure  \ref{fig:Lgal_sptype_fractions}-b, exhibits a similar effect.   Complementing the rising PAS fraction, the CSF 
fraction declines steadily over the \Lgal\ range, from a population that dominates PAS galaxies by many-to-one in the field, a 
few-to-one in poor and moderate groups, one-to-one in rich groups and clusters, and dropping to only one-in-five of the 
PAS in cluster cores.  (Even this small remaining CSF population is likely exaggerated, since some apparent CSF types in the
core are likely projections along the line-of-sight to the cluster core.)  More to the point, like the PSB and PAS fractions, the SB 
track the CSF population, within the errors, by a factor of 4, that is, SB are about 4 times less populous than CSFs 
\emph{in all environments}.  Comparing starbursts and poststarburst galaxies, the fraction of starbursts greatly exceeds 
the fraction of poststarburst galaxies for isolated galaxies and small- and moderate-sized groups, but rapidly drops to 
match the PSB fraction in rich groups and cluster cores.

The first conclusion to be drawn from Figure \ref{fig:Lgal_sptype_fractions} is that SBO + SBH starbursts, in 
environments running from the isolated field galaxies to modest-sized groups, cannot be a common path to PAS 
galaxies --- there are far too many.  The lifetime of these bursts is likely to be no more than 1 Gyr --- indeed, 
this is long for a starburst.\footnote{We have, however, argued in Oemler \etal\ (2009) that the degree to which A stars 
dominate the light necessitates a minimum lifetime $\tau >$ 100 Myr, the lifetime of an early A star, and sufficient time for 
such a stars to migrate from the dusty sites of their birth.}  In environments like the field and small-to-moderate groups, 
the SB/PAS ratio is near unity.  If most of SB turned to PAS galaxies, the fraction of these would more than double in 
a Gyr or less.  In fact, the PAS fraction is only growing by at most some tens of percent over the several Gyr that the 
ICBS spans.  It would seem that most of the SB galaxies in these environments must return to the pre-starburst 
CSF state, as was also concluded by P99.  We note, however, that this conclusion weakens considerably 
for rich groups and cluster populations.  For these environments, SB fraction falls to $\sim$10\% while the PAS fraction 
has risen to 40\% or more, and --- again for these environments --- modest growth in the PAS fraction from $z\sim0.5$ to the 
present-epoch is observed (Li \etal\ 2009; 2012).  So, when combined with the fact that the active starburst population is 
declining from $z\sim0.5$ to the present \emph{in all  environments} (Dressler \etal\ 2009a), it appears that active starbursts 
could be significant contributors to the PAS population only in the densest intermediate-redshift environments.

While most active starbursts --- members of the field population --- cannot be linked one-to-one with the quenching of 
starforming galaxies to form PAS galaxies, the situation seems more favorable for PSB galaxies.  Indeed, a PSB/PAS fraction 
of $\sim$10-20\% in all environments --- very different from the varying  SB/PAS fraction across environments) in fact urges 
a direct connection of the PSB phase to a quenching event that produces a new PAS galaxy.  The decay time for this phase, 
$\tau\ls500$ Myr (more certain in this case because the absence of star formation constrains the spectral evolution), suggests 
that --- if the PSB fraction remained constant from $z\sim0.5$ to the present --- the PAS fraction would approximately double, 
probably more growth than the observations will support.  However, since there is also good evidence for a steep decline in the 
fraction of PSBs in the field, from the ICBS value of $\sim$1\% to a level of $\sim$0.1\% at the present epoch (Zabludoff \etal\ 
1996; Wild \etal\ 2009), so it would appear that decaying PSBs would not overproduce PAS galaxies, even with the slow growth 
of the passive galaxy population in the field since $z\sim0.5$ found by, for example, Faber \etal\ (2007) and Brown \etal\ (2007).  
The situation is much the same for rich groups and clusters: the same 10-20\% of PSB/PAS is found, and the poststarburst 
fraction is known to decline substantially with time (the PSB class was essentially unknown until intermediate-redshift clusters 
were studied).  

However, demonstrating that PSBs could be the primary channel to PAS galaxies since $z<0.5$ is not sufficient to
show that PSBs are a primary path to passive galaxies.  The problem is that extrapolating back in time from $z\sim0.5$ 
to $z\sim1$ in both the field and in clusters similar fractions of PSB galaxies are found, for example, in the extensive study 
of the CL1604 supercluster (Lemaux \etal\ 2010), a poststarburst fraction of $\sim10-15$\% is found in the lower density 
environments outside the virialized clusters (B.C. Lemaux, private communication).  In the EDisCS Survey, Poggianti \etal\ 
(2009a) find a $3-6$\% PSB fraction for the field and poor groups and Yan \etal\ (2009) find a PSB fraction of $2\pm1$\% at 
$z\sim0.8$ in the field-dominated Deep2 Survey.  It would appear, then, that PSB rates were at least as high, perhaps higher, out 
to redshift $z\sim1$.  In contrast to this, results from the Carnegie-Spitzer-IMACS prism survey (Kelson, private communication) 
show a growth in the passive population in the field of only $\sim$25\% from $z=0.8$ to $z=0.5$. Therefore, while the overproduction 
of PAS galaxies by the decay of PSB galaxies since $z\sim0.5$ might not be a problem (because of the substantial decline 
of PSBs after $z=0.3$), it would likely be a serious problem for the earlier epoch $z\sim1$ to $z\sim0.5$. 

In this conclusion, that PSB galaxies cannot be the dominant mechanism for producing the PAS population between
$z\sim0.5$ and the present --- certainly in the field and probably in rich clusters as well, we are in agreement with De Lucia 
\etal\ (2009), whose argument includes the issue of generally lower masses of PSB compared to PAS galaxies, 
which is illustrated by the ICBS sample in Figure \ref{fig:sptype_BVo_SFR}.  The small sample in De Lucia \etal\ is an issue, 
however --- the two clusters in their study show very different PSB/PAS fractions, undoubtedly due to the short time scales 
of the PSB phenomenon and resultant statistical uncertainties.  Wild \etal\ (2009) come to a conclusion that could also be 
consistent with the ICBS result, finding a possible production of $\sim$40\% of PAS galaxies in another small sample at 
$z=0.5-1.0$.

In summary, while it is likely that some fraction of starbursts and poststarbursts are phases on the path to passive galaxies, a model 
in which this is the dominant path is troubled by the short timescale of the phenomena and the commonness of these types  
compared to the relatively slowly changing populations of passive galaxies --- a point made by De Lucia \etal\ (2012) for quenching 
mechanisms in general.  The fact that the starburst phenomenon is in rapid decline since $z=0.3$ helps, but observations of galaxies 
at higher redshift ($0.6<z<1.0$) show only a small change of the starburst fraction,  so overproduction during this earlier time is likely 
to be problematic.  Furthermore, there are clearly important trends that do not seem to flow easily from such a model: the field 
population of PAS galaxies changes very slowly to the current epoch while the fraction of PAS in rich groups and clusters grows 
substantially, suggesting an environmentally sensitive quenching method, while the PSB/PAS fraction is near-constant from the 
field to rich clusters at $z\sim0.5$ and rapidly declining for all environments to the present day.  Likewise, SB/CSF is 
roughly constant over all environments, but the timescale argument indicates that only in rich groups and clusters could these 
be major contributors to the PAS population, and only a very small fraction can be funneled to PAS galaxies in the lower-density 
field.  As both McGee \etal\ (2009) and De Lucia \etal\ have suggested, the mild trend of PAS growth points to quenching 
mechanisms with long time scales, $\tau>1$\,Gyr, and these are not compatible with the starburst signature.  Some 
fraction of PAS galaxies could be the result of starbursts in CSF galaxies, but it appears that most cannot.  By implication,
this suggests that most PSBs must come from PAS galaxies, and not the other way around.

%Figure 20:  starburst cycles cartoon

\begin{figure*}[t]

\vspace{0.3in}
\centerline{
\includegraphics[width=2.8in, angle=90.]{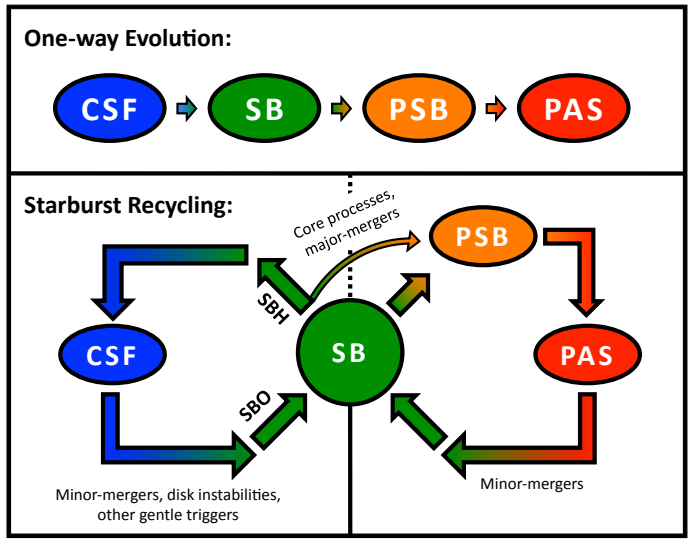}
}

\caption{Cartoon showing the proposed starburst cycles described in this paper.  The conventionalevolutionary sequence for
starbursts converts starforming galaxies to passive galaxies through an active- and post-starburst phase --- we believe
this path is followed by some galaxies, in particular, major mergers of starforming galaxies, and galaxies subject to the 
extreme environment in the cores of rich clusters.  However, for most, we believe that two cycles are operating, largely
independently, one beginning and completing with starforming disk galaxies, the other beginning and completing with 
at least one spheroidally-dominated passive galaxy.  As explained in the \S4.7, this alternate picture is motivated by the 
fractions and timescales associated with starbursts, the spatial distribution and structural properties that link SB to CSF 
galaxies and PSB to PAS galaxies, and the near-constant ratios of PSB/PAS and SB/CSF across environment shown 
in Figure \ref{fig:Lgal_sptype_fractions}.  The locations of "SBO" and "SBH" on the starforming-galaxy cycle are notional,
suggesting a possible sequence for the two types, and recognizing that the SBH class includes both in-progress starbursts
and those in a post-starburst phase as the burst itself subsides.
\label{fig:starburst_cycles_cartoon}}

\end{figure*}

\subsection{A Different Picture: starbursts are a signature of mergers across all environments
\label{discussion_starburst_mergers}}

As we commented in \S2.5 in our discussion of the spatial distribution of different spectral types, the simplest 
relation between spectral types that involves the starburst phenomenon,

\vspace{0.03in}
\hspace{0.7in}CSF $\Rightarrow$ SB  $\Rightarrow$ PSB $\Rightarrow$ PAS
\vspace{0.03in}

\noindent{seems to be inconsistent with the cross correlation functions of these types, at least for the metacluster samples 
of our four ICBS fields (see Figure \ref{fig:cross_correlation})}.  In the previous section, we confirmed with the ICBS data 
the basic argument of P99 that this simple sequence cannot be the dominant path outside of clusters because 
SB galaxies outnumber PSB galaxies by a factor of $\sim$5 in all lower-density environments, including moderate-sized
groups.  We conclude that

\vspace{0.03in}
\hspace{0.7in}CSF $\Rightarrow$ SB $\Rightarrow$ CSF
\vspace{0.03in}

\noindent{is the unquestionable fate of most starbursts.  It is still possible, of course, that the PSB galaxies that
are found outside of clusters are the third step in the above sequence, for example, the small fraction that are
produced in major mergers.  But, even for this case we showed a likely contradiction with observations --- that the 
shear numbers of PSB galaxies, combined with their relatively short lifetimes, $\tau \ls$ 0.5\,Gyr, implies a too-rapid 
growth in the fraction of PAS galaxies since $z=1$.  On the other hand, locating PSB galaxies on the alternate sequence

\vspace{0.03in}
\hspace{0.7in}PAS $\Rightarrow$ SB $\Rightarrow$ PSB $\Rightarrow$ PAS
\vspace{0.03in}

\noindent{would solve the ``PAS overproduction" problem and explain the strong correlation between PAS and 
PSB galaxies seen in Figure \ref{fig:cross_correlation}.

Figure \ref{fig:Lgal_sptype_fractions}, which shows further that the ratios CSF:SB and PAS:PSB are roughly constant
over two orders-of-magnitude of galaxy clustering scale and galaxy local density, points the way to such alternative
relationships between these spectral types.  We suggest that --- rather than indicative of a common CSF-to-PAS
evolution --- starbursts are events in which the PAS and CSF galaxies are \emph{progenitors} and \emph{end 
points}.  In other words,  we imagine two mostly independent \emph{starburst cycles}, one that starts and ends with PAS 
galaxies, and one that starts and ends with CSF galaxies.  The general picture is illustrated in cartoon form in 
Figure \ref{fig:starburst_cycles_cartoon}.

The approximately constant ratios of 15\% for PSB/PAS and 25\% for SB/CSF over all  environments motivates such a 
picture, one in which SBO + SBH starbursts, and PSB poststarbursts, are commonly the  result of minor mergers and 
accretions of gas-rich satellites onto PAS and CSF galaxies, respectively.  Minor mergers are the mechanism of choice 
because (1) a rapid change in SFR is expected, explaining the starburst signature that cannot be obtained in slow processes 
such as starvation; (2) the basic morphology of the host galaxy remains unchanged; and (3) merging is a local process, to 
first-order independent of global conditions, which further accounts for the near-constant ratios of SB:CSF and PSB:PAS 
over the full range of environment.  We note also that, for the CSF $\Rightarrow$ SB $\Rightarrow$ CSF cycle, strong 
tidal interactions share thesesame three features.

In this picture, a CSF galaxy accreting a smaller gas-rich satellite (or, more rarely, a major merger with a gas-rich peer), 
or a strong tidal encounter, would result in a substantial rise in SFR, appearing as an SBO or SBH, perhaps both in sequence.  
Because a minor merger or tidal encounter would not change the basic morphology of the CSF galaxy, especially the survival 
of the gas disk, the system would subsequently return to the CSF population.  Likewise, when the principal galaxy is a passive 
galaxy, in particular, a galaxy with a large spheroidal component, accreting a gas rich companion would produce a starburst
(perhaps a red SBO) leading to a poststarburst before the system returns to a passive state.  In this case, not only minor 
mergers but also major mergers would be viable candidates: for these rarer events the conventionally assumed CSF 
$\Rightarrow$ SB $\Rightarrow$ CSF $\Rightarrow$ PAS process could in fact carry through.  These separate starburst 
cycles would explain both the spatial correlations of CSF--SB and PAS--PSB shown in Figure \ref{fig:cross_correlation} 
and avoid an overproduction of PAS galaxies in any environment.  The basic structure/morphology of the galaxy hosting the 
starburst should in general not change, with the caveats that (1) major mergers can be a leak in the CSF cycle that does lead 
to more PAS galaxies (and a change of morphology), and (2) a CSF galaxy with a larger bulge may result from a minor merger.  

This prediction of the starburst cycles model --- that basic galaxy structure remains unchanged --- is supported by an analysis 
of near-infrared images of the ICBS galaxies by Abramson \etal\ (2013).  By fitting Sersic models to the CSF, SB, PSB, and 
PAS galaxies in field and supercluster galaxies, Abramson \etal\ show that CSF and SB galaxies are well described as disk 
systems and quite distinct from PAS and PSB systems, which have the steeper profiles of spheroidally dominated galaxies. 

%Figure 21: sSFR and SFR vs. (B-V)o
\begin{figure*}[t]

\centerline{
\includegraphics[width=3.3in]{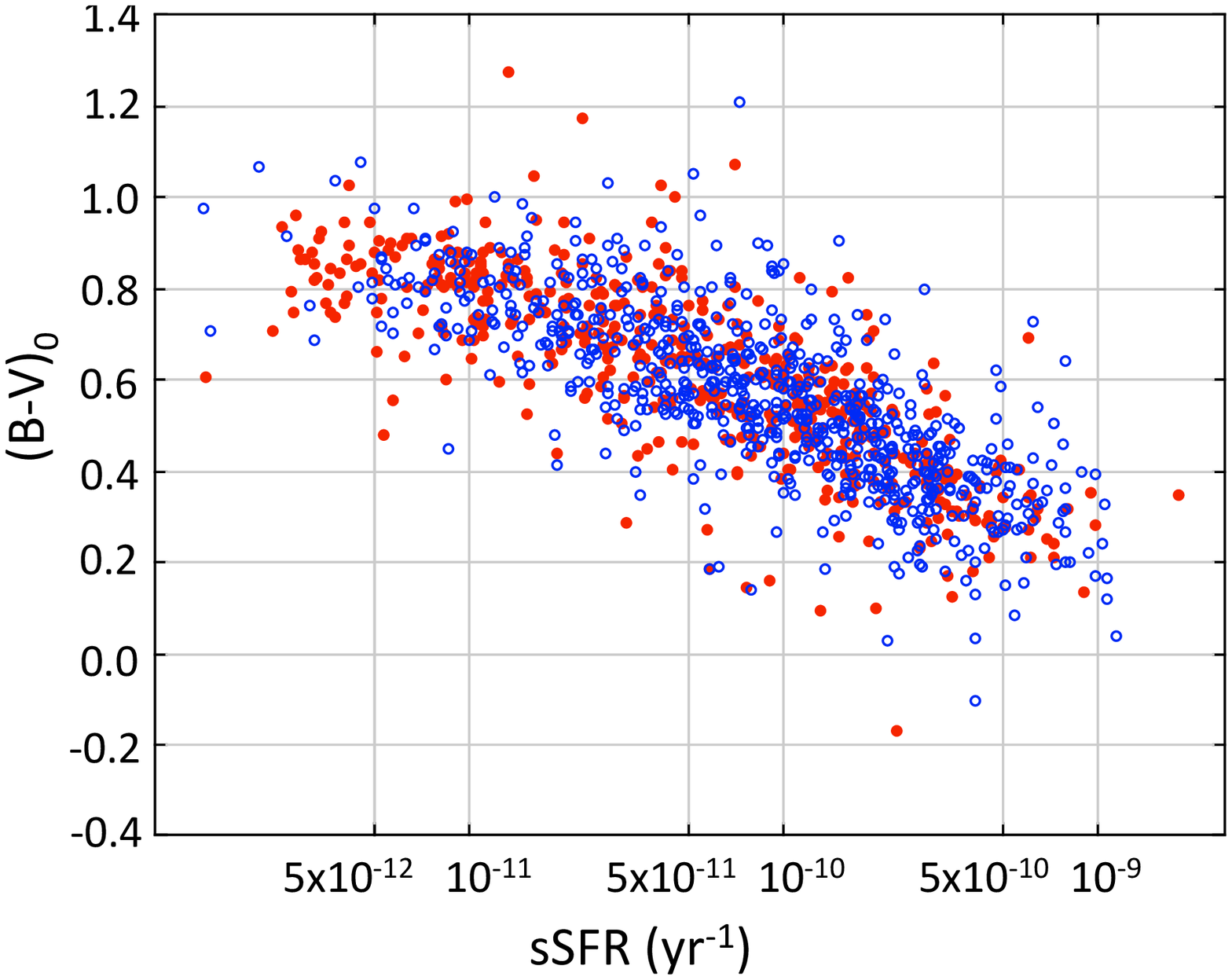}
\qquad
\includegraphics[width=3.3in]{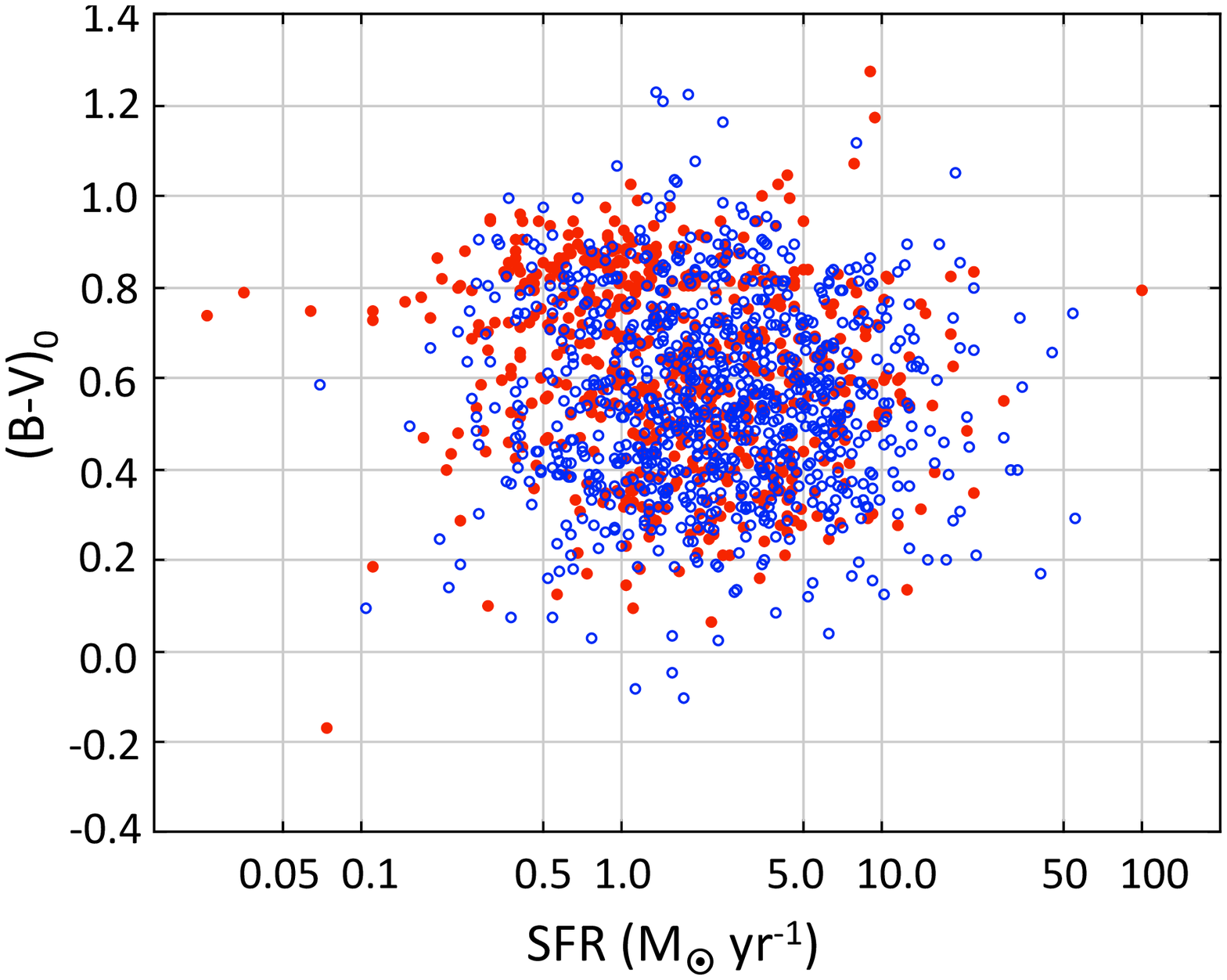}
}

\caption{(left) The specific star formation rate (sSFR) vs rest-frame B-V color for the same 4 metacluster sample (red
points) and the cl\_field (blue circles) as in Figures \ref{fig:sptype_MB_Mass} \& \ref{fig:sptype_BVo_SFR}. (right) The 
star formation rate vs rest-frame B-V color.   There is a excellent correlation of sSFR with color, but this relationship 
obscures the fact that many galaxies that are dominated by an old population have high star formation rates, as seen 
from the scatter plot at right.  These objects could be early-type spirals forming stars continuously, or formerly passive 
galaxies that are experiencing a starburst due to a minor-merger or accretion event, as discussed in \S4.7.
\label{fig:SFR_vs_BV}}

\end{figure*}

The nominal SB galaxy would be the intermediate stage of the CSF cycle, but what would the starburst phase of an 
accreting passive galaxy look like?  Because the SB galaxies greatly outnumber PSB galaxies in all but the densest 
environments, the active starburst phase of the PAS $\Rightarrow$ SB $\Rightarrow$ PSB $\Rightarrow$ PAS cycle could be 
difficult to distinguish from the CSF $\Rightarrow$ SB $\Rightarrow$ CSF cycle.   Alternatively, a spheroid-dominated galaxy 
accreting a gas rich companion should produce a spectrum resembling an early type spiral before returning to the PAS 
spectral class, one that might be difficult to distinguish as a starburst against the continuum light of an old stellar population.  
In Figure \ref{fig:SFR_vs_BV} we see that --- in addition to an very good correlation of specific star formation rate with rest-frame 
B-V color, there are a significant number of galaxies with the red color of old stellar populations with very high SFRs.  In a future 
paper we will model whether such progenitors can match our criteria for SBH and SBO starbursts --- anomalously 
strong \Hd and/or \OII --- and evolve into PSB galaxies with their strong Balmer absorption but little ongoing star formation.  
A third possibility is that some starbursts can be hidden in the dust obscured nuclei of bulge-dominated galaxies, as 
was found through Spizter \24m observations of the intermediate-redshift clusters A851 by Dressler \etal\ (2009b).

An attraction of the two-cycle starburst model is that it explains ICBS observations as well as the results of many 
other studies. The starburst phenomenon is pervasive among high redshift galaxies, especially when considering the 
relatively short duty cycle, but the connection of this to popular mechanisms such as ram-pressure stripping or starvation 
is forced, at best.  Such mechanisms are expected to act on a longer timescale --- an attractive feature when trying
to account for quenching of starforming galaxies and the growth of the PAS population, but ill-matched to the
starburst signature.  In contrast, mergers undoubtedly play a significant role in the evolution of many galaxies,
and their connection to the starburst phenomenon --- first elucidated by Zabludoff \etal\ (1996) in their study
of poststarbursts in the low-redshift field --- has also cropped up in morphological studies of active starbursts 
and poststarbursts in rich clusters (see Oemler 1997, \S6).  Mergers are known to lead to rapidly increasing and 
decreasing star formation rates, for which the spectral types SBH, SBO, and PSB are easily associated.  Indeed, 
Hogg \etal\ (2006) were drawn to this association of starbursts and poststarbursts with mergers, through a 
comprehensive analysis of the environments of present-epoch examples in the SDSS: 

\begin{quotation}
\small
The remaining hypothesis for the triggering of the starburst (or, more properly, star formation truncation) events
that precedes the poststarburst phases of these galaxies are: some kinds of random internal catastrophes or some 
kinds of galaxy-galaxy mergers. This latter possibility, which is consistent with all of the results here, is directly
supported by the discovery of post-merger morphological signatures (e.g., tidal arms) in many poststarburst
galaxies (Yang \etal\ 2004; Goto 2005).  It is also exciting, because merging is one of the fundamental processes
of cosmogony, and holds great promise for providing precise connections between cosmological observations
and theory at small scales.
\normalsize
\end{quotation}

It is this ubiquity of the starburst phenomenon over the whole range of galaxy environments that is one of the most 
attractive features of minor mergers and accretions, and strong tidal encounters, as the important mechanisms.  
The dependence of such processes on local as opposed to the global environment provides a natural explanation to the 
commonness of  starburst phenomenon in intermediate-redshift field, groups, and clusters.  Minor mergers and accretions
in particular offer an explanation for the marked decline in starburst activity over the last $\sim$5 Gyr.   While the focus 
has been on major mergers as a method of producing spheroidal stellar systems from disk-dominated systems, particularly 
at early times ($z\gs2$), minor mergers and accretions, and strong tidal interactions, are far more common events and 
offer the possibility of leaving the basic structure and starforming character of the primary galaxy intact.  Thus, identifying 
the precursors of  most starbursts as normal starforming galaxies, and the precursor of poststarbursts as passive galaxies, 
can explain the tracking of PSB to PAS and SB to CSF seen in Figure \ref{fig:Lgal_sptype_fractions} in a natural way.

\subsection{Pieces for the big picture}

Is there a notional model that approximately describes the histories of star formation and structure evolution of 
galaxies with redshift and environment, and identifies the key processes?  Despite great progressin the last 
few decades in quantifying the characteristics of galaxies over most of cosmic time, and the identification of 
many mechanisms thought to influence galaxy evolution, there is no consensus picture that explains the basic data.  
In this paper we have used a uniform data set that includes photometric and spectroscopic observations of 
intermediate-redshift galaxies over the full range of galaxy environment, from isolated field galaxies to the cores of 
rich clusters.  We have presented evidence that suggests that the ``quenching'' that turns starforming galaxies into 
passive galaxies is for the most part the result of slow processes such as starvation that are particularly effective in 
galaxy groups.  The other notable process at higher redshift --- the increasing frequency of starbursts --- does not, we 
argue, contribute very much to this dominant quenching process, but instead is a signal of an increasing merger/interaction 
rate at higher redshifts.  These starbursts may have a small but important effect on the evolution of galaxy structure/morphology, 
and a detailed understanding of the starburst phenomenon is of course necessary for a complete description of the star formation 
history of many if not most galaxies.

Combined with some speculation about the evolution of galaxies before $z=2$, these two basic programs
of starvation quenching and merging starbursts could help frame a picture of galaxy evolution that accounts for
much of what is observed.  Considering our results and voluminous literature on the subject of galaxy evolution,
we see seven principal ingredients that we believe inform a ``big picture" description of galaxy evolution: 
1) nature -- a very early formation processes for a substantial fraction of today's passive galaxies, those in low-density
environments  2) hierarchical clustering -- widening the density range of galaxy environments over cosmic time; 
3) quenching -- processes that \emph{slowly} transform starforming galaxies into passive galaxies; 4) galaxy 
merging --- local acquisition of a neighboring galaxy;  5) starbursts --- relatively rapid rise and fall in a galaxy's 
star formation rate; 6) preprocessing --- increase in the fraction of passive galaxies in the group environment; 
7) rich cluster environment --- unique and extreme conditions in the core of a rich cluster that are hostile to 
ongoing star formation.

\subsubsection{Nature before nurture -- the early birth of some passive galaxies} 

We consider $z\sim2$ as the beginning of galaxy evolution and identify the epoch $2<z<6$ as the time of galaxy \emph{formation}.  
The components of galaxy building as observed at $z>2$ are not readily comparable to the mature galaxy types we see today.   
For the ``isolated galaxy" component of the ICBS cl\_field sample at $0.31<z<0.54$, we find a $\sim$10\% (luminosity-limited) 
fraction of passive galaxies ($\sim$20\% for a mass-limited fraction, $M>3\times10^{10}$\,\Msun), in an environment where only 
processes that are internal to a galaxy (secular evolution) or local (strong interactions or merging with neighbor galaxies, or 
accreting major satellites) can be credited with ending star formation.  With the significantly greater leverage of intermediate-redshift 
compared to local samples, it is certain that these passive galaxies formed the bulk of their stars early, at $z\gs2$.  By our 
definition, then, passive systems are more the result of ``nature" than ``nurture." 

Although this population has remained passive most of the time since $z\sim2$ to the present epoch,  they have not 
been undisturbed or even free of star formation at all times.  Indeed, the "red nuggets" -- small, relatively massive, passive 
galaxies at $z\sim2$ (see, e.g., van Dokkum \etal\ 2008; Williams \etal\ 2010) are thought to evolve to somewhat more 
massive and significantly larger spheroidally-dominated systems by the present day, and mergers and some star formation
is likely to be part of the picture.   Nurture follows nature, but later evolution of the stellar population of such 
galaxies looks to be modest.

Today, elliptical and S0 galaxies are the morphologies associated with this base population of passive galaxies.  It 
it is worth remembering that, even though passive galaxies dominate in very dense environments, those passive galaxies 
found in relative isolation or in loose-groups are the majority of the population, since this is the environment of more 
than 90\% of galaxies.  At the present epoch S0 galaxies are roughly twice as common as  elliptical galaxies in these 
sparsest environments; this could be the result of structural evolution of, or addition to, a population once dominated by 
ellipticals.

As hierarchical clustering proceeds since $z\sim2$, major mergers add to the passive population, but the merger rate is 
declining with time, and major mergers --- because they involve two roughly equal masses --- are the least common.  These 
events, if involving gas rich galaxies, can result in luminous infrared sources (LIRGs and ULIRGs), and ``dry" mergers (PAS) 
when they are not.  

\subsubsection{Nurture after nature -- adding passive galaxies to the legacy population}

Our ICBS sample across environment indicates that only a modest increase in galaxy density through hierarchical clustering 
is needed to add again as much to the ``legacy population" of passive galaxies.  We have argued in this paper, as have others 
elsewhere, that starbursts and poststarbursts are not signposts for this transformation from actively starforming to passive 
galaxies; such a relationship is not compatible with the short timescale for the starburst phenomenon and their $\sim$10\% 
frequency, which is enough to overproduce the growth in the passive population since $z=1$.  The increase in the passive fraction 
over time in groups is referred to as ``preprocessing" to distinguish it from the processes that occur in the dense environments 
of clusters.  Therefore, we look to other, relatively slow ``quenching" mechanisms as responsible for the starforming-to-passive 
transformation.  

Boselli \etal\ (2006) have comprehensively reviewed the environmental mechanisms that can turn starforming galaxies into 
passive ones.  `Starvation' (or strangulation) is currently the most-cited mechanism for turning off star formation (Larson \etal\ 
1980; Bekki \etal\ 2002).   Starvation is also the quenching mechanism that is most effective in groups, as the possibility of a 
galaxy being incorporated into a large halo and/or losing its circumgalactic gas through tidal stripping is highest in this environment.  
The starvation quenching mechanism is undoubtedly too slow to result in even the weakest PSB spectral signatures, those 
that do not require a burst but do require a relatively rapid truncation of star formation (discussed above).  This slower time 
scale, $\tau\gs2$ Gyr, is a virtue in explaining the growth of the passive population, as explained in the previous section.

We find for the ICBS groups, cluster and field, a monotonic increase in passive fraction that is proportional to the mass 
of the group, which is essentially the halo mass.  However, we do not identify the group halo mass \emph{per se}  
as responsible for promoting the starvation (or similar) mechanism, partly because the halo mass is in fact growing 
in the process of hierarchical merging.  That is, the growth of the halo, rather than its size at any given time, is related
to the efficacy of  the mechanism.  The fact that we do not see a correlation with the encounter time between group galaxies 
also suggests to us that the mergers of smaller groups, and not simply galaxy-galaxy interactions, factors in the loss of 
the fuel supply for one or two group members, begining their slow transition from starforming to passive. 

\subsubsection{Starbursts trace the history of minor mergers and acquisitions}

Starbursts and poststarburst galaxies are an order-of-magnitude more prevalent at intermediate redshift than today.
We identify minor mergers and accretions of smaller satellites as the reason for this higher incidence; the rapid decline 
in merger rate since $z\sim1$ is at least qualitatively consistent with the rapid decline in the starburst frequency, as might
a general decline in gas available for star formation since $z=1$.  While major mergers may fundamentally change the 
character of involved galaxies, we expect that these smaller mergers and acquisitions do not substantially alter the basic
character of the host galaxy.

Since poststarburst galaxies are observed across the whole range of intermediate-redshift environments, it is necessary
in our picture that minor mergers and accretions are as well.  Certainly major and minor mergers are viable in the lowest density 
environments. However, the group environment has long been touted as the most favorable environment because of a ``sweet spot'' 
of higher galaxy density and a moderate encounter speed, when compared to rich clusters where galaxy densities are very high 
but encounter velocities are sufficiently high to discourage mergers.  Just \etal\ (2010, 2011) have highlighted the role of groups 
as preprocessing sites, specifically in the production of S0 galaxies, which have been suggested as a common outcome of  
mergers of galaxies with mass ratios $\gs$3 (Bekki 1998, 2001;  Bekki and Couch 2011).  A study by Wilman \etal\ (2009) of groups 
at intermediate redshift also supports the idea that preprocessing in groups is a major component of  the production of S0 
galaxies. 

However, the importance of minor mergers in the production of S0 galaxies may turn on the question of starbursts.  If the
observations of such events, and numerical models, point to starbursts as a usual result of the process, then our 
argument based on the frequency of starbursts and the growth of passive galaxies suggests that minor mergers 
and accretions are not the primary channel for S0 production. Given that slow quenching is certainly going to  produce 
passive disk galaxies identifiable as S0 galaxies, it may be that a preferred model involves a combination of mechanisms.  
In particular, there are two characteristics that must be reproduced: (1) S0 galaxies have systematically larger 
bulge-to-disk ratios and thicker disks than spirals, and (2) the fraction of S0 galaxies has increased rapidly since $z=1$ while
the fraction of ellipticals has changed little or not at all (Dressler \etal\ 1997; Postman \etal\ 2005; Desai \etal\ 2007; Poggiant 
\etal\ 2008).  Starvation is, of course, a mechanism for producing S0 galaxies, not ellipticals, and it is one that works
effectively down to the present epoch, while minor mergers and accretions are strongly decreasing since $z\sim0.4$.  
However, the minor mergers and accretions we identify as starbursts and poststarbursts are likely to heat up the disk 
and raise the bulge-to-disk ratio of the host disk galaxy (Bekki and Couch 2011).  So, even though minor mergers and accretions may not
be a primary quenching mechanism, structural changes in the host galaxy may be important to producing the S0 galaxy 
population we see today.   Put another way, a synthesis of what these studies have found, and what we have proposed 
here, is that mechanisms such as starvation are responsible for the evolution of the stellar population, but that minor mergers 
and accretions that we observe as active starbursts and particularly poststarbursts produce the structural evolution that
also distinguish S0 galaxies as a class.

Although we have no morphological information from HST imaging to look for merging associated with the SBH, SBO, and 
PSB galaxies in the ICBS sample, there are abundant examples for the environment of the rich clusters of the Morphs study.  
Figure 8 of Dressler \etal\ (1999) shows minor mergers (M1/M2 = 3:1 to 10:1) and perhaps even accretion events (M1/M2 $>$ 10:1) 
for more than half of the examples of e(a), e(b), and k+a or a+k, the categories that correspond to the SBH, SBO, and PSB of 
this study.  Unlike the transformational processes that we have identified as increasing the passive population, 
mergers and accretions are not strongly influence by global environment with the possible exception of the cores of
rich clusters (but see Mihos reference, below).  In one of the clearest case of subgroups in a rich environment, Oemler 
\etal\ (2009) used \emph{HST} images to investigate starburst and poststarburst galaxies in Abell\,851.  That study found 
a large population of such galaxies, some of them cases of hidden starbursts that could only be detected through their \24m 
emission.  The latter were identified as the youngest systems, and mostly had disturbed morphologies, including tidal 
signatures of major mergers.  Oemler \etal\ show that --- throughout the cluster --- disturbed morphologies (indicative of tidal 
encounters or mergers) are common, particularly for the youngest (most recent) starbursts, although some poststarburst galaxies 
appear quite normal compared to a present-epoch early-type spiral.  Similar morphologies have been identified for lower
redshift field samples of starburst and poststarburst galaxies (Yang \etal\ 2004; Goto 2005).

The near-constant ratio of starbursts/starforming and poststarburst/passive galaxies we find over the full range of environment 
partially motivates our conclusion that starbursts are not indicative of the production of passive galaxies.  In our picture, SBO
and SBH starbursts happen mostly in continuously starforming disk galaxies, to which they return after the burst.  A poststarburst, 
on the other hand, is only observable when a bulge- or spheroid-dominated host galaxy accretes a starforming and/or gas-rich 
system, and the system returns to a passive state following the burst.

\subsubsection{Agents of galaxy transformation in the extreme environment of cluster cores}

With both field and groups identified as fertile ground for galaxy interaction and mergers, it seems that rich clusters would 
be the only unfavorable locations, and this view --- that high velocity dispersions would suppress the merger rate --- held 
sway for decades. However, Mihos (2004) pointed out that building galaxy clusters through hierarchical clustering changes 
the situation dramatically:  even in the cluster environment (and especially in subgroups falling into rich clusters), a wide range 
of slow and fast tidal encounters, and even mergers, should occur at sufficiently high rates to play a major role in galaxy 
evolution.  In fact, Mihos was the first to use the term ``preprocessing" to describe the role that mergers and interactions 
might play in the group phase of cluster evolution, and he predicted a range of starburst phenomena, from moderate and 
over the full galactic disk, to central and strong, that close encounters of group and cluster galaxies could produce.  

The dense environment of rich clusters has traditionally been identified as the environment where galaxy star formation 
rates and galaxy morphology could be drastically affected.  The hot intracluster medium and high velocity dispersion of 
cluster cores means that \emph{fully} stripping a galaxy's intergalactic and circumgalactic gas through ram-pressure 
stripping is expected to be prevalent here, and here alone.

Tidal stripping of a galaxy's halo, including the gas supply of a starforming galaxy, should be very efficient in the cluster 
environment.  However, in respect to quenching through starvation or stripping, the main location of such 
transformations is likely to be galaxy groups.   We have argued here that the hierarchical merging of groups are events 
that initiate quenching for more and more galaxies as time progresses, and that such ``preprocessing" is able to 
produce a majority of the passive systems that dominate the rich clusters, well before these groups are incorporated 
into the clusters.  However, our data also suggest the possibility of additional quenching mechanisms that would
take the passive population from the 40-50\% level found in the richest groups to the 70-80\% found in extreme
environments.   Ram pressure stripping (Gunn \& Gott 1972; Balsara \etal\ 1994; Abadi \etal\ 1999; Quilis \etal 2001; 
Bekki 2009), and harassment (Richstone 1976; Moore \etal\ 1998) are processes that should ``come into their 
own" only in cluster cores.

It is possible that ram-pressure stripping can result in a sufficiently sudden end to star formation to produce at least the minority 
of the PSB galaxies \emph{in cluster cores} --- those cases where truncation of star formation on a timescale $\tau<1$ 
Gyr --- without a starburst --- is sufficient to produce a PSB spectrum. There is evidence, however, that although ram-pressure 
stripping is able to clear spiral disks of HI gas in central cluster regions, the denser molecular gas towards the galaxy's 
center is more resistant (Kenney \& Young 1986; see also Boselli \& Gavazzi (2006).  If true, ram-pressure stripping should only curtail 
star formation on a longer time scale ($\tau\gs10^9$ yr), causing the SFR to decline relatively slowly as disk gas is used up by
astration.  The theoretical modeling in these studies implies that the effects of starvation and ram-pressure stripping may both be 
effective in stopping star formation, but not abruptly, and therefore unable to  populate the PSB class.

However, recent work suggests that ram-pressure might do more than this.   As first suggested by Dressler \& Gunn (1983), 
the hydrodynamic interaction of a gas-rich disk with a hot, high-pressure intracluster medium (ICM) might trigger a burst of star 
formation in an infalling cluster galaxy.  This idea has been more explored recently (Kapferer \etal\ 2008, 2009) along with 
the idea that the static pressure alone of the ICM could trigger a starburst in a gas-rich disk galaxy (Bekki \& Couch 2003; 
Bekki \etal\ 2010).  There is at least one case where busts of star formation induced by ram-pressure may have been 
observed (Cortese \etal\ 2006), and the effects appear to be dramatic.  Of course, even if ram-pressure or a static 
high-pressure ICM can induce starbursts, this mechanism only functions over a small range of the environments 
studied here --- in the cores of rich clusters. However, this is exactly where the fraction of poststarbursts is highest, 
so it could be that the few active starbursts that are found here, and some of the many poststarbursts, are the result of 
ram-pressure-induced star formation that adds to the passive population in this unique environment.  We note from our
own data (Figure \ref{fig:Lgal_sptype_fractions}) that the fractions of starbursts and poststarbursts are equal in the cluster
cores, suggesting that most or all of the active starbursts in this environment should pass through the poststarburst
phase as their star formation is shut off permanently --- a very different outcome than what we infer for lower-density 
environments where most starbursts return to starforming galaxies.  

In other words, the cluster core environment is perhaps the only one where the conventional 
CSF $\Rightarrow$ SB  $\Rightarrow$ PSB $\Rightarrow$ PAS path (see Figure \ref{fig:starburst_cycles_cartoon}) is in 
fact the common one.  Moran \etal\ (2008), find examples of ``passive spirals'' in two intermediate-redshift clusters that 
they believe are transitioning from active spirals to S0 galaxies through the agency of ram pressure stripping --- they claim 
 such objects are only found in regions where the intracluster gas is sufficiently dense.  Their sample includes both objects 
undergoing a slow evolution (undetected in UV light), and those with more recent signs of star formation --- detected in 
the UV, and possibly a burst.  In the study of the structural morphology of ICBS galaxies of different spectral types, Abramson 
\etal\ (2013) also find that, for the cluster cores alone, the PAS-PSB (spheroid) and CSF-SB (disk) symmetry is broken,
as the PSB distribution shifts to disk-dominated structures.

Lastly, harassment, as developed by Moore \etal\ (1998), is a cluster-specific process in which a galaxy is whittled down by
a combination of encounters with other galaxies and through the tidal field of the cluster as it traverses the cluster core.  
Moore \etal\ did in fact, suggest that harassment of a gas-rich galaxy could produce starbursts, in a manner similar to
the possibility of ram-pressure-induced star formation.  If so, harassment could also contribute to the population of the
conventional channel CSF to PAS through starbursts, and only in the unique environment of dense cluster cores.

\subsubsection{The thumbnail picture} 
We conclude, as others have, that starvation, and tidal- and ram-pressure- stripping are likely the dominant mechanisms 
for turning star forming galaxies into passive ones.  Because these are strongly dependent on environment, and operate 
more slowly than mergers, they are consistent with the basic fact that the passive fraction is growing mostly in denser
environments since $z<1.5$, and slowly.  On the other hand, mergers and strong tidal interactions are likely to be primarily 
responsible for the starburst phenomenon (with a possible contribution from ram-pressure stripping and harassment in 
cluster cores).  These mechanisms can operate effectively over a wide range of environments --- even in clusters.  The
two distinctive features of this model are (1) starbursts are not an important quenching mechanism, and (2) poststarburst 
galaxies are for the most part more common in dense environments not because they are produced by a environmentally
sensitive mechanism, but because they are tied to the large population of passive galaxies in such environments.
Both conclusions are at-odds with many previous studies, including our own.

In the Introduction of this paper, we highlighted two unanticipated discoveries, the rapid decline in the cosmic SFR 
since $z=1$,  and the dramatically higher incidence of starburst galaxies from $0.3 \ls z \ls 1.0$ compared to today.  
Because the population of passive galaxies grew over this period, it seemed reasonable to think this a substantial 
contributor to the decline. When we began the ICBS, we imagined we would confirm that starbursts are one of the 
quenching mechanisms that contribute to the growth of passive galaxies, instead, we found that starbursts are 
minor contributors, plausibly dominant only in the cores of rich clusters.  But, we also found clear evidence for the growth of 
the passive population as groups are built --- \emph{preprocessing} --- and concluded that a relatively slow process such 
as starvation must be responsible.  However, this substantial growth of the passive population has, it turns out, little to 
do  with the rapid decline of the cosmic SFR density since $z=1$.  In Paper 3 we explore the the histories of star 
formation we observe over that epoch, and in Paper 4, the assembly of SFHs that are required to explain that decline.  
We will show in a future paper that the growth of passive galaxies through quenching in groups and clusters 
is playing only a minor role as the universe heads for a second dark age. 

\section{Summary}

Our spectroscopic and photometric study of 4 fields of $\sim$0.5 deg diameter has produced high quality data
for some 2200 galaxies in 5 rich clusters and the field, $0.31<z<0.54$.  From these data we have measured galaxy
magnitudes, colors, line strengths, and velocities and computed galaxy star formation rates and masses.  Using these 
basic data we have separated galaxies into 5 spectral types: passive, continuously star forming, two types of
starbursts, and poststarburst.

For 4 of the 5 clusters in our sample, we find substantial infall of moderate-sized, cold groups with typically
10-20 spectroscopic, (20-40 photometric) members; these groups contribute roughly half of the infall into the clusters, 
and the total infall within $R<5$ Mpc is sufficient to double the mass of the virialized cluster.  The ICBS clusters are 
more representative of clusters at intermediate redshift than those selected by, for example, strong X-ray emission: 
the one rich, regular cluster of this type in the ICBS sample shows much less infall and is presumably in a more 
advanced dynamical state.

The groups infalling into the clusters have been compared to field groups and filaments of similar size, mass, and
velocity dispersion.  For all three samples we find a factor of 2-3 growth in the fraction of passive galaxies from
the smallest to the largest groups, indicating that preprocessing in groups is substantial. However, there is also
evidence that in rich cluster cores additional quenching mechanisms ``kick in" and further elevate the passive
fraction.  Cluster groups that are projected along the cluster center, presumably infalling or exiting from the
cluster core, show this effect strongly: their fraction of passive galaxies is $\sim$70-80\%.

Together, starbursts and poststarbursts make up about 15-20\% of the intermediate galaxy population, so common that
--- given the $\tau\ls500$ Myr the timescale of the starburst phenomenon --- it is unlikely that they are a major component
in the growth of the passive galaxy population in any environment, with the possible exception of rich cluster cores.  We find in 
addition new relationships for starburst galaxies: the poststarburst/passive fraction is approximately constant at 10-20\% 
over all environments, from isolated galaxies, through groups, to cluster cores, and the active-starburst/continuously-starforming 
fraction near constant at $\sim$25\%.  From this we suggest that mild-to-moderate starbursts in this era are the result of mergers, 
mostly minor mergers and accretions.  Specifically, we suggest that readily identified active starbursts are primarily events in 
previously continuously star forming galaxies, to which they will generally return, and poststarbursts are events occurring in 
previously passive galaxies, to which they will usually return.  These events are thought not to fundamentally change the galaxy, 
except perhaps in the building the larger bulges and thick disks of S0 galaxies.  

By combining this explanation of the starburst phenomenon with the popular notion that starvation and stripping --- comparatively slow, 
environmentally sensitive quenching mechanisms --- are mainly responsible for building the higher passive fractions of groups and 
clusters, we complete a picture which is plausibly consistent with the major features of structural and star formation features for 
galaxy populations at $z<1$, over the full range of environment.

\section{Acknowledgments}

Dressler and Oemler acknowledge the support of the NSF grant AST-0407343.  All the authors thank NASA for its support 
through NASA-JPL 1310394.  Vulcani and Poggianti acknowledge financial support from ASI contract I/016/07/0 and 
ASI-INAF I/009/10/0.  Gladders thanks the Research Corporation for support of this work through a Cottrell Scholars Award. 
The authors thank Dr. Lori Lubin for helpful discussions and, in particular, Dr. Kristian Finlator, for a provocative comment 
that led to the interpretation of starburst cycles presented in this paper.

%%BEGIN TABLES

\begin{table*}
\begin{center}
{\scriptsize
\caption{Metacluster Properties}\label{Clusters}}

\begin{tabular}{lcccccccccccc}

\tableline\tableline

\\     ID  &   RA  &  DEC    &     z   &   N   &  $\sigma_0$  & R$_{200}$    & PAS & CSF & SBH & PSB & SB0 
\\        &         &              &          &        &       \kms        &        Mpc         &         &         &          &         &         \\      
\tableline

\\ RCS0221A  &  35.41530  &  -3.77685  &  0.430924 & 247 &  896 & 1.27 & 85  &  121 &   27  &  9  &  5   
\\ RCS1102B  & 165.67800 &  -3.66588  &  0.385657 & 275 &  698  & 1.39 & 107  & 137 &   19 &   8  &   3    
\\ SDSS0845A  & 131.36600  &  3.45924 &  0.329637 & 278 & 1436 & 1.43 & 132 & 109 &  17  &  12  &  7    
\\ SDSS1500A  & 225.14300  & 1.89275 &  0.419252  & 113  & 637  &   1.17 & 47  &   43 &   16 &   3  &   4   
\\  SDSS1500B  & 225.09400  & 1.85731 &  0.517742  & 160  & 1398 &  1.23 & 53 &   83  &  14  &   5  &   5    \\
\\
\tableline
\end{tabular}

\tablecomments{spectral-types are in numbers of galaxies}
\end{center}
\end{table*}

\begin{table*}
\begin{center}
{\scriptsize
\caption{Sample Incompleteness}\label{incompleteness_table}}

\begin{tabular}{lcccccccccc}

\tableline\tableline

\\    spectral samples  &   N  &  WTmag  &  WTrad  &    WTtot   &   sample  &  N  &  WTmag  & WTrad   & WTtot \\ 
\tableline

\\ all cluster  &    913  &    1.98    &    1.94   &   1.89  &   all cl\_field   &  1090  &   2.00  &   1.90  &   1.98 
\\ all (normalized)  &  913   &   1.00    &    1.00   &    1.00   &  all (normalized)  &  1090  &   1.00  &    1.00  &   1.00 
\\
\\ PAS  &  371   &   0.97    &    1.01   &    0.97   &    PAS    &  193   &  0.94   &   0.95   &   0.89  
\\ CSF  &  410  &   1.00    &   1.00   &   1.01   &    CSF   &    692    &  1.02   &   1.02   &   1.03  
\\ SBH  &  79   &  1.11    &   0.91   &   1.08   &    SBH    &  126   &   0.91   &    0.94   &   0.94   
\\ PSB  &   32   &   0.91   &   1.15    &  1.00   &   PSB    &   29   &    1.00   &     0.96   &    0.98  
\\ SBO  &   14   &  1.44   &   090    &   1.17   &    SBO    &    29   &  1.00    &   0.96  &   0.98  \\

\\    group samples  &   N  &  WTmag  &  WTrad  &    WTtot   &   sample  &  N  &  WTmag  & WTrad   & WTtot \\ 
\tableline

\\ all metacluster     &     795   &    1.81   &    1.91  &    1.89   &   all  cl\_field   &    1090  &   2.00  &   1.90   &    1.97
\\ all (normalized)    &    795  &   1.00   &    1.00   &   1.00   &     all (normalized)    &   1090  &   1.00   &    1.00   &    1.00
\\
\\ groups  &   257  &    0.97   &    0.89   &   0.91   &    groups   &   425   &  0.99   &    1.00   &     1.00
\\ nongroup  &   466   &   1.02    &   1.10   &   1.09   &      &        &          &         &       
\\
\\ cluster   &   539   &      1.08    &    1.09    &   1.01    \\   
\tableline

\end{tabular}

\tablecomments{Incompleteness of spectral-types with different samples. Average weight for spectroscopic compared to 
photometric sample is a factor of $\sim$2.0, as shown in the first lines of the table.  Values for subsets are normalized to 
the appropriate weight for each type of incompleteness for whole sample.}
\end{center}
\end{table*}

\begin{table*}
\begin{center}
{\scriptsize
\caption{Sample Mass Limits}\label{mass limited samples}}

\begin{tabular}{lcccccccccc}

\tableline\tableline

\\    metacluster   &  N  & $ M > 10^9$   &  $ M > M_{lim}^{(a)} $ &  $f^{(b)}$ &  cl\_field  &  N  &  $M > 10^9$  &  $ M > M_{lim}  $   & $f$ \
\\    \,\,\,\,sptype     &   &   \Msun   &   &   & \,\,sptype &   &   \Msun  &        &   \\
\\
\tableline
&&&
\\  PAS   &   371   &    343     &   292   &   0.851  &   PAS    &   193    &    175    &   146    &   0.834
\\  CSF   &   410   &    381     &   200   &   0.525  &   CSF    &   692    &    604    &    297   &   0.492
\\  SBH   &     79   &      73     &      34   &   0.466  &  SBH    &   126    &    110    &      52   &    0.473 
\\  PSB   &     32   &      28     &      19    &   0.679   &  PSB   &     29    &      23    &      19   &    0.826
\\  SBO   &     29   &      19     &        6    &   0.316   &  SBO   &     47    &      46    &      15   &    0.326  \\
\\
\tableline
\end{tabular}

\tablecomments{Numbers and fractions of mass limited samples by spectral types.  Not all N galaxies have measured masses.}
\tablenotetext{a}{$ M_{lim} = 2.5 \times 10^{10} $ \Msun }
\tablenotetext{b}{$f$ is fraction of sample with masses above $ M_{lim}$ }
\end{center}
\end{table*}

\begin{table*}
\begin{center}
{\scriptsize
\caption{Cluster Group Properties\label{Cluster_Groups}}
\resizebox{18cm}{!} {
\begin{tabular}{lccccccccccccccccc}
\tableline\tableline
\\ Group ID  &  Ntot & RA & DEC &  N$_{tot}^{(a)} $ & L$_{gal}^{(b)}$ & R$_{pair}^{(c)}$  & $\Delta V_0$ & $\sigma_0$  & PAS &  CSF &  SBH & PSB & SBO & \  Nex &   Prob  & $\tau_{enc}$
\\                 &          &  deg     &  deg   &     &     &   Mpc     &   \kms    &       \kms        &   \%      &     \%    &     \%      &    \%     &    \%      &       &         &  \\      

\tableline
&&& 
\\ RCS0221A-1A  & 20 &  35.3304 & -3.7006 & 41.35 & 11.42 & 2.18  & 541.9 & 362.0 & 25.0 & 70.0 &  0.0 &  0.0 &  5.0  &  6 &  3.2E-3  & 1.4490
\\  RCS0221A-1B & 16 &  35.3493 & -3.6811 & 29.19 &  6.46  & 1.65  & -1499.3 &189.4 & 25.0 &  64.3 & 12.5 & 0.0 & 6.3 & 5 &  $<$ E-4 & 1.5949 
\\  RCS0221A-2A  & 14 &  35.2734 & -3.8977 & 36.48 & 7.27 &  1.20  & 433.3  &  244.8 & 21.4 &  57.1 & 14.3 & 0.0 & 0.0 & 0 & 1.3E-3  & 0.3904 
\\  RCS0221A-2B  & 5  &  35.2905 & -3.8953 & 12.66 & 3.02 & 0.98  &  1331.3 & 77.7 & 0.0 & 80.0 & 20.0 & 0.0 & 0.0 & 0 & 3.0E-4 & 1.6917  
\\  RCS0221A-4    & 24 &  35.5511 & -3.9102 & 51.99 & 15.82 & 2.14 & 297.5 & 349.9 & 20.8 & 50.0 & 29.2 & 0.0 & 0.0 & 0 & 6.0E-4 & 1.0781 \\
\\  RCS1102B-1A  & 20 &  165.845 & -3.605 & 64.38 & 12.59 & 1.08 & -42.5 & 181.3 & 20.0 & 60.0 & 10.0 & 5.0 & 5.0 & 1 &  6.0E-4 & 0.2452  
\\  RCS1102B-1B & 4  &  165.829 & -3.6788 & 11.64 & 3.07 & 0.49 & -985.8 & 158.3 & 0.0 & 100.0 & 0.0 & 0.0 & 0.0 & 4 &  1.1E-2 &  0.3897 
\\  RCS1102B-2  & 15 &  165.499 & -3.5894 & 53.82 & 13.62 & 1.75 & 897.0 &  385.0 & 13.3 & 60.0 & 20.0 & 6.7 & 0.0 & 6 & $<$ E-4 & 0.5650 
\\  RCS1102B-3  & 12 &  165.502 & -3.7916 & 48.80  & 10.70 & 0.86 & 268.9 & 162.7 & 16.6 & 75.0 &  8.3 &  0.0 &  0.0 & 6 &  6.2E-3 & 0.1654
\\  RCS1102B-4  & 13 &  165.725 & -3.8076 & 36.28  & 9.29 & 1.14 & -679.6 & 508.4 & 9.1 & 81.8 &  9.1 & 0.0 &  0.0 & 5 &  2.0E-4 & 0.3353 
\\  RCS1102B-5  & 10 &  165.703 & -3.6828 & 18.73  & 5.04 & 0.16 & -843.1 & 335.8 & 60.0 & 20.0 &  0.0 & 20.0 & 0.0 & 7 & 1.8E-3 &  0.0014 \\
\\  SDSS0845A-1 & 18 & 131.178 & 3.5366 & 41.79 & 17.19 & 0.89 & 244.8 &  280.0 & 22.2 & 77.8 & 0.0 & 0.0 & 0.0  & 4 & 3.0E-4 & 0.1298
\\  SDSS0845A-2 & 6  &  131.244 & 3.4757 &13.01  & 3.81  & 0.48 & -3273.8 & 125.2 & 16.7 & 33.3 & 33.3 & 0.0 & 16.7 & 5 & $<$ E-4 & 0.3084  \\  
\\  SDSS1500A-1 & 8  & 225.239 & 1.9881 & 15.33 & 7.10 & 0.67 & 398.2 & 169.7 & 25.0 & 50.0 & 25.0 & 0.0 & 0.0 & 2  & 1.1E-2 & 0.2269  
\\  SDSS1500A-2 & 6  & 225.331 & 1.8750 & 11.17 & 6.72 & 1.08 & 140.2 & 199.8 & 16.6 & 66.7 & 16.7 & 0.0 & 0.0 & 0 & 1.6E-1 & 1.0781 
\\  SDSS1500A-3 & 13 &  225.102 & 1.9919 & 26.02 & 12.06 & 1.00 & 224.5 & 305.3 & 38.5 & 38.5 & 15.4 & 0.0 & 7.6 & 2 &  5.5E-2 & 0.2530
\\  SDSS1500A-4A & 14 & 225.153 & 1.8936 & 24.46 & 10.18 & 0.70 & -791.1 & 255.9 & 50.0 & 21.4 &  7.1 & 14.3 & 7.2 & 14 & 2.0E-4 &  0.1104 
\\  SDSS1500A-4B & 6 &  225.147 & 1.8926 &13.76 & 4.09 & 0.90 & 467.7 &  236.3 & 83.3 & 0.0 & 16.7 & 0.0 & 0.0 & 5 & 8.8E-2 & 0.4217 \\
\\  SDSS1500B-1A & 3 & 225.053 & 1.7293 & 13.30 & 5.58 & 1.63 & 2143.3 & 116.2 & 33.3 & 66.7 & 0.0 & 0.0 & 0.0 & 4 & 5.6E-3 & 0.0096  
\\  SDSS1500B-1B & 8 & 225.026 & 1.7496 & 21.84 & 7.96 & 1.13  & -2895.3 & 401.2 & 0.0 & 75.0 & 25.0 & 0.0 & 0.0 & 3 & $<$ E-4 & 0.3187 
\\  SDSS1500B-2  & 29 & 224.952 & 1.802 & 107.45 & 38.67 & 0.86 & -99.8 & 915.4 & 37.9 & 55.2 &  0.0 &  6.9 & 0.0 & 0 & 2.3E-1 & 0.0132
\\  SDSS1500B-3  & 8 & 225.126 & 1.8688 & 41.17  & 13.72 & 0.58 & 1497.0 & 476.8 & 62.4 & 25.0 & 12.5 & 0.0 & 0.0 & 10 & 2.4E-2 & 0.0195  
\\  SDSS1500B-4A & 6 & 225.143 & 1.738 & 22.59 &  5.83 & 0.88 & 1960.9 & 319.1 & 16.7 & 66.6 & 0.0 & 0.0 & 16.7 & 2 & $<$ E-4 & 0.1836  
\\  SDSS1500B-4B & 3 & 225.178 & 1.727 & 10.08 & 2.74 & 1.84 & -2440.9 & 34.0 & 0.0 & 66.7 & 33.3 & 0.0 & 0.0  & 2 &  1.5E-3  & 0.0039  \\

&&&&\\
\tableline
\end{tabular}
}
}
\tablenotetext{a,b}{N$_{tot}$ \& L$_{gal}$ are values for each group of the total number of galaxies, after adjusting 
for sampling depth, and the total luminosity, from fitting a Schechter function, as explained in the text.}
\tablenotetext{c}{R$_{pair}$ is a radius calculated by as the mean of the separations of all possible pairs}

\end{center}
\end{table*}

\begin{table*}
\begin{center}
{\scriptsize
\caption{Field Group Properties\label{Field_Groups}}
\begin{tabular}{lccccccccccccccc}
\tableline

\\ Group & N & RA  &  DEC  &  z &  Ntot  &  Lgal  &   $\sigma_{0}$  & Rpair & PAS & CSF & SBH & PSB & SBO & $tau_{enc}$            
\\            &     &       &            &     &          &           &      \kms            &   Mpc  &  \%    &    \%      &    \%     &    \%     &       \%      &    Gyr   \\
\tableline
& 
\\ RCS0221 1A & 13 & 35.30412 & -3.85102 & 0.3157 & 21.10 &  4.05 &  238.0 & 1.17 & 23.1 &  53.8 & 15.4 & 0.0 &  7.7 & 0.6677 
\\ RCS0221 1B & 41 & 35.27720 & -3.80787 & 0.3257 & 55.44 & 16.32 &  317.0 &  2.08 & 24.4 & 56.1 & 12.2 & 2.4 &  4.9  &  1.0720 
\\ RCS0221  2  & 17 & 35.55209 &  -3.81320 & 0.3487 & 25.81 &  7.21 &  151.0 & 1.77  & 17.6 & 76.5 &  5.9 &  0.0 &  0.0   &   2.9787
\\ RCS0221 3  &  7 & 35.55083 &  -3.87912 & 0.3663 & 11.88 &  2.69 &  174.0 & 0.83  & 0.0  & 71.5 &  28.5 & 0.0 & 0.0  &   0.5791
\\ RCS0221 4  &  8 & 35.53390 &  -3.69391 & 0.4970 & 22.50 &  8.19 &  223.0 & 2.12 & 25.0 & 50.0 &  25.0 & 0.0 & 0.0 &  3.5795
\\ RCS0221 5  & 10 & 35.46751 &  -3.63309 & 0.3969 & 20.11 &  4.24 & 240.0 & 0.78 & 10.0 & 60.0 & 10.0 & 0.0 & 20.0  &  0.2058
\\ RCS0221 6A &  8 & 35.60182 &  -3.82601 & 0.4970 & 22.50 &  8.19 & 223.0 & 2.12 & 12.5 & 50.0 &  25.0 &  0.0 & 12.5  & 3.9755 
\\ RCS0221 6B & 13 & 35.49378 & -3.88737 & 0.5002 & 35.15 &  14.51 & 123.0 &  2.02 & 15.4 & 76.9 & 0.0 &  7.7 & 0.0   &   3.9911
\\ RCS0221  7   &  7  & 35.54801 & -3.75429 & 0.5159 & 26.01 &   7.06 &  295.0 & 0.85 & 14.3 & 85.7 &  0.0 & 0.0 & 0.0 &   0.1676 \\
\\ RCS1102  1  & 13 & 165.53380 & -3.60421 & 0.3424 & 45.53  &  9.25 &  123.0 &  2.07 & 15.4 & 69.2 & 15.4 & 0.0 & 0.0  &  3.3157 
\\ RCS1102  2  & 15 & 165.65413 & -3.83397 & 0.3481 & 38.44  &  7.65 &  172.0 &  2.07 & 20.0 & 53.3 & 20.0 & 0.0 & 6.7  &  2.8085
\\ RCS1102  3  & 16 & 165.75301 & -3.61600 & 0.3617 & 46.32 &  17.23 & 204.0 & 1.47 & 18.8 & 62.4 & 12.5 & 6.3 & 0.0   &  0.7038
\\ RCS1102  4 &  21 & 165.50757 & -3.74132 & 0.3667 & 83.06 &  22.24 & 291.0 & 1.52 &  23.8 & 66.7 & 4.7 &  4.8 &  0.0  &   03042 
\\ RCS1102  5 & 24 & 165.66133 & -3.66528 & 0.3993 & 86.30 &  20.82 & 145.0 & 2.65 &  25.0 & 45.8 & 25.0 & 0.0 & 4.2  &  3.1133 
\\ RCS1102  9 & 16 & 165.74455 & -3.77131 & 0.4755 & 87.68 &  23.01 & 109.0 & 3.11 &  12.5 & 68.8 & 18.7 & 0.0 & 0.0  & 6.5981
\\  RCS1102 10 & 48 & 165.67100 & -3.73191 & 0.4992 & 255.38 & 82.89 & 341.0 & 4.23 & 35.4 & 50.0 &  2.1 & 10.4 & 2.  & 1.8195 \\
\\  SDSS0845  2B & 10 & 131.38068 & 3.34949 & 0.3811 & 30.39 & 10.33 & 269.0 & 2.09 &  0.0 &  80.0 & 10.0 & 0.0 & 10.0  &  2.3379
\\  SDSS0845  3  & 14 &  131.25034 & 3.50936 & 0.4438 & 48.28 & 16.33 & 157.0 & 2.39 & 21.4 &  64.4 &  0.0 & 7.1 & 7.1  &   3.7705
\\  SDSS0845  4  &  7 &  131.44269 &  3.57199 & 0.4545 & 27.93 &  6.20 &  447.0 &  2.42 & 14.2 & 85.8 & 0.0 & 0.0 & 0.0  &   2.3765 \\ 
\\  SDSS1500  1  & 11 &  225.20789 & 1.79321 &  0.3719 & 19.16 &  7.51 &  214.0 & 1.44 &  54.6 & 45.4 &  0.0 & 0.0 &  0.0 &   1.5246
\\  SDSS1500  3  &  8 &   225.31461 & 1.90860 & 0.3955  & 12.90 &  4.43 &  302.0 & 1.60 & 25.0 & 37.5 & 25.0 & 12.5 & 0.0 &  2.2011  
\\  SDSS1500  5  & 14 &  225.22456 & 2.00235 & 0.4577 & 32.26 &  11.65 &  212.0 & 2.25 &  21.4 & 42.8 & 21.4 & 7.2 & 7.2 &   3.4867
\\  SDSS1500  6  &  9 &   225.13403 & 1.95897 & 0.4799 & 19.80 &   4.05  &  262.0 &  2.85 & 11.1 &  55.6 & 22.2 & 11.1 & 0.0 &  9.3420 \\

&\\
\tableline
\end{tabular}
}
\end{center}
\end{table*}

\begin{table*}
\begin{center}
{\scriptsize
\caption{Field Filament Properties\label{Field_Filaments}}
%\vspace{1.0cm}
\begin{tabular}{lccccccccccccccc}
\tableline

\\ Group & N & RA  &  DEC  &  z &  Ntot  &  Lgal  &   $\sigma_{0}$  & Rpair & PAS & CSF & SBH & PSB & SBO & $tau_{enc}$           
\\            &     &       &            &     &          &           &      \kms             &   Mpc  &    \%     &  \%       &  \%       &   \%       &  \%   &  Gyr   \\

\tableline
& 
\\  RCS1102   6  &  8  & 165.72307 & -3.84014 & 0.4265 &  30.24  &  9.96 &  287.0 & 1.88 & 25.0 &  62.5 & 12.5 & 0.0 & 0.0 &  1.6028
\\  RCS1102   7  & 15 & 165.73518 & -3.65126 & 0.4424 &  48.53 & 19.02 & 165.0 &  2.75 &  20.0 &  80.0 & 0.0 &  0.0 & 0.0 &  5.4372  
\\  RCS1102   8  & 11  & 165.69684 & -3.56506 & 0.4741 &  50.53 & 10.28 & 278.0 &  3.14 &  9.1 &  81.8 &  9.1 &  0.0 & 0.0 &     4.6139 \\
\\  SDSS0845  1 & 11 &  131.4595 &  3.52035  & 0.3580 &  25.00 &   5.18 & 169.0 &  1.73 &  9.1 &   72.7 & 18.2 & 0.0 & 0.0 &   2.5656 
\\  SDSS0845  2A  & 8 & 131.36554 & 3.35107 & 0.3647  & 17.97  &   5.51 & 203.0 &  2.10 &  0.0 & 100.0 &  0.0 &  0.0 & 0.0 &  5.3148 \\
\\  SDSS1500   2   & 14 & 225.10097 & 1.83667 & 0.3775 & 25.59 & 11.28 & 187.0  & 2.59 &  57.1 &  42.9 &  0.0 & 0.0 & 0.0  &   7.6008
\\  SDSS1500   7   & 15 & 225.08087 & 1.72892 & 0.4835 & 40.31 & 11.43 & 345.0 &  2.48 &  26.7 &  53.3 & 13.3 & 0.0 & 6.7 &   2.2961 \\ 

&\\
\tableline
\end{tabular}
}
\end{center}
\end{table*}

\begin{table*}
\begin{center}
{\scriptsize
\caption{Properties of Other Samples\label{Other_Samples}}
\resizebox{18cm}{!} {
\begin{tabular}{lccccccccccc}
\tableline\tableline
\\ Sample  &  N &  N$_{tot}^{(a)}$ & L$_{gal}^{(b)}$  & $\sigma_0$ & R$_{pair}^{(c)}$ & PAS &  CSF &  SBH & PSB & SBO 
\\               &      &                           &                          &    \kms         &       Mpc              &   \%  &    \% &    \%  &  \%   &   \%   \\      

\tableline
&&& 
\\ RCS0221A cluster    &     91   & 210.46  & 62.66  &  918  &  1.43   &   55.0  & 27.5 & 10.9 & 6.6 &  0.0 
\\ RCS1102B cluster   &   169    & 485.59  & 125.21 &  705 &  1.49   &   50.9  &  38.5 &  6.5 &  3.6 &  0.6 
\\ SDSS0845A cluster  &  180    & 452.85  & 199.57 & 1273 & 1.27   &    61.1 &  27.2 &  3.9 &  6.1 &  1.7 
\\ SDSS1500A cluster    &   46   & 108.12  &  30.22  &  764 &  1.03   &    54.3  &  26.1 &  8.7 &  6.5 &  4.3 
\\ SDSS1500B cluster    &   53   & 205.02  &  57.40  &  828 &  1.19   &   49.1  &  34.0  & 13.2 & 1.9 &  1.9   
\\
\\ RCS0221A supercluster  & 156  &        &           & 888 &  3.38  &  22.4 & 61.6 & 10.9 &  1.9 &  3.2 
\\ RCS1102B supercluster   & 106  &       &            & 688   &  2.62  &  19.8 & 68.9  &  7.5  & 1.9 &  1.9   
\\ SDSS0845A supercluster   &  98  &      &            & 1676  &  3.24  &  23.5  & 61.2 & 10.2 &  1.0 &  4.1
\\ SDSS1500A supercluster   &  67  &      &            &   513   &  3.72  &  34.3  & 44.8 &  17.9 &  0.0 &  3.0 
\\ SDSS1500B supercluster    & 101 &     &            & 1385  &   4.25  &  25.7  &  61.4 &  5.9  &  3.0 &  4.0
\\
\\ RCS0221A core                   &   20  &     &           &   1142  &            &  70.0  &  10.0  & 5.0  &  15.0 &  0.0
\\ RCS1102B  core                    &   37  &     &          &    761   &           &   67.6  &  21.6  & 5.4  &    5.4 &  0.0
\\ SDS0845A  core                    &   70  &     &           &  1227   &          &   74.3  &  14.5  &  2.9  &   8.6  &  0.0
\\ SDSS1500A core                  &   20  &      &           &   912   &          &   75.0  &   15.0  &  5.0  &   5.0  &  0.0
\\ SDSS1500B core                  &   11  &      &           &   718   &          &   72.9  &    18.2  &  0.9  &   0.0  & 0.0  \\

&&&&\\
\tableline
\end{tabular}
}
}
\tablenotetext{a,b}{N$_{tot}$ \& L$_{gal}$ are extrapolated values of the number of galaxies and total luminosity, as described in the text.}
\tablenotetext{c}{R$_{pair}$ is a radius calculated by as the mean of the separations of all possible pairs}

\end{center}
\end{table*}

\end{document}